\documentclass[aps,10pt,pra,twocolumn,showpacs,showkeys,groupedaddress,nofootinbib]{revtex4-1}
\usepackage[colorlinks,linkcolor=red,citecolor=blue,urlcolor=blue]{hyperref}
\usepackage{amsmath}
\usepackage{graphicx}
\usepackage{bm}
\usepackage{xcolor}
\usepackage{comment}
\usepackage[normalem]{ulem}
\usepackage{enumitem}
\usepackage{natbib}
\usepackage{textcomp}
\usepackage{microtype}

\newcommand{\be}{\begin{equation}}
\newcommand{\ee}{\end{equation}}
\newcommand{\bea}{\begin{eqnarray}}
\newcommand{\eea}{\end{eqnarray}}

\def\bfB{\mathbf{B}}
\def\bfr{\mathbf{r}}
\def\mum{\rm\mu m}
\def\mus{\rm\mu s}

\def\muK{\rm\mu K}

\def\Tc{T_{\rm c}}
\def\ie{{\it i.e.,\/}}
\def\eg{{\it e.g.,\/}}

\def\via{{\it via\/}}

\def\insitu{{\it in~situ\/}}

\def\Li{$\rm^{7}Li$}
\def\Na{$\rm^{23}Na$}
\def\K{$\rm^{40}K$}
\def\Rb{$\rm^{87}Rb$}
\def\Sr{$\rm^{88}Sr$}

\def\posskip {\vskip   \baselineskip}
\def\negskip {\vskip  -\baselineskip}
\setlist[enumerate]{wide=-4pt,nolistsep}
\begin{document}

\title{Fifteen Years of Cold Matter on the Atom Chip: \\ Promise, Realizations, and Prospects}

\author{Mark Keil}
	\thanks{\footnotesize Corresponding authors:  
	\href{mailto:mhkeil@gmail.com}{\tt mhkeil@gmail.com}, \href{mailto:folman@bgu.ac.il}{\tt folman@bgu.ac.il}}
	\affiliation{Department of Physics, Ben-Gurion University of the Negev, Be'er Sheva 84105, Israel}
\author{Omer Amit}	
	\affiliation{Department of Physics, Ben-Gurion University of the Negev, Be'er Sheva 84105, Israel}
\author{Shuyu Zhou}	
	\thanks{\footnotesize Present address: Laboratory for Quantum Optics, Shanghai Institute of Optics and Fine Mechanics, Chinese Academy of Sciences, P.O. Box 800211, Shanghai 201800, China.}
	\affiliation{Department of Physics, Ben-Gurion University of the Negev, Be'er Sheva 84105, Israel}
\author{David Groswasser}
	\affiliation{Department of Physics, Ben-Gurion University of the Negev, Be'er Sheva 84105, Israel}	
\author{Yonathan Japha}
	\affiliation{Department of Physics, Ben-Gurion University of the Negev, Be'er Sheva 84105, Israel}	
\author{Ron Folman}
	\thanks{\footnotesize Corresponding authors:  
	\href{mailto:mhkeil@gmail.com}{\tt mhkeil@gmail.com}, \href{mailto:folman@bgu.ac.il}{\tt folman@bgu.ac.il}}
	\affiliation{Department of Physics, Ben-Gurion University of the Negev, Be'er Sheva 84105, Israel}

\makeatletter
\renewcommand\frontmatter@abstractwidth{\dimexpr\textwidth-27mm\relax}
\makeatother

\begin{abstract}
Here we review the field of atom chips in the context of Bose-Einstein Condensates~(BEC) as well as cold matter in general. Twenty years after the first realization of the BEC and fifteen years after the realization of the atom chip, the latter has been found to enable extraordinary feats: from producing~BECs at a rate of several per second, through the realization of matter-wave interferometry, and all the way to novel probing of surfaces and new forces. In addition, technological applications are also being intensively pursued. This review will describe these developments and more, including new ideas which have not yet been realized. 
\end{abstract}

\date{\today}

\maketitle

\twocolumngrid

\makeatletter			
\renewcommand\tableofcontents{\@starttoc{toc}}
\makeatother
{\hypersetup{linkcolor=black}\tableofcontents}

\vfill

{\onecolumngrid
\begin{center}
\begin{figure}[b!]
   \vskip-4.0\baselineskip	
   \begin{minipage}[t!]{0.70\textwidth}	
      \centering
      \includegraphics[width=\textwidth]{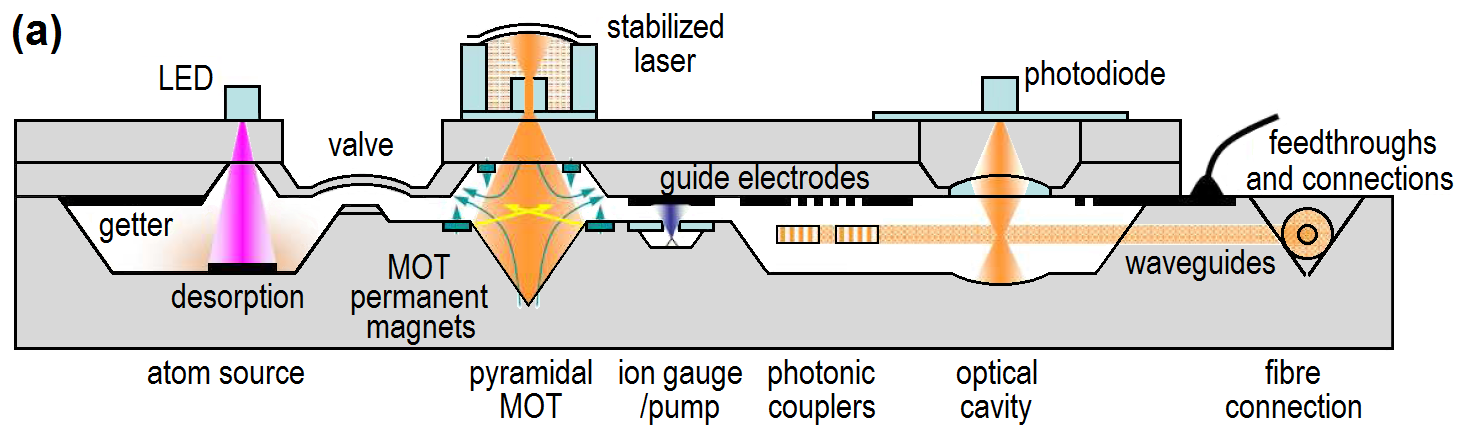}	
   \end{minipage}
   \hfill
   \begin{minipage}[t!]{0.28\textwidth}	
      \centering
      \includegraphics[width=\textwidth]{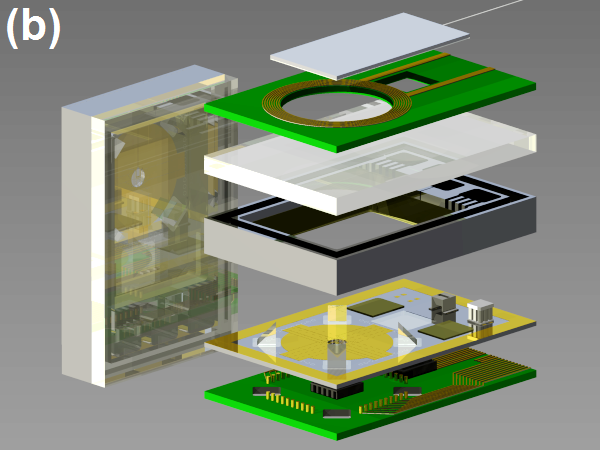}
   \end{minipage}
   \vskip-0.0\baselineskip	
   \caption{Futuristic visions of highly integrated atom chips. (a)~A single substrate (gray shading) holds all the light sources (yellow), atom sources (purple), photonics, and micro-magnetic traps. Successive stages of cooling and experimentation proceed in a series of miniature vacuum chambers (unshaded) from left to right. Conceptual sketch courtesy of Tim Freegarde (Southampton) and adapted from~\cite{Folman2011a}, with permission from Springer Science+Business Media. (b)~An integrated design currently being realized incorporates the vacuum system, atom source, and optical components in a permanently sealed micro-liter system~\cite{Rushton2014}. The device is designed for magneto-optical trapping and cooling and is capable of passively maintaining ultra-high vacuum for~$>\rm1000\,days$. The composite chip size will be~$20\times24\times5\rm\,mm^3$. Schematic courtesy of Matt Himsworth (Southampton). Compact modular systems developed by Dana Anderson's group (Boulder) are now capable of producing chip-based~BECs~\cite{Salim2011,Farkas2014x}.}  
   \label{fig:Folman2011afig1}
   \vskip-0.5\baselineskip	
\end{figure}
\end{center}}

\quad\clearpage\newpage 

\twocolumngrid
\negskip

\section{Introduction}	
\label{sec:introduction}

The atom chip is a device intended to miniaturize setups for quantum optics with cold matter. This means an integrated device that can create cold matter, trap it, manipulate it, and measure it, while controlling both internal and external degrees of freedom. The enhanced accuracy and versatility is expected to enable not only new technology, but also a variety of novel fundamental studies.

The original vision of the atom chip is simple: combine the mature technology of the semiconductor industry with the dramatic progress made in atom optics. This combination creates a type of ``solid state'' device for matter-wave optics with the best of both worlds: on the one hand, accuracy, scalability, complexity, miniaturization and integration, and on the other hand, the long coherence times of isolated cold matter. The atom chip is expected to operate with atom-surface distances in the range of~$\rm100\,nm-100\,\mum$, thus enabling the isolation required for atom optics, while simultaneously achieving the noted advantages of solid-state devices. Such a device could enable trapping and guiding potentials with virtually arbitrary architecture and the ability to individually address specific traps (\eg\ in a high-density lattice of traps) and control interactions between them (\eg\ with dynamic control over tunneling barriers). The hope was~--~and still is~--~that such a level of miniaturization and integration can bring to cold matter and matter waves the same revolution that it brought to electronics and optics. In Fig.~\ref{fig:Folman2011afig1} we show a schematic representation of such a vision.

This vision of the atom chip has already achieved several very important milestones. First it was shown that spin flips due to noise from the nearby surface are at a level where the lifetime of atoms trapped in a magnetic potential is long enough. Hence atom loss is under control. It was then shown that the heating rate is low enough that the~BEC can survive very close to a surface. This observation is truly remarkable as we remind the reader that atom chips, typically at room temperature, create a temperature ratio of~9 orders of magnitude over a distance of a few~$\rm\mum$. Next, it was shown that spin coherence survives for a very long time, and finally, it was shown that even spatial coherence is very robust. All these findings are essential for the above vision. A more detailed description of these milestones, all of which are essential for the above vision, appears in the next sections.

\begin{figure}[t!]
   \centering
   \includegraphics[width=0.40\textwidth]{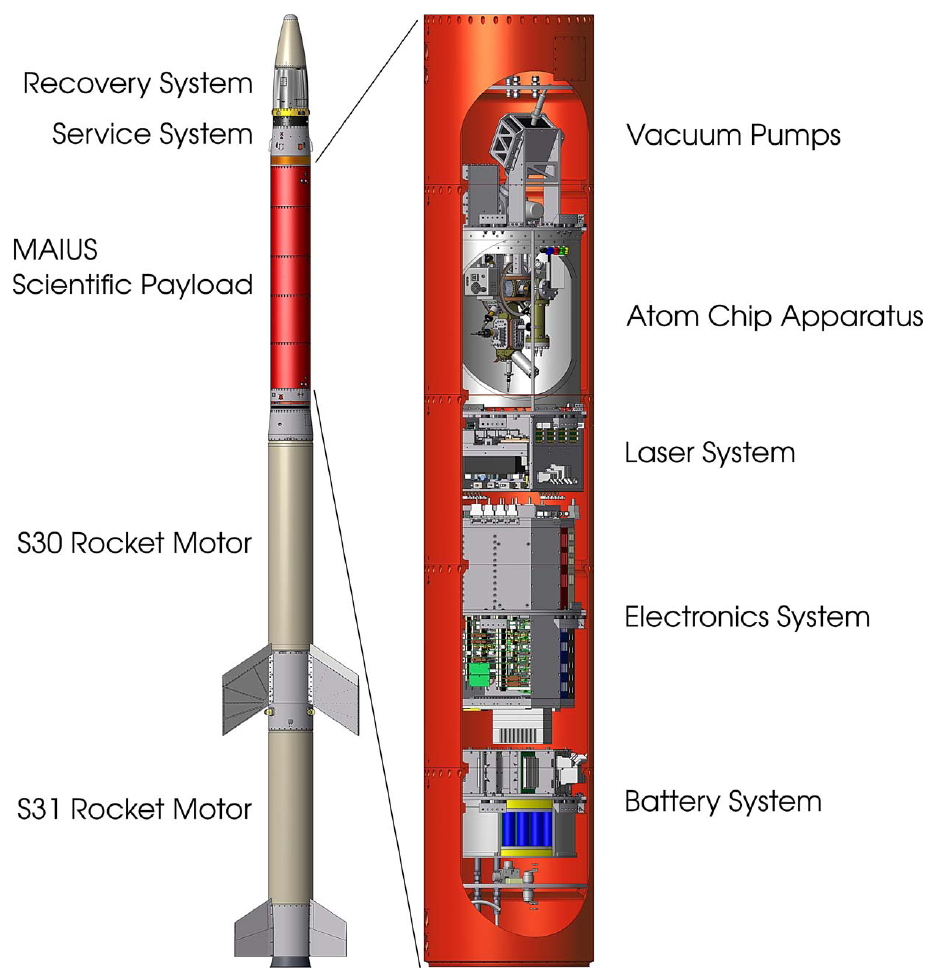}
\caption{Vehicle and payload of the~DLR sounding rocket MAIUS-1 in launch configuration. The scientific payload 
developed by the consortium led by Ernst Rasel consists of an autonomous atom chip device for interferometry employing~BECs, including the vacuum system housing a three-layer atom chip, solid-state laser system, electronics, control system and batteries. Courtesy of Stephan Seidel (Hannover), Jens Gro{\ss}e (DLR-Bremen), and DLR-MORABA~\cite{DLR2015}.}
   \label{fig:rocket}
\end{figure}

The atom chip was first realized at the turn of the century~\cite{Reichel1999,Folman2000,Dekker2000}, and the first~BECs on atom chips were realized shortly thereafter~\cite{Ott2001,Hansel2001}. Since its birth, the atom chip has continued to surprise and delight us, both in terms of fundamental physics and as a technological device. We note that atom chips are expected to go into space in 2016~\cite{Altschul2015} (Fig.~\ref{fig:rocket}) and to be installed in the International Space Station in~2017~\cite{JPL2015,Bongs2015}, whereby fundamental tests of physics in micro-gravity (\eg~the equivalence principle) and at extremely low temperatures (down to~$\rm1\,pK$) will take place. Quite a number of reviews have been written on the extraordinary conception and implementation of the atom chip~\cite{Hinds1999,Folman2002,Reichel2002,Fortagh2007,Reichel2011x,Folman2011a}. Here we briefly present previous results while mainly aiming to update the reader on the state-of-the-art and on new ideas. It is our aim to hint at possible future directions and roadmaps, and in this way help young atom chip researchers who have just begun their journey.

If we may summarize the last~15 years, it may be said that while enormous development has occurred, technological applications of the method are still in need of further progress. Work is being done on applications such as magnetic sensors, atomic clocks, inertial acceleration sensors, atom transport dynamics analogous to electronics (``atomtronics''), building blocks for quantum communication (\eg\ memory and repeaters), and devices for quantum simulation and computing. Just as important, in parallel, a mesmerizing amount of exciting fundamental physics has been done with atom chips and this in itself has made the effort worthwhile. 

Focusing on experimental work with neutral atoms, such fundamental studies include the probing of 
Johnson (thermally-induced) noise~\cite{Jones2003,Lin2004,Emmert2009}, 
electron transport~\cite{Aigner2008}, 
the testing of the Casimir-Polder potential near surfaces~\cite{Lin2004,Harber2005,Obrecht2007a}, 
quantum reflection from surfaces~\cite{Pasquini2006}, 
investigations of degenerate Bose gases in low dimensions~\cite{Esteve2006,Yuen2015,Rauer2016}, 
statistical many-body features such as thermalization~\cite{Hofferberth2007,Gring2012,Langen2015}, entanglement~\cite{Riedel2010,Haas2014,Barontini2015} 
and squeezing~\cite{Maussang2010} effects in an interacting Bose gas, 
self-rephasing through collisions~\cite{Deutsch2010}, 
complementarity in the context of general relativity~\cite{Margalit2015},
Fermi gases~\cite{Trotzky2015}, and much more. Additional fundamental experiments are underway, such as testing the equivalence principle~\cite{vanZoest2010} and 
searching for the hypothesized fifth force~\cite{Wolf2007,Sorrentino2009}. 	
The starting point of many of these experiments is the~BEC that we now celebrate in this Special Issue.

Theoretical work concerning the atom chip is overwhelming in scope and depth. In this review we will be focusing on experimental work. Nevertheless, concerning neutral atoms, let us briefly mention here some of the extensive work done on 
Johnson noise and single-atom decoherence~\cite{Henkel1999,Henkel2003}, 
electron transport~\cite{Wang2004,Japha2008,Dellabetta2012}, 
Casimir-Polder interactions~\cite{Haakh2014},			
decoherence in low dimensions~\cite{Burkov2007},	
loss, heating, and decoherence of a many-body system near a surface~\cite{LlorenteGarcia2013,Japha2016x},
quantum gates~\cite{Charron2006,Treutlein2006,Muller2011}, 	
hybrid devices~\cite{Petrosyan2009},	
superconductors~\cite{Scheel2007,Hohenester2007,Sokolovsky2014a,Sokolovsky2014b},
exotic materials and geometries affecting everything from Johnson noise to Casimir-Polder 
forces and vacuum modes~\cite{Dikovsky2005,Fermani2007,David2008,Petrov2009,Jha2015},
double-well potentials and matter-wave 
interferometry~\cite{Japha2007,Guarrera2015,Stevenson2015,Ammar2015},			
matter-wave pulse shaping~\cite{Nest2010}, 	
atomtronics~\cite{Stickney2007,Pepino2009,Chow2015},
and so on. Two more examples consist of a simulator for quantum pumping of electrons in mesoscopic circuits~\cite{Das2009}, and the investigation of bosonic superflow~\cite{Simpson2014}. The last three topics are good examples of the possible uses of tunneling barriers on atom chips.

\begin{figure}[t!]
   \centering
   \includegraphics[width=0.40\textwidth]{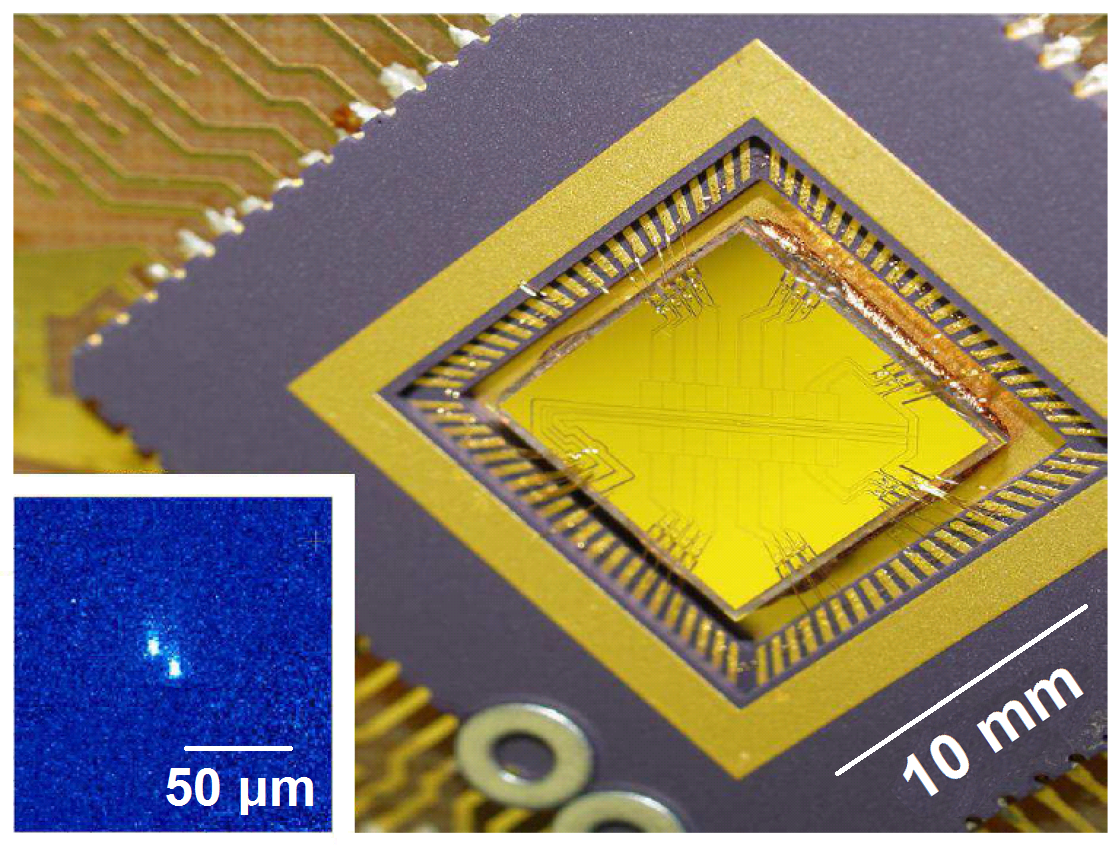}
   \caption{Ion chip produced at Ben-Gurion University before installation in the vacuum chamber at Mainz. Inset: Two ions trapped on the chip. Courtesy of Ferdinand Schmidt-Kaler (Mainz) and adapted from~\cite{Folman2011a}, with permission from Springer Science+Business Media.}
   \label{fig:Folman2011afig2}
\end{figure}

This paper is structured as follows: after this brief introduction, Sec.~\ref{sec:particles} focuses on the wide range of particles that are now used with atom chips. While the first chips were planned for ground state neutral bosons, atom chips have since been designed to trap and manipulate Rydberg atoms, fermions, molecules, ions, and electrons. Figure~\ref{fig:Folman2011afig2} presents an example of an atom chip operated with cold ions. Atom chips are now being proposed for trapping antiprotons, positrons, and antihydrogen. Our review does not deal with atom chips for room temperature atomic vapor although impressive work has been done here as well~\cite{Wu2010x,Stern2013,Knappe2005,Jimenez-Martinez2014}, nor with atom chips for solid-state ``atom-like'' systems~\cite{Zhu2014,You2014}. We mainly focus on cold neutral atoms where, again, the starting point of most experiments is the~BEC.

Section~\ref{sec:technology} deals with the technology. While the first atom chips were mainly based on a surface with current-carrying wires, they have diversified significantly and now include permanent magnets, electrodes for electric fields, antennas for radio-frequency and microwave potentials, and photonics. Superconductors are also being used now. Proposals exist to make use of exotic materials such as alloys, crystalline materials, nano-wires, carbon nano-tubes and so on. The vacuum chamber has also evolved. Some atom chips are utilized as a facet of the chamber itself such that the substrate holds the vacuum. Other atom chips have the vacuum within the substrate. These are important steps towards the vision of a fully self-contained atom chip which will have within the substrate its own vacuum (including pumps), particle sources, light sources, and magnetic and electric field sources. Integrated readouts \via\ photodiodes and cavities will be transmitted from the chip by way of electronics and fibers (Fig.~\ref{fig:Folman2011afig1}). 

In Sec.~\ref{sec:surfaces}, we focus on interactions between the atoms and the surface. These interactions include forces such as the Casimir-Polder force, and effects due to noise sources such as Johnson noise. Here we also describe limitations of atom chips due to imperfections in material or fabrication. We emphasize that while the atom chip was designed to serve as a base for matter-wave optics, it may also be used to provide new insights on materials, surfaces, and searches for such effects as the hypothesized fifth force.

As an example of matter-wave optics made possible by the atom chip, we focus on interferometry in Sec.~\ref{sec:interferometry}. We discuss free-space, guided, and trapped interferometry, including Mach-Zehnder interferometers as well as Sagnacs. 

In Sec.~\ref{sec:applications} we describe potential applications such as clocks, sensors, and quantum information processing. In addition, we expect that progress in atom chips will open new possibilities for the nascent field of atomtronics. We also discuss the use of atom chips for rapid production of~BECs.

We give some concluding remarks and perspectives in Sec.~\ref{sec:conclusion}.

Last but not least, let us note that we explicitly mention the names of those who led many of the atom chip and related efforts. This is done in order to honor these people and to make available a more intimate perspective. However, it is of course possible that, although we have made considerable efforts, we may have inadvertently neglected to mention some additional people. We apologize for any such oversights.

\addtocontents{toc}{\protect\vspace{0pt}}

\section{Chip-Trappable Species} 
\label{sec:particles}

In its earliest days, the atom chip was used for the control and manipulation of ground-state neutral bosons only, and the alkali metals were the bosons of choice. Although it remains true that the workhorse of atom chip experiments and technology is~\Rb, more recent developments have expanded to include a much wider range of ultracold neutral particles, including Rydberg atoms, fermions, and molecules. Atom chips are even being proposed for the capture and manipulation of antimatter. 

This section gives a few detailed examples for each of these classes. In addition, we briefly discuss charged particles (ions and electrons), particularly in the context of quantum information processing~(QIP), but a more detailed review of this very wide field (see, \eg~\cite{Blatt2008,Amini2011x,Monroe2013,Cho2015}) is beyond the scope of the present survey. Except as explicitly noted, it is indeed~\Rb\ that is used elsewhere in this review.

\subsection{Rydberg atoms} 
\label{subsec:rydberg} 

Rydberg atoms are atoms excited to a very high principal quantum number~$n$~\cite{Gallagher1994}. The excited valence electron is loosely bound and its radius is large~($\propto n^2$), making Rydberg atoms highly polarizable (electric polarizability~$\propto n^7$) and extremely sensitive to external fields. 

Since the alkali atoms so commonly used for ultracold experiments are precisely those most easily excited to one-electron Rydberg states, it is natural to marry atom chip studies of surface potentials with this great sensitivity. For example, Rydberg energy levels are very sensitive to residual electric fields and they have therefore been used as probes for weak fields emanating from the chip surface~\cite{Carter2012}. Disruptive electrostatic fields near surfaces have been shown to arise mainly from surface adsorbates~\cite{McGuirk2004,Tauschinsky2010,Hattermann2012}, which cannot be avoided entirely since the alkali atoms are chemically very reactive, and once released from the atom trap may interact and be absorbed by the surface. Nevertheless, it has recently been possible to reduce the stray fields and thereby to bring the Rydberg atoms much closer to the surface ($\approx\rm10\,\mum$~\cite{Naber2016}). 

Rydberg electromagnetically induced transparency is a common method to probe the electrostatic field of adsorbed and polarized adatoms. A cryogenic atom chip was used to measure the temperature dependence of atom-surface physisorption on a high-$\Tc$ YBCO/YSZ\,\footnote{$\rm YBa_2Cu_3O_{7-\delta}$/yttrium-stabilized~$\rm ZrO_2$} surface~\cite{Chan2014}. It was shown that the van der Waals interaction is the dominant factor at cryogenic temperatures. 

The interaction between ground-state neutral atoms is fairly weak and short. Two-body interactions for Rydberg atoms can however, be vastly stronger, since their dipole-dipole interactions scale as~$n^4$. Dipole blockade, a candidate for building quantum gates~\cite{Lukin2001,Saffman2010}, can occur at distances~${>\rm10\,\mum}$ between Rydberg atoms~\cite{Urban2009,Gaetan2009}. This is the same length scale achieved for permanent-magnet lattice traps recently implemented on atom chips~\cite{Leung2014}, with much shorter length scales currently being implemented (Sec.~\ref{subsec:lattices}) in the same laboratory.

\begin{figure}[t!]
   \centering
   \includegraphics[width=0.45\textwidth]{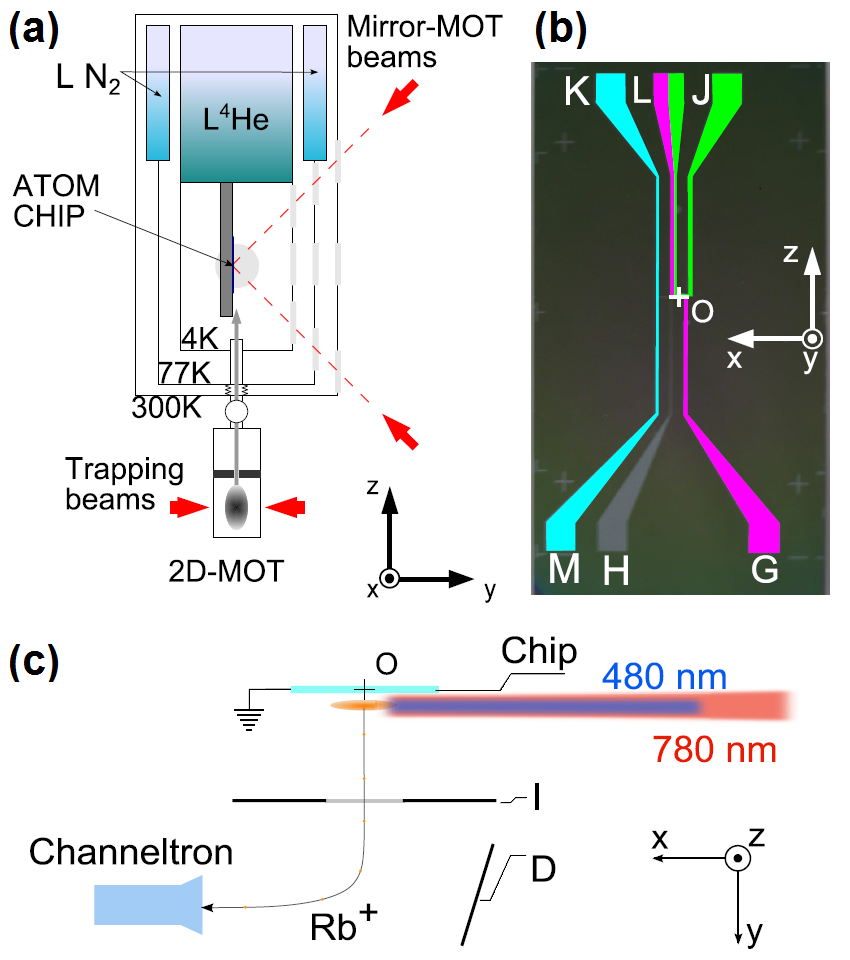}
   \caption{Microwave spectroscopy of Rydberg atoms on a superconducting atom chip. (a)~Scheme of the experimental setup. (b)~Scheme of the superconducting atom chip. (c)~Scheme of the field-ionization detection system. Note the different co-ordinate axis definitions in the three panels. The origin~{\it O} is taken at the center of the horizontal segment of the~Z-wire connecting pads~{\it G} and~{\it L}. Adapted from~\cite{Hermann-Avigliano2014}, with permission \textcopyright~2014 by the American Physical Society.}
   \label{fig:Hermann-Avigliano2014fig1}
\end{figure}

Recent experiments have demonstrated several promising features of Rydberg atoms for~QIP. While coherence lifetimes of cold Rydberg atoms were measured near the surface of an atom chip and shown to be sensitive to static electric field inhomogeneities, spin-echo and spin-locking methods extended the measured coherence times~\cite{Carter2013}. Coherence of Rydberg qubits on a superconducting atom chip was maintained for~$>\rm600\,\mus$ by the Haroche group, exceeding even the lifetime of the~$n=60$ Rydberg level used~\cite{Hermann-Avigliano2014}. The same superconducting atom chip (shown in Fig.~\ref{fig:Hermann-Avigliano2014fig1}) was used for direct measurements of the dipole blockade regime, revealed by microwave spectroscopy of a dense Rydberg gas~\cite{Teixeira2015}. These experiments are advancing the feasibility of quantum simulations and hybrid atom chip quantum information architecture based on Rydberg atoms. 

\subsection{Fermions} 
\label{subsec:fermions}

\begin{figure}[t!]
   \centering
   \includegraphics[width=0.45\textwidth]{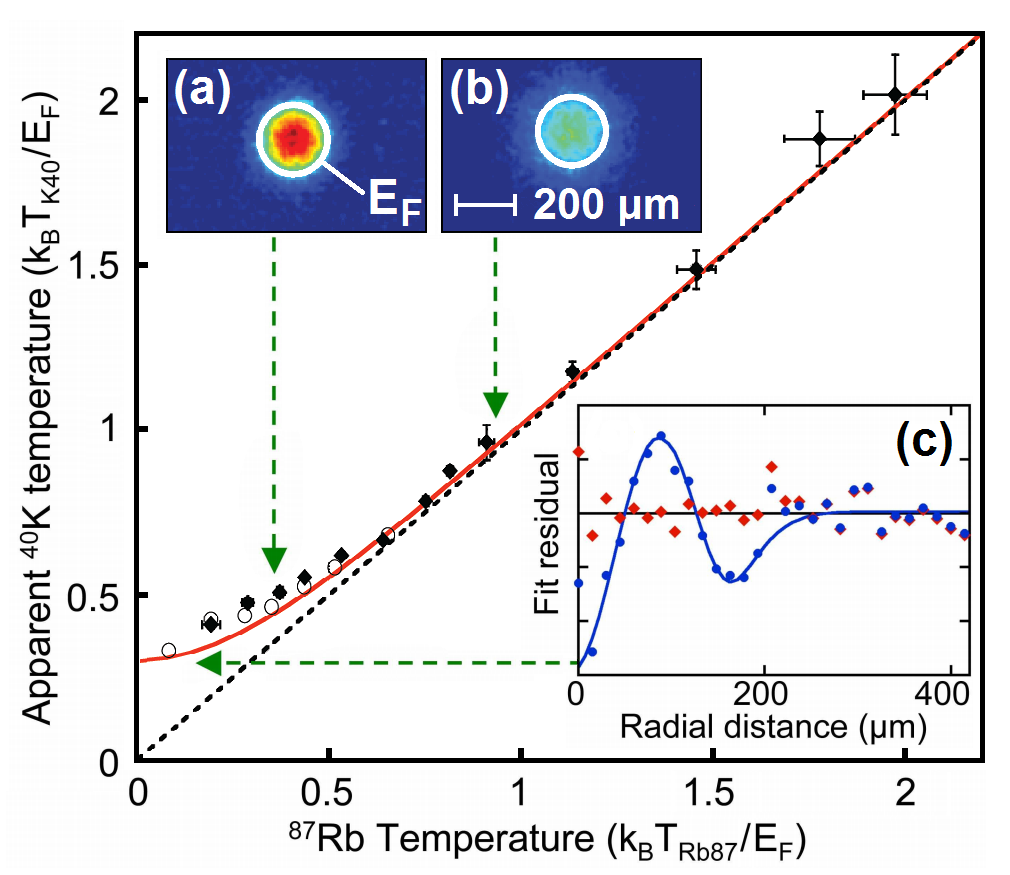}
   \caption{Appearance of a degenerate Fermi gas of~\K\ on an atom chip. Owing to Pauli pressure, Fermi degenerate~\K\ clouds seem to stop getting colder, even when the reservoir temperature approaches zero. Data is compared with its classical expectation (dashed black line) and with a Gaussian fit of a theoretically generated ideal Fermi distribution (solid red line). Absorption images are shown for~(a) $k_{\rm B}T/E_{\rm F} = 0.35$ and (b)~0.95, where the white circle indicates the Fermi energy~$E_{\rm F}$. Fit residuals in~(c) of a radially averaged cloud profile (taken after~$\rm9\,ms$ time-of-flight) show a strong systematic deviation when assuming Boltzmann (blue circles) instead of Fermi (red diamonds) statistics. Courtesy of Marcius Extavour (Toronto)~\cite{Extavour2009} and adapted from~\cite{Aubin2006}, with permission by Macmillan Publishers Ltd.}
   \label{fig:Aubin2006fig3}
\end{figure}

Forced evaporative cooling is a common method for achieving ultracold Bose gases, but it does not work for a single-component Fermi gas because binary elastic collisions of identical spin-polarized fermions are prohibited at ultra-low temperatures. This problem can be solved by using two different spin states of fermionic \K~\cite{DeMarco1999} or by sympathetic cooling of fermionic~$\rm^{6}Li$ by bosonic~\Li~\cite{Truscott2001}. Fermi degeneracy of~\K\ on an atom chip was achieved by the Thywissen group by applying sympathetic cooling with~\Rb~\cite{Aubin2006} (Fig.~\ref{fig:Aubin2006fig3}). The strong confinement and large inter-species collision rate afforded by the atom chip trap permitted cooling to Fermi degeneracy in just~$\rm6\,s$, faster than previously possible in conventional magnetic traps. 

Since anisotropic magnetic potentials with high aspect ratios are easily produced by atom chips, they serve as excellent platforms for studies of fermionic physics~\cite{Extavour2011x}. Ideal Fermi gas splitting has been demonstrated with a radio-frequency (RF)-dressed double well on an atom chip~\cite{Extavour2006}. This kind of potential is species-selective and may be useful for studying tunable boson-fermion interactions in ultracold atomic mixtures and the effect of fermions on the coherence of bosons in the double well. Optical potentials can trap the spin mixtures which are necessary for investigating strong inter-particle interactions in fermions. 

Bringing optical traps near the surface of an atom chip provides opportunities to combine strongly interacting degenerate Fermi gases with near-field~RF and microwave probes~\cite{Bardon2014a}. This has become a powerful tool for investigating quantum dynamics of fermions. The dynamics of a transversely magnetized unitary Fermi gas in an inhomogeneous magnetic field has been studied experimentally, and shows how a transversely spin-polarized Fermi gas decoheres and becomes strongly correlated at a Feshbach-tuned interaction resonance~\cite{Bardon2014b}. In another experiment on spin diffusion, an unambiguous signature of the Leggett-Rice effect in a strongly interacting Fermi gas was observed~\cite{Trotzky2015}. Recently, ``$p$-wave contacts'' in non-equilibrium dynamics was also investigated~\cite{Luciuk2016x}. In this experiment, the $s$-wave interactions are suppressed by polarizing an ultracold Fermi gas of~\K, while $p$-wave interactions are enhanced by working near a scattering resonance.

Atom chip traps for~\K\ and~\Rb\ were compared as simulators for quantum pumping~\cite{Das2009}; this theoretical analysis showed both differences and common features for geometric behavior and resonance transmission of bosons and fermions across a~1D channel between magnetic reservoirs.

\subsection{Molecules} 
\label{subsec:molecules}

With their numerous internal degrees of freedom and strong long-range interactions, ultracold molecules are another frontier in the investigation of fundamental phenomena, such as measurements of the electron electric dipole moment and parity violation, and tests of the hypothesized fifth force. Due to their complicated rotational-vibrational-electronic energy level manifolds, molecules generally lack the closed two-level cycling systems that have proven so useful for efficient laser cooling. So far, only alkaline-earth mono-halides, with their near-unity Franck-Condon factors, have been successfully cooled and slowed with lasers (SrF in~\cite{Shuman2010,Barry2012} and~CaF in~\cite{Zhelyazkova2014}). Temperatures on the order of a few~mK have been achieved

More generally, polar molecules have a high enough sensitivity to electric fields (due to Stark shifts) that electrostatic trapping can be accomplished once the molecules have undergone initial cooling. This has been achieved using time-varying inhomogeneous electric fields for a molecular beam of~$\rm NH_3$~\cite{Bethlem2000}. Filtering of slow molecules for loading into electrostatic traps has been demonstrated for~$\rm ND_3$ using an electrostatic quadrupole guide~\cite{Rieger2005}, and has also been demonstrated for~Rb using a magnetic octupole guide as a prototype for paramagnetic molecules~\cite{Nikitin2003}. Regarding chip potentials, storage and adiabatic cooling of~$\rm CH_3F$ has recently been demonstrated using a micro-structured electric trap, and Sisyphus cooling has been applied \via\ spontaneous emission between vibrational states to effect further reductions in temperature to the~mK range~\cite{Zeppenfeld2012}.

\begin{figure}[t!]
   \centering
   \includegraphics[width=0.4\textwidth]{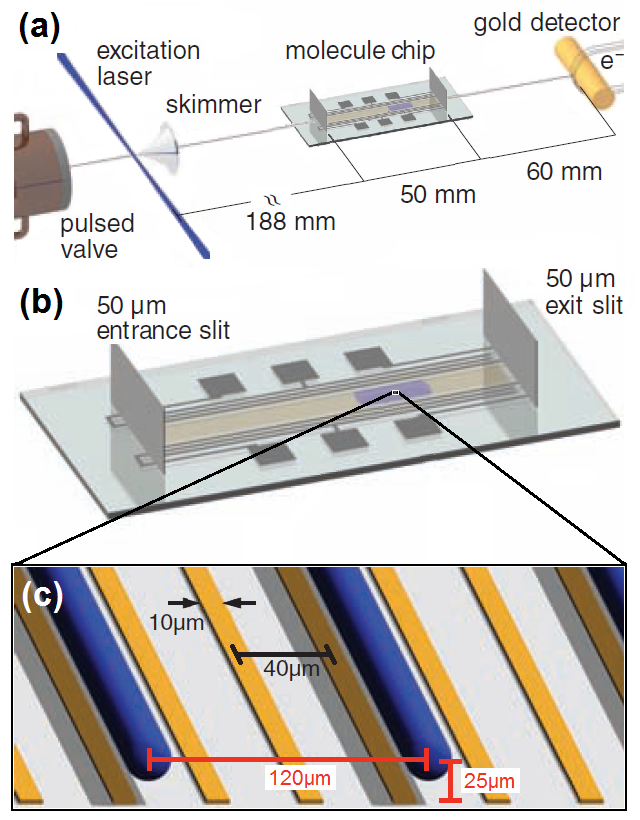}
   \caption{Trapping molecules on a chip. (a)~A pulsed beam of~CO molecules is prepared in the upper $\Lambda$-doublet level of the~$a^3\Pi_1$ ($v'=0, J'=1$) state by direct laser excitation from the electronic ground state. (b)~The molecules are collimated and then travel closely above the ``molecule chip'' over its full~50-mm length. Upon arrival above the chip, the molecules are confined in tubular electric field microtraps  centered~$\rm25\,\mum$ above the chip that move with the molecular beam at a velocity of several hundred~m/s. (c)~An array of these miniaturized moving traps (blue tubes) is brought to a standstill over a distance of only a few centimeters by applying phase-shifted MHz-range potentials to the micro-structured electrode array. After a certain holding time, the molecules are accelerated off the chip again for detection. Adapted from~\cite{Meek2009b}, with permission from~AAAS.}
   \label{fig:Meek2009bfig1}
\end{figure}

Strong electric field gradients can be produced by an atom chip. This has been used to demonstrate trapping of~Sr atoms using time-varying inhomogeneous electric fields generated by electrodes on an atom chip~\cite{Kishimoto2006} (spinless~\Sr\ cannot be trapped using just magnetic fields from the atom chip). Similarly, an array of time-varying electric fields can be created by a series of electrodes arranged along the length of a ``molecule chip''~\cite{Meek2008}. The local minima of this array of electric fields can be made to move parallel to the atom chip axis, and very close to its surface, by sinusoidally varying the electrode potentials at~MHz frequencies. The resulting ``supersonic conveyor belt'' can be moved smoothly over the surface of the chip at a speed commensurate with molecules emerging from a supersonic beam; chirping the frequency of these electrode potentials slows down the molecules over the length of a few~cm~\cite{Meek2009a}. Figure~\ref{fig:Meek2009bfig1} shows the design and implementation of this chip~\cite{Meek2009b,Santambrogio2015}.

But these are molecules after all, and besides their mechanical manipulation, all their internal degrees of freedom can be manipulated too, using light of the appropriate wavelength while the molecules are on the chip. Light from the~UV to the~IR and microwave ranges can be coupled to the molecules on the chip to induce electronic, vibrational, and rotational transitions. The method can be used for a wide variety of polar molecules and has been applied most extensively for~CO~\cite{Santambrogio2015}. Detection is also needed for the molecules on the chip. A suitably general and sensitive imaging technique is provided by multi-photon resonance-enhanced ionization~\cite{Marx2013}. This method is quantum state-selective and generally applicable, adding the final fundamental component to the molecule chip and offering a new and promising route for investigating cold molecules. Incidentally, the analogous chip-based Stark decelerator method has also been used for trapping Rydberg-state~H atoms~\cite{Hogan2012}. We remark that these molecule chips are essentially miniaturized versions of Stark decelerators consisting of hundreds of discrete electrode assemblies over lengths exceeding~$\rm1\,m$~\cite{Schnell2007}. 

Instead of directly trapping polar molecules with electric fields, another route for producing ultracold molecules relies on tuning the interactions between ultracold atomic gases with Feshbach resonances~\cite{Chin2010}. A proposal for shifting magnetic Feshbach resonances with~RF fields from atom chips~\cite{Tscherbul2010}, rather than with~DC magnetic fields alone, promises selective tuning of the scattering length in ultracold mixtures, much faster control, and the capability of addressing individual pairs of atoms for~QIP applications. Alternatively, ultracold alkali diatomics can be formed by photoassociation of their atomic precursors. We refer the reader to recent reviews for other methods and protocols that are being used to generate ultracold molecules, and for detailed discussions of their physics and chemistry~\cite{Carr2009,Quemener2012,Lemeshko2013}. 

\subsection{Antimatter} 
\label{subsec:antimatter}

The Weak Equivalence Principle claims that the trajectory of a particle is independent of its composition and internal structure if the only force acting upon it is gravity. But does antimatter obey this rule? In fact, the Weak Equivalence Principle has never been examined experimentally for gravitational interactions of antimatter with matter. Some theoretical work has discussed the possibility that the gravitational interaction could be significantly different from the analogous matter-matter interaction, or even that it is repulsive~\cite{Villata2011}. A recent review summarizes experimental progress for measuring the effect of gravity on cold neutral antimatter atoms~\cite{Dufour2015}.

Charged test particles are not well suited for determining gravitational forces although they are much easier to trap. Since the Coulomb force is orders-of-magnitude stronger than the gravitational force, any stray electric fields could mask small gravitational effects. The~GBAR project (Gravitational Behaviour of Antihydrogen at Rest) proposes a cold-atom scheme to solve this problem by neutralizing an antiproton with a positron to create antihydrogen~\cite{Indelicato2014}. 

Experiments with antihydrogen are also important for tests of~CPT violation thought to be responsible for the lack of anti-matter in the observed natural universe~\cite{Kellerbauer2015}. Other interesting uses of antihydrogen include the study of quantum reflection of antimatter from a matter surface~\cite{Dufour2013}.

Further into the future, a new international chip-based consortium is now developing ideas for replacing the macroscopic traps of the~ALPHA~\cite{Amole2014} and AEgIS~\cite{Scampoli2014x} collaborations at~CERN with atom chips. Such a chip would enable the combined operations of trapping antiprotons and positrons, cooling them, recombining them, and finally trapping the resulting antihydrogen atom. The atom chip surface is essential for this idea, \eg~to cool the charged antiparticle constituents by way of resistive cooling~\cite{Itano1995} before recombining them. The nearby surface may also enable efficient multi-species traps as required for this experiment, while the high gradients may allow efficient recombination processes~\cite{Leefer2016x}. One of the challenges concerns the final temperature of the antihydrogen. Materials engineering focusing on new high current density conductors (Sec.~\ref{subsec:materials}) could enable much deeper atom chip traps, thereby improving prospects for efficient antihydrogen capture. 

\subsection{Charged particles} 
\label{subsec:ions}

Trapped ions are advantageous for~QIP because of their long lifetimes, long coherence times relative to gate times, strong inter-ion interactions, high reproducibility, and efficient detection~\cite{Monroe2013,Eltony2015x}. Ideally, the motional modes for ions in a multi-ion trap can be laser-cooled to the ground state of their motion, thereby providing a well-defined initial quantum state. This delicate preparation can however, be disturbed by even one quantum of motion absorbed from the environment~\cite{Hite2013}.

Scaling quantum processors to hundreds or thousands of qubits in order to outperform their classical counterparts in certain applications remains a central challenge. Micro-fabricated ion traps provide an opportunity for significant advances in~QIP, and to realize quantum simulations, cavity quantum electrodynamics, quantum hybrid systems, precision measurements, and many other interesting topics.  

Most ion chip traps use planar electrodes, unlike the traditional Paul trap and the Penning trap. For multi-layer chip designs, the ion is trapped between electrodes located in two or more planes~\cite{Stick2006}. There are also single-layer chips with surface electrode traps that make use of mature micro-fabrication technologies~\cite{Seidelin2006}. In addition to~1D trapping arrays, 2D~structures such as~Y-shaped junctions~\cite{Moehring2011} have been developed. 

Micro-fabricated ion traps bring a greater level of precision and density to ion-based systems for~QIP and quantum simulation. But trap miniaturization also worsens electric field noise as the ions are brought closer to the trap surface. Single ions are usually confined~$\rm30-200\,\mum$ from the electrodes, where measured heating rates are much faster than expected from Johnson (thermally-induced) noise and technical noise. The dependence on the distance~$d$ from the ion to the nearest electrode is very strong, and is consistent with a~$d^{-4}$ ``anomalous heating'' scaling law~\cite{Daniilidis2011}.

Cooling the trap electrodes to cryogenic temperatures has been shown to drastically reduce heating rates~\cite{Deslauriers2006,Labaziewicz2008}. Implementing \insitu\ $\rm Ar^+$-beam cleaning produced a~100-fold reduction in the heating rate~\cite{Hite2012}. Pulsed laser cleaning of the trap surface has been reported to reduce the electric-field noise spectral density by about~50\%~\cite{Allcock2011}. Heating rates of trapped single~\Sr$^+$ ions in a superconducting micro-fabricated ion trap~\cite{Wang2010a} showed no significant change across the superconducting transition. These experimental results suggest that anomalous heating is caused primarily by noise sources on the surface, and not in the bulk metal. Monolayer graphene, acting as a field-transparent coating and free of surface charges, was synthesized directly on the copper electrodes of an ion chip, but the measured heating rate was about~100 times {\it faster} than typical for an uncoated trap operated under similar conditions at~$\rm4\,K$~\cite{Eltony2014}. Given the beneficial results for~$\rm Ar^+$-cleaning, it was suggested that graphene fails because of hydrocarbon contamination adsorbed during the preparatory baking for ultra-high vacuum.

\begin{figure}[t!]
   \centering
   \includegraphics[width=0.4\textwidth]{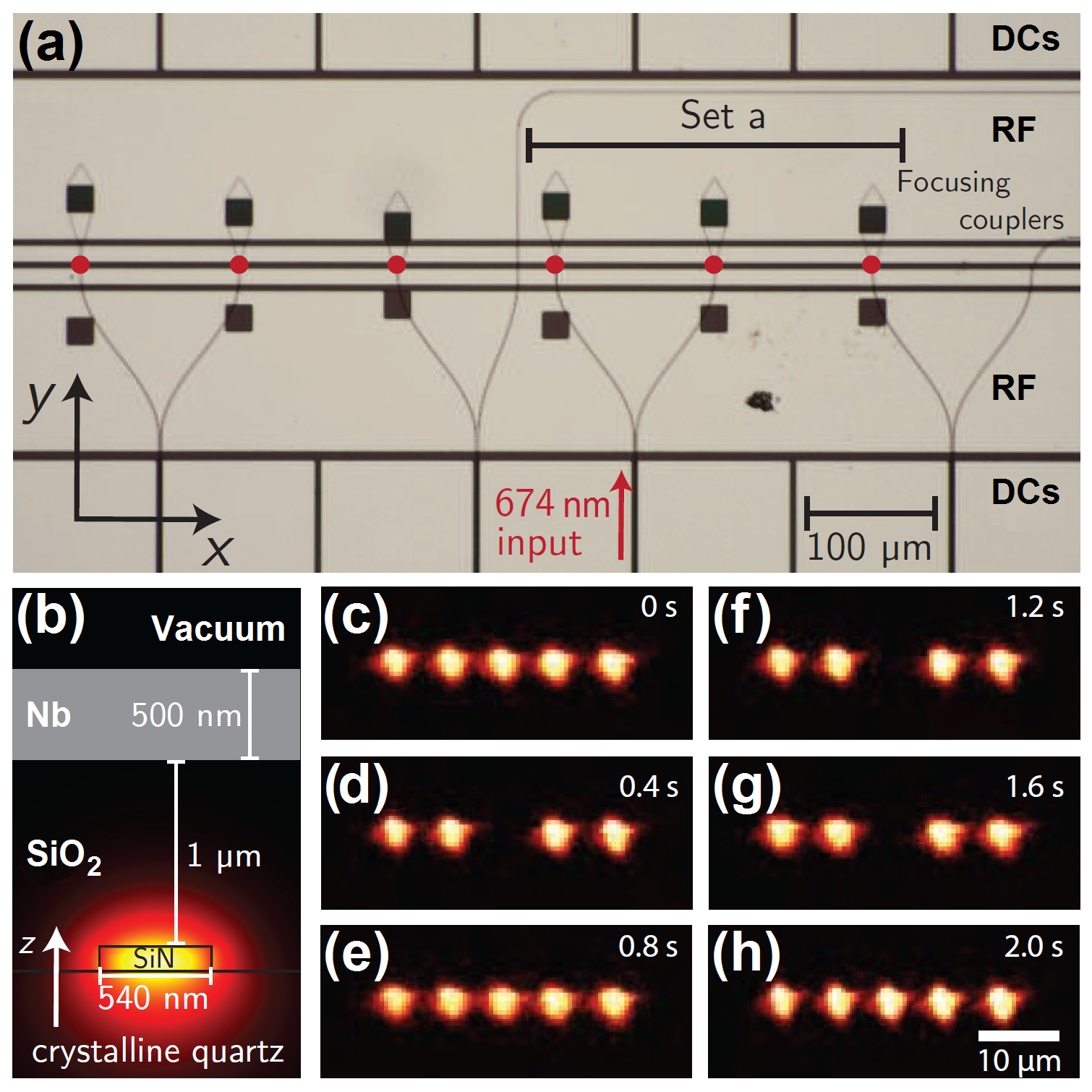}
   \caption{Integrated ion chip. (a)~Optical micrograph of the ion chip with integrated waveguides and couplers underneath at multiple trap zones; optical waveguides and couplers are visible \via\ topography transfer to the metal. Ions are trapped at one of the positions marked by the red dots,~$\rm50\,\mum$ above the electrodes, with appropriate potentials applied to the~DC and~RF electrodes. (b)~Simulated electric field mode profile of the single quasi-TE mode (field oriented predominantly horizontally) waveguide used for routing. Quantum coherent operations are performed on the optical qubit transition in individual~\Sr$^+$ ions by visible light routed in and emitted from the~SiN waveguides and couplers. (c-h) Sequence of images of~$\rm422\,nm$ fluorescence from a chain of five~\Sr$^+$ ions, with the middle ion aligned to the grating coupler's focus and occasionally entering a dark state; the sequence spans~2\,s with frames spaced evenly. Courtesy of Karan Mehta~(MIT) and MIT~Lincoln Laboratory~\cite{Mehta2015x}.}
   \label{fig:Mehta2015xfig1and3}
\end{figure}

As an obstacle to high-fidelity two-qubit gate operations in large-scale trapped-ion~QIP, anomalous heating still requires great efforts to be understood and further reduced, as recently reviewed in~\cite{Hite2013}. Alongside efforts to suppress the origin of these stray electric fields, another approach could be taken by compensating for them using efficient measurement methods~\cite{Narayanan2011}. 

Delivering light to multiple individual ions in surface-electrode traps at various wavelengths is required for qubit manipulations and readout in~QIP. The first realization of optical functionality for trapped ions with nano-fabricated integrated optics has been reported very recently~\cite{Mehta2015x}, with nano-photonic dielectric waveguides integrated within a linear surface-electrode ion chip. Qubit addressing at multiple locations was realized by focusing light from grating couplers emitted through openings in the trap electrodes to specific ions trapped~$\rm50\,\mum$ above the chip (Fig.~\ref{fig:Mehta2015xfig1and3}). This high level of optics integration within the chip trap presents a scalable implementation for large-scale~QIP.

A collective ion-photon interface was demonstrated by a micro-fabricated planar trap array of ion chains coupling to an optical cavity~\cite{Cetina2013}. The optical cavity serves as a quantum information bus between ions and achieves long-lived sub-wavelength localization of ions~\cite{Karpa2013}. This system was also used for investigating friction with trapped ions in an optical lattice and demonstrates extensive control achieved at the atomic scale~\cite{Gangloff2015}.

\begin{figure}[t!]
   \centering
   \includegraphics[width=0.5\textwidth]{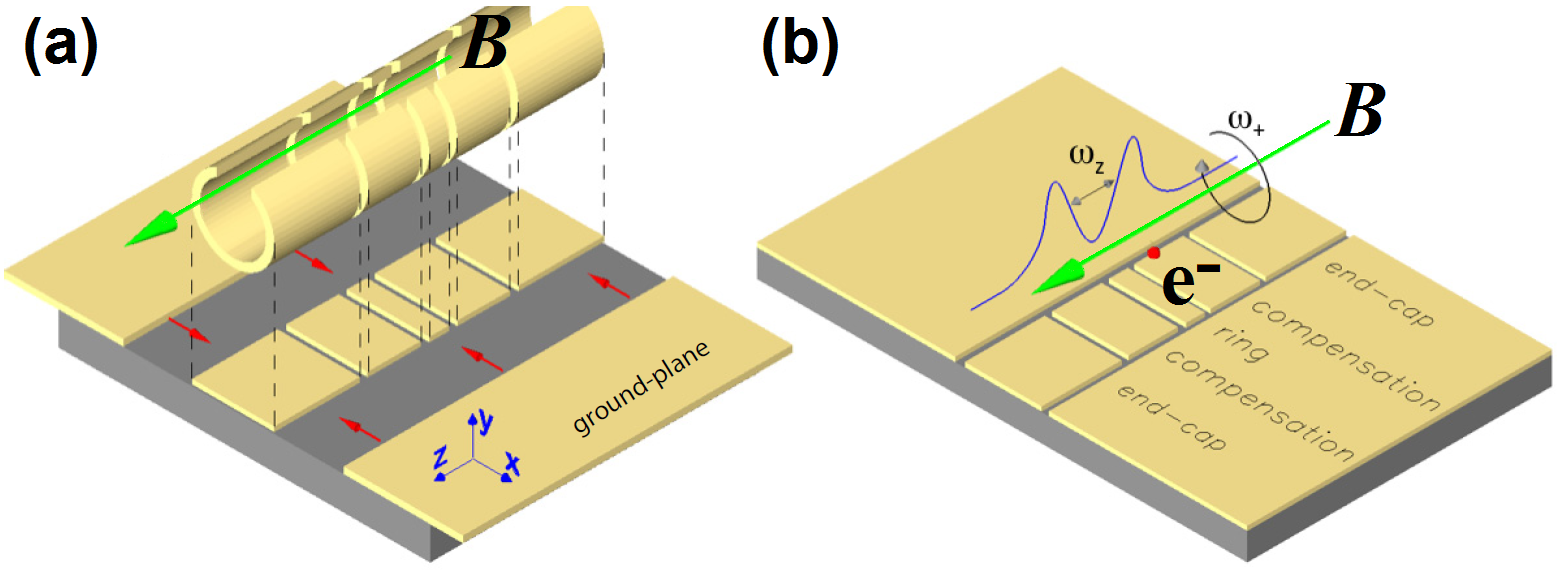}
   \caption{Chip trap for electrons. (a)~Genesis of the coplanar-waveguide~(CPW) Penning trap. The figure shows the projection of a standard cylindrical five-pole Penning trap onto a plane. The projected segments are shielded with two outer ground-planes. (b)~Sketch of the trap, with the resulting cyclotron and axial motions of an electron (red dot). Adapted from~\cite{Verdu2011}, with permission~\cite{CCBY} \textcopyright~IOP Publishing~\& Deutsche Physikalische Gesellschaft.}
   \label{fig:Verdu2011fig1}
\end{figure}

Chip traps can also trap electrons. A trapped single electron (a geonium-atom~\cite{VanDyck1976x}) is an outstanding system for testing the laws of physics with ultra-high precision~\cite{Wineland1973,Brown1986}. Figure~\ref{fig:Verdu2011fig1} shows a planar Penning trap on a chip that has been designed and analyzed for very accurate control of the dynamics of a trapped single electron~\cite{Verdu2011,AlRjoub2012}. Recently, microwave guiding~\cite{Hoffrogge2011} and beam splitting~\cite{Hammer2015} of electrons was demonstrated by using micro-structured guiding potentials created above the surface of planar microwave chips, providing novel tools for electron-based quantum matter-wave optics.

We have seen that the atom chip has proven useful for a variety of particle types and this spectrum is still growing. As this Special Issue concerns~BECs, we will retain our focus on neutral atoms. We now describe a range of enabling technologies related to atom chips.

\addtocontents{toc}{\protect\vspace{0pt}}

\section{Enabling Technologies} 
\label{sec:technology}

The original atom chip configurations were based on single-layer current-carrying wires deposited on top of substrates. These designs were simple but quite ingenious. For example,~U- and~Z-shaped wires were found to create quadrupole fields for a magneto-optical trap~(MOT) and Ioffe-Pritchard fields for holding cold atoms. Many nice anecdotes may be related concerning those early days. For example, in Innsbruck the external bias coils were held by hand and moved around the chamber while the experiment was running until the right geometry could be found. Furthermore, contrary to expectations, it was surprising to find that the shadows in the cooling beams created by the etchings defining the wires did not hurt the reflection~MOT. Nobel laureate Theodor H{\"a}nsch tells the story of how he drew the idea for an atomic conveyor belt on a napkin in a restaurant~\cite{HanschBio2005}. 

However, it soon became clear that many more novel ideas are required in order to make progress towards the ultimate vision of the atom chip. In recent years multitudes of new technologies and techniques have been conceived and implemented, allowing impressive progress towards the realization of the atom chip vision. Here we present some of these new technologies and techniques.

\subsection{Materials}
\label{subsec:materials}

Proposals for using new materials include novel ideas for using molecular conductors such as carbon nano-tubes~(CNTs)~\cite{Fermani2007,Petrov2009} and graphene. CNTs are expected to sustain current densities up to~100 times those of gold, where current density is the figure of merit for potential gradients. These materials also decrease the magnitude of Johnson noise and the Casimir-Polder~(CP) potential and may even have less electron scattering, thereby reducing potential roughness. They may also have the advantage of sharp absorption peaks, allowing them to be placed next to high-Q photonic devices without degrading the finesse. Since the deterministic directional growth of~CNTs and electrically contacting them are still difficult procedures, chemical processes are also being developed to etch wires from graphene~\cite{Geim2009,Wang2010b}. We expect that additional types of high current-density materials and composites will be developed that can more easily be utilized for fabrication of current-carrying wires~\cite{Subramaniam2013,Mehta2015a}, thereby enabling the creation of tighter and deeper magnetic traps. 

Anisotropic conductors have been suggested as current-carrying wires on the chip, since the decoherence rate may drop by several orders of magnitude, even at room temperature~\cite{David2008}. This is due to the fact that conductance can be suppressed in the directions perpendicular to the current, precisely the directions that produce the magnetic fields inducing decoherence. 

Utilizing alloys should reduce Johnson noise, especially at low temperatures. The alloy composition strongly affects the resistivity~$\rho$ and its temperature dependence becomes non-linear, thereby reducing~$T/\rho$, the dominant term in the noise~\cite{Dikovsky2005}. This may be advantageous compared to superconductors, which also exhibit very low noise, since alloys such as~Au-Ag are easily deposited and do not suffer from limitations due to critical currents, sensitivity to external magnetic fields, or the appearance of vortices that may generate complex~DC fields or increased noise~\cite{Fruchtman2012}.

\begin{figure}[t!] 
   \centering
   \includegraphics[width=0.45\textwidth]{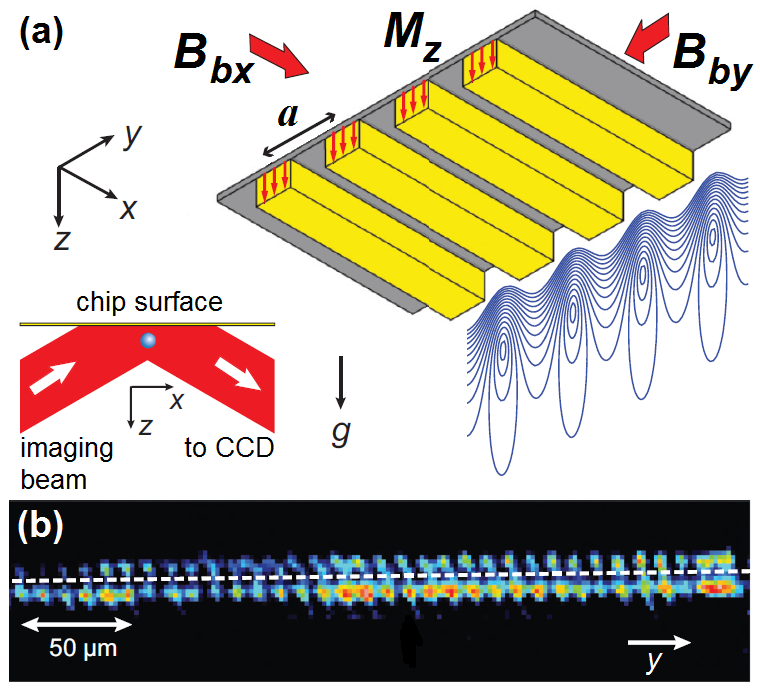}
   \caption{Permanent magnet chips. (a)~Schematic of the magnetic micro-structure used to create a periodic~1D lattice of magnetic microtraps. Contour lines are calculated equipotentials with contour intervals of~$\rm0.5\,G$. (b)~Part of the absorption image for an array of clouds of~\Rb\ $|F=1,m_F=-1\rangle$ atoms trapped in the~$\rm10\,\mum$-period magnetic lattice, after evaporative cooling to below the critical temperature. Adapted from~\cite{Jose2014}, with permission \textcopyright~2014 by the American Physical Society.} 
   \label{fig:Jose2014fig1} 
   \end{figure}

The control of~2D electron gases has been considered by the Fromhold and Kr{\"u}ger groups as an alternative to current-carrying wires~\cite{Sinuco-Leon2011}. Nano-wires~\cite{Salem2010} and nano-bridge wires~\cite{Chuang2011a} have also been suggested for suppressing Johnson noise, increasing current density, and decreasing finite-size effects whereby the magnetic field gradients suffer when atom-surface distances approach the width of the wire.

Permanent magnets have been used as another alternative to current-carrying wires for creating~1D and~2D magnetic lattices by the Hannaford~\cite{Jose2014} (Fig.~\ref{fig:Jose2014fig1}) and Spreeuw groups~\cite{Leung2014}, respectively. They have the advantages of strong fields, less Johnson noise (due to their low conductivity), and no technical noise, but an obvious disadvantage is their lack of dynamical control. Crucially, permanent magnet lattices can be fabricated with more complex symmetries and with smaller periods than their optical counterparts~\cite{Schmied2010,Herrera2015}, which may also be advantageous for the goals of quantum simulations and~QIP (Sec.~\ref{subsec:QIP}). 

The interaction with electric fields is also of interest for trapping on chips~\cite{Kishimoto2006,Kruger2003} or for other manipulations through Stark shifts. Here novel materials may be used, such as transparent indium-tin oxide, with advantages for photonics as well as low Johnson noise. 

New setups using atom chips cooled to cryogenic temperatures have allowed superconductors to be  incorporated~\cite{Nirrengarten2006,Mukai2007,Roux2008,Kasch2010,Muller2010a,Imai2014,Minniberger2014}.
To the best of our knowledge, the first cold atomic cloud~\cite{Nirrengarten2006} as well as the first~BEC~\cite{Roux2008} on a superconducting atom chip were achieved by the Haroche group. The first trapping by a persistent current was done by the Shimizu group~\cite{Mukai2007}. These developments open the door for cryogenic material science~\cite{Chan2014}, suppressing noise~\cite{Scheel2005,Skagerstam2006,Dikovsky2009,Nogues2009,Kasch2010}, and attaining enhanced coherence~\cite{Bernon2013,Hermann-Avigliano2014}. Rapid sample exchange, as an alternative to single-use atom chips, has recently been implemented in a cryogenic apparatus~\cite{Naides2013} and may enable probing exotic materials, as well as shedding new light on, for example, high-$T_c$ superconductivity or the evolution of domains in magnetic materials. 

Cryogenic setups require special care to reduce heat loads. An electron beam source was developed to minimize the heating required to vaporize atoms for loading the~MOT~\cite{Haslinger2011}. Initial~MOT loading also requires magnetic fields that can heat the chip environment. Most systems completely remove the heat load of these initial steps by cooling the atomic ensemble in one chamber, and then transferring it to an adjoining science chamber using either magnetic or optical dipole transport~\cite{Naides2013}. Lenses with tunable focal lengths~\cite{Mishra2014} allow precise dipole transport to be performed without physically moving any optics~\cite{Leonard2014}. A magnetic conveyor-belt has also been used for transferring the atomic ensemble both horizontally and vertically, thereby allowing improved radiation shielding for reaching~mK chip temperatures~\cite{Minniberger2014}. This technique however, requires a large number of overlapping coil pairs to create the traveling quadrupole magnetic trap, and its efficiency drops for colder atoms due to Majorana spin-flip losses.

\begin{figure}[t!] 
   \centering
   \includegraphics[width=0.45\textwidth]{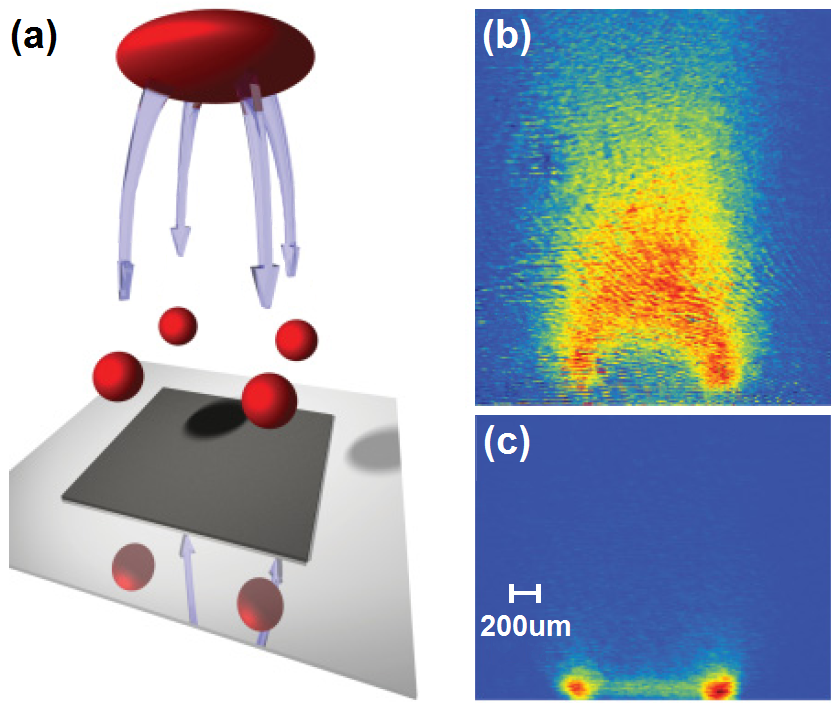}
   \caption{Superconducting atom chip. (a)~Magnetic trap geometries formed above an~YBCO superconducting square and an external bias magnetic field. (b)~For low values of~$B_{\rm bias}$ there exists one field zero above the square center. (c)~Higher values of $B_{\rm bias}$ bring the atoms closer to the chip surface and split the initial, central trap into four separate traps (two of the atom clouds are hidden due to the imaging angle). Adapted from~\cite{Siercke2012}, with permission \textcopyright~2012 by the American Physical Society.}
   \label{fig:Siercke2012fig7} 
\end{figure}

The persistent current carried in a superconductor by vortices can be utilized to create traps, as well as a magnetic lattice close to the surface~\cite{Muller2010b,Siercke2012}. Such a system is shown in Fig.~\ref{fig:Siercke2012fig7}, where the location of the lattice sites is determined by the geometry of the superconducting surface. Corresponding theoretical proposals have been made in~\cite{Zhang2012,Sokolovsky2014a,Sokolovsky2014b,Sokolovsky2016}. Let us also note that cryogenic surfaces hold the promise of a ``quantum surface'' (Sec.~\ref{subsec:hybrid}). 

\subsection{Substrates}
\label{subsec:substrates} 

The substrate of an atom chip is just as important as the structures on it. For example, its thermal conductivity is crucial for determining the currents that may be driven through the chip wires~\cite{Folman2011a}. Its electrical insulation and breakdown point are paramount for enabling high electric fields. For specific applications one requires additional features such as low tangent loss for chips with high-frequency radiation (\eg\ ion chips). Substrates should have the right crystal unit structure and thermal expansion coefficient to accommodate unique materials (\eg\ superconductors) and be strong enough for use as a facet of the vacuum chamber, in which case through-wafer etchings (vias) should also be vacuum compatible~\cite{Chuang2014a}. Substrates should also be able to easily accommodate multi-layer chip designs~\cite{Trinker2008,Chuang2011a,Chuang2011b}. 

\begin{figure}[t!]
   \centering
   \includegraphics[width=0.40\textwidth]{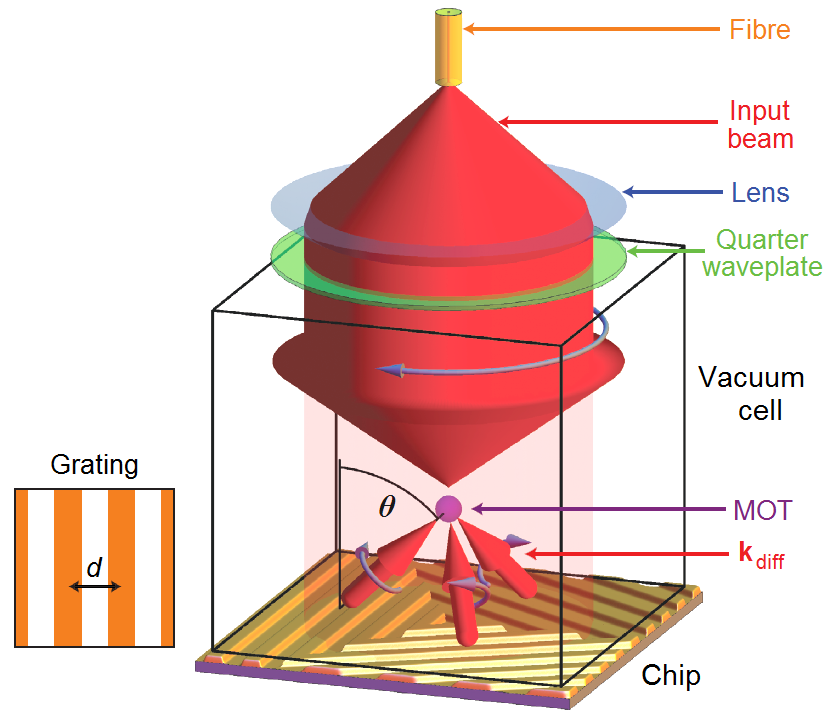}
   \caption{Concept of the grating chip~MOT. Linearly polarized light diverging from the output of an optical fiber is collimated and circularly polarized by the combination of a lens~(grey arrow) and~$\lambda/4$ waveplate. This single input beam diffracts from micro-fabricated gratings on the chip to produce the additional beams (small red arrows) needed to form a~MOT. Three linear gratings (the inset shows one pattern) diffract the light into~$n_x=\pm1$ orders to form a four-beam~MOT (only trapping beams are shown). A square array of cylindrical indentations (not shown in this adaptation) could instead be used to diffract the input into the~$n_x=\pm1$ and~$n_y=\pm1$ orders to form a five-beam MOT. Magnetic quadrupole coils are omitted for clarity. Adapted from~\cite{Nshii2013}, with permission from Macmillan Publishers Ltd.}
   \label{fig:Nshii2013fig1}
\end{figure}

Atom chip substrates have, almost since their inception, been envisioned by the Birkl and Ertmer groups as integrating optical elements and interactions~\cite{Birkl2001}. Chip fabrication processes~\cite{Folman2011a,Folman2011x} are developing continuously and we have seen, for example, the fabrication of waveguides within the substrate (Fig.~\ref{fig:Mehta2015xfig1and3}), as well as high-Q devices, which are also of paramount importance (Sec.~\ref{subsec:photonics}). Fabrication processes can now accommodate an optical window embedded within the substrate~\cite{Salim2013,Caliga2016} as well as transparent substrates~\cite{Huet2012,Chuang2014b}, allowing high-resolution imaging of the atomic ensemble. Let us also touch briefly upon the topic of the~MOT, which is the first stage of cooling and collecting the atoms. Atom chips generally cannot utilize the standard~6-beam configuration since the chip blocks some of the beams. Chip~MOTs have evolved from using a flat mirror surface so that the reflected beams create the~MOT, to a pyramidal micro-mirror where the facets create the~MOT beams~\cite{Trupke2006}, and recently to a grating etched into the silicon wafer whereby the refracted light is used to create the~MOT beams~\cite{Nshii2013,Lee2013a} (Fig.~\ref{fig:Nshii2013fig1}). The above-mentioned embedded optical window has recently enabled hybrid magnetic and optical potentials for confinement and controlled tunneling~\cite{Caliga2016}, while transparent substrates have allowed the direct transmission of the~MOT laser beams through the substrate~\cite{Huet2012}.

As~3D fabrication becomes more common, substrates allowing high aspect-ratio etching are also required. Tasks include etching channels for conductors in multi-layer chips, vias for back-side electrical contacts~\cite{Chuang2014a}, and holes for loading atoms or ions from the back side. More elaborate tasks include vacuum chambers embedded within the substrate. Eventually, all the complex surrounding infrastructure, such as the vacuum system, optics, and lasers, will be incorporated directly onto the chip~\cite{Rushton2014} (Fig.~\ref{fig:Folman2011afig1}), with the substrate playing a crucial role in this effort. We may foresee a future in which a fully autonomous atom chip will look like a regular electronics chip from the outside.

\subsection{Photonics}
\label{subsec:photonics}

\begin{figure}[t!]
   \centering
   \includegraphics[width=0.35\textwidth]{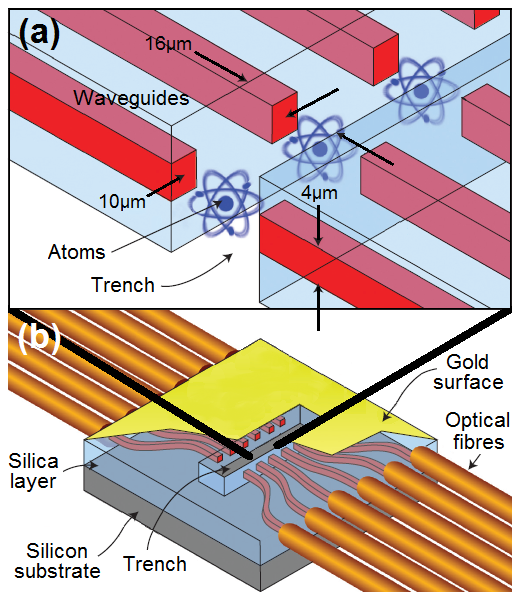}
   \caption{On-chip photonics. Schematic of an integrated-waveguide atom chip. An expanded view of the trench at the center of~(b) is shown in~(a). A silicon substrate supports a layer of silica cladding, within which~$\rm4\,\mum$-square doped silica waveguide cores are embedded. There are~12 parallel waveguides (for clarity, only~6 are shown) spaced at the center of the chip by~$\rm10\,\mum$. These flare out at the edges of the chip so that optical fibers can be connected. The top layer of the chip is coated with gold to reflect the laser light used for cooling the atoms. Current-carrying wires below the chip provide magnetic fields to trap and move the atoms. Adapted from~\cite{Kohnen2011}, with permission by Macmillan Publishers Ltd.}
   \label{fig:Kohnen2011fig1}
\end{figure}

The first truly integrated photonics device in which current-carrying wires were joined with optical waveguides was realized by the Hinds group~\cite{Kohnen2011} (Fig.~\ref{fig:Kohnen2011fig1}). Their implementation used a trench cutting across the waveguides that was narrow enough for most of the light entering the trench to be collected by the waveguides on the far side (in the absence of atoms). An atom in the trench affects the phase and the intensity of the transmitted light. Conversely, the light affects the state of the atom. Thus, each waveguide comprises a microscopic atom-photon junction. These highly integrated photonic devices may form the basis for a quantum network where photons (``flying qubits'') transfer information to atoms (``memory qubits'').

Additional attempts included on-chip Fabry-P{\'e}rot~(FP) cavities by the Stamper-Kurn group~\cite{Purdy2008} and the detection of atoms with concave mirrors etched into the substrate, as reported in~\cite{Goldwin2011}. Gluing fibers to the chips has also proven to be useful~\cite{Heine2010}. Fibers have even enabled strong coupling of ultracold atoms to a single mode of a high-finesse~FP cavity by the Reichel group~(\cite{Haas2014} and references therein). As explained in Sec.~\ref{subsec:methods}, this has proven to be extremely successful. 

Chip-based high-Q micro-disks were first coupled to ultracold atoms by the Kimble~\cite{Aoki2006} and Mabuchi groups~\cite{Barclay2006}. They can be used as a means of simultaneously trapping and detecting atoms~\cite{Rosenblit2006,Rosenblit2007}, and the near-field optical interactions they enable are being pursued intensively~\cite{Shomroni2014,Rosenblum2016} (Sec.~\ref{subsec:QIP}).

\begin{figure}[t!] 
   \centering
   \includegraphics[width=0.5\textwidth]{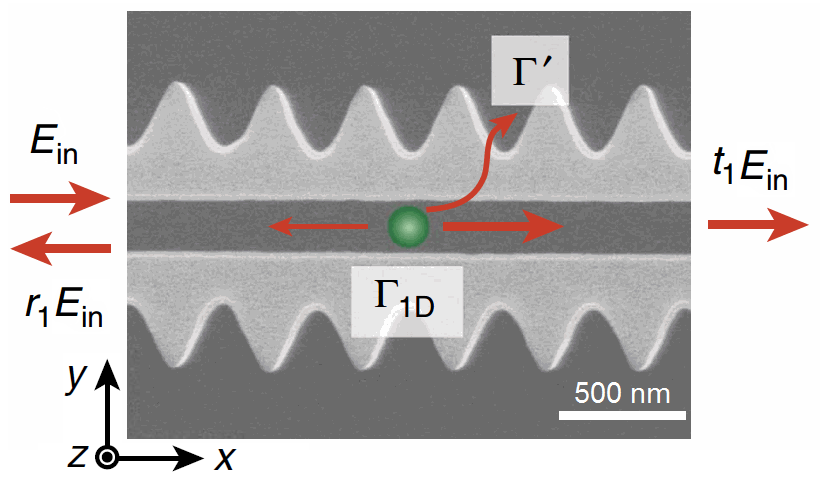}
   \caption{On-chip integrated photonic crystal waveguides. The scanning electron microscope image shows a photonic crystal waveguide made from~$\rm200\,nm$-thick~SiN. Arrows indicate radiative processes of an atom (green circle) coupled to an incident electric field~$E_{\rm in}$. Adapted from~\cite{Goban2014}, with permission by Macmillan Publishers Ltd.}
   \label{fig:Goban2014fig1a} 
\end{figure}

The idea of creating cavities from band-gap materials was analyzed in~\cite{Lev2004}. Single atoms have been brought close~($\approx\rm250\,nm$) to a photonic crystal cavity using optical tweezers in~\cite{Thompson2013b}, where the near-surface trap was made by reflecting the optical tweezer light from the photonic crystal itself. 
Figure~\ref{fig:Goban2014fig1a} shows an example of an integrated device where atoms have been coupled into a photonic crystal waveguide~\cite{Goban2014}. Such nanophotonic systems have recently been proposed for enabling a novel quantum interface in which atomic spin degrees of freedom, motion, and photons are strongly coupled over long distances~\cite{Douglas2015}.

Arrays of tunable micro-cavities have also been successfully integrated on chips and may enable the creation of large numbers of tunable light-matter interfaces~\cite{Derntl2014}. A proposal for realizing~QIP building blocks has also been made in~\cite{Abdelrahman2014}, where arrays of cavities connected by silica waveguides and integrated with magnetic traps form long-range interactions.

\subsection{Lattices} 
\label{subsec:lattices}

\begin{figure}[t!]
   \begin{minipage}[t!]{0.45\textwidth}	
      \centering
      \includegraphics[width=\textwidth]{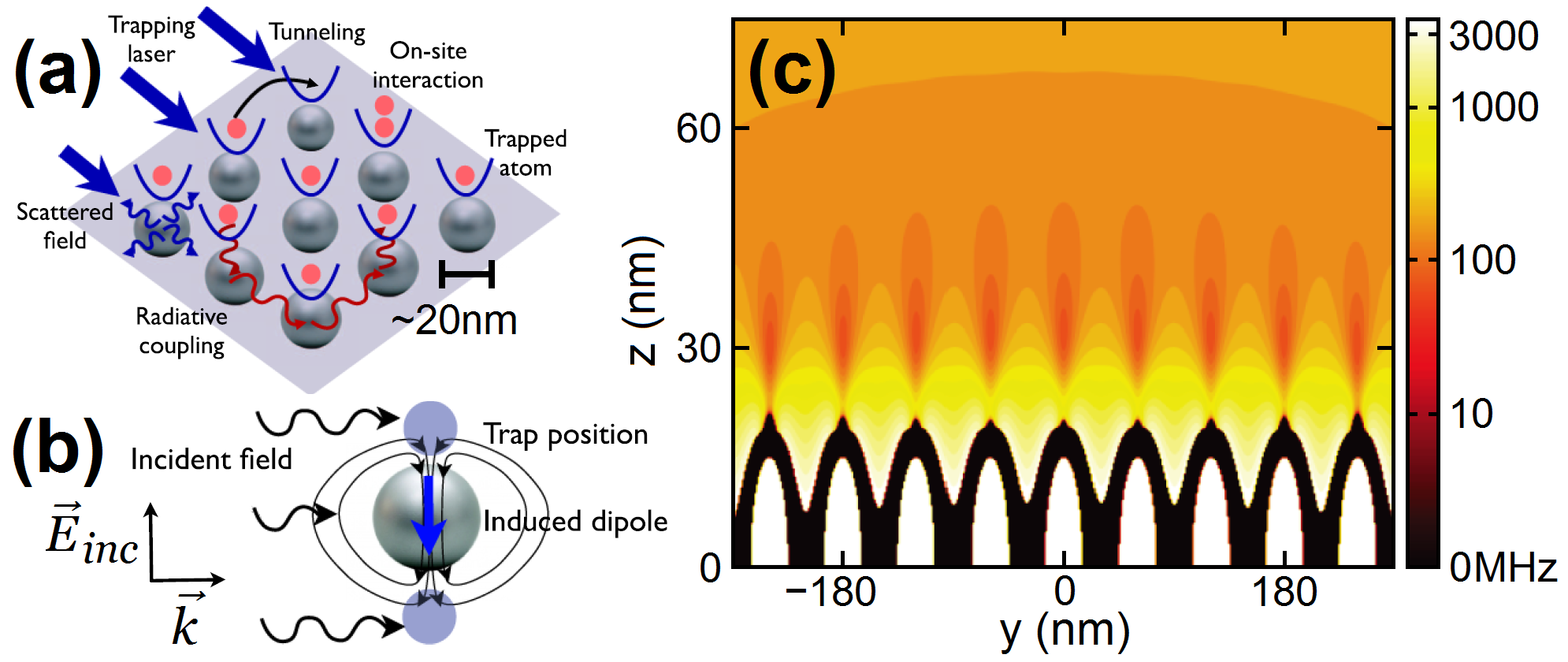}	
      \\[12pt]
      \includegraphics[width=\textwidth]{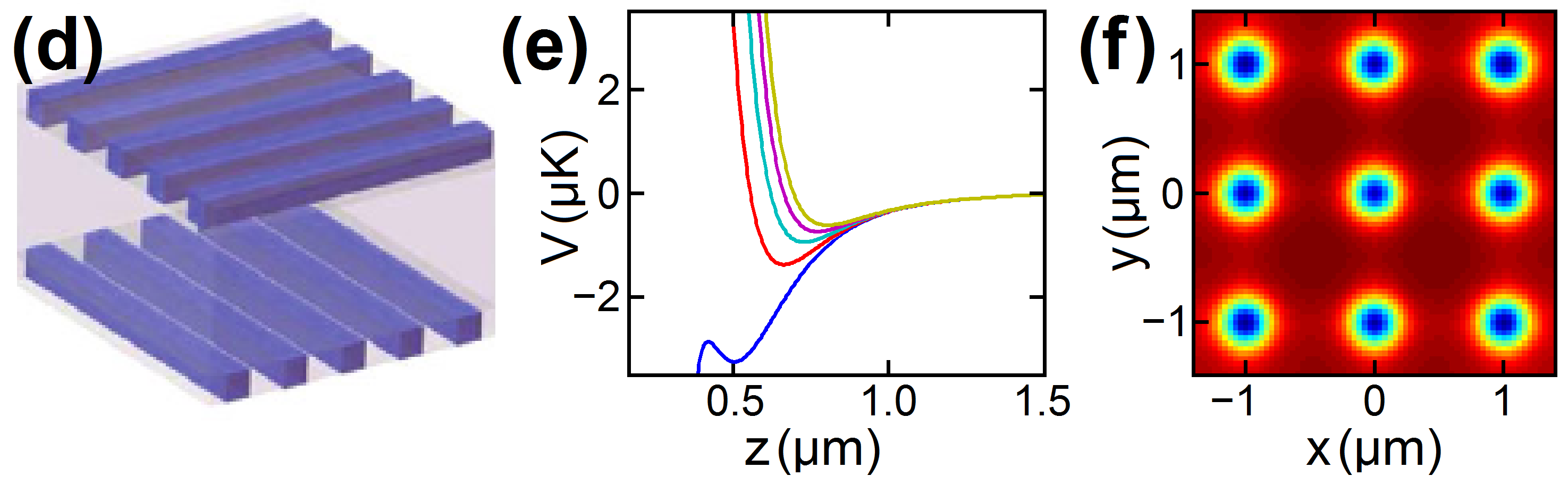}
   \end{minipage}
   \caption{On-chip lattice proposals.\\[0pt]
(a-c)~Plasmonic lattice. (a-b)~Illustration of how to engineer an optical dipole trap by driving on the blue side of the plasmon resonance. (c)~$y$-$z$ contours ($z$:$y$~axis expanded by~6:1) of the calculated potential for~Rb near a line of nine spheres in the center of a~$45\times45$ square lattice with a~$\rm60\,nm$ lattice spacing. Black regions are where the potential is negative due to van der Waals attraction, and spheres are shown in white. Adapted from~\cite{Gullans2012}, with permission \textcopyright~2012 by the American Physical Society.\\[0pt]
(d-f)~Capacitive lattice. (d)~A~2D array of point capacitors can be created when each layer of wires (purple) is connected to an opposite voltage. A prism delivering a blue-detuned evanescent repulsive potential is located below these two layers and the atoms are above. Wire spacings are~$\rm1\,\mum$ horizontally and vertically. (e-f)~Trap potentials calculated assuming that each capacitor is charged with one electron in the top layer and one positive hole in the bottom layer. The Casimir-Polder force is included. (e)~Trap potentials as a function of surface light intensity (blue to yellow): $\rm20,..,60\,W$. (f)~Simulated potential wells (blue) at a distance of~$\rm644\,nm$ from the chip. Adapted from~\cite{Folman2011a}, with permission by Springer Science+Business Media.}
   \label{fig:Gullans2012fig1} 
\end{figure}

Lattices are of great interest both for fundamental science, where they may simulate solids, and for technological applications, such as clocks. Optical standing waves generated by counter-propagating lasers are most commonly used. Optical lattices are a very mature technology~(\cite{Straatsma2015} and references therein) and have even become a commercial product for atom chips~\cite{ColdQuanta2015}(Sec.~\ref{subsec:clocks}). Plasmons have been suggested as an enabling technology for lattices that would allow sub-optical wavelength resolution~\cite{Gullans2012} (Fig.~\ref{fig:Gullans2012fig1}). A proposal for nano-engineering a vortex array in a thin-film type-II superconductor has been analyzed as a magnetic lattice for ultracold atoms~\cite{Romero-Isart2013}. Engineering a row of micro-cantilever oscillators coupled to atoms in an atom chip optical lattice has also been proposed and analyzed~\cite{Geraci2009}.

Another example of a lattice is also presented in Fig.~\ref{fig:Gullans2012fig1}. This proposal is based on single electrons trapped in miniature capacitors, whereby their electric field attracts the induced electric dipole of the atom~\cite{Folman2011a}. Repulsive balance is provided by a blue-detuned evanescent optical field. The wires would be fabricated from indium-tin oxide to avoid interfering with the evanescent fields. A similar atom chip architecture is suggested for generating sub-wavelength magnetic lattices using an~RF dressed potential approach~\cite{Sinuco-Leon2015}. In this proposal, the two layers of crossed-array conductors carry~DC and~RF currents with each successive parallel wire alternating in polarity and phase, respectively. The resulting combination of static and~RF magnetic fields produces a~2D array with minima~$<\rm2\,\mum$ from the surface. The
spatial periodicity of these minima depends on the wire spacing, reported as~$\rm1.5\,\mum$ for the~DC wires and~$\rm1.0\,\mum$ for the~RF wires.

Permanent magnet lattices are also a realistic option, such as a~2D lattice that has been loaded with~\Rb\ (Sec.~\ref{subsec:QIP}). While this lattice has a period of~$\rm10\,\mum$~\cite{Leung2014}, it is now being replaced by a lattice having a~$\rm250\,nm$-period. General methods for designing tailored lattices of magnetic microtraps for ultracold atoms on the basis of patterned permanently magnetized films have been introduced~\cite{Schmied2010,Herrera2015}. 

Near-surface lattices have the potential advantage of single-site addressability even for very small periodicities, \eg\ with near-field optics. One may instead envision a nano-scale charged electrode to create local Stark shifts next to each lattice site such that qubit rotations could be done in parallel on many selected sites. Existing lithographic techniques allow this to be done even for lattice periods as small as a few tens of~nm.

For lattices based on fields emanating from the surface, one typically needs to trap the atoms at an atom-surface distance similar to the required period. Hence, very small periods require positioning the atoms very close to the surface. Consequently, all limitations must be considered, such as losses, heating and decoherence from technical and Johnson noise, and from effects arising from the~CP potential, including ways in which these disruptive limitations can be reduced. To the best of our knowledge, the state-of-the-art for trapped atoms near an atom chip surface (where the trap is formed by fields from the surface) stands at~$\rm500\,nm$~\cite{Lin2004} (Sec.~\ref{sec:surfaces}). 

\subsection{Hybrid systems and quantum surfaces}
\label{subsec:hybrid}

\begin{figure}[t!]
   \centering
   \includegraphics[width=0.5\textwidth]{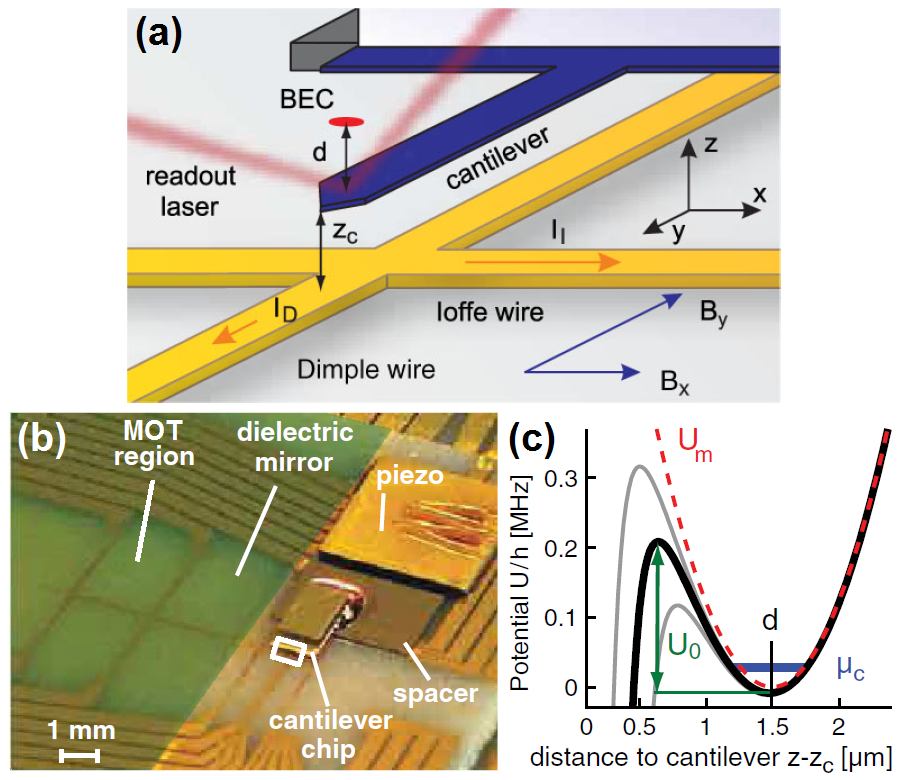}
   \caption{On-chip~MEMS. (a)~Micro-cantilever with wires for magnetic trapping of atoms. Cantilever vibrations can be independently probed with a readout laser. (b)~Photograph of the atom chip showing the~MOT loading region and the cantilever sub-assembly with a piezo for cantilever excitation. Rectangle: region shown in~(a). (c)~Potential~$U=U_m+U_s$ for trap frequency~$\omega_z/2\pi=\rm10.5\,kHz$ at~$d=\rm1.5\,\mum$ from the driven cantilever. Dashed red line: magnetic potential~$U_m$. The surface potential~$U_s$ reduces the trap depth to~$U_0$. 
Gray lines:~$U$ during the extremum positions of the cantilever for an oscillation amplitude~$a=\rm120\,nm$. Blue line:~BEC chemical potential~$\mu_c$ for~600 atoms. Adapted from~\cite{Hunger2010a}, with permission \textcopyright~2010 by the American Physical Society.}
   \label{fig:Hunger2010afig1}
\end{figure}

Atom chips offer a rich ground for hybrid systems. In this section we briefly discuss some suggestions and implementations for such systems. Several recent reviews have given detailed accounts of the great variety of possibilities~\cite{Treutlein2014,Kurizki2015}.

Integration of moving elements such as micro-electro-mechanical systems~(MEMS) exemplify yet another field holding the promise of multiple applications on the atom chip. Work performed by the Treutlein group is a first demonstration of a mechanical coupling between a resonator and ultracold atoms~\cite{Hunger2010a} (Fig.~\ref{fig:Hunger2010afig1}). Cantilever oscillations modulate the potential, thereby coupling to atomic motion. Here high-quality fabrication is required at several levels. The resonant interaction of trapped cold atoms with a magnetic cantilever tip has been demonstrated in a joint project of the Kitching and Geraci groups~\cite{Montoya2015}. A proposal evaluating the interaction between a vibrating~CNT and a~BEC found a coupling that is strong enough to sense quantum features of the~CNT current noise spectrum~\cite{Kalman2011}. 

\begin{figure}[t!] 
   \centering
   \includegraphics[width=0.5\textwidth]{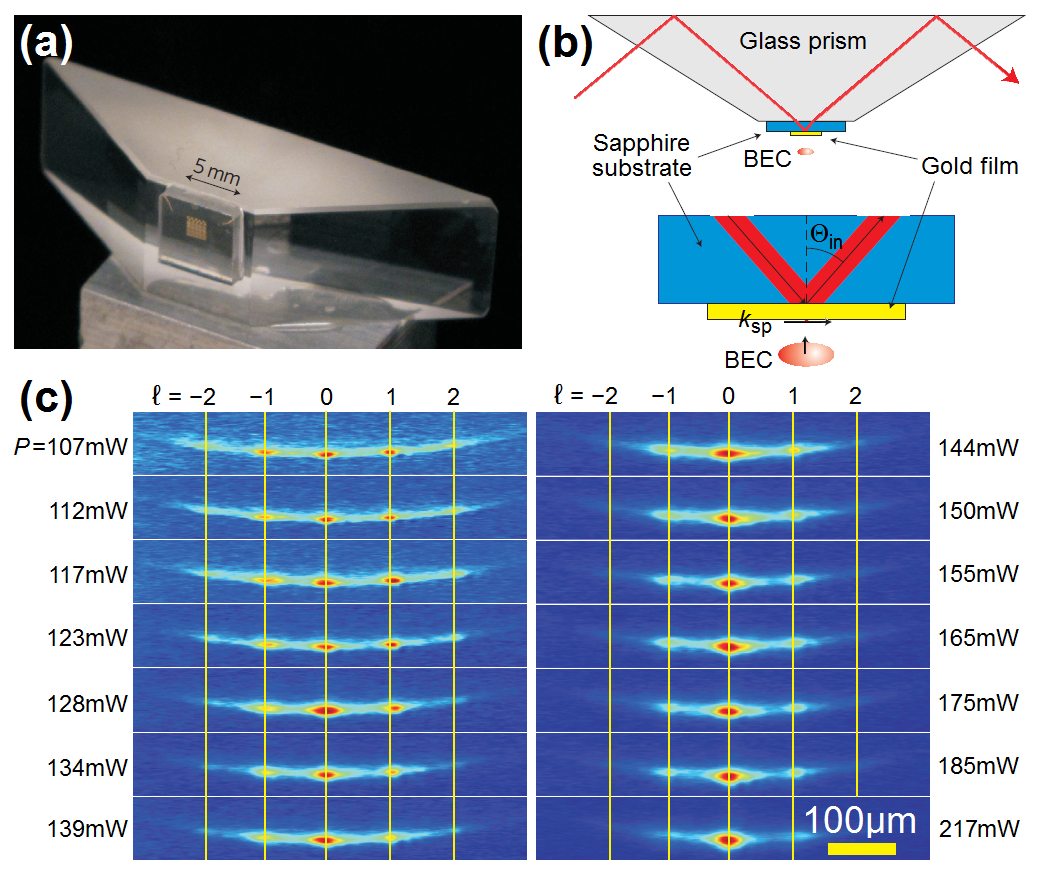}
   \caption{On-chip plasmonics. (a)~Photograph of the dielectric prism with sapphire substrate and fabricated gold structures. (b)~Schematic side view of the prism. A laser beam is internally reflected with an adjustable angle~$\Theta_{\rm in}$. In this Kretschmann configuration, plasmons are excited at the angle~$\Theta_{\rm in}=\Theta_{\rm pl}$. (c)~Diffraction images of cold atoms, which are reflected from the nominally~$\rm500\,nm$ gold stripes for laser powers of the evanescent wave~$107<P<\rm217\,mW$. Diffraction orders~$l$ are indicated by the vertical yellow lines. Adapted from~\cite{Stehle2011}, with permission by Macmillan Publishers Ltd.}
   \label{fig:Stehle2011fig1and5} 
\end{figure}

We have briefly mentioned plasmons in connection with the short-period lattice proposal shown in Fig.~\ref{fig:Gullans2012fig1}. Surface plasmons are a type of surface polariton that results from the coupling between an electromagnetic field and collective oscillations of the conduction electrons in a metal. They can be created using light and they propagate along the surface of a thin metal layer. Such surface waves can be focused and engineered to generate intense sub-wavelength features because their velocity is much slower than light in free space~\cite{Shaffer2011}. Plasmons can also be made to create light, \eg\ for near-field dipole traps or single-site addressability. Experiments done by the Zimmermann group~\cite{Stehle2011,Bender2014} (Fig.~\ref{fig:Stehle2011fig1and5}) have shown how plasmon potentials interact with atoms and have opened the door to yet another new kind of atom-surface interaction. 

Metasurfaces are a fascinating new field which may have implications for atom chips. For example, metasurfaces may be designed to influence vaccum modes and in this way affect the emission properties of atoms~\cite{Jha2015}.

Standard atom chip surfaces are classical in nature, including their on-chip sources of magnetic, electric, and optical fields that prepare, manipulate and measure quantum states of the atomic system. An interesting challenge is to create a surface that is also in a well-defined quantum state, whereby the surface quantum state and the atomic quantum state can interact with each other and be regarded as one quantum system. 

The advent of cryogenic setups has made this a feasible prospect, opening the door to many fascinating atom-surface interactions for matter-wave optics and for probing the surface,~\eg\ trapping with the magnetic field from vortices~\cite{Muller2010b,Romero-Isart2013} and shedding new light on high-$\Tc$ superconductivity, respectively.

Superconductor persistent currents have been used to trap atoms~\cite{Mukai2007,Bernon2013,Imai2014}, and a recent experiment demonstrated the sensitivity of a cold atomic cloud to single flux quanta in a~$\rm10\,\,\mum$-radius superconducting ring of~Nb~\cite{Weiss2015}. High-Q microwave cavities may also serve as buses mediating long-range interactions between atoms~\cite{Verdu2009}. Going to even lower surface temperatures~\cite{Jessen2014,Minniberger2014}, one may imagine superconducting qubits interacting with atoms~\cite{Petrosyan2009}, or perhaps even flux qubits in a superposition of counter-propagating currents coupled to polarized atoms, thereby creating a highly entangled, so-called ``high-$N00N$'' state~\cite{Lee2002}. The direct interaction between superconducting qubits and atoms may enable an optimal system whereby the atoms serve as memories while the qubits give rise to fast quantum computing gates. Such hybrid devices are currently at the center of intense work.

A quantum surface may also include quantum dots, wherein single electrons enable the entanglement of the electronic spin with the atomic spin. Surface plasmons and polaritons (Sec.~\ref{subsec:photonics}) may also provide an interface between atoms and quantum states of the surface. A quantum surface may also generate squeezed currents, thereby allowing chip operations below shot-noise limits. These are amongst many possibilities just beginning to be explored.

This section on enabling technologies has only briefly touched upon the vast field of atom chip technology. It is truly impressive to see the level of creativity and ingenuity that appears in such a variety of novel realizations. We have also seen how photonics, plasmonics, MEMS, lattices, and quantum systems are being integrated with cold atoms. The atom chip has really brought material engineering together with quantum optics and this interface is sure to produce many more interesting results in the future.

\addtocontents{toc}{\protect\vspace{0pt}}

\section{Close Encounters with the Surface}
\label{sec:surfaces}

We now move on to discuss what is perhaps the most crucial element of the atom chip: atom-surface interactions. These interactions are extremely interesting intrinsically, since they allow the study of fundamental fluctuations and forces. In addition, they enable utilizing atoms as ultra-sensitive surface probes. Nevertheless, atom-surface interactions also exhibit disruptive effects that need to be understood and overcome. For example, one challenge is to moderate the~CP force, since it lowers the potential barrier to the surface and provides an escape route for the atoms.

Controlled proximity to the surface promises numerous advantages such as tunable tunneling barriers, small-periodicity lattices, high gradients, low power consumption (for portable devices), and near-field optics or local fields (for single-site addressability). Tunneling barriers may be used for atomic circuits (atomtronics), \eg\ for interferometery, or for~QIP gates. High gradients may ensure guided propagation with less excitation to higher vibrational states due to imperfections and noise, and they may also enable reaching the Lamb-Dicke regime to the point of using side-band cooling for single atoms, recently achieved in a (non-chip) 3D~optical lattice~\cite{Reiserer2013}.

Bringing cold atoms closer to surfaces offers an ever-widening variety of possibilities for applying the atom chip method to surface and solid-state science. The investigation of topics such as Johnson noise, the~CP force, surface quantum phenomena, and electron transport are being realized and are discussed in this section. In relation to close encounters with the surface, we also discuss the achievement of spin and spatial coherence, as well as work on the hypothesized fifth force. Additional topics, including plasmons, vortices and high-$\Tc$ superconductors, as well as domain formation in exotic materials, have already been noted in Sec.~\ref{sec:technology}. Even near-contact biophysical measurements may be projected for atom chips (Sec.~\ref{subsec:fields}).

\subsection{Noise and stray fields}
\label{subsec:noise}

As atom traps are brought closer to the surface, the surface increasingly acts as a disruptive environment. The effect of noise on cold atoms in magnetic traps, due to technical sources and random thermally-induced currents (Johnson noise), has been theoretically analyzed in detail 
in~\cite{Henkel1999,Henkel2003,Dikovsky2005,Scheel2005,David2008}. For neutral atoms, both sources of noise scale as~$d^{-2}$ for typical atom-surface distances~$d$~\cite{Scheel2005,Emmert2009} and may cause losses from atom chip traps due to spin-flip transitions (when the noise frequency is on resonance), as well as heating due to position and frequency instability of the traps or guides~\cite{Henkel2003,LlorenteGarcia2013}. 

Technical noise is the simplest form of disruptive noise, originating in the power supplies which drive the atom chip currents, or from electromagnetic noise that is picked up from the environment by the cables feeding the chip. By shifting the energy levels~(Zeeman shifts), technical noise (at all frequencies) may also cause spin decoherence. Technical noise has a long correlation length and is therefore not expected to give rise to spatial decoherence directly. However, as shown in~\cite{Japha2016x} in the context of atom-atom interactions, long correlation-length noise may cause spatial decoherence indirectly by giving rise to atom loss. Technical noise is typically the strongest type of noise encountered on most types of atom chips, except those with persistent supercurrents or with permanent magnets. 

\begin{figure}[t!]
   \centering
   \includegraphics[width=0.4\textwidth]{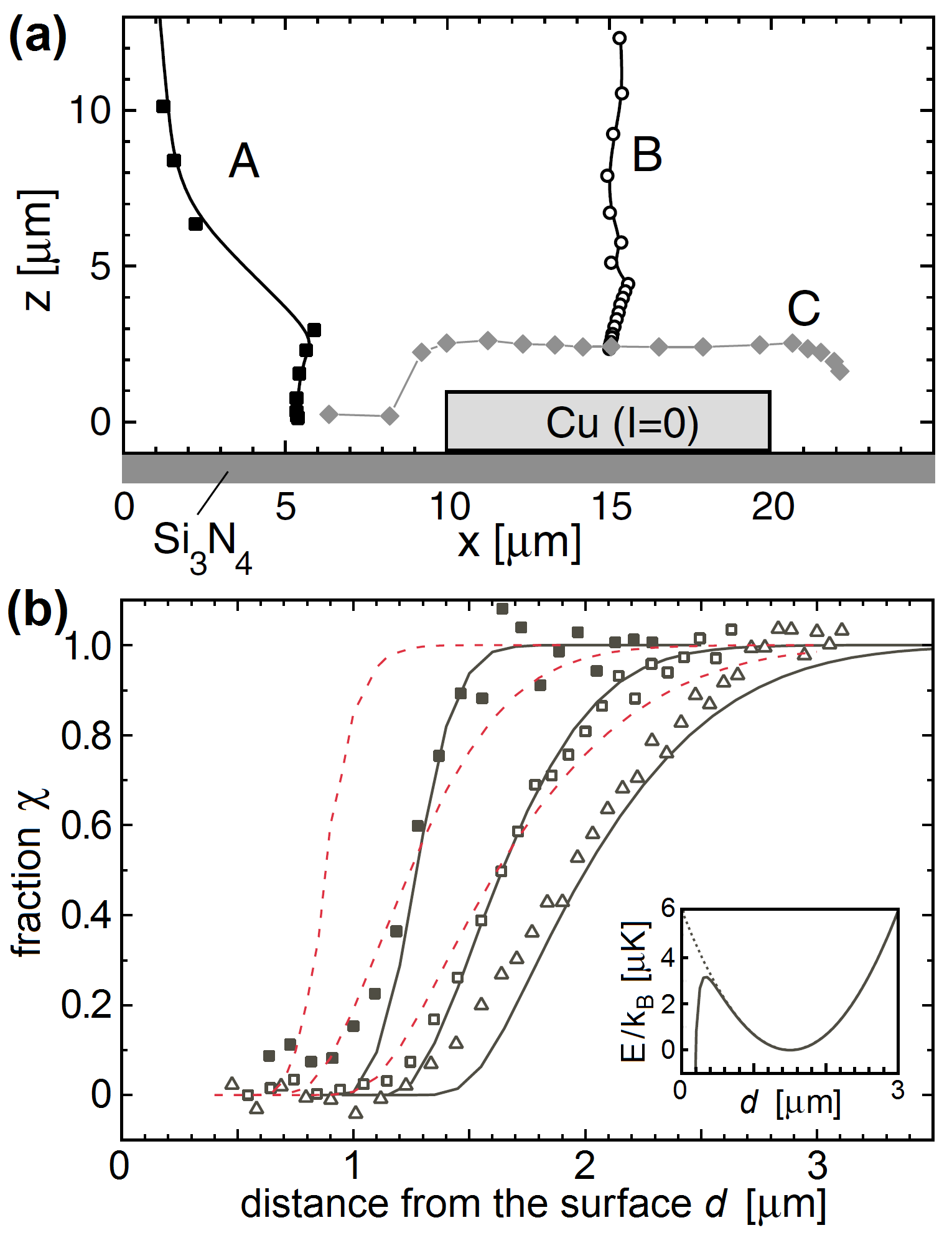}
   \caption{Probing Johnson noise and the Casimir-Polder~(CP) potential on a chip. (a)~Paths chosen for trap lifetime measurements above a dielectric surface (A) and above a copper film (B). Line C is the measured contour line of~$\rm22\,ms$ lifetime near the metal. (b)~Remaining atom fraction~$\chi$ in a trap at distance~$d$ from the dielectric surface for a condensate (solid squares), and for thermal clouds at~$\rm2.1\,\muK$ (open squares) and~$\rm4.6\,\muK$ (triangles). The solid (dashed) lines are calculated with (without)~CP potential for the condensate, $\rm2.1\,\muK$, and~$\rm4.6\,\muK$ clouds (left to right). The inset shows the trapping potentials for~$C_4=\rm8.2\times10^{-56}\,J\,m^4$ (solid line) and~$C_4=0$ (dotted line). Adapted from~\cite{Lin2004}, with permission \textcopyright~2004 by the American Physical Society.}
   \label{fig:Lin2004fig3and5}
\end{figure}

Johnson noise is another type of noise. Its origin lies with virtual electron currents due to thermal energy. Johnson noise becomes progressively more important relative to technical noise as atoms are brought closer to the surface; at small distances the power requirements are much lower and this typically enables the use of ultra-low-noise current sources. The first experimental study of this noise in atom chips was done by the Hinds group~\cite{Jones2003}. Results from the Vuleti{\'c} group followed shortly thereafter~\cite{Lin2004}, as shown in Fig.~\ref{fig:Lin2004fig3and5}. A systematic experimental study of the difference between the magnitude of Johnson noise and technical noise was done in~\cite{Emmert2009}. Superconducting surfaces are expected to have very low levels of noise~\cite{Scheel2005,Scheel2006,Skagerstam2006,Hohenester2007,Dikovsky2009} and this has indeed been confirmed by experiments measuring the effect of noise on spin-flips~\cite{Nogues2009,Kasch2010}. Additional experiments by the Fort{\'a}gh group extended their previous studies to include the effect of noise on the coherence of superposition states~\cite{Bernon2013}.

Johnson noise may also be minimized by using materials with lower conductance for the nearby surface, as accomplished with permanent-magnet chips. Thinner conducting layers reduce Johnson noise through the use of less material, which may also be achievable with~CNTs or nano-wires. Additional proposals have also been made for using exotic materials such as electrically anisotropic conductors~\cite{David2008}, or by utilizing alloys at low temperatures~\cite{Dikovsky2005}, as noted in Sec.~\ref{subsec:materials}.

Electric fields caused by surface chemistry are also of great concern, especially for ultracold Rydberg atoms. Effects due to alkali adsorbates have been measured by the Cornell~\cite{McGuirk2004}, Spreeuw~\cite{Tauschinsky2010}, and Fort{\'a}gh~\cite{Hattermann2012} groups, with recent results enabling their reduction and the consequent lowering of Rydberg atoms much closer to the chip surface~\cite{Naber2016} (Sec.~\ref{subsec:rydberg}). Patch potentials (forming~DC fields) have recently been analyzed experimentally by the Dumke group using Rydberg atoms as sensitive probes~\cite{Chan2014}. Surface chemistry is also implicated in the ``anomalous heating'' that plagues ion chips (Sec.~\ref{subsec:ions}), where dynamic patch potentials or surface dipoles are hypothesized as the source of high-frequency noise.

Disruptive fields in the~DC domain may also be caused by magnetic impurities as well as by scattered electron currents from rough wire edges, polycrystalline domain walls, and internal geometric imperfections or material impurities. Such fabrication imperfections can become especially damaging at small atom-surface distances and may cause potential roughness, thereby damaging or destroying the desired trapping or guiding of atoms. In the early days of the atom chip, these static magnetic fields sometimes became strong enough to cause breakup of the ultracold atomic cloud. This important ``fragmentation'' phenomenon was first reported by the Ketterle and Pritchard groups~\cite{Leanhardt2002} and by the Zimmermann group~\cite{Kraft2002}. It has been further characterized by many subsequent studies~\cite{Jones2003,Esteve2004,Wang2004,Schumm2005b,Kruger2007} and is reviewed in~\cite{Fortagh2007}. Several ideas for combating such effects have arisen, including time-dependent potentials suggested by the Westbrook and Bouchoule group~\cite{Trebbia2007}. Fragmentation effects may be somewhat overcome by using more advanced and careful fabrication techniques, such as those reviewed in~\cite{Folman2011x}. Further progress is still required. For example, it has been shown that significant effects come from internal bulk scattering~\cite{Japha2008}. It would be interesting to check if single-crystal gold exhibits the same magnitude of internal scattering. Similarly, it would be interesting to check other crystalline materials such as~CNTs and graphene. 

The recent advent of spatial interferometry close to the surface~\cite{Zhou2015x} could enable the measurement of correlation lengths, a topic of considerable interest extending beyond atom optics on atom chips. For example, what is the actual correlation length of Johnson noise? Is it on the same order as the distance to the surface~--~as predicted~\cite{Henkel2003} but never measured directly? Similarly, what is the correlation length of the current shot-noise? There are many other interesting noise sources to be studied, another example being the noise peak produced by superconductors at~$T_{\rm c}$~\cite{Jiang1993,Khrebtov1993}.

\subsection{Coherence close to an atom chip}
\label{subsec:coherence}

One of the major challenges of atom interferometry with magnetically sensitive atoms is the preservation of coherence. This problem increases when trapping and manipulating the atoms is performed very close to the chip surface. Decoherence mechanisms, \eg\ due to noise sources discussed above, were already analyzed in the early days of the atom chip (see \eg~\cite{Henkel2003}).

Internal-state coherence (spin coherence) has been shown to be preserved for a long time for atoms trapped near a chip surface when the two spin components are realized with magnetically trappable ``clock states'' at the ``magic value'' of the magnetic field ($\rm3.23\,G$ for the~$|F,m_F\rangle=|1,-1\rangle$ and~$|2,1\rangle$ states of~\Rb)~\cite{Treutlein2004}, and most dramatically when spin self-rephasing occurs~\cite{Deutsch2010}. However, spatial coherence, which is essential for spatial interferometry, has usually been observed only for distances of at least a few tens of~$\mum$ from the chip~\cite{Wang2005,Schumm2005a,Jo2007,Baumgartner2010}. Diffraction has been observed for atoms dynamically reflected from close encounters with surfaces~\cite{Gunther2005,Bender2014}, but the atoms were not trapped by the atom chip potential. Preliminary proof of robust spatial coherence at about~$\rm5\,\mum$ from the chip surface was recently provided by experiments with atoms trapped in a magnetic lattice generated by engineered potential corrugations in a chip wire~\cite{Zhou2015x}. In addition to long-lived coherence, these experiments showed a coherence length of~$\gtrsim\rm15\,\mum$, considerably longer than expected if decoherence arises from, for example, Johnson noise, whose correlation length is expected to be about~$\rm5\,\mum$ in this experiment.

Theoretical understanding of spatial coherence in a~BEC trapped in a confined configuration would involve both the effects of atom-atom interactions, which may be the major limitation in systems at large distances from the surface of the chip, and the effects of external noise. These effects are usually non-additive and the interplay between them may yield counter-intuitive results. A discussion of this interplay in the context of atom chip interferometry has recently been presented~\cite{Japha2016x}. Using a double-well model, it was shown that interactions may partially suppress decoherence for external noise causing phase fluctuations between the two wells. For external noise causing atom loss with relative number fluctuations however, as is likely the case in the spatial coherence experiment, decoherence may be enhanced by the atom-atom interactions. However, the interplay of these effects during a non-adiabatic time evolution has yet to be investigated. All these factors affecting coherence are also highly relevant for the field of atomtronics.

\subsection{Casimir-Polder effects}
\label{subsec:CP}

Casimir-Polder and van der Waals forces become measurable for atom-surface distances below about~$\rm10\,\mum$. Pioneering experiments measured losses from a thermal-energy atomic beam of~Na passing through a parallel-plate cavity at distances of~$0.7-7\,\mum$ and were able to quantitatively distinguish between~CP and (non-retarded) van der Waals interactions~\cite{Sukenik1993}. Atomic beams transmitted through~50\,nm-wide slits of a~SiN nano-grating acquire a phase shift due to the surface~$\rm<25\,nm$ away. These phase shifts have been measured interferometrically in experiments sensitive to the non-retarded regime~\cite{Perreault2005,Lepoutre2009}. Experiments with cold atoms, also sensitive to interactions in this range, have been performed using evanescent-wave ``atomic mirrors'' that add a repulsive force to the~CP attraction~\cite{Landragin1996}.

Probing the~CP force by balancing evanescent-wave and surface potentials has been refined considerably by using ultracold atoms~\cite{Bender2010,Bender2014}. In the first of two experiments conducted by the Zimmermann group, the position of the potential barrier is adjusted from~160 to~$\rm230\,nm$ by varying the laser power. Reflection probabilities are then measured for a range of initial velocities, controlled by switching magnetic potentials, that span a range from complete reflection below the classical barrier to complete transmission above it, the latter corresponding to complete atom loss. Quantum reflection (discussed at the end of this sub-section) is negligible for the velocity range chosen. Because the evanescent field can be accurately characterized, these experiments provide direct measurements of the~CP force without requiring assumptions about the potential shape. The results suggest that the best agreement is reached with a full quantum electrodynamics calculation of the force~\cite{Bender2010}. The second of these experiments refined the surface to include an overlayer of parallel stripes of gold, adjacent to which the~CP force is maximal due to the absence of the evanescent wave. The surface therefore acts as a diffraction grating for a~BEC (incident at~$\rm3.4\,cm/s$) with a landscape that can be adjusted by varying the laser power. The data analysis explicitly accounts for the lateral periodicity of the potential, enabling a significant advance in our understanding of how~CP forces are modified by surface structures~\cite{Bender2014}. It should be noted that these experiments have been extended by the same group, again using only optical and CP forces emanating from the surface, to the coupling of ultracold atoms and surface plasmons~\cite{Stehle2011,Stehle2014} (see Sec.~\ref{subsec:photonics}).

Ultracold atoms brought close to the surface of an atom chip by using purely magnetic potentials provide a potentially ideal platform for measuring~CP interactions in the~$\rm0.1-10\,\mum$ range that spans the retarded and temperature-dependent regimes (reviewed in detail in~\cite{Intravaia2011}). The Vuleti{\'c} measurements of atom loss due to Johnson noise (Sec.~\ref{subsec:noise}) near a copper conductor also included measurements near the dielectric substrate of the atom chip~\cite{Lin2004}. This allowed CP-induced surface evaporative losses to be measured exclusively, and clearly showed the effects of the~CP potential (Fig.~\ref{fig:Lin2004fig3and5}). This seminal study concluded that the~CP force limits the atom chip trap depth at distances~$<\rm2\,\mum$ from the dielectric surface. 

\begin{figure}[t!]
   \centering
   \includegraphics[width=0.4\textwidth]{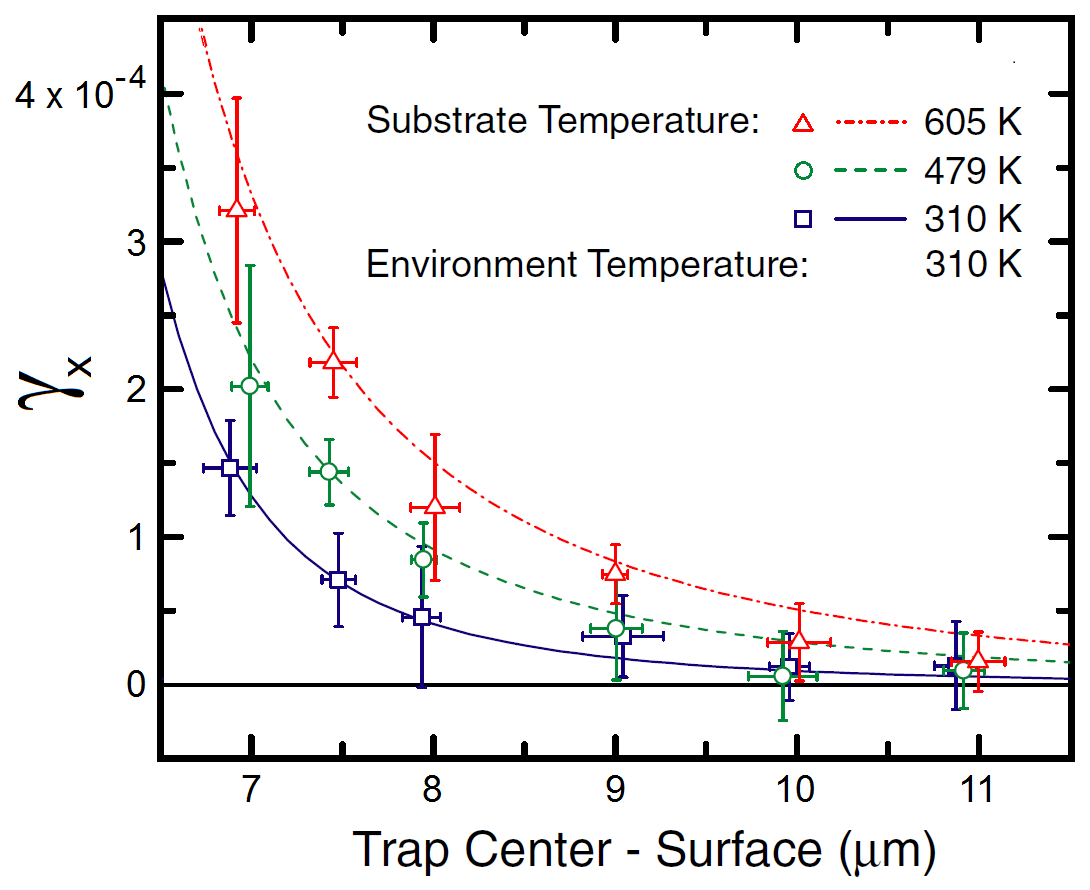}
   \caption{Probing the temperature dependence of the~CP force. The fractional change in the trap frequency due to the~CP force is shown as a function of the distance to the chip. Three sets of data are presented, with accompanying theoretical curves having no adjustable parameters. Error bars represent the total uncertainty (statistical and systematic) of the measurement. Adapted from~\cite{Obrecht2007a}, with permission \textcopyright~2007 by the American Physical Society.} 
   \label{fig:Obrecht2007afig4a}
\end{figure} 

While the~CP potential can destroy the trap at short atom-surface distances, it also affects the trap at larger distances. To quantify this weak effect at large distances, sensitive measurements of the trap frequency can be made as a function of the atom-surface distance~\cite{Harber2005}. With measurement times on the order of~$\rm1\,s$ and an oscillation frequency~$>\rm200\,Hz$, fractional changes~$\lesssim\rm10^{-4}$ can be measured, thereby rendering the measurements sensitive to the~CP force at distances up to~$\approx\rm10\,\mum$. Subsequent experiments were extended to measure temperature-dependent~CP effects characteristic of these distances~\cite{Obrecht2007a} (Fig.~\ref{fig:Obrecht2007afig4a}). Both sets of results agreed very well with theoretical predictions based on the substrate being at the same~\cite{Antezza2004} or elevated~\cite{Antezza2005} temperatures relative to the environment.  The agreement between theory and experiment is particularly impressive as it was obtained with no adjustable parameters. For completeness, we also mention (non-chip) high-temperature experiments conducted at thermal equilibrium that show a~CP force for Cs-sapphire interactions enhanced by about~50\% at~$\rm100\,nm$ and~$\rm1000\,K$~\cite{Laliotis2014}.

\begin{figure}[t!]
   \centering
\includegraphics[width=0.5\textwidth]{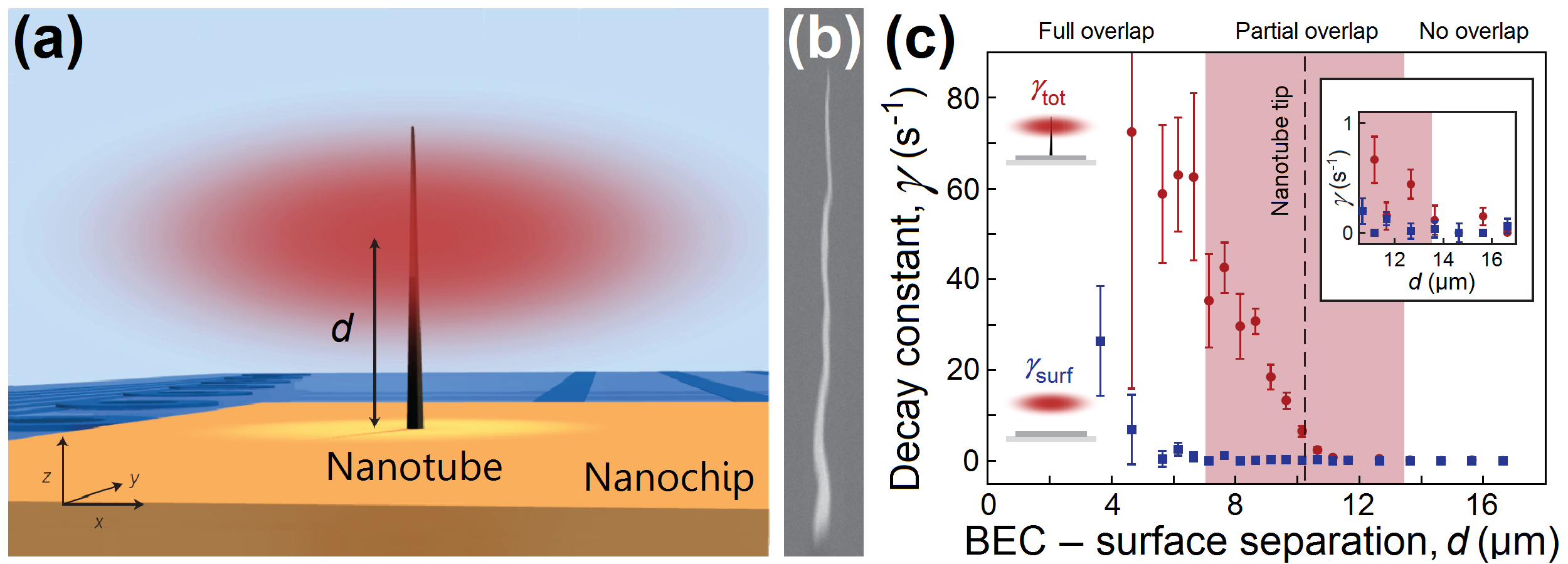}
   \caption{Probing dispersion forces using a chip-mounted carbon nano-tube~(CNT). (a)~A multi-walled~CNT (length,~$\rm10.25\,\mum$), immersed in an ultracold quantum gas, stands on a silicon substrate (nano-chip). (b)~Scanning electron microscope image of the~CNT used for the experiments. (c)~Exponential decay constant of atom number at the~CNT (red points) and at the plain surface (blue points) plotted against distance between the surface and the center of the condensate,~$d$. The vertical dashed line indicates the position of the~CNT tip. The red shaded area denotes the regime where the condensate is in partial overlap with the~CNT. Inset: detailed view of the sharp onset of scattering losses once the~BEC touches the~CNT. Adapted from~\cite{Schneeweiss2012}, with permission by Macmillan Publishers Ltd.}
   \label{fig:Schneeweiss2012fig1and2and4}
\end{figure}

A recent example of measuring dispersion interactions for controlled small distances~$d$ between the cold atoms and a surface, or in this case an object on the surface, is shown in Fig.~\ref{fig:Schneeweiss2012fig1and2and4}. Here a single~CNT, standing vertically on an atom chip surface, is immersed in a~\Rb\ BEC~\cite{Schneeweiss2012}. Atom losses caused by attractive forces between the atoms and the~CNT were measured and used to characterize the atom-CNT interaction. In a related implementation, cold atoms were used in a scanning probe microscopy configuration to measure~CNT surface structures~\cite{Gierling2011}. In addition, because the~BEC and the nano-tube are comparable in size and mass, it may be possible to use the atoms to cool the nano-tube, leading ultimately to its vibrational ground state~\cite{Weiss2013}.

Detailed features of~CP forces continue to be investigated intensively including, for example, the influence of surface geometry~(\eg~\cite{Halif2011}) and in the design of nano-wire or CNT-based atom chips that could enable purely magnetic atom chip traps at sub-$\mum$ distances~\cite{Fermani2007,Petrov2009,Salem2010,Chuang2011a}. Experimental and theoretical work has however, been much more extensive for Casimir forces than for~CP interactions~(for a review, see~\cite{Rodriguez2011}), largely related to searches for repulsive Casimir forces that would allow, for example, the construction of frictionless~MEMS devices. Casimir forces have been demonstrated to be very dependent on surface geometry, and measurements show significant deviations from a pairwise additive formalism~\cite{Chan2008}. Structured materials including metamaterials and layered substrates have been shown theoretically to exhibit tunable Casimir forces, including repulsion~\cite{Henkel2005,Leonhardt2007,Yannopapas2009,Zhao2009}. Repulsive Casimir forces have been measured experimentally, but only with a very specific choice of materials and a liquid medium~\cite{Munday2009}. It remains to be seen what kind of manipulations of the~CP force are possible through geometry and materials on the atom chip~\cite{Scheel2011x,Eberlein2011,Milton2012}.

Intimately related to the~CP attractive potential is the phenomenon of quantum reflection. As an atom approaches a surface, its classical trajectory simply accelerates towards the surface until the atom is either adsorbed or reflected by repulsive forces operating at the atomic-scale distances of surface chemical potentials. Quantum mechanically however, if the atom is moving sufficiently slowly, the~CP potential can instead cause reflection at much larger distances, with a probability related to the abruptness of the attractive potential~\cite{Friedrich2002}. This quantum reflectivity was first observed in experiments using slow~H atoms~\cite{Berkhout1989}, and somewhat later using metastable~Ne*~\cite{Shimizu2001}, followed by observations for~BECs of~\Na~\cite{Pasquini2004}. Grazing-incidence reflectivity was considerably enhanced for~Ne* when the flat surface was replaced with a ridged structure~\cite{Shimizu2002}. It should be noted that reflection typically occurs at sub-$\mum$ distances from the surface. Quantum reflection can be firm but gentle, not even disrupting the extraordinarily fragile bond of~He$_2$~\cite{Zhao2011b}. These reflection characteristics may also be relevant for experiments on antimatter~\cite{Dufour2013} (Sec.~\ref{subsec:antimatter}).

Quantum reflection may be used to study~CP interactions with different surface configurations. A recent suggestion for introducing atom chip technology envisions periodically doping the surface; the electric field generated by the surface dopants provides a force in addition to the~CP interaction that can locally suppress quantum reflection~\cite{Stickler2015}. The surface, though flat, then acts as a diffraction grating for matter waves and may even be extended to realize further atom optics with flat substrates. 

\subsection{Electron transport}
\label{subsec:electron-transport}

\begin{center}
\begin{figure}[t!]
   \begin{minipage}[t!]{0.23\textwidth}	
      \centering
      \includegraphics[width=\textwidth]{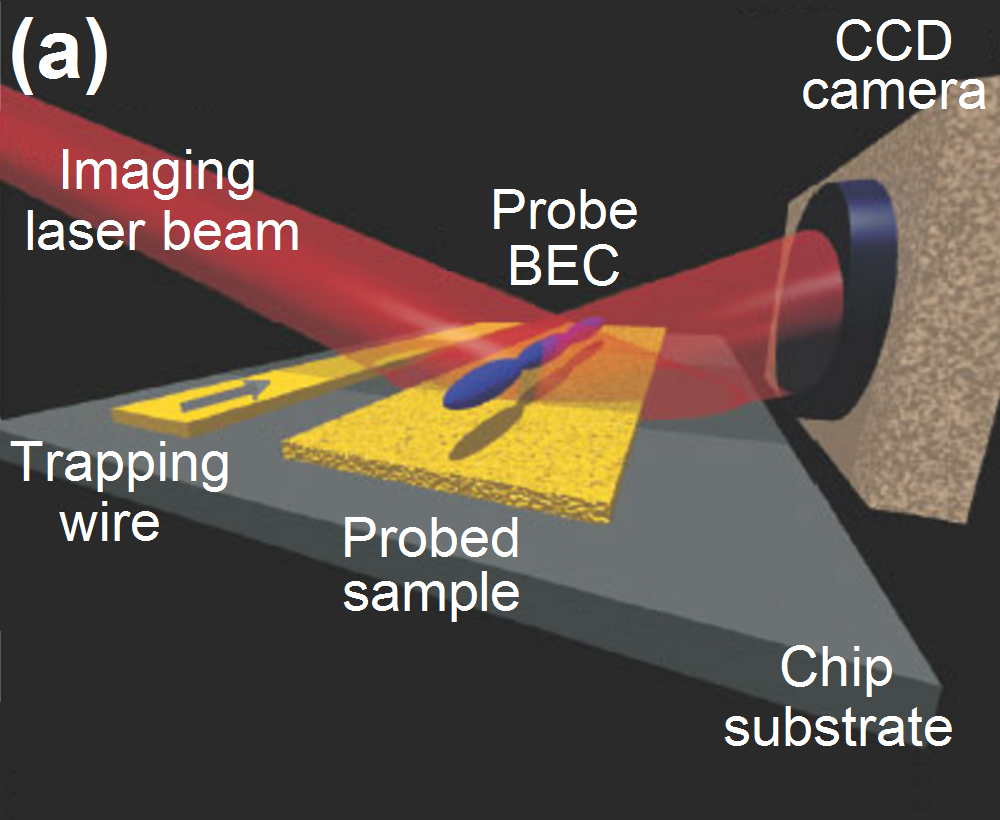}	
   \end{minipage}
   \hfill
   \begin{minipage}[t!]{0.23\textwidth}	
      \centering
      \includegraphics[width=\textwidth]{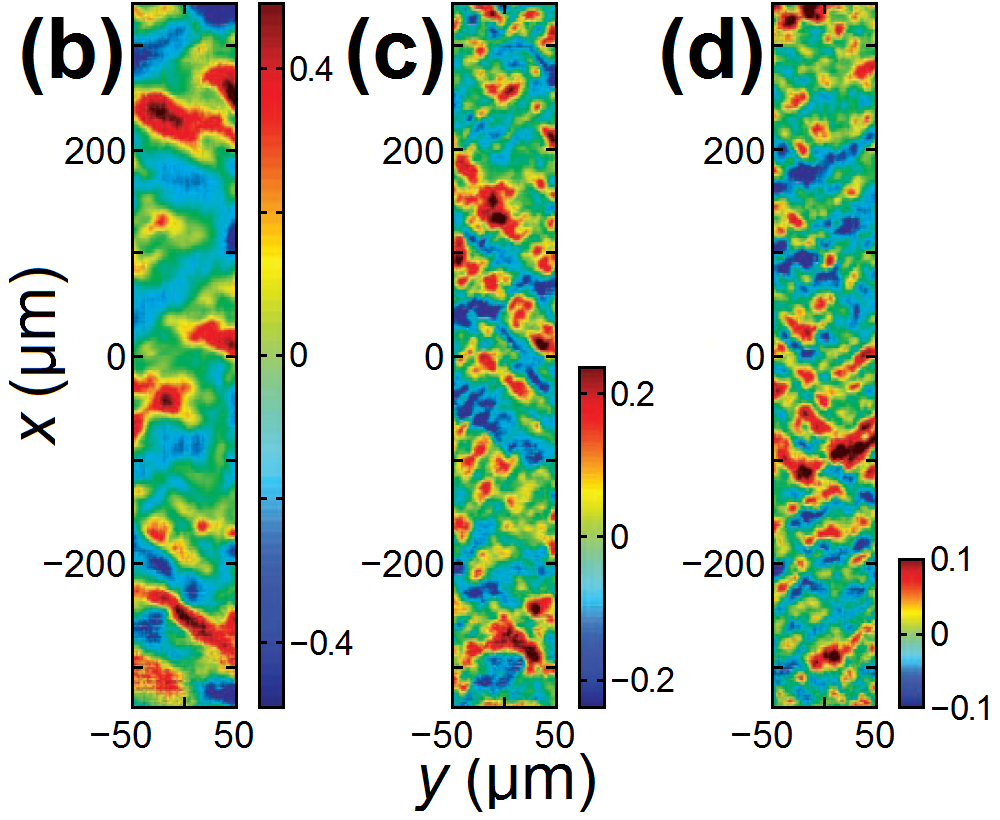}
   \end{minipage}
   \caption{Probing electron transport with cold-atom magnetometry on a chip. (a)~Readout of the information by measuring the atomic density with reflected absorption imaging. (b-d)~Magnetic field angle fluctuations (color scale bars, mrad) above (b)~$\rm2.08\,\mum$-thick and (c-d)~$\rm0.28\,\mum$-thick polycrystalline gold films [grain sizes are (b)~60-$\rm80\,nm$, (c)~30-$\rm50\,nm$, and (d)~150-$\rm170\,nm$]. These fluctuations are due to variations in the direction of the current flow, nominally along the~$x$ axis. The appearance of~$\pm45^\circ$ patterns is clearly observable and reflects a correlated scattering of the electrons. (a)~Adapted from~\cite{Wildermuth2005}, with permission by Macmillan Publishers Ltd.\@ and (b-d)~adapted from~\cite{Aigner2008}, with permission by~AAAS.}
   \label{fig:Aigner2008fig1a}  
\end{figure}
\end{center}
\negskip\negskip

Here we briefly describe electron transport, a fundamental topic in solid-state physics, as a specific example of surface probing accomplished with atom chips. Several additional prospects for interactions with nearby surfaces, including hybrid devices and quantum surfaces, have been discussed in Sec.~\ref{subsec:hybrid}.

Although electron transport has been studied for decades, probing by ultracold atoms has offered new possibilities, including for example, the discovery of previously unknown effects at a microscopic level by the Folman and Schmiedmayer groups~\cite{Aigner2008,Japha2008} (Fig.~\ref{fig:Aigner2008fig1a}). It was found that electrons scatter collectively with a correlation length extending over tens of~$\mum$ in a well-formed evaporated conductor. This exceeded by three orders of magnitude anything expected from the conductor itself, whose structural spatial correlations are no longer than a few tens of~nm (typical of the grain size). It was explained by Fourier transforming the random scattering centers (due to impurities, grain boundaries, and geometrical imperfections), and showing that specific scattering waves, namely those forming a~$\rm45^\circ$-angle with the wire axis, have the strongest scattering probability. This is a good example of how cold atoms in the vicinity of a surface may act as a novel probe. An additional suggestion for cold-atom microscopy of electron transport in topological insulators has been proposed in~\cite{Dellabetta2012}.

Different types of electron transport microscopy probes based on cold atoms could address many interesting questions. For example, can one detect and characterize different current regimes, \eg~ballistic, small mean-free-path, turbulent (chaotic)? Can one detect and characterize non-classical currents, \eg~squeezed currents~\cite{Strekalov2011}? Or relativistic corrections~\cite{Folman2013}? In addition, better understanding of electron transport may enable completely new types of conductors for atom optics on atom chips, as discussed in Sec.~\ref{subsec:materials}. 

\subsection{Searching for non-Newtonian gravity}
\label{subsec:non-Newton}

Ultracold atoms are used to measure gravitational acceleration~$g$ and the universal gravitational constant~$G$, with interferometric measurements yielding the highest accuracy and precision~(see, \eg~\cite{Dickerson2013,Rosi2014}). Although these experiments do not use atom chips, the latter measurements in particular may soon be conducted in micro-gravity environments, for which the compactness afforded by atom chip-based platforms is crucial for experimental implementation~\cite{Altschul2015,JPL2015}. Such micro-gravity experiments are also designed as tests of fundamental physics such as the Weak Equivalence Principle. Here we briefly discuss progress in experiments designed to improve known limits on non-Newtonian gravity by structuring the atom chip as a source of measurable gravitational fields in the $\rm100\,nm-100\,\mum$ range~\cite{Dimopoulos2003,Wolf2007}.

Most experimental work to date has been based on the use of a vertical optical lattice, typically created by retro-reflecting an off-resonant laser from the atom chip surface~\cite{Dimopoulos2003,PereiraDosSantos2009}. The lattice site closest to the surface would then be only~$\lambda/2$ away, well into the sub-$\mum$ range that is difficult to access using purely magnetic potentials. A variety of methods, using optical tweezers for example~\cite{Thompson2013b}, have been proposed to place the atoms into just one particular lattice site, with a known location.  

\begin{figure}[t!] 
   \centering
   \includegraphics[width=0.5\textwidth]{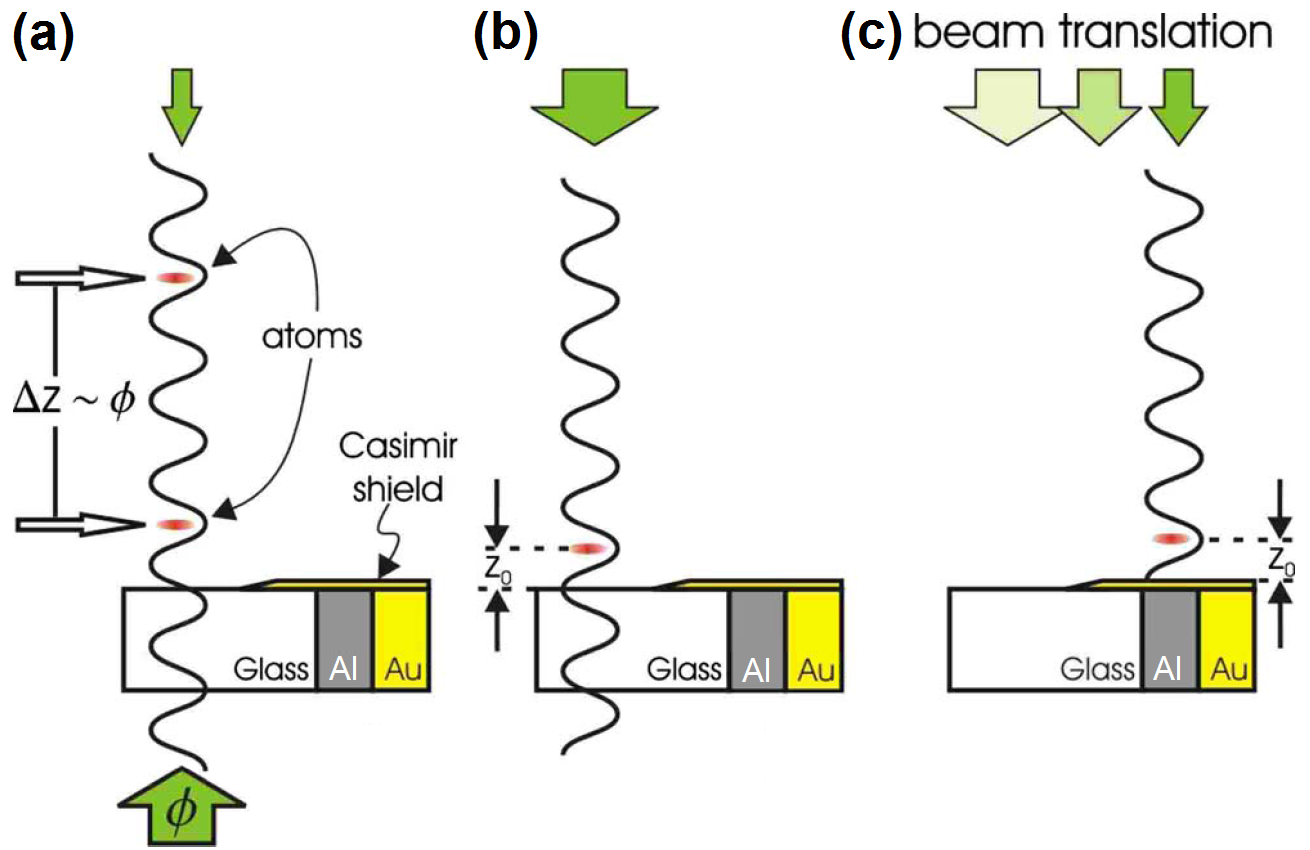}
   \caption{Probing the surface for non-Newtonian gravitational effects. An optical lattice is used to position cold~\Sr\ atoms near two adjacent test masses. (a)~The atoms are first placed close to the transparent part of the test surface; a vertical translation of~$\Delta z=(\lambda/4\pi)\phi$ is accomplished by adjusting the relative optical phase~$\phi$ accumulated between the two laser beams (green arrows). (b)~The counter-propagating (lower) lattice beam is then switched off adiabatically, and the atoms remain trapped in the standing wave made of the co-propagating beam and the weak reflected beam. (c)~The lattice beam is translated laterally, placing the atoms just above the Casimir shield and close to the~Al and~Au masses. The width of the arrows represents the relative intensity of laser beams. Adapted from~\cite{Sorrentino2009}, with permission \textcopyright~2009 by the American Physical Society.}
   \label{fig:Sorrentino2009fig1} 
\end{figure}

An ``optical elevator'' has been developed, taking advantage of transparency built into an area of the atom chip surface. A counter-propagating (rather than retro-reflected) laser is frequency-chirped to generate a traveling wave~\cite{Sorrentino2009} that can bring the atoms to the desired distance from the surface, as illustrated in Fig.~\ref{fig:Sorrentino2009fig1}. The laser is then displaced to move the trapped cloud laterally over different regions of the atom chip, which is structured as a layered sandwich of metals with different densities. The gravitational field strength emanating from the surface is then a periodic function of the lateral position of the atoms. Since differences in the gravitational attraction can easily be masked by the much stronger~CP interaction, a ``Casimir shield'' is interposed directly above the surface~\cite{Dimopoulos2003,Chiaverini2003,Ribeiro2013}. Useful strategies for minimizing limitations imposed by surface patch potentials have also been discussed~\cite{Behunin2014}.

Alternatively, magnetic potentials from the atom chip may be used to prepare and transport atoms and, in particular, to bring them close to the surface for initial loading into an optical lattice integrated into the atom chip (Sec.~\ref{subsec:clocks}). Although these proof-of-principle experiments~\cite{Gallego2009,Straatsma2015} do not state the atom-surface distances achieved, they may in the near future enable precision measurements of atom-surface interactions and other short-range forces. 

As discussed above, a multi-layer hybrid atom chip was used to magnetically transport cold atoms to the vicinity of a micro-cantilever~\cite{Montoya2015} (Sec.~\ref{subsec:hybrid}). In addition to atom loss measurements for atom-cantilever separations of about~$\rm100\,\mum$, the system was analyzed as a detector for the force exerted on the cantilevers by the atoms at a range of a several~$\rm\mum$. It is interesting to note that several micro-cantilever experiments (not using cold atoms) have provided some of the best limits on Yukawa-type deviations in the~$\rm3-15\,\mum$ range~\cite{Geraci2008}.

Pure magnetic traps may also be used for the complete experiment. As noted, the state-of-the-art for such traps stands at~$\rm500\,nm$~\cite{Lin2004}, and analyses of nano-wire and~CNT traps~\cite{Fermani2007,Petrov2009,Salem2010,Chuang2011a} suggest that even smaller distances may be achieved.

An alternative orientation of an atom chip has been proposed for measuring short-range interactions with a nano-sphere~\cite{Geraci2015}. Orienting the atom chip vertically, the optical lattice may be created by a retro-reflecting laser aimed horizontally. The test mass consists of alternating vertical stripes behind the vertical wall of the atom chip; the ballistic trajectory of the falling nano-spheres then depends on the density of the stripe closest to its initial position. An interferometric scheme is expected to improve the sensitivity of the ballistic experiment. This scheme may also be applicable for measuring falling atomic trajectories adjacent to alternating gravitational fields. 

For completeness, we note here that a variety of experiments (reviewed, \eg\ in~\cite{Tino2014}) have been conducted at considerably greater distances from the retro-reflecting surface used to create the vertical optical lattice. Combined with the gravitational gradient, the optical lattice produces a Wannier-Stark potential manifold, allowing very long-lived Bloch oscillations to be observed. This has improved the precision for corresponding measurements of~$g$~\cite{Poli2011,Beaufils2011,Zhou2013}, though not to the level achieved in the interferometric measurements referred to above.

It remains to be seen if the atom chip can also contribute to improved limits for a variety of related searches, \eg\ the so-called chameleon field and dark energy~\cite{Burrage2015}. Accurate control of small distances and wavepacket size, as well as exotic surfaces (\eg\ very rough or even porous surfaces to increase surface area) may enable establishing better limits. Recent realizations of atom interferometers close to the surface~\cite{Machluf2013a} suggest the possibility of sensitive probing with ``matter-wave homodyning'' (Sec.~\ref{subsec:fields}). In addition, measuring correlation lengths of different effects may be possible using wavepackets separated parallel to the surface~\cite{Zhou2015x}. 

\addtocontents{toc}{\protect\vspace{18pt}}

\section{Interferometry on an Atom Chip} 
\label{sec:interferometry}

Atom interferometry on a chip is much younger and less mature than conventional free-space interferometry using atomic beams, fountains, and more recently, BECs. Nevertheless, its development offers many potential advantages for the future. Beyond the fact that atom chips provide a compact miniature platform for precise manipulation of atoms, they also provide a unique environment for new kinds of interferometric schemes, \eg\ high-finesse configurations and making use of tunneling barriers. In addition to traditional applications such as measuring acceleration and gravitation, inerferometry with ultracold atoms near a chip surface allows precise measurements of atom-surface interactions and thorough investigations of the fundamental physics of a Bose gas as a many-particle system, \eg\ using a single sample of a~1D gas. 

In this section we review spatial interferometry, in which each atom is split into a superposition of two locations or two momenta. At the end of this section we will also briefly mention some interesting recent schemes of interferometry with superpositions of internal or discrete motional levels that are not separated in space. 

Specific aspects of interferometry on atom chips have been discussed in previous reviews of atom interferometry~\cite{Cronin2009,Schaff2014}. Here we attempt to encompass most of the important aspects of this interferometry, with emphasis on new developments during the last few years~--~and focusing on achievements and proposals for future advancement. 

\subsection{Bragg splitting} 
\label{subsec:Bragg}

The first demonstration of spatial interferometry on an atom chip was the Michelson interferometer realized jointly by the groups of Cornell, Anderson, and Prentiss~\cite{Wang2005}. A~BEC was released at the center of a magnetic waveguide~$\rm250\,\mum$ from the chip surface and split by a laser pulse into a superposition of two wavepackets counter-propagating along the waveguide with momenta~$\pm2\hbar k$, where~$k$ is the wave vector of the standing-wave light. The latter was formed by a tightly focused laser beam reflected by a pair of mirrors on the chip. The splitting laser pulse consisted of a pair of sub-pulses adjusted to ensure a splitting efficiency near~100\%. The wavepackets propagated in the magnetic waveguide for up to about~$\rm10\,ms$, whereupon a Bragg pulse reversed their direction. They were recombined upon arrival back at the center of the waveguide using a second pair of laser sub-pulses. The wavepackets reached a maximum separation of about~$\rm120\,\mum$, so far unsurpassed in a coherent interferometer of trapped or guided atoms on a chip. The relative number of atoms in the output momentum components showed clear interference fringes as a function of the relative phase between the wavepackets, which was engineered by imparting an initial velocity to the atom cloud before the splitting.

\begin{figure}
   \centering
   \includegraphics[width=0.40\textwidth]{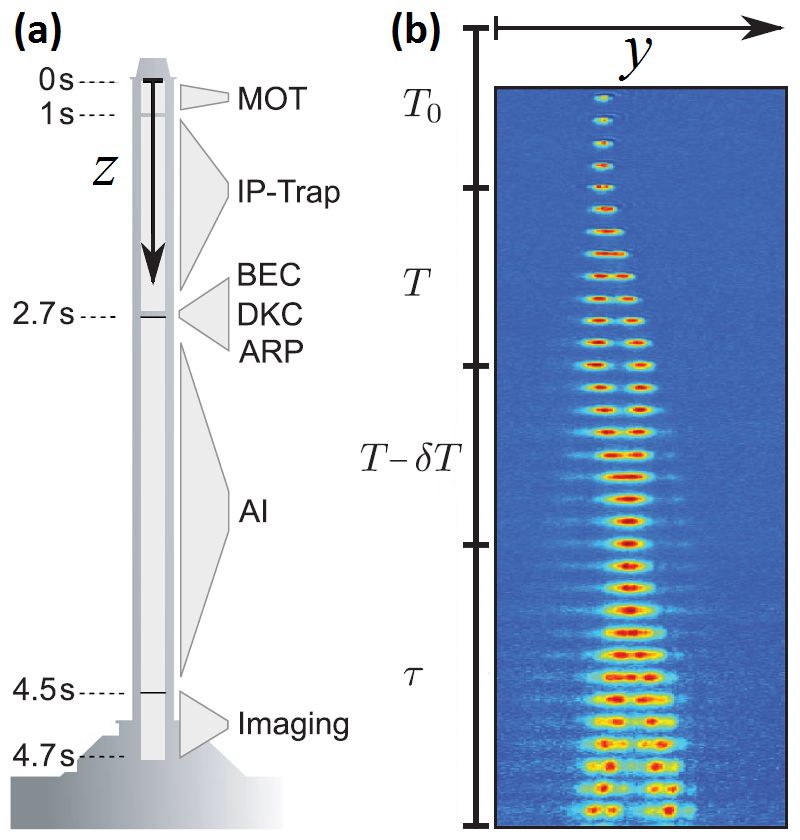}
   \caption{Bragg interferometery in micro-gravity on an atom chip at the Bremen drop tower. (a)~The experimental sequence includes capturing cold atoms in a~MOT, loading a Ioffe-Pritchard trap, creating a~BEC, and applying delta-kick cooling~(DKC) followed by adiabatic rapid passage~(ARP). The remaining time before the capture of the capsule at the bottom of the tower is used for atom interferometry~(AI) and imaging of the atoms. (b)~The evolution of the~BEC and the asymmetric Mach-Zehnder interferometer is visualized by a series of absorption images of the atomic densities separated by~$\rm1\,ms$. A~$\pi/2$ pulse from two counter-propagating light beams of frequency~$\omega$ and~$\omega+\delta$ creates (time~$T_0$) a superposition of two wave packets that drift apart with a two-photon recoil velocity~$v_{\rm rec}=\rm11.8\,mm/s$. After~$T$ they are redirected by a~$\pi$ pulse and partially recombined after~$T-\delta T$ by a second~$\pi/2$ pulse. A nonzero value of~$\delta T$ leads to a spatial interference pattern after~$\tau=\rm53\,ms$ in free-fall. The fringe spacing is inversely proportional to the separation~$d=v_{\rm rec}\ \delta T$ of the wave packets. Reprinted from~\cite{Muntinga2013}, with permission \textcopyright~2013 by the American Physical Society.}
   \label{fig:Muntinga2013fig1and2}
\end{figure}

The magnetic waveguide in this experiment is important for preventing the expansion of the~BEC, which would occur if it was released into free space. An atom chip as a platform for~BEC inteferometry has a few additional advantages even if the interferometric sequence itself is performed with a freely propagating~BEC, including simple and fast~BEC formation. The atom chip also enables better control of the~BEC after it is formed in order to optimize its shape and expansion rate. These advantages were exploited by the Rasel group in a device designed to perform interferometry in micro-gravity~\cite{Muntinga2013}. The interferometry sequence of three Bragg pulses was applied after releasing the~BEC from a magnetic trap on an atom chip during free-fall in a drop tower (Fig.~\ref{fig:Muntinga2013fig1and2}). This group is now preparing a dual-species~$\rm^{85}Rb$/\Rb\ atom chip interferometer for space flight~\cite{Schuldt2015}.

\subsection{Double-well interferometry}
\label{subsec:doublewell}

\subsubsection{Adiabatic dressed-state potentials}

A second kind of interferometry, which is unique to atoms and has no direct analog in light waves, is based on splitting a~BEC in a double-well potential. Coherent splitting of a~BEC in a double-well potential was first demonstrated using an optical potential~\cite{Shin2004}. The first demonstration of coherent double-well interferometry on an atom chip was achieved by the Schmiedmayer group by a splitting scheme based on an adiabatic dressed potential of~RF and static magnetic fields~\cite{Schumm2005a}. In the presence of a static field~${\bf B}_{\rm st}({\bf r})$ and an oscillating magnetic field~${\bf B}_{\rm ac}({\bf r})\cos(\omega t+\phi)$, the adiabatic eigenvalues of the magnetic energy in the rotating-wave approximation are given by 
\begin{equation} V_{\rm dressed}({\bf r})=\mu_F\sqrt{[|{\bf B}_{\rm st}({\bf r})|-\hbar\omega/\mu_F]^2+|{\bf B}_{\rm ac,\perp}({\bf r})|^2}, 
\end{equation}
where~$\mu_F=m_Fg_F\mu_B$ is the magnetic moment of the specific Zeeman sub-levels in the hyperfine state~$F$, and~${\bf B}_{\rm ac,\perp}$ is the component of the~RF field that is perpendicular to the direction of the static field. When the frequency~$\omega$ of the~RF field is ramped up above the Larmor frequency~$\omega_L=|\mu_F{\bf B}_{\rm st}|/\hbar$ of the atom at the trap minimum of the static field, new trap minima appear at the points where~$\hbar\omega=\mu_F|{\bf B}_{\rm st}|$ so that the minimum of the static magnetic field may become a potential barrier. The details of the potential depend on the specific form of the vectorial static and~RF fields, which are determined by the current configuration on the chip. 

In the original experiment~\cite{Schumm2005a}, the double-well potential split the~BEC into two parallel elongated clouds separated by~$\rm3-80\,\mum$. After trap release and a short time-of-flight, interference fringes were measured for small cloud separations (they could not be optically resolved for cloud separations~$\gtrsim\rm5.5\,\mum$). The fringe patterns were repeatable (deterministic), as shown by a very narrow distribution of phases over many experimental realizations (\ie\ high contrast of the averaged fringes), thereby proving that the splitting process is phase-preserving and that the beam splitter is coherent. The relative phase between the two condensates was locked at zero whenever the chemical potential of the~BEC was larger than the barrier height. However, once tunneling was fully inhibited, for a cloud separation~$>\rm3.4\,\mum$, the phase difference began to evolve deterministically.

Non-random phase differences were maintained for up to~$\rm2\,ms$, during which time the phase spread broadened, as measured by decreased fringe contrast. It was suggested that this rapid loss of coherence was due to the finite coherence length of the quasi-1D~BEC along the axial direction, and this was later investigated in a study of many-body effects in a~1D Bose gas~\cite{Hofferberth2007}. Using their double-well interferometer, the Schmiedmayer group conducted additional studies of fundamental effects, such as relaxation and pre-thermalization in isolated quantum systems~\cite{Gring2012}, as well as measurements of a variety of non-equilibrium effects in a~1D Bose gas~\cite{Langen2015,Steffens2015,Rauer2016}. 

An interferometric measurement of a non-magnetic (gravitational) potential difference between the locations of the two wells was performed by the Hinds group~\cite{Baumgartner2010}. The~BEC was split into a double-well trap~$130\,\mum$ from the chip by using two~RF currents whose relative magnitude determines the rotation of the splitting axis relative to the vertical axis. The phase of the observed interference fringes allowed the determination of the energy difference between the chemical potentials of the two condensates. 
The precision of this measurement was limited by the chemical potential uncertainty, which determines the coherence time due to phase diffusion (about~$\rm10\,ms$ in this experiment). 

\begin{figure}
   \centering
   \includegraphics[width=0.45\textwidth]{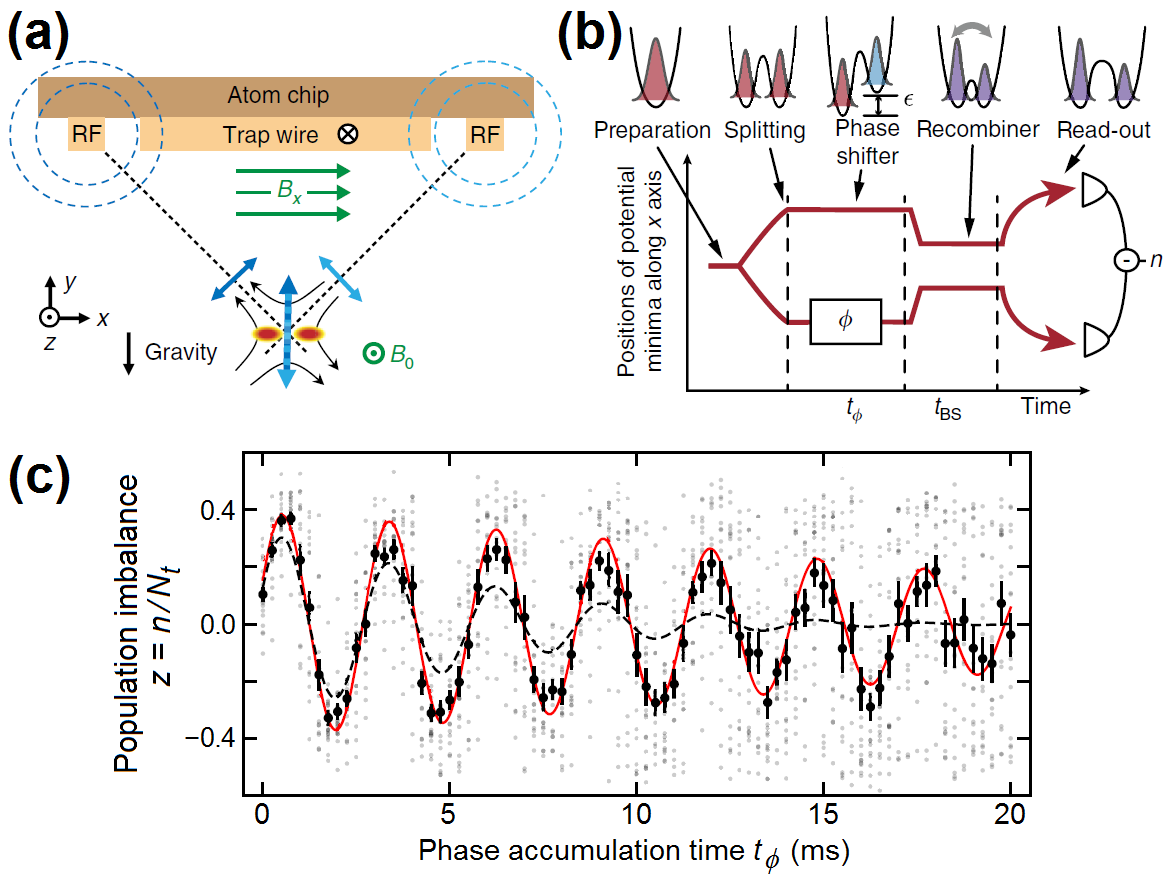}
   \caption{An~RF potential Mach-Zehnder interferometer on an atom chip. (a)~Schematic of the atom chip: DC~currents in the trap wire and in two perpendicular wires (not shown), together with a uniform external field, create an elongated Ioffe-Pritchard trap~$\rm60\,\mum$ below the chip. RF~currents with a relative phase of~$\pi$ are applied on the dressing wires to perform the splitting along~$x$. (b)~A relative phase between the two arms is imprinted by tilting the double well during a time~$t_{\phi}$; the spacing between the two wells is then abruptly reduced and the potential barrier acts as a beam splitter for both wave packets, transforming the relative phase into a population imbalance read out after the two clouds are separated again. (c)~The normalized population difference~$z=n/N_{\rm t}$ as a function of $t_{\phi}$ exhibits interference fringes damping due to phase diffusion. Grey dots: imbalance of individual experimental realizations; black dots: ensemble average~$\langle z\rangle$; red curve: theoretical prediction taking into account phase diffusion; dashed black line: expected signal for a classical coherent state. Adapted from~\cite{Berrada2013}, with permission by Macmillan Publishers Ltd.}
   \label{fig:Berrada2013fig1}
\end{figure}

A more recent experiment~\cite{Berrada2013} (Fig.~\ref{fig:Berrada2013fig1}) reported a coherence time of more than~$\rm20\,ms$, a factor of~3 longer than the expected phase diffusion time for a coherent state. This was shown to be directly related to number squeezing, as also found in~\cite{Jo2007}. The experiment demonstrated a novel beam recombiner that does not require a long time-of-flight to read out the relative phase; the barrier is reduced non-adiabatically and the two condensates are allowed to overlap for a given time and then they are separated again. The resulting population imbalance between the two wells is then proportional to the sine of the relative interferometric phase before the recombination. 

The double-well method has so far been used mainly to explore many-body effects, but the fact that phase coherence can be maintained has yet to be exploited in practical interferometry. Atom-atom interactions play a crucial role in interferometry with a~BEC in a trapping potential, as discussed extensively in the literature. For example, in addition to the work discussed above, superfluidity and hydrodynamics in a Josephson junction of a~BEC were studied experimentally in an~RF dressed double-well potential on a chip~\cite{LeBlanc2011}. Theoretical studies of~BEC coherence in a double-well potential typical of atom chips were performed using various approaches, including stochastic methods~\cite{Scott2009,Judd2008}. A theoretical study of a full interferometric process in a Mach-Zehnder configuration with interacting atoms~\cite{Grond2011} examined the effects of number-squeezing on the phase stability, showing that the standard quantum limit~(SQL) for phase sensitivity can be overcome and the Heisenberg limit reached. 

\subsubsection{Static potentials}

Tunneling barriers for splitting atoms on a chip by static magnetic fields created by current-carrying wires may have the advantage of better integrability with other atom chip circuits to be used as platforms for more complex interferometric schemes. However, such potentials, based on magnetic fields obeying the stationary Maxwell's equations in the form of the Biot-Savart law, require the atoms to be very close (a few~$\rm\mum$) to the chip surface in order to allow fine control of the potential shape over a range comparable to the barrier penetration length (typically $\approx\rm1\,\mum$ for a chemical potential a few~nK below the barrier). This may make the potential sensitive to imperfections in the structure of the wire, Johnson noise, time-dependent fluctuations of the currents, and environmental magnetic noise that is transmitted by the metallic wires on the surface, as discussed in Sec.~\ref{subsec:noise}. 

\begin{figure}
   \centering
   \includegraphics[width=0.45\textwidth]{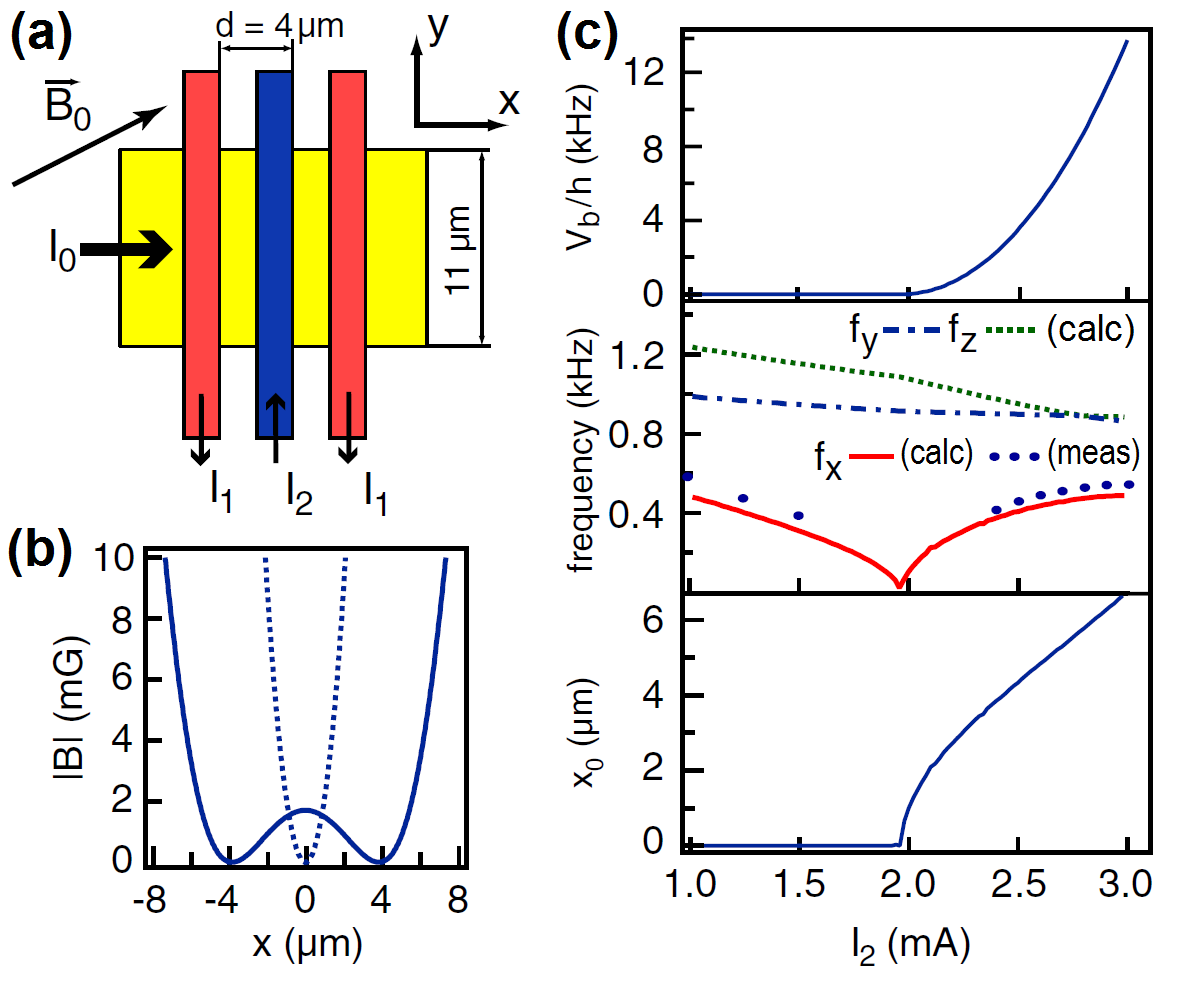}
   \caption{A static field double-well potential on an atom chip. (a)~The magnetic field is generated by~DC currents superposed with a static homogeneous field~$B_0$. The minimum is located at a distance~$z_0=\mu_0 I_0/2\pi B_y$ from the~$I_0$ wire and the field at the minimum is~$B_x$. This minimum is modified by~$I_1=\rm2.0\,mA$ and~$I_2$, whose fields are predominantly along~$x$ on the trap axis. For~$z_0\geq d$, the two~$I_1$ wires together provide harmonic confinement along~$x$. $I_2$,~of opposite polarity, creates the barrier of adjustable height and also determines the spacing between the two resulting wells. (b)~Profile of the trapping potential along the splitting axis~$x$, for~$I_2=\rm0\,mA$ (cooling trap, dashed line) and~$I_2=\rm2.4\,mA$ (solid line). (c)~Barrier height~$V_b$~(top), trap frequencies~$(f_x,f_y,f_z)$~(center), and position~$x_0$ of the right minimum~(bottom) as functions of the current~$I_2$. Adapted from~\cite{Maussang2010}, with permission \textcopyright~2010 by the American Physical Society.}
   \label{fig:Maussang2010fig1}
\end{figure}

Splitting a~BEC using the double-well potential of a static magnetic field on an atom chip was realized in~\cite{Shin2005}. Interference fringes were observed, but the phase of these fringes was random. A double-well potential closer to the chip surface was later realized (Fig.~\ref{fig:Maussang2010fig1}) and coherent properties such as enhanced or reduced number fluctuations were observed~\cite{Maussang2010}.

\begin{figure}[t!]
   \centering
   \includegraphics[width=0.45\textwidth]{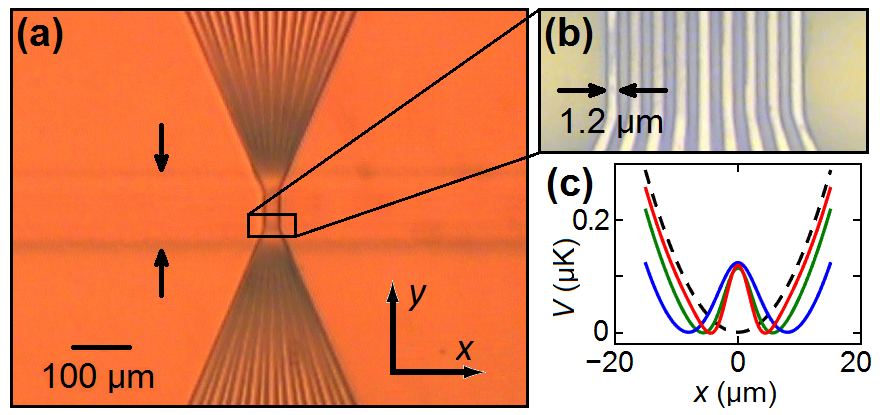}
   \caption{Atom chip for a tight double-well potential. (a-b)~Central region of the multi-layer chip; wire layers are separated by a~$\rm15\,\mum$-thick layer of polyimide insulation, and the entire chip is covered by an additional gold mirror layer. (a)~Arrows show the faint outline of the trapping structure in the lower layer. (b)~Up to~9 central wires ($\rm1\,\mum$ wide, $\rm1\,\mum$ gaps) in the upper layer can be used for barrier control. (c)~Magnetic potentials produced~$\rm5.5\,\mum$ from the chip surface, shown along the longitudinal axis~($\hat{x}$). The harmonic potential~(dashed) is modified by using (blue)~a single central wire and (green, red)~adding opposing currents in successive pairs of adjacent wires. The barrier becomes progressively narrower as wire pairs are added while the barrier height is maintained at~$\rm100\,nK$ and the transverse frequency (along~$\hat{y}$) is maintained at~$\rm2.4\,kHz$. Potential calculations and design courtesy of Shimon Machluf (Be'er Sheva)~\cite{Machluf2013b}.} 
   \label{fig:IFMchip}
\end{figure}

An example of how the atom chip allows the sculpturing of a variety of potential shapes is given in Fig.~\ref{fig:IFMchip}. Here~9 parallel $\mum$-scale wires can be used to fine-tune the shape of a tunneling barrier. This offers a method for studying topological effects in tunneling. Another example of engineering $\rm\mum$-scale wires specifically for tunneling studies is given in~\cite{Chuang2011a}. While the tunneling experiments envisioned for these chips have not yet been tested with atoms, preliminary proof of spatial coherence from atoms trapped in a modulated potential just a few~$\mum$ from the chip has been demonstrated by the Folman group~\cite{Zhou2015x} (Sec.~\ref{subsec:coherence}). 

\subsection{Splitting with state-selective potentials}
\label{subsec:state-selective}

A state-selective dressed potential is formed by applying an inhomogeneous (on-chip) microwave field somewhat detuned from a specific transition between two Zeeman sub-levels in two hyperfine levels~\cite{Bohi2009}. Interaction with the microwave field shifts the energy of the two coupled levels by an amount which is proportional to the intensity of the field (AC~Stark shift). This creates an effective potential for the two levels, whose shape is determined by the intensity and may be either attractive or repulsive.

The two states that are usually used for state-selective potential splitting are the two ``trapped clock states''; for~\Rb\ these are~$|F,m_F\rangle=|1,-1\rangle$ and~$|2,1\rangle$. These two states have the same magnetic susceptibility at the ``magic field'' of~$\rm3.23\,G$ and two photons~(microwave+RF) are required for transitions between them. One of the two states can be shifted by a dressed microwave potential or they may both be shifted by a different potential. Splitting is done by applying a microwave~$\pi/2$ pulse to create an equal superposition of the two states and then applying the state-dependent potential to make them move in different directions.

\begin{figure}
   \centering
   \includegraphics[width=0.5\textwidth]{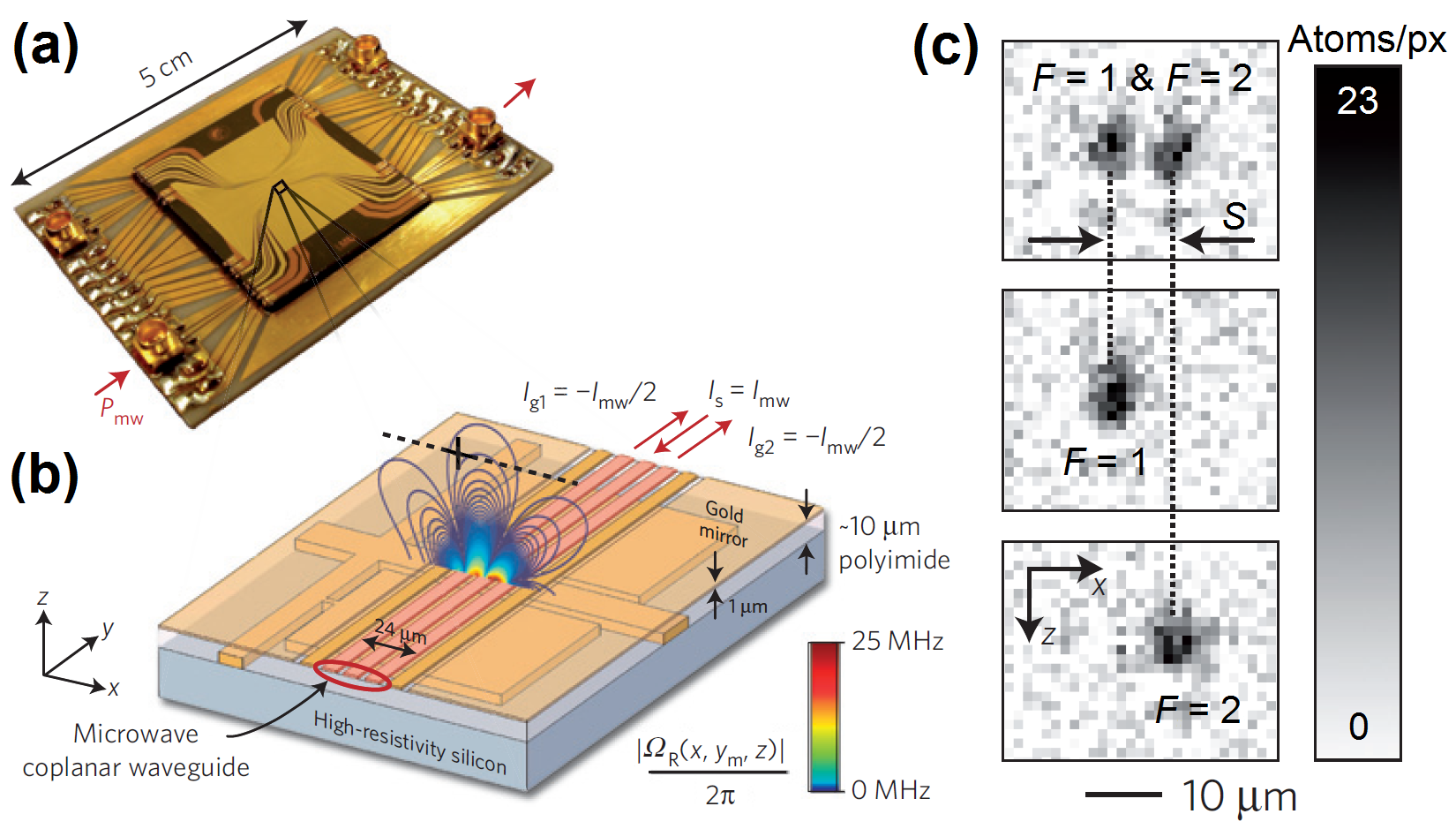}
\caption{Double well on an atom chip using potentials created by microwave co-planar waveguides~(CPWs). (a)~Photograph of the chip assembly. The~Si experimental chip has two layers of gold wires separated by a thin polyimide layer, with~CPWs on the upper layer. It is glued and wire-bonded to an~AlN carrier chip with a single gold layer. (b)~Schematic close-up of the experimental region. The three central wires~(red) form a~CPW. All wires (including the~CPW) can carry stationary currents for the generation of static magnetic traps. The position of the minimum of the static trap is indicated by the black cross. (c)~State-selective splitting of the~BEC. Absorption images of the adiabatically split~BEC. By imaging both hyperfine states simultaneously~(top), only~$F=1$~(middle) or only~$F=2$~(bottom), the state selectivity of the splitting is established. Adapted from~\cite{Bohi2009}, with permission by Macmillan Publishers Ltd.}
   \label{fig:Bohi2009fig2and3}
\end{figure}

A~BEC on a chip was first split with state-selective potentials by the Treutlein group in order to create an entangled two-component~BEC, allowing the~SQL of interferometry to be overcome~\cite{Bohi2009,Riedel2010}. The microwave near-field was created by a~CPW on the chip surface, as shown in Fig.~\ref{fig:Bohi2009fig2and3}. 

Next, we discuss some recent proposals for spatial interferometry based on splitting by state-selective potentials.

A scheme for a double-well beam splitter was proposed by the Reichel group and some of its components were demonstrated experimentally~\cite{Guarrera2015}. The idea is to use the pair of states~$|F,m_F\rangle=|1,-1\rangle$ and~$|2,1\rangle$ in a state-dependent microwave-dressed potential which creates a single harmonic well for one of the states and a double-well potential for the other. A two-photon Rabi pulse can be tuned to transfer the ground state of the single well to the ground state of the double-well potential with minimal excitation of other trap modes in the two wells. A numerical simulation based on mean-field calculations showed a high splitting efficiency. It would be interesting to investigate the evolution of the many-body state during this splitting and compare it to the experimentally studied double-well systems discussed in~Sec.~\ref{subsec:doublewell}.  

A scheme for a Sagnac interferometer for inertial sensing based on state-selective RF-dressed potentials for the two trapped clock states has been proposed~\cite{Stevenson2015}. It is based on quasi-adiabatic transport of the two states in two corresponding~3D harmonic traps, moving in opposite directions along a ring-shaped trajectory. The magnetic configuration for these potentials had been proposed earlier~\cite{Fernholz2007}, but the new proposal also investigates the dependence of interference visibility on the adiabaticity of the transport dynamics for a~BEC as well as thermal atoms. 

Another proposal for an interferometer based on state-selective microwave potentials was analyzed in~\cite{Ammar2015}. Symmetric splitting would be accomplished by two~CPWs producing two different microwave near-field frequencies. This allows splitting a thermal cloud of atoms into two clouds at positions shifted symmetrically from the original position and then reversing the splitting process to get an interferometric signal. A thorough analysis of this scheme shows that symmetry allows a high-contrast signal even for thermal atoms and that it is robust against magnetic field fluctuations. Robustness due to time-reversal symmetry was previously found in relation with guided atom interferometers~\cite{Japha2007} and will be further discussed in Sec.~\ref{subsec:guided}.

\subsection{Stern-Gerlach interferometry}
\label{subsec:FGBS}

Stern-Gerlach interferometry, in which a magnetic field gradient coherently splits a beam of spin-$\frac{1}{2}$ particles into two spatially separated beams of different spin projections, was one of the first proposals for interferometry of massive particles and was made shortly after the discovery of the Stern-Gerlach effect~\cite{Gerlach1922a,Gerlach1922b}. However, it was predicted that coherent operation of such an interferometer would require extremely precise magnetic field gradients that were beyond reach~(\cite{Scully1989} and references therein). Nevertheless, it was recently demonstrated by the Folman group that coherent Stern-Gerlach interferometry is possible for a~BEC on an atom chip~\cite{Machluf2013a}. To the best of our knowledge, this is the first realization of a coherent spatial Stern-Gerlach beam splitter.

This realization is based on the general idea of a field-gradient beam splitter~(FGBS): first, a~$\pi/2$ Rabi pulse creates an equal superposition of two atomic internal states~$(|1\rangle+|2\rangle)/\sqrt{2}$. Then a field gradient pushes the two components~$|1\rangle$ and~$|2\rangle$ with a different force and they acquire a state-dependent momentum~$p_1$ and~$p_2$, respectively. A second~$\pi/2$ pulse creates the atomic state 
\begin{equation}
{\textstyle\frac{1}{2}}\left[(|1,p_1\rangle-|1,p_2\rangle)+(|2,p_1\rangle+|2,p_2\rangle)\right], 
\end{equation}
such that each of the internal states is in a superposition of the two momentum states and may be used as a spatial interferometer.  An interference signal can be obtained either by bringing the two wavepackets back to the same point and recombining them in a similar beam splitter, or by stopping their relative motion and observing interference fringes once they expand and overlap, as was done in the actual experiment. 

A proof-of-principle experiment for interferometry based on an~FGBS was demonstrated using~\Rb\ atoms in the states~$|1\rangle\equiv|F=2,m_F=1\rangle$ and~$|2\rangle\equiv|F=2,m_F=2\rangle$. A two-state system was engineered by keeping a relatively strong homogeneous magnetic field that induced nonlinear Zeeman splitting and pushed the transitions to other Zeeman sub-levels away from resonance with the~RF field used for~$\pi/2$ pulses. A magnetic field gradient generated by a current pulse in a chip wire was used to create the differential potential gradient. It was shown that differential momentum kicks of~$100\,\hbar k$ (in units of a~$\rm1\,\mum$-wavelength photon recoil) are possible in the short splitting time of a few~$\mus$. However, the recombination and imaging method allowed proof of the coherence only for a wavepacket separation of a few~$\mum$. New schemes for recombination of the split wavepackets, such as using a second beam splitter based on the same principle, would allow future experiments to benefit from the high momentum splitting achieved. 

Compared to standard Stern-Gerlach interferometery, an important advantage of the~FGBS for interferometry is that its output contains a superposition of pairs of wavepackets with an indistinguishable state, such that long-wavelength magnetic fluctuations in space do not affect the relative phase between the two paths. This makes interferometry with atoms split with the~FGBS insensitive to stray magnetic fields during propagation through the interferometer arms, since the atoms are in a superposition of different spin states only during the very short time of the splitting. On the other hand, momentum transfer by magnetic gradients cannot achieve the precision offered by splitting with optical beams and would therefore be more suitable for applications using an atom guide, where momentum precision may not be as crucial (Sec.~\ref{subsec:guided}). It should be noted that the~FGBS may be realized with various kinds of state-selective potentials: magnetic, electric, or optical.   

\begin{figure}[t!]
   \centering
   \includegraphics[width=0.5\textwidth]{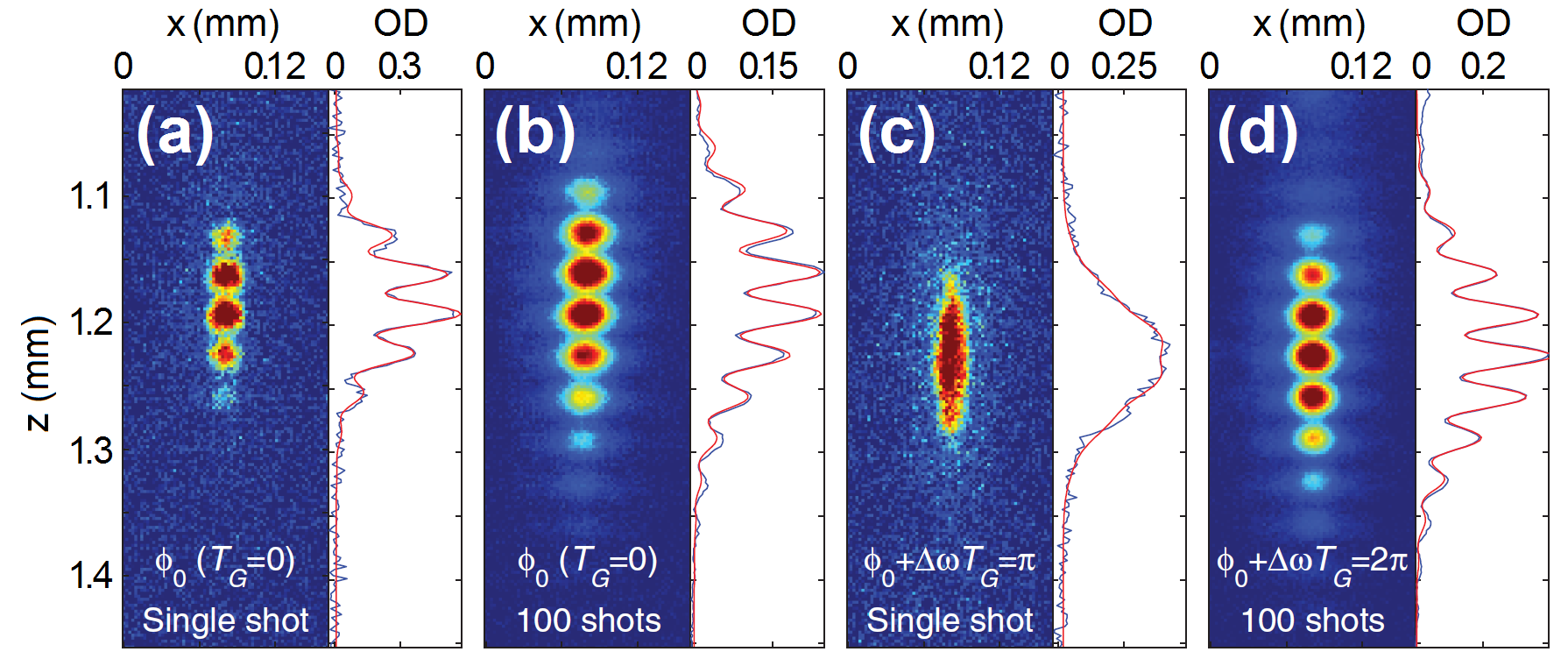}
   \caption{Clock interferometry on a chip. A magnetic gradient pulse of duration~$T_G$ is applied to induce a relative angle of rotation between two clock wave packets (inducing a clock ``tick'' rate difference~$\Delta\omega$ simulating a gravitational red shift). (a-b)~For~$T_G=0$, the clock rate is approximately the same in the two wave packets, and interference is visible. (c-d)~Visibility of the clock spatial interference may be destroyed~($\Delta\omega T_G=\pi$) or restored~($\Delta\omega T_G=2\pi$) merely by adjusting~$T_G$. Optical density curves are data (blue) and fits (red) to a simple combination of a sine with a Gaussian envelope. The vertical axis~$z$ is relative to the chip surface. Adapted from~\cite{Margalit2015}, with permission by~AAAS.}
   \label{fig:Margalit2015fig2}
\end{figure}

The main limitation of splitting based on magnetic field gradients is the shot-to-shot stability of these magnetic fields, which are determined by current instabilities in the wires. An improved scheme based on a three-wire configuration was implemented to reduce the phase accumulation during the splitting due to the quadrupole field. This enabled considerable improvement of the phase stability in a later experiment, which demonstrated complementarity between interference visibility and distinguishability of two paths when the atom is used as a quantum clock~\cite{Margalit2015} (Fig.~\ref{fig:Margalit2015fig2}). 

\subsection{Guided atom interferometry}
\label{subsec:guided}

As mentioned in Sec.~\ref{subsec:Bragg}, the first realization of interferometry on an atom chip was a Michelson interferometer with a magnetically guided~BEC~\cite{Wang2005}. One of the obvious applications of guided interferometry is an area-enclosing loop, which would enable Sagnac interferometry for inertial sensing of rotation. The sensitivity of such a Sagnac interferometer is proportional to the area enclosed by the interferometer loop, which would be small on an atom chip as compared to free-space implementations. High sensitivity could be recovered however, by constructing a high-finesse loop that would enable the atoms to perform many revolutions. In addition, closed-loop interferometer configurations may have symmetry properties that allow a high interferometric sensitivity even for thermal atoms and in the presence of potential imperfections~\cite{Japha2007}, as also shown for another symmetric interferometric scheme~\cite{Ammar2015}.

Prospects for guided closed-loop interferometry have encouraged several proposals for overcoming some major technical obstacles. The first attempt to realize a guided Sagnac interferometer was probably done by the Prentiss group~\cite{Wu2004}. One detailed proposal for a full scheme of a Sagnac interferometer with a~BEC on a chip is based on optical Bragg splitting and a magnetic atomic guide generated by an array of~3 or~4 parallel current-carrying wires~\cite{Baker2009}. The proposed magnetic guide would prevent perturbations caused by the input and output currents on the loop by exchanging the active wires during the round trip of the atoms in the loop. Another suggestion for preventing the problem of input and output for the loop current is a superconducting loop that maintains a persistent current~\cite{Muller2008}. Many very creative ideas frequently emerge, such as the recently suggested potential loop generated by the current in a (single-layer) Archimedean spiral of two interleaved wires~\cite{Jiang2015}. However, care has to be taken to ensure that the potential is completely smooth as any roughness, due for example to the current leads, would hinder the operation of the Sagnac interferometer.

Adiabatic dressed potentials have also been suggested as a method to create a variety of shapes of traps and guides for ultracold atoms. Designs of ring traps for atoms were suggested~\cite{Morizot2006,Lesanovsky2006,Fernholz2007,Lesanovsky2007} as loops for Sagnac interferometry as well as structures for studying superfluidity in confined geometries. Inductive dressed ring traps have also been suggested~\cite{Vangeleyn2014,Sinuco-Leon2014}, as well as rings based on moving actuators~\cite{West2012b}. Ring traps have been loaded~\cite{Sherlock2011,Pritchard2012}, but not on an atom chip. Finally, a proposed scheme for splitting and guiding by a~3D state-selective RF-dressed potential~\cite{Berrada2013} was briefly discussed in Sec.~\ref{subsec:state-selective}. To the best of our knowledge, coherent splitting and a measurement of the Sagnac effect has not yet been demonstrated on any kind of ring potential.

While the coherence and precision of free-space atom interferometers based on splitting by laser pulses is ensured by the high momentum precision of such splitting, the precision of guided-atom interferometers may be ensured by their guiding potentials. Sensitivity to the precision of the beam splitter is reduced since the guiding potential fixes the exact path of the atoms and hence the interferometer area. In addition, for Sagnac interferometers where the atoms propagate in two opposite directions along the same closed loop, the interferometric signal is expected to be insensitive to the exact value of the momentum transfer~$k$ at the splitting, as long as the splitting is symmetric (\ie\ to~$+k$ and~$-k$). In addition to high interferometric precision for beam splitters with very precise momentum transfer such as with Bragg pulses (Sec.~\ref{subsec:Bragg}), this enables high interferometric precision  also for splitting with state-selective potentials (Sec.~\ref{subsec:state-selective}) or~FGBS (Sec.~\ref{subsec:FGBS}).

\subsection{Internal and vibrational state interferometry} 
\label{subsec:internal}

This section has concentrated on spatial interferometry, in which atoms are split into a superposition of spatially separated states. We end the section by briefly mentioning some interesting schemes that use quantum interference between internal or motional atomic states that are typically not separated in space.

\begin{figure}[t!]
   \centering
   \includegraphics[width=0.45\textwidth]{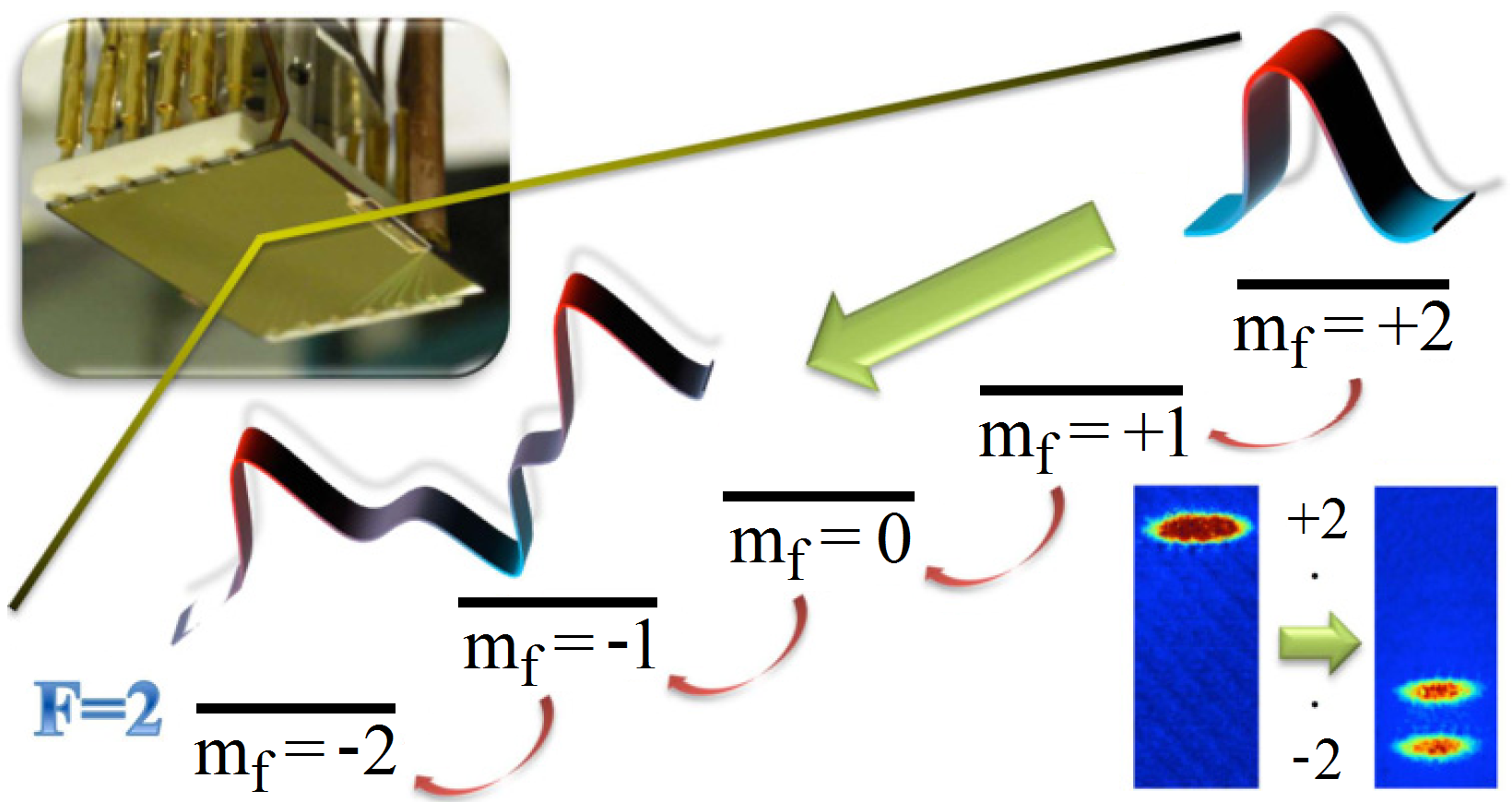}
   \caption{Preparation of the internal states of a~BEC realized on an atom chip. Optimal control theory is used to design frequency-modulated~RF pulses to prepare an arbitrary coherent superposition of states starting from a given initial state (fixed by the~BEC preparation). Shown in the bottom right corner are the experimentally observed clouds of atoms that are initially in sub-level~$+2$ and then split coherently into the sub-levels~$-1$ and~$-2$ \via\ optimal control. Adapted from~\cite{Lovecchio2016}, with permission \textcopyright~2016 by the American Physical Society.}
   \label{fig:Lovecchio2016fig1}
\end{figure}

Manipulation of atomic internal states, including Rabi flips and Ramsey interferometry, is a common exercise in atomic physics in general and in ultracold atoms and chip-based~BECs in particular. One of the most important examples of internal state interferometery is that of a clock~\cite{Treutlein2004}. Coherence times up to a minute have been observed on a chip~\cite{Deutsch2010}.

\begin{figure}[t!]
   \centering
   \includegraphics[width=0.5\textwidth]{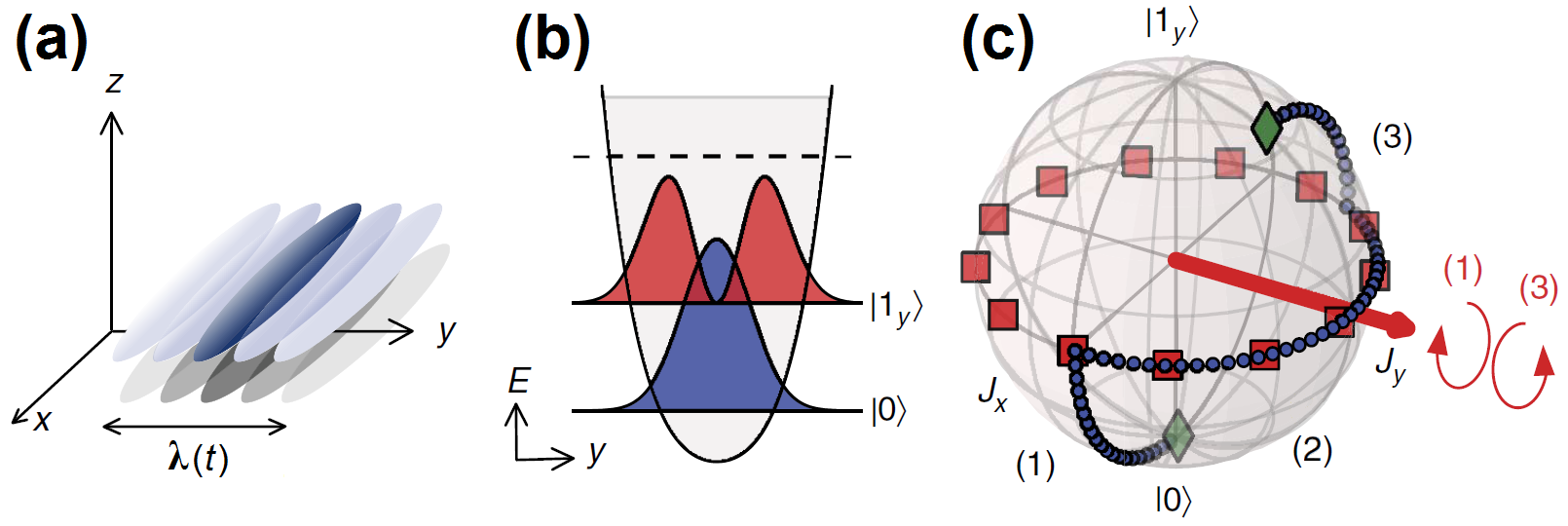}
   \caption{Vibrational state interferometry: schematic of the Ramsey interferometric sequence. (a)~Representation of the~BEC subjected to a fast displacement~$\lambda(t)$ in the~$y$-direction. (b)~Trapping potential and effective two-mode system. The anharmonicity in the~$y$-direction leads to a unique transition frequency between (blue)~the ground state~$|0\rangle$ and~(red) the lowest-lying excited state~$|1_y\rangle$, effectively almost isolating the two-level system~~$|0\rangle-|1_y\rangle$. The other states~(dashed line) have higher energies. (c)~Example of an interferometric trajectory~(blue dots) on the Bloch sphere representation of the two-level system. 
(1)~is the first pulse that prepares a balanced coherent superposition; (2)~is the phase accumulation time corresponding to a rotation around the vertical axis; (3)~is the second pulse, which is equivalent to a~$\pi/2$ pulse for the states on the equator and corresponds to a~$90^\circ$ counter-clockwise rotation around~$J_y$. The red squares show the~15 points on which the second pulse was optimized. Adapted from~\cite{vanFrank2014}, with permission~\cite{CC}.}
   \label{fig:vanFrank2014fig1}
\end{figure}

An example of interferometric signals which reveal the effect of many-body interactions on two-state interference is the observation of periodic coherence revivals in a~BEC consisting of the two trapped clock states~\cite{Egorov2011}.

Another example is the demonstration, by the Cataliotti group, of Ramsey interferometry with all~5 levels of the~$F=2$ hyperfine state of~\Rb~\cite{Petrovic2013}. Later the same group developed a scheme for preparing any internal state at will with~\Rb\ atoms in the~$F=2$ hyperfine state manifold~\cite{Lovecchio2016} (Fig.~\ref{fig:Lovecchio2016fig1}). Optimal control theory was used for achieving this goal. 

Interferometry with motional states of trapped atoms is based on the ability to control the occupation of higher-energy vibrational states in the trap. Transitions into such excited states can be induced by shaking an anharmonic trap~\cite{LlorenteGarcia2013}. A procedure for exciting transverse vibrational modes in an elongated trap by using adiabatic RF-dressed potentials has enabled the creation of twin-atom beams with opposite momenta~\cite{Bucker2011}. The same kind of transition was later used for performing vibrational state interferometry analogous to Ramsey interferometry of internal states~\cite{vanFrank2014} (Fig.~\ref{fig:vanFrank2014fig1}). Optimal control theory was again used in order to enable this interferometry. Indeed, it was suggested that control of vibrational states of trapped atoms may be beneficial in~QIP~\cite{Martinez-Garaot2013,Charron2002}, as well as in loop interferometry~\cite{Marti2015,Ryu2013}.

\addtocontents{toc}{\protect\vspace{4pt}}

\section{Atom Chip Applications}	
\label{sec:applications}

In this section we review several applications for which the atom chip may provide an advantageous platform. 

In terms of technological applications, atom chips not only offer robustness and low prices for mass production by utilizing fabrication techniques borrowed from semiconductor manufacturing, they also offer increased accuracy and reduced power consumption owing to their favorable scaling laws, whereby higher gradients are produced by lower currents. Prospects for precise integration of different elements also promise new possibilities, \eg\ by integrating photonics for high-Q devices and high-efficiency collection (Fig.~\ref{fig:Folman2011afig1}). 

The versatility of particles that can be trapped and manipulated has been described in Sec.~\ref{sec:particles} and the available technologies have been detailed in Sec.~\ref{sec:technology}. In Secs.~\ref{sec:surfaces} and~\ref{sec:interferometry} we presented some of the fundamental scientific topics which may be studied with the atom chip. Here we extend our discussion of the above-mentioned advantages for fundamental research to advantages that the atom chip offers for technological applications. For example, several groups are currently using it as the platform of choice for rapidly producing a~BEC~\cite{Farkas2014x,Rudolph2015}. Concurrently, atom chips are perhaps the best available technology for the emerging field of  atomtronics~\cite{Charron2006,Stickney2007,Japha2007,Pepino2009,Ryu2013,Eckel2014,Sinuco-Leon2014,Chow2015,Japha2016x,Caliga2016,NJP_SpecialIssue}, for which dynamic tunneling barriers are required. Such barriers may be formed on atom chips at distances of several~$\mum$~\cite{Salem2010,Chuang2011a}. The atom chip offers the ability to realize guides and traps with virtually arbitrary architecture.

Crucially important to the realization of atom chip applications are of course imperfections and hindering interactions. Progress on overcoming these inhibiting factors, including issues related to fabrication, noise, and coherence, were also presented in Sec.~\ref{sec:surfaces}. It therefore seems that the door is indeed open for the realization of quantum technology applications based on atom chips.

In the following, we start with specific applications, namely acceleration and field sensors, clocks, and devices for~QIP, and we end with several examples of general performance-enhancing techniques.

\subsection{Inertial sensors}
\label{subsec:sensors}

Inertial sensors are particularly important for high-accuracy navigation that is independent of the global positioning system~(GPS), and for detecting gravitational fields for geological searches, topography-based navigation, and underground detection. These sensors are also important for fundamental research. To date, very good inertial sensors are mechanical~--~MEMS~--~while optical ring-laser and fiber-gyroscopes also exhibit high performance. Ultracold atoms for inertial sensing promise orders-of-magnitude improvement over existing technologies (\eg~\cite{Dickerson2013}). The use of atom chip technology for an inertial sensing device based on cold atoms offers many possible advantages, such as integration and miniaturization, robustness, and a high bandwidth due to fast~BEC production. In fact, the atom chip may enable continuous-mode operation by utilizing a thermal beam with tunneling barriers~\cite{Japha2007}. It was also shown that a guided high-finesse device may compensate for the loss of area due to miniaturization. However, to date, guided atom interferometry for inertial sensing has yet to be realized (Sec.~\ref{subsec:guided}).

\begin{figure}[t!]
   \centering
   \includegraphics[width=0.5\textwidth]{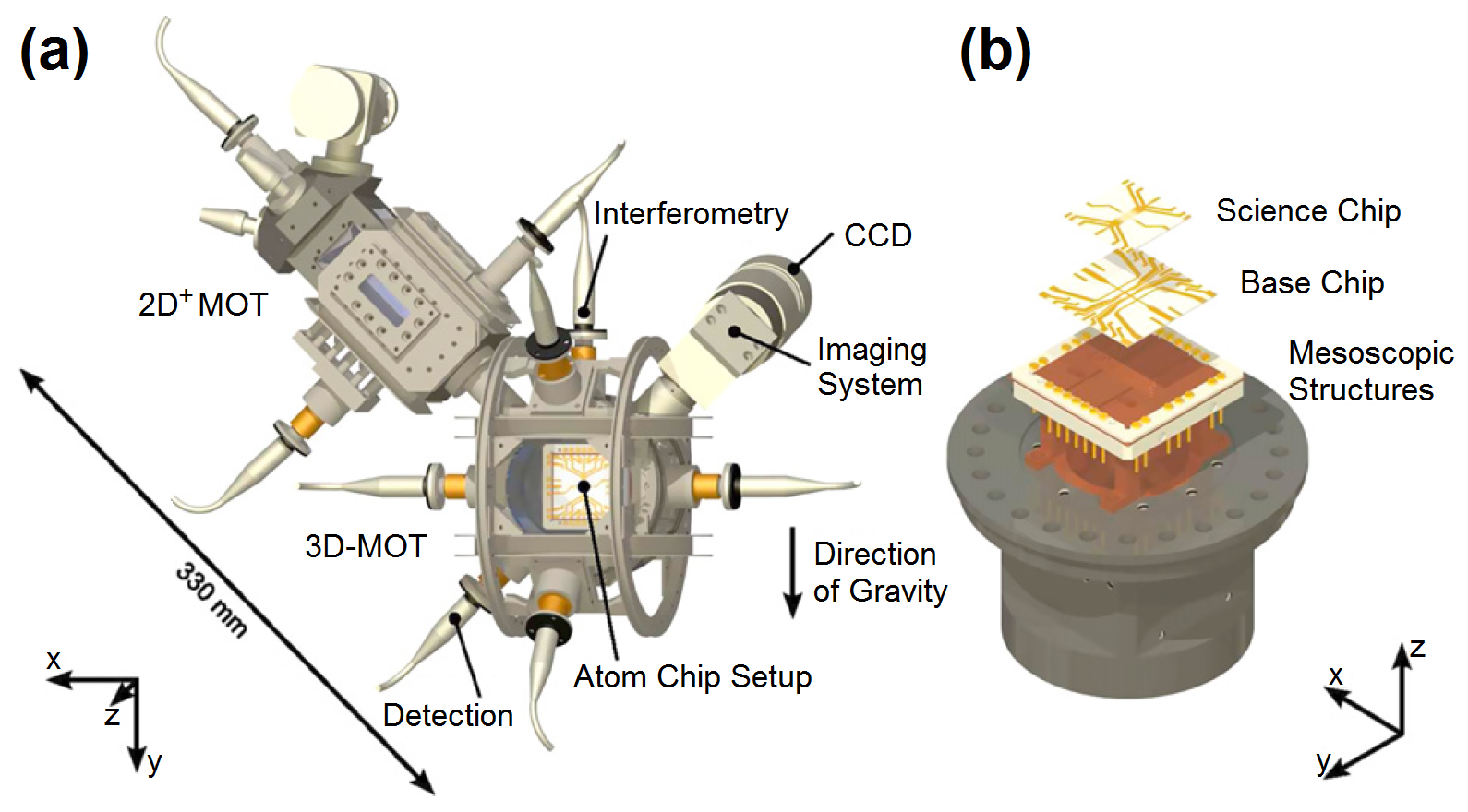}
   \caption{High-flux atom chip-based BEC source. Schematic drawings of (a)~the 2-chamber vacuum setup and (b)~the atom chip setup. A beam of pre-cooled~Rb atoms is formed in a~2D$^+$~MOT and injected into the~3D chamber. 
Atom interferometry, as well as detection of the atoms, is carried out in the~3D chamber. The three layers of the atom chip setup are shown in an exploded view. Adapted from~\cite{Rudolph2015}, with permission~\cite{CC} \textcopyright~IOP Publishing~\& Deutsche Physikalische Gesellschaft.} 
   \label{fig:Rudolph2015fig2}
\end{figure}

The atom chip is already proving useful for inertial sensors in the preparatory cooling stage of the cycle. Atomic inertial sensors typically use laser-cooled atomic ensembles at temperatures of a few~$\rm\muK$. Lower temperatures improve the precision both on the Earth and in micro-gravity due, for example, to longer evolution times. A high level of precision is necessary not only for technological applications, but also for fundamental studies, \eg\ for verifying predictions of violations of general relativity postulates at different accuracy levels. One of the key properties of the atom chip is to enable high magnetic gradients at relatively low currents near the chip surface for fast evaporative cooling and production of~BECs. Low power consumption is particularly important for mobile systems that operate on batteries. The~``QUANTUS'' project~\cite{vanZoest2010} started as a feasibility study of a compact, robust and mobile experiment for the creation of a~BEC in a micro-gravity environment. The latest result of this effort reports on the realization of a miniaturized setup, able to generate a flux of~$4\times10^5$ quantum degenerate~\Rb\ atoms every~$\rm1.6\,s$.  Ensembles of~$1\times10^5$ atoms can be produced at a~$\rm1\,Hz$ rate under micro-gravity conditions~\cite{Rudolph2015}. Rates of~$\rm1.2\,Hz$ were also reached for smaller numbers of atoms, with a projection of rates up to~$\rm10\,Hz$ being attainable with improved vacuum. The entire setup was placed in a capsule, which was released from the top of a drop tower (Fig.~\ref{fig:Muntinga2013fig1and2}). A schematic illustration of the physical package is presented in Fig.~\ref{fig:Rudolph2015fig2}. In this context, it is also important to mention the work on a miniaturized atom chip setup that can also generate~BECs at a~$\rm1\,Hz$ rate~\cite{Farkas2014x,ColdQuanta2015}. 

Although guided interferometry on the atom chip using coherent splitting into a magnetic or an optical waveguide remains unrealized, the potential impact of such a device for various applications of inertial sensing is so important that engineering a working device continues to drive many efforts. Two such efforts are the~``OnACIS''~\cite{OnACIS2013} and~``iSense''~\cite{deAngelis2011,Malcolm2016} projects. Another effort is taking place at SYRTE~\cite{GarridoAlzar2012}. A recent proposal is based on~RF potentials and analyzes the possibility of utilizing the atom chip for the creation of~3D traps which move on a loop in opposite directions~\cite{Stevenson2015}. The latter is part of yet another effort to realize a chip-based sensor~\cite{MatterWave}. A recent review of~20 years of development of the Sangac effect for atomic inertial sensing is in~\cite{Barrett2014}.

It stands to reason that the atom chip will eventually enable the right combination of methods for coherent guided interferometry to be realized. For example, integrating a mirror on the chip to create the necessary two Bragg beam splitters, together with the source of a magnetic guide, is important because any relative movement between the potential and the Bragg fringes will destroy the sensitivity. In another example, one may think of incorporating a bright or dark ring~\cite{Ramanathan2011,Marti2015,Turpin2015} for guiding, while two Stern-Gerlach type beam splitters~\cite{Machluf2013a} close a Ramsey-type sequence. It also remains to be seen whether guided interferometry allows one to relax the current stringent requirements regarding the~$\Delta p/p$ in interferometry, since the guides determine the path (Sec.~\ref{subsec:state-selective}).

\subsection{Field sensing}
\label{subsec:fields}

One of the well-known applications of the atom chip is to probe magnetic fields originating from the surface~(\eg\ from electron transport). Since atom-surface interactions were presented in Sec.~\ref{sec:surfaces}, here we briefly describe this topic with an emphasis on applications.

Cold-atom magnetic microscopy has the advantage of both high sensitivity~($\rm10^{-10}\,T$) and high resolution (several~$\rm\mum$). These properties were applied to discover long-range order in electronic transport in polycrystalline metal (gold) films~\cite{Aigner2008}. This technology is particularly attractive for exploring current transport and vortex formation in superconducting wires, as well as domain formation in different magnetic materials. The Lev group developed an advanced atom chip setup that is designed to cool the chip to below~$\rm4\,K$ and enable an optical resolution of~$\rm0.7\,\mum$ at~$\rm780\,nm$~\cite{Naides2013}. The atom chip may thus potentially give rise to new insight in the race to understand high-$\Tc$ superconductivity. An alternative method to investigate magnetic fields through the corrugations they form in magnetic traps is based on phase-space tomography after displacement from equilibrium and evolution of the atom cloud in the trap~\cite{Zhou2014}. It was shown that the phase-space distribution is extremely sensitive to small anharmonicities in the potential that nanoscale imperfections can cause. 

\begin{figure}[t!]
   \centering
   \includegraphics[width=0.45\textwidth]{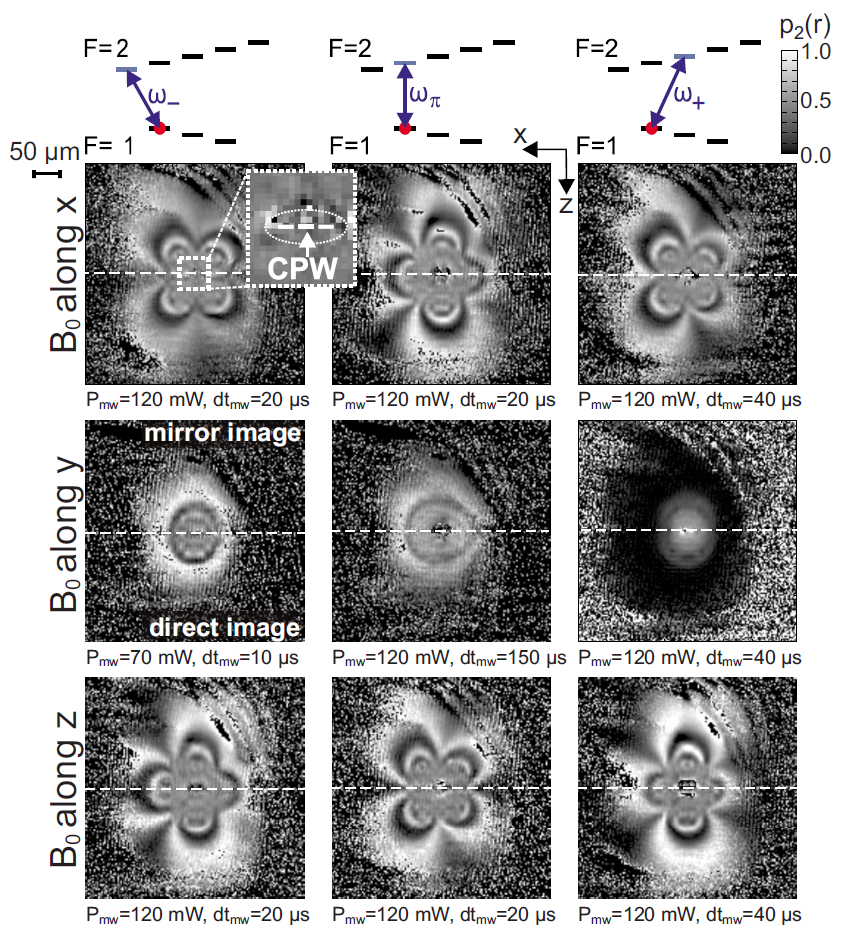}
   \caption{Imaging of microwave magnetic field components near the atom chip microwave source (a~CPW structure, inset). The images show the measured probability~$p_2(\bfr)$ to find an atom in~$F=2$ after applying the microwave pulse. Columns correspond to measurements on the three different transitions~$\omega_\gamma$ ($\gamma=-,\pi,+$); rows to three different orientations of~$\bfB_0$. The imaging beam is reflected from the chip surface at a small angle. As a result, on each picture, the direct image and its reflection on the chip surface are visible. The dashed line separates the two.  Adapted from~\cite{Bohi2010}, with permission \textcopyright~2010 by AIP Publishing~LLC.} 
   \label{fig:Bohi2010fig2}
\end{figure}

In addition to probing~DC magnetic fields, microwave fields can also be investigated with microscopic resolution. In~\cite{Bohi2010} it was shown how the magnetic field component of a microwave pulse drives different transitions in~Rb at Rabi frequencies that depend on the local strength of the field. Imaging the atomic cloud allows a complete reconstruction of the microwave field, as demonstrated in Fig.~\ref{fig:Bohi2010fig2}.

Other schemes use a~BEC as an interferometric scanning probe for mapping the microwave field with similar microscopic resolution but with improved sensitivity due to entanglement between the atoms~\cite{Ockeloen2013}. This enabled surpassing the~SQL for interferometry by~$\rm4\,dB$ and reaching a magnetic field sensitivity of~$\rm77\,pT/\sqrt{Hz}$ in a probe volume of~$\rm20\,\mum^3$. It was suggested that this development could help improve integrated microwave circuits for~QIP and in applications to communication technology.

Atom chip technology can also be used to detect electric field potentials originating from the surface~\cite{Chan2014}. This is carried out with Rydberg states that are extremely sensitive to electric dipole interactions.

Looking into the future, it seems likely that atom chips will continue to achieve increased sensitivity through interferometric schemes. For example, the atom chip group at Ben-Gurion University is working on the concept of ``matter-wave homodyning'', whereby one wavepacket is sent close to the surface for interaction with a system of interest, while the other wavepacket is maintained further away as a reference.  

As an extension of cold-atom microscopy, where ultracold atoms in an elongated trap scan an area close to a metallic surface, a~BEC in a vertical optical lattice trap can cover a large area parallel to the surface and allow simultaneous sensing of field patterns near it. For example, a~BEC of~$10^6$ \Rb\ atoms with a ``pancake'' shape of horizontal radius~$\rm100\,\mum$ can be formed in a potential with a horizontal frequency of~$\rm14\,Hz$ and a vertical frequency of~$\rm25\,kHz$~\cite{Gallego2009} (note that tunneling to other sites of the optical lattice may be suppressed by self-trapping due to atom-atom interactions). The necessary hardware has already been integrated onto an atom chip, as illustrated in the next section (Fig.~\ref{fig:Straatsma2015fig1and2}). Cold-atom magnetometry may even be applied to biophysical systems under ambient conditions, such as neural networks, using~$\rm\mum$-thick membranes capable of holding ultra-high vacuum over small areas. 

\subsection{Clocks}
\label{subsec:clocks}

High-end applications that use secondary atomic frequency standards based on vapor cells must currently rely on frequent synchronization with the~GPS system to correct long-term drifts. This is a major drawback and a significant problem for various military and civilian applications in which the~GPS signal may be temporarily or permanently inaccessible. The~Cs beam clock is a possible solution that has been available commercially for many years.  

Progress in the field of ultracold atoms and the maturity of the technology has motivated various projects to develop more compact clocks with improved long-term stability. One approach is to release the trapped atoms and use a Ramsey sequence during a time-of-flight of several msec. This requires a very uniform microwave and magnetic c-field over several~mm and the physical package is not very compact. The~HORACE cold atom clock is one example of such an effort~\cite{Esnault2011}. 

A contrasting approach achieves long coherence times~--~and therefore long measurement times and high accuracy~--~by trapping the atoms instead of releasing them. In~2004 it was demonstrated that atoms close to an atom chip can retain long coherence times for internal degrees of freedom~\cite{Treutlein2004}. Confining the atoms to the small volume of an atom chip also relaxes the homogeneity conditions for the microwave radiation~\cite{Vuletic2011}. In addition, it was subsequently discovered that the coherence lifetime of the atoms can be extended to almost~$\rm1\,min$ due to spin rephasing induced by the identical spin rotation effect~\cite{Deutsch2010}. Together with the vision of the atom chip platform presented in Fig.~\ref{fig:Folman2011afig1}, these developments are contributing to chip-based atomic clocks with improved accuracy and further miniaturization. 

\begin{figure}[t!] 
   \centering
   \includegraphics[width=0.35\textwidth]{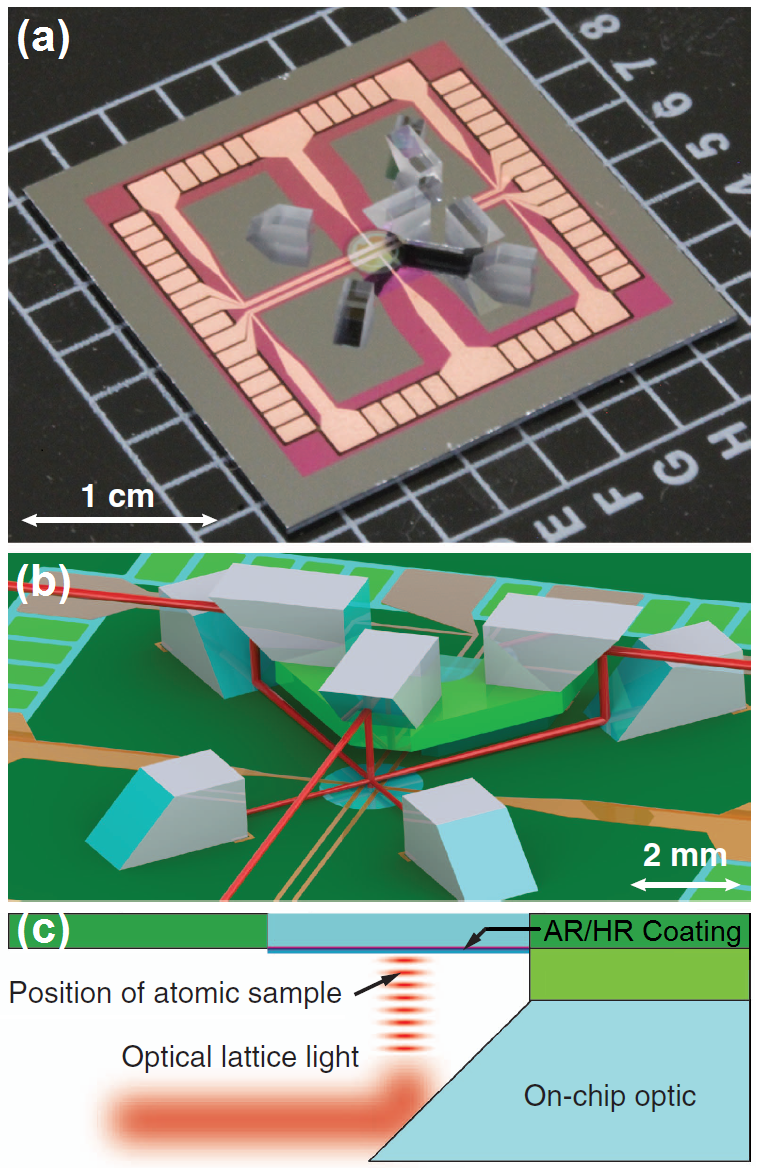}
   \caption{On-chip lattices. (a)~Photo of an atom chip showing the optical lattice system. (b)~Model of the on-chip optical lattice system. The two mirrors at the bottom of the image retro-reflect two of the incoming lattice beams, while the window is used to retro-reflect the third beam. (c)~Schematic illustration of the vertically oriented~1D lattice used to demonstrate Landau-Zener tunneling. Adapted from~\cite{Straatsma2015}, with permission from the Optical Society of America.}
   \label{fig:Straatsma2015fig1and2} 
\end{figure}

The first complete realization of an atom chip clock was reported in~\cite{Szmuk2015}. Aside from the fact that the sequence of trapping atoms is more complex than that of free-falling experiments, there remain various noise sources that degrade the stability of this clock. The most dominant are shot-to-shot fluctuations in the atom temperature and in the offset magnetic field. Performance is also degraded by the local oscillator. Next-generation versions of this clock are expected to incorporate non-destructive detection and shorter cycle times.

Another approach would be to implement designs that incorporate different optical elements such as mirrors enabling, for example, a compact and robust on-chip optical lattice system for trapping and detection. One such design, where prisms embedded directly onto the chip create a~3D standing wave, has been developed by the Anderson group~\cite{Straatsma2015} (Fig.~\ref{fig:Straatsma2015fig1and2}) and is now a commercial product~\cite{ColdQuanta2015}. This chip also features a through-chip optical window enabling high-resolution imaging (the through-chip optical window of this design has also enabled hybrid magnetic+optical transistor-like potentials for atomtronics~\cite{Caliga2016}). Optical lattice clocks currently hold the record for the best stability and accuracy~\cite{Nicholson2015}, and precisely integrating the optical elements directly onto a chip surface to control all aspects of the potentials and of the detection elements promises improved robustness and miniaturization.

In addition to work on atom chips, there are also various efforts to use ion chips for miniature frequency standards. The ``MITICC'' project is an example of such an attempt~\cite{Yb_ion_clock2014}. It is based on trapping a single~$\rm Yb^+$ ion using a micro-fabricated circuit, with the aim of exhibiting a relative frequency stability of at least~$10^{-14}$.

\subsection{Quantum information processing}
\label{subsec:QIP}

Key operations needed for the realization of~QIP systems include qubit preparation, measurement, arbitrary single-qubit rotation, and entangling gates for~2-qubit operations~--~all while achieving clock times faster than any loss, heating, or decoherence time scales. Cold, isolated particles on chips constitute excellent platforms for accomplishing these operations and there is consequently great interest in them, due also to their accuracy, scalability, and single-site addressability.

Quantum information processing with trapped ions currently seems to be quite a bit ahead of neutral atoms~\cite{Monroe2008,Blatt2008,Harty2014}, with the most successful multi-particle quantum information processors to date being based on trapped ions \eg~by the Blatt group~(\cite{BellPrize2015} and references therein). Ion chips are also poised to perform simulations, \eg~by the Wineland group~\cite{Wilson2014}. There are still numerous difficult challenges that the field is facing, such as fidelity, scale-up, and material engineering aspects to reduce heating and decoherence from the environment. Here we discuss a few examples of current projects.

\begin{figure}[t!]
   \centering
   \includegraphics[width=0.45\textwidth]{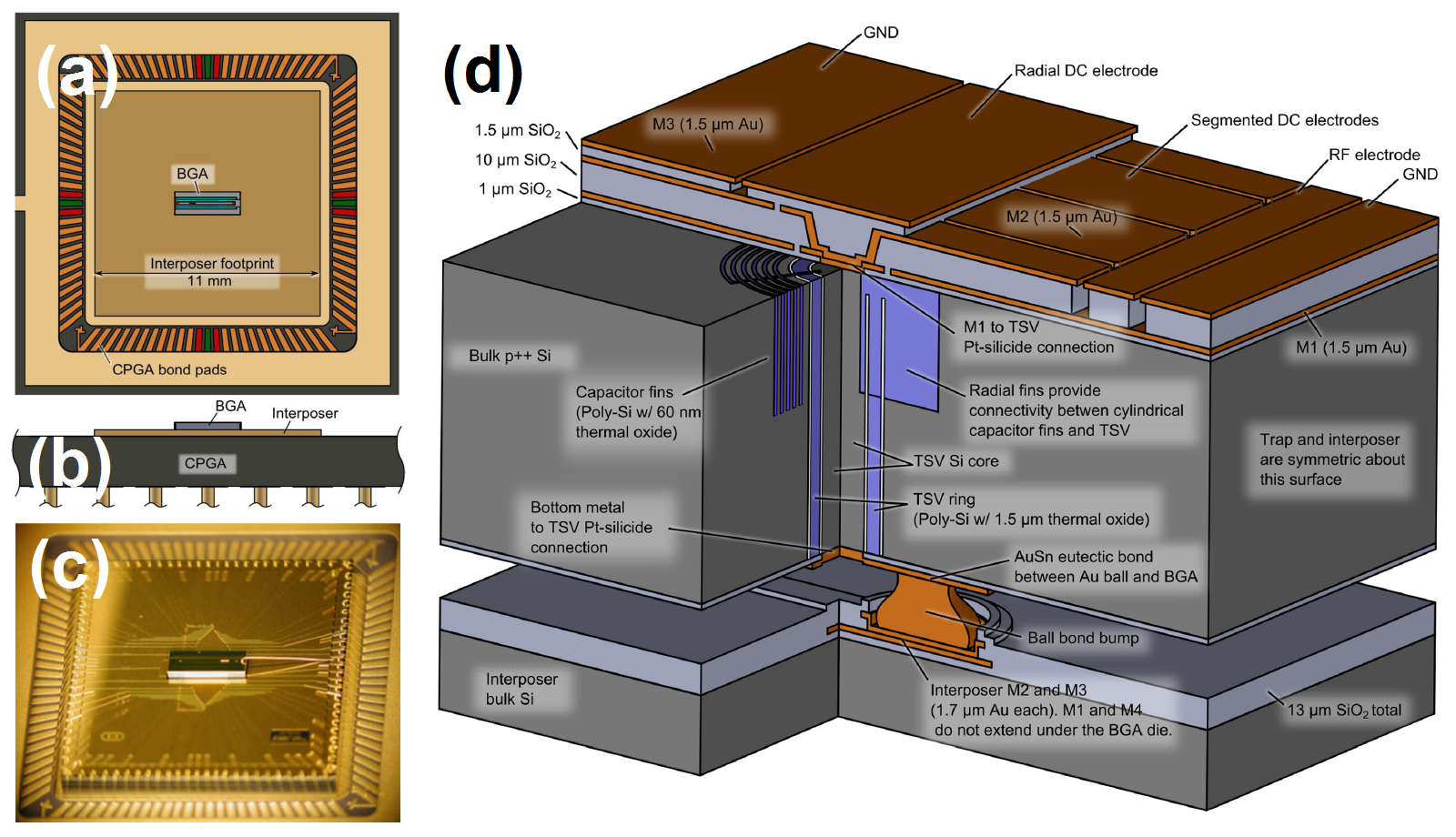}
   \caption{High-density electrode chip. (a)~Die bond region showing the ball-grid array~(BGA) trap and the interposer footprint; (b)~side view; (c)~fully packaged~BGA trap. The long bond wire supplies the trap~RF signal. (d)~Schematic cross section through a~BGA trap die and interposer (not to scale).  Adapted from~\cite{Guise2015}, with permission \textcopyright~2015 by AIP Publishing~LLC.}
   \label{fig:Guise2015fig1and2}
\end{figure}

Existing fabrication techniques for micro-fabricated ion chips limit the number of surface electrodes. The capacitors and bond pads can consume a majority of the overall chip area and perimeter, strongly constraining the layout of electrode structures and~DC/RF lead traces which may obstruct laser access (the lasers must interact with ions confined~$\rm\approx100\,\mum$ above the chip surface). This scale-up problem is the subject of major research efforts, with many ideas for new ion trap architectures. One such effort builds the chip around a ball-grid array~(BGA) with connections that allow through-substrate vias to bring electrical signals from the back side of the trap die to the surface trap structure on the top side~\cite{Guise2015} (Fig.~\ref{fig:Guise2015fig1and2}). This kind of advance is an important step for increasing qubit densities and bringing the vision of a quantum computer closer to reality. A scalable architecture for an ion chip quantum simulator has also been suggested in~\cite{Mielenz2015x}.

Regarding scalability, (neutral) atom chips have an advantage because it is possible to trap large numbers of atoms in an optical or magnetic lattice, or in a combined potential, with single-site addressability. Additional ideas for lattices include plasmonic, RF~dressing, and static electric lattices. Single-site addressability may come from near-field optics or local electric and magnetic fields emanating from the nearby surface. A commercially-available integrated setup for optical lattices on atom chips was discussed earlier in this section, indicating the maturity of this technology.

\begin{figure}[t!]
   \centering
   \includegraphics[width=0.4\textwidth]{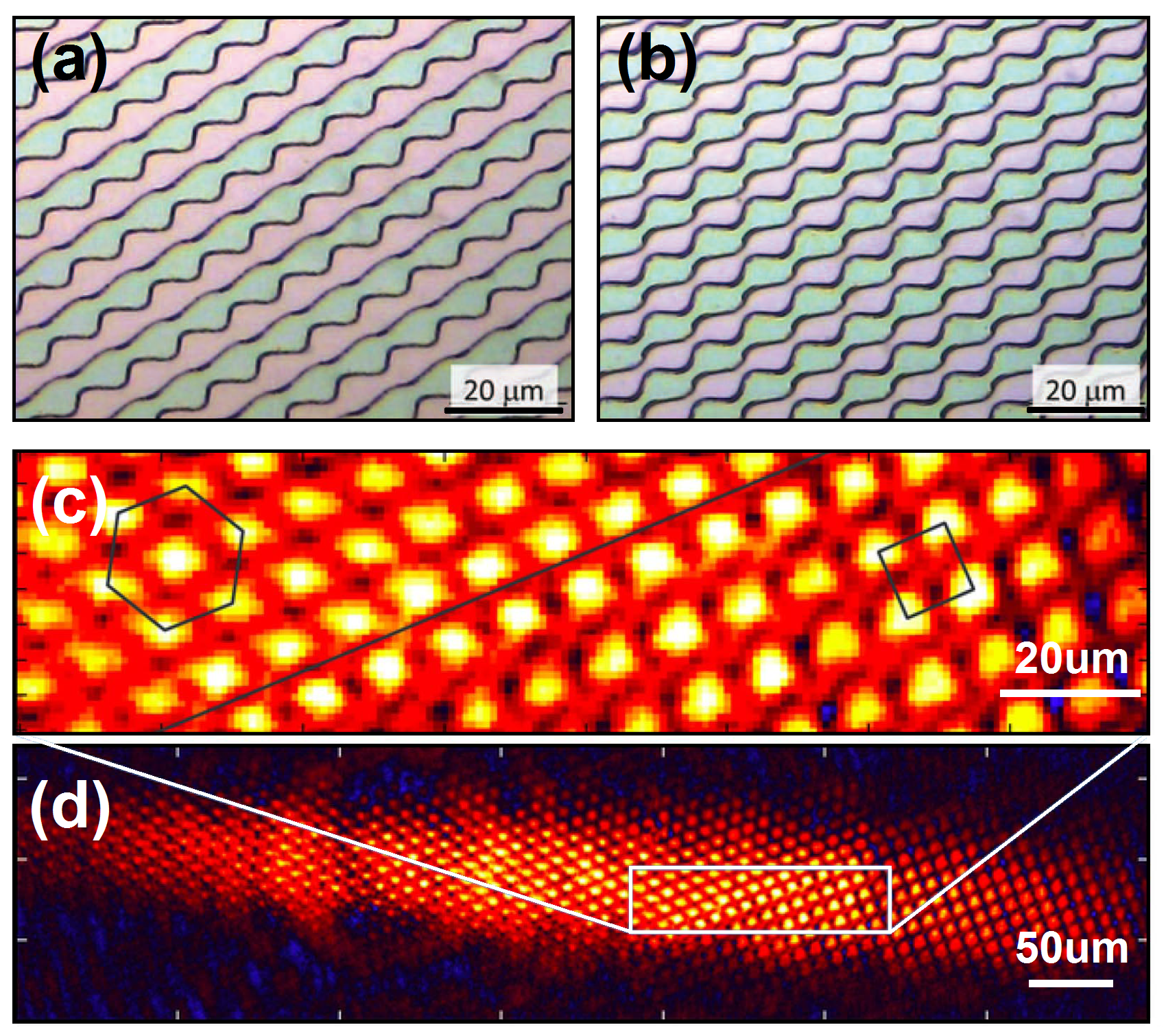}
   \caption{2D array and trapped lattice atoms on a permanent-magnet chip. The upper images show alternating strips of etched and non-etched regions in a~$\rm200\,nm$-thick layer of~FePt that generates a magnetic lattice with (a)~square translational symmetry and (b)~hexagonal (equilateral triangular) symmetry. An absorption image of~\Rb\ is shown in~(c) for atoms loaded simultaneously into the hexagonal (upper left) and square (lower right) lattices, with (d)~showing the entire extent of the lattice. Atom chip fabricated at Ben-Gurion University. Adapted from~\cite{Leung2014}, with permission \textcopyright~2014 by AIP Publishing~LLC.} 
   \label{fig:Leung2014fig2and7}
\end{figure}

Magnetic lattices may be made by current-carrying wires or by a lithographically patterned permanent magnetic film (Sec.~\ref{subsec:lattices}). Beyond the~1D permanent magnet lattice presented in Fig.~\ref{fig:Jose2014fig1}, a~2D permanent magnet lattice has been realized for a~BEC of~\Rb~\cite{Leung2014}. This experiment demonstrated such an atom chip with a lattice constant of~$\rm10\,\mum$, fabricated by selective etching of an~FePt thin-film permanent magnet (Fig.~\ref{fig:Leung2014fig2and7}). This approach enables a large variety of trapping geometries over a wide range of length scales and is also suitable for experiments in quantum information science based on interactions between atoms in highly excited Rydberg energy levels. Plans to load a~$\rm250\,nm$-period lattice are now underway in the same group. Such a lattice can enable Hubbard-model physics with ground state atoms, but in new geometries and regimes. Geometries will be more flexible in terms of lattice types, including interfaces between different lattices, both~1D and~2D and everything in between. Disorder and designer defects can be built in. Tapered lattices with a gradually changing period across the chip may also be of interest. 

An advantage of a permanent magnet lattice over optical lattices is that any lattice period and cell structure may be designed and fabricated. Reducing the lattice period may even enable direct interaction between ground-state atoms. The nearby surface may also be engineered to render specific atoms resonant or non-resonant, thereby enabling parallel control over many qubits using, for example, nm-scale electrodes near each lattice site that would Stark-shift atomic energy levels.  

Quantum simulations date back to the ideas of Richard Feynman, who famously expounded that the proper way to  simulate quantum physics would be to use a quantum computer instead of a classical computer~\cite{Feynman1982}. 
Such simulations, based on small-scale quantum systems that are relatively easy to control, could be used to efficiently simulate other quantum systems that are difficult or impossible to describe analytically or numerically on classical computers. 

Prospects for quantum simulations using systems of trapped ions are discussed in~\cite{Blatt2012}, where the available set of quantum operations and first proof-of-principle experiments for both analog and digital quantum simulations with trapped ions are reviewed. Deterministic tuning of the Coulomb interaction between two ions was demonstrated in~\cite{Wilson2014}, independently controlling their local wells. The scheme is suitable for emulating a range of spin-spin interactions. It was suggested that extension of this building block to a~2D network, which is possible using ion trap micro-fabrication processes, may provide new quantum simulator architecture with broad flexibility in designing and scaling the arrangement of ions and their mutual interactions. All of these prospects of course rely on the ability to accurately control and manipulate systems of trapped ions, to engineer a large variety of interactions with high precision, and to measure relevant observables with nearly~100\% efficiency. 

Neutral atoms in an optical lattice on an atom chip also constitute a possible system for a quantum simulator~\cite{Bloch2012}. Quantum simulations using neutral atoms retain their scalability advantage as discussed earlier~\cite{Bakr2009,Gillen2009}. Building blocks for quantum simulations with ensembles of trapped atoms were demonstrated with bosons~\cite{Simon2011} and fermions~\cite{Haller2015}. Permanent magnet atom chips with Rydberg atoms constitute another possible system for quantum simulators~\cite{Leung2014}, with the benefit of long-range atom-atom interactions.

Building blocks for quantum simulations have also been demonstrated with magnetically trapped neutral atoms on an atom chip. The experimental generation of multi-particle entanglement on an atom chip was achieved in~\cite{Riedel2010} by controlling elastic collisional interactions with a state-dependent potential, a significant step forward for atom chip applications in quantum simulations and~QIP. 

\subsection{Methods for atom chip applications}
\label{subsec:methods}

General methods for simpler and more efficient atom chip techniques continue to be developed for practical operations, as noted throughout this review. For example, an efficient procedure for direct evaporative cooling, whereby the~RF radiation is kept constant while the magnetic trap parameters are changed, was recently developed~\cite{Farkas2013}. In the remainder of this section, we discuss several additional experimental techniques that may become useful for technological applications based on atom chips. 

Efficient and high-fidelity internal state manipulation is essential. A crucial milestone in~QIP is the ability to prepare any initial state at will, an example of which was recently demonstrated \via\ optimal control on an atom chip~\cite{Lovecchio2016} (Fig.~\ref{fig:Lovecchio2016fig1}).  Optimal control allows one to accelerate the
state preparation process to the ultimate bound imposed by quantum mechanics, the so-called quantum speed limit. With control pulses as short as~$\rm100\,\mus$, transfer to the desired final coherent superposition was achieved with only a few percent error. Another example is the realization of stimulated Raman adiabatic passage with microwave frequencies on an atom chip by the Westbrook group, thereby enabling highly efficient coherent population transfer~\cite{Dupont-Nivet2015}. A~\Rb~BEC in its maximal magnetic moment state~$|F=2,m_F=2\rangle$ was prepared and coherently transfered to the~$|F=2,m_F=1\rangle$ state, a gateway for addressing the $|F=2,m_F=1\rangle\leftrightarrow|F=1,m_F=-1\rangle$ two-photon transition. This magnetically trappable ``clock'' transition is very robust against magnetically-induced decoherence at the ``magic field'' of~$\rm3.23\,G$ (Sec.~\ref{subsec:state-selective}).  New ideas concerning the manipulation of the external degrees of freedom are also continuing to surface. For example, using adiabatic passage to coherently move atoms from one magnetic guide to another is suggested in~\cite{Morgan2013}.

Further initial-state preparation may include squeezing, which is particularly important for metrological applications. Naturally, the precision of a measurement is limited by the projection noise of the system under test, which scales according to the~SQL as~$1/\sqrt{N}$. This is a consequence of the Poissonian statistics of~$N$ independent particles without any correlations between them. Squeezing and entanglement can overcome the~SQL, enabling~$1/N$ scaling in a system of~$N$ entangled particles~\cite{Giovannetti2004}. Attaining this fundamental Heisenberg limit is one of the goals of the emerging field of quantum metrology and has motivated many efforts to develop systems and schemes for generating entanglement~\cite{Gross2010}. 

\begin{figure}[t!]
   \centering
   \includegraphics[width=0.5\textwidth]{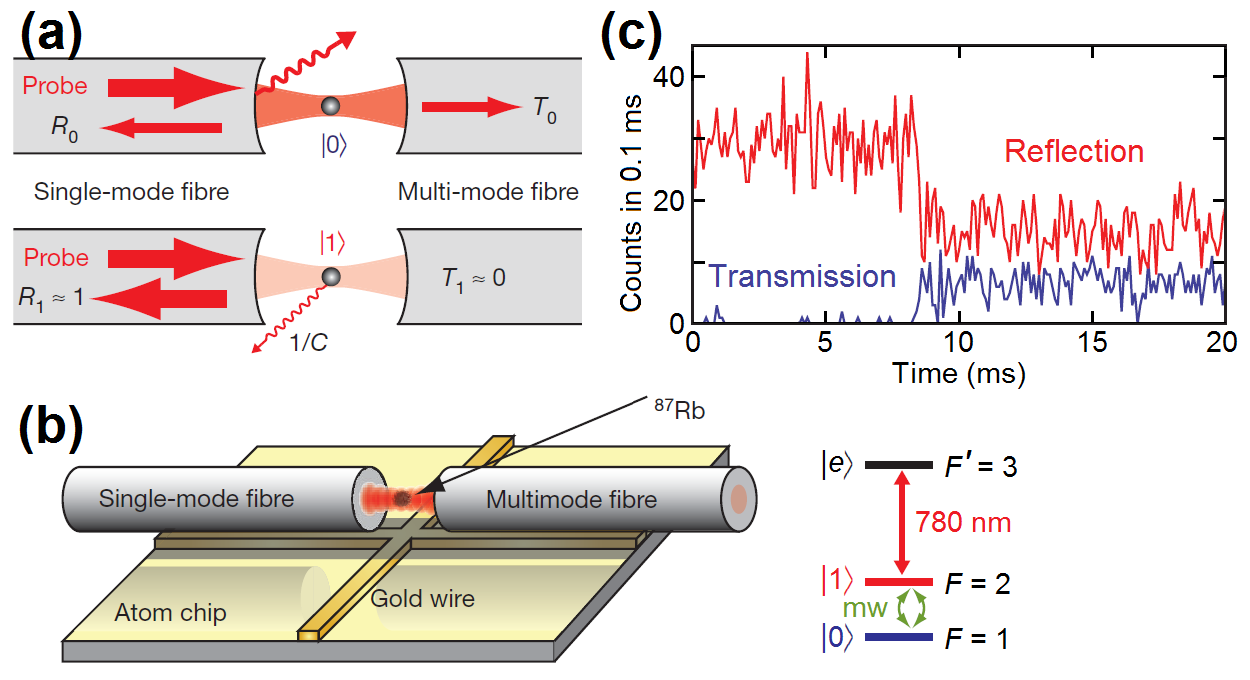}
   \caption{Optical fiber-based high-finesse cavity mounted on an atom chip, enabling cavity-assisted detection of an atomic qubit. (a)~For an atom in the dark state~$|0\rangle$~(top), probe light is either transmitted, reflected or lost by mirror imperfections. For the bright state~$|1\rangle$~(bottom), most incident photons are reflected. In both cases, only a small fraction is scattered by the atom. (b)~The cavity is formed by the coated end facets of two optical fibers. The qubit states~($F=1,m_F=0$ and~$F=2,m_F=0$) can be coupled by a resonant microwave pulse. The cavity and the atomic transition $|1\rangle\rightarrow|e\rangle$  are resonant with the $\pi$-polarized probe laser at~$\rm780\,nm$ wavelength. (c)~Typical detection trace, showing cavity transmission (blue) and reflection (red) for an atom initially in~$|1\rangle$  performing a quantum jump to~$|0\rangle$ owing to spontaneous emission. Adapted from~\cite{Volz2011}, with permission by Macmillan Publishers Ltd.}
   \label{fig:Volz2011fig1}
\end{figure}

A state-dependent microwave potential from an atom chip was used to generate spin-squeezed states of a two-component~BEC~\cite{Riedel2010}. The observed reduction in spin noise demonstrated multi-particle entanglement on the atom chip and metrologically useful squeezing. Subsequent experiments by the Reichel group demonstrated entanglement by strongly coupling~$N$ atoms to a single mode of a high-finesse optical cavity integrated on the atom chip~\cite{Haas2014}. This integrated atom chip cavity was also used to deterministically entangle~36 qubit atoms in less than~$\rm5\,\mus$~\cite{Barontini2015}. The high degree of entanglement utilizes quantum Zeno dynamics. The states created in the presence of the measurement are highly entangled in general, and their purity depends on the measurement being strong enough and, at the same time, sufficiently nondestructive.

Looking into the future, there are many ideas on how to use atom chips for enhanced entanglement. For example, having polarized atoms in a superposition of magnetic fields (created by a superposition of currents in a flux qubit) would create a highly entangled state, the so-called ``high-$N00N$'' state~\cite{Lee2002}. Such states are advantageous for metrology based on interferometry. It remains to be seen what methods eventually become mature enough to be used in actual applications.

\begin{figure}[t!]
   \centering
   \includegraphics[width=0.5\textwidth]{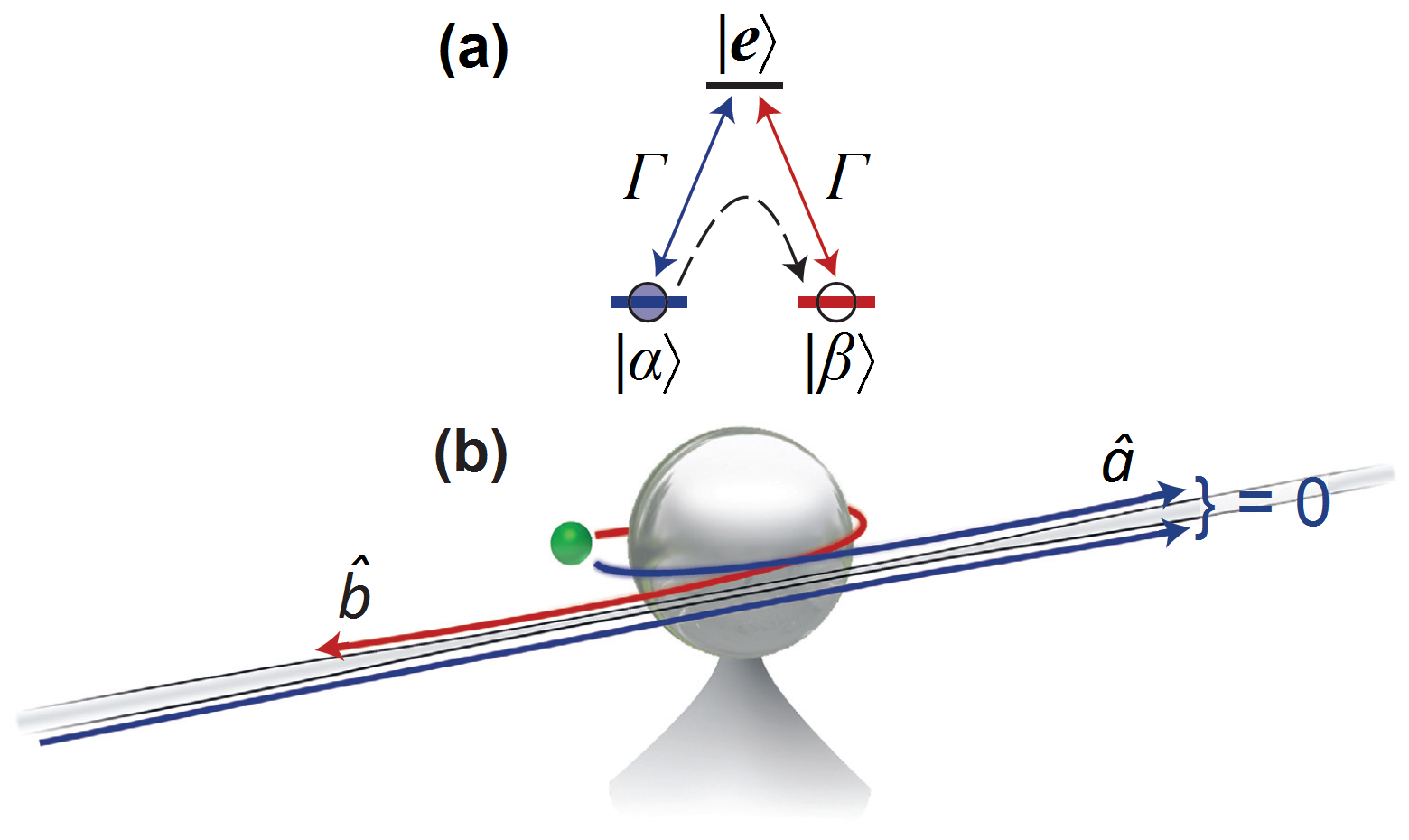}
   \caption{On-chip integrated resonator. (a)~Schematic depiction of the single-photon Raman interaction. The two transitions in a three-level~$\Lambda$-system (a single atom of~\Rb\ in this case) are coupled (b)~\via\ a micro-resonator to different directions of a nano-fiber waveguide. A photon coming from the left is deterministically reflected~(red arrows) due to destructive interference in the transmission~(blue arrows), resulting in the Raman transfer of the atom from ground state~$\alpha$ to~$\beta$. The atom then becomes transparent to subsequent photons, which are therefore transmitted. Adapted from~\cite{Rosenblum2016}, with permission by Macmillan Publishers Ltd.}
   \label{fig:Rosenblum2016fig1}
\end{figure}

As a final example, let us note that strong light-matter interaction is essential for many tasks, such as qubit readout, non-destructive detection, and photon switching. Such atom-photon interactions are also required for quantum communications (\eg\ repeaters~\cite{Briegel1998,Li2015}) or for links between quantum computers. A high level of control (including physical stability) is required to transferring  information from photons (``flying qubits'') to atoms or ions (``trapped qubits'')~\cite{Kimble2008,Northup2014}.  Such couplings have been realized in a fiber~FP cavity on an atom chip~\cite{Volz2011} (Fig.~\ref{fig:Volz2011fig1}). In another experiment, light evanescently coupled by a fiber into an on-chip ultrahigh-Q micro-resonator was used to couple photons to single~\Rb\ atoms falling past the micro-resonator~\cite{Rosenblum2016} (Fig.~\ref{fig:Rosenblum2016fig1}), where a further step could be taken by trapping the atoms. Photonic crystals have also been used for localizing and interfacing atoms with guided photons~\cite{Goban2014,Douglas2015} (Fig.~\ref{fig:Goban2014fig1a}).

\addtocontents{toc}{\protect\vspace{4pt}}

\section{Conclusion}
\label{sec:conclusion}

In this review, we have tried to give the reader a glimpse of the atom chip in~2015. We have attempted not only to describe the state-of-the-art, but also to hint at possible future directions and roadmaps, and in this way help young atom chip researchers who have just begun their journey.

The atom chip has been around for~15 years. In that time it has transformed continuously: from current-carrying wires and permanent magnets to dipole traps and plasmonics; from macroscopic structures to micro- and even nano-scales; from simple materials to superconductors, alloys, anisotropic conductors,~CNTs and graphene; from external detection to on-chip single-atom detection; from single-layer to multi-layer chips, and all the way to~3D fabrication. It has progressed from slow experimental cycles and bulky machinery to~BECs produced at a rate faster than~$\rm1\,Hz$ in stand-alone portable packages~--~and it would not be surprising if this rate reaches~$\rm10\,Hz$ in the near future.

This stunning transformation of the atom chip includes a wide variety of particles now routinely trapped, cooled, and manipulated, from neutral ground-state atoms, to Rydberg atoms, molecules, ions, electrons, and perhaps even antimatter in the near future.

The goals have also expanded: a wide variety of fundamental studies from single particles to many-body physics; from Casimir-Polder to searches for hypothesized forces; from surface science to foundations of quantum mechanics and all the way to quantum simulations. Applications have also been intensively pursued: from guided matter-wave interferometry, through clocks and all the way to quantum information processing.

This widening of the techniques and scientific scope go hand-in-hand with the ever-growing number of groups, from just three at the turn of the century to a few dozen at the present time. If one takes into account all types of particles, the number of groups is very large indeed. 

It is very hard, and perhaps even foolhardy, to guess what this field will achieve in the future or even what directions it will take. Given enough time and funding, it is clear that the vision presented in Fig.~\ref{fig:Folman2011afig1} is achievable. Will atom chips really enable an applicative revolution, a new kind of 21st century technology? On the front of fundamental science, will more integration, miniaturization, accuracy (fidelity), complexity, scalability and control, enable new discoveries? The answer to both questions is~--~of course~--~that no-one knows. We can say with a high level of likelihood however, that if we extrapolate from the past, we should expect many fascinating novelties in the future.

Looking back at the last~15 years, one cannot but admire how dynamic this field has become. So much so that one is reminded of the saying attributed by some to Heraclitus of ancient Greece: ``There is nothing permanent except change''. This is perhaps the most fitting summary and outlook one may propose for this field of surface interfaces with cold matter.

We hope that the reader has enjoyed this glimpse and remains eager to see what additional surprises await.

\posskip	

\addtocontents{toc}{\protect\vspace{0pt}}	

\section*{List of abbreviations}

{\parindent=0pt
\begin{tabular}{ l l }
BEC		& Bose-Einstein condensate (or condensation)				\\
BGA		& ball-grid array																		\\	
CNT		& carbon nano-tube																	\\
CP		& Casimir-Polder																		\\
CPT		& charge-parity-time																\\
CPW		& co-planar waveguide																\\
FGBS	& field-gradient beam splitter											\\
FP		& Fabry-P{\'e}rot																		\\
GBAR 	& Gravitational Behaviour of Antihydrogen at Rest		\\
GPS		& global positioning system													\\
MEMS	& micro-electro-mechanical system										\\
MOT		& magneto-optical trap															\\
RF		& radio-frequency																		\\
SQL		& standard quantum limit														\\
QIP		& quantum information processing										\\
YBCO	& yttrium barium copper oxide $\rm(YBa_2Cu_3O_{7-\delta})$	\\
YSZ	 	& yttrium-stabilized~$\rm ZrO_2$										\\[0pt]
\end{tabular}
}

\posskip\centerline{\bf Disclosure Statement}

The authors declare that they have no competing financial interests.

\posskip\centerline{\bf Acknowledgments}

We wish to warmly thank all the members~--~past and present~--~of the Atom Chip Group at Ben-Gurion University of the Negev, and the team of the~BGU nano-fabrication facility for designing and fabricating innovative high-quality chips for our laboratory and for others around the world. We are grateful to Shimon Machluf and Judy Kupferman for their helpful criticism of the manuscript. We also wish to express our thanks to members of the atom chip scientific and technical communities for their enthusiastic responses to our requests for figures and other materials.

\posskip\centerline{\bf Funding}

This work is funded in part by the Israel Science Foundation, the~EC ``MatterWave'' consortium~(FP7-ICT-601180), and the German DFG through the DIP program (FO 703/2-1). We also acknowledge support from the~PBC program for outstanding postdoctoral researchers of the Israeli Council for Higher Education and from the Ministry of Immigrant Absorption (Israel).

\addtocontents{toc}{\protect\vspace{0pt}}

\bibliography{Keil2016}

\begin{thebibliography}{100}

\bibitem{Folman2011a}
R.~Folman.
\newblock {Material science for quantum computing with atom chips, in: \it
  Special Issue on Neutral Particles, \rm R. Folman, ed.}
\newblock
  \href{http://link.springer.com/article/10.1007%2Fs11128-011-0311-5#/page-1}{\it
  Quantum Inf. Process. \bf10\rm, 995-1036 (2011).}

\bibitem{Rushton2014}
J.~A. Rushton, M.~Aldous, and M.~D. Himsworth.
\newblock {The feasibility of a fully miniaturized magneto-optical trap for
  portable ultracold quantum technology}.
\newblock
  \href{http://scitation.aip.org/content/aip/journal/rsi/85/12/10.1063/1.4904066}{\it
  Rev. Sci. Instrum. \bf85\rm, 121501 (2014).}

\bibitem{Salim2011}
E.~A. Salim, J.~DeNatale, D.~M. Farkas, K.~M. Hudek, S.~E. McBride,
  J.~Michalchuk, R.~Mihailovich, and D.~Z. Anderson.
\newblock {Compact, microchip-based systems for practical applications of
  ultracold atoms, in: \it Special Issue on Neutral Particles, \rm R. Folman,
  ed.}
\newblock \href{http://link.springer.com/article/10.1007/s11128-011-0300-8}{\it
  Quantum Inf. Process. \bf10\rm, 975-994 (2011).}

\bibitem{Farkas2014x}
D.~M. Farkas, E.~A. Salim, and J.~Ramirez-Serrano.
\newblock {Production of Rubidium Bose-Einstein Condensates at a 1 Hz Rate}.
\newblock \href{http://arxiv.org/pdf/1403.4641v2.pdf}{\tt arXiv:1403.4641v2
  \rm(2014).}

\bibitem{Reichel1999}
J.~Reichel, W.~H\"{a}nsel, and T.~W. H\"{a}nsch.
\newblock {Atomic Micromanipulation with Magnetic Surface Traps}.
\newblock
  \href{http://journals.aps.org/prl/pdf/10.1103/PhysRevLett.83.3398}{\it Phys.
  Rev. Lett. \bf83\rm, 3398-3401 (1999).}

\bibitem{Folman2000}
R.~Folman, P.~Kr{\"u}ger, D.~Cassettari, B.~Hessmo, T.~Maier, and
  J.~Schmiedmayer.
\newblock {Controlling Cold Atoms using Nanofabricated Surfaces: Atom Chips}.
\newblock
  \href{http://journals.aps.org/prl/pdf/10.1103/PhysRevLett.84.4749}{\it Phys.
  Rev. Lett. \bf84\rm, 4749-4752 (2000).}

\bibitem{Dekker2000}
N.~H. Dekker, C.~S. Lee, V.~Lorent, J.~H. Thywissen, S.~P. Smith,
  M.~Drndi{\'c}, R.~M. Westervelt, and M.~Prentiss.
\newblock {Guiding Neutral Atoms on a Chip}.
\newblock
  \href{http://journals.aps.org/prl/pdf/10.1103/PhysRevLett.84.1124}{\it Phys.
  Rev. Lett. \bf84\rm, 1124-1127 (2000).}

\bibitem{Ott2001}
H.~Ott, J.~Fortagh, G.~Schlotterbeck, A.~Grossmann, and C.~Zimmermann.
\newblock {Bose-Einstein Condensation in a Surface Microtrap}.
\newblock
  \href{http://journals.aps.org/prl/pdf/10.1103/PhysRevLett.87.230401}{\it
  Phys. Rev. Lett. \bf87\rm, 230401 (2001).}

\bibitem{Hansel2001}
W.~H\"{a}nsel, P.~Hommelhoff, T.~W. H\"{a}nsch, and J.~Reichel.
\newblock {Bose-Einstein condensation on a microelectronic chip}.
\newblock
  \href{http://www.nature.com/nature/journal/v413/n6855/full/413498a0.html}{\it
  Nature \bf413\rm, 498-501 (2001).}

\bibitem{Altschul2015}
B.~Altschul, Q.~G. Bailey, L.~Blanchet, K.~Bongs, P.~Bouyer, L.~Cacciapuoti,
  S.~Capozziello, N.~Gaaloul, D.~Giulini, J.~Hartwig, L.~Iess, P.~Jetzer,
  A.~Landragin, E.~Rasel, S.~Reynaud, S.~Schiller, C.~Schubert, F.~Sorrentino,
  U.~Sterr, J.~D. Tasson, G.~M. Tino, P.~Tuckey, and P.~Wolf.
\newblock {Quantum tests of the Einstein Equivalence Principle with the
  STE-QUEST space mission}.
\newblock
  \href{http://www.sciencedirect.com/science/article/pii/S0273117714004384}{\it
  Adv. Space Res. \bf55\rm, 501-524 (2015).}

\bibitem{JPL2015}
I.~Hahn, N.~Yu, U.~Israelsson, and J.~Goree.
\newblock {Cold Atom Laboratory}.
\newblock
  \href{http://www.nasa.gov/sites/default/files/atoms/files/np-2015-04-021-jsc_fundamental_physics-iss-mini-book-508.pdf}{\it
  International Space Station \rm(Houston, TX, 2015) pp.~24-26.}

\bibitem{Bongs2015}
K.~Bongs, Y.~Singh, L.~Smith, W.~He, O.~Kock, D.~{\'S}wierad, J.~Hughes,
  S.~Schiller, S.~Alighanbari, S.~Origlia, S.~Vogt, U.~Sterr, C.~Lisdat, R.~{Le
  Targat}, J.~Lodewyck, D.~Holleville, B.~Venon, S.~Bize, G.~P. Barwood,
  P.~Gill, I.~R. Hill, Y.~B. Ovchinnikov, N.~Poli, G.~M. Tino, J.~Stuhler, and
  W.~Kaenders.
\newblock {Development of a strontium optical lattice clock for the SOC mission
  on the ISS}.
\newblock
  \href{http://www.sciencedirect.com/science/article/pii/S1631070515000602}{\it
  Comptes Rendus Phys. \bf16\rm, 553-564 (2015).}

\bibitem{Hinds1999}
E.~A. Hinds and I.~G. Hughes.
\newblock {Magnetic atom optics: mirrors, guides, traps, and chips for atoms}.
\newblock
  \href{http://iopscience.iop.org/article/10.1088/0022-3727/32/18/201/pdf}{\it
  J. Phys. D \bf32\rm, R119-R146 (1999).}

\bibitem{Folman2002}
R.~Folman, P.~Kr{\"u}ger, J.~Schmiedmayer, J.~Denschlag, and C.~Henkel.
\newblock {Microscopic atom optics: from wires to an atom chip}.
\newblock
  \href{http://www.sciencedirect.com/science/article/pii/S1049250X02800118}{\it
  Adv. At. Mol. Opt. Phys. \bf48\rm, 263-356 (2002).}

\bibitem{Reichel2002}
J.~Reichel.
\newblock {Microchip traps and Bose-Einstein condensation}.
\newblock \href{http://link.springer.com/article/10.1007/s003400200861}{\it
  Appl. Phys. B \bf74\rm, 469-487 (2002).}

\bibitem{Fortagh2007}
J.~Fort\'agh and C.~Zimmermann.
\newblock {Magnetic microtraps for ultracold atoms}.
\newblock \href{http://journals.aps.org/rmp/pdf/10.1103/RevModPhys.79.235}{\it
  Rev. Mod. Phys. \bf79\rm, 235-289 (2007).}

\bibitem{Reichel2011x}
{\it Atom Chips, \rm J. Reichel and V. Vuletic, eds.}
\newblock
  \href{http://eu.wiley.com/WileyCDA/WileyTitle/productCd-3527407553,subjectCd-PHG1.html}{(Wiley-VCH:
  Hoboken, NJ, 2011).}

\bibitem{DLR2015}
\href{http://www.dlr-gsoc.de/moraba/}{\tt http://www.dlr-gsoc.de/moraba/}.

\bibitem{Jones2003}
M.~P.~A. Jones, C.~J. Vale, D.~Sahagun, B.~V. Hall, and E.~A. Hinds.
\newblock {Spin Coupling between Cold Atoms and the Thermal Fluctuations of a
  Metal Surface}.
\newblock
  \href{http://journals.aps.org/prl/pdf/10.1103/PhysRevLett.91.080401}{\it
  Phys. Rev. Lett. \bf91\rm, 080401 (2003).}

\bibitem{Lin2004}
Y.~Lin, I.~Teper, C.~Chin, and V.~Vuleti{\'c}.
\newblock {Impact of the Casimir-Polder Potential and Johnson Noise on
  Bose-Einstein Condensate Stability Near Surfaces}.
\newblock
  \href{http://journals.aps.org/prl/pdf/10.1103/PhysRevLett.92.050404}{\it
  Phys. Rev. Lett. \bf92\rm, 050404 (2004).}

\bibitem{Emmert2009}
A.~Emmert, A.~Lupa{\c s}cu, G.~Nogues, M.~Brune, J.-M. Raimond, and S.~Haroche.
\newblock {Measurement of the trapping lifetime close to a cold metallic
  surface on a cryogenic atom-chip}.
\newblock
  \href{http://link.springer.com/article/10.1140/epjd/e2009-00001-5}{\it Eur.
  Phys. J. D \bf51\rm, 173-177 (2009).}

\bibitem{Aigner2008}
S.~Aigner, L.~{Della Pietra}, Y.~Japha, O.~Entin-Wohlman, T.~David, R.~Salem,
  R.~Folman, and J.~Schmiedmayer.
\newblock {Long-Range Order in Electronic Transport Through Disordered Metal
  Films}.
\newblock \href{http://www.sciencemag.org/content/319/5867/1226.full.pdf}{\it
  Science \bf319\rm, 1226-1229 (2008).}

\bibitem{Harber2005}
D.~M. Harber, J.~M. Obrecht, J.~M. McGuirk, and E.~A. Cornell.
\newblock {Measurement of the Casimir-Polder force through center-of-mass
  oscillations of a Bose-Einstein condensate}.
\newblock \href{http://journals.aps.org/pra/pdf/10.1103/PhysRevA.72.033610}{\it
  Phys. Rev. A \bf72\rm, 033610 (2005).}

\bibitem{Obrecht2007a}
J.~M. Obrecht, R.~J. Wild, M.~Antezza, L.~P. Pitaevskii, S.~Stringari, and
  E.~A. Cornell.
\newblock {Measurement of the Temperature Dependence of the Casimir-Polder
  Force}.
\newblock
  \href{http://journals.aps.org/prl/pdf/10.1103/PhysRevLett.98.063201}{\it
  Phys. Rev. Lett. \bf98\rm, 063201 (2007).}

\bibitem{Pasquini2006}
T.~A. Pasquini, M.~Saba, G.-B. Jo, Y.~Shin, W.~Ketterle, D.~E. Pritchard, T.~A.
  Savas, and N.~Mulders.
\newblock {Low Velocity Quantum Reflection of Bose-Einstein Condensates}.
\newblock
  \href{http://journals.aps.org/prl/pdf/10.1103/PhysRevLett.97.093201}{\it
  Phys. Rev. Lett. \bf97\rm, 093201 (2006).}

\bibitem{Esteve2006}
J.~Est{\`e}ve, J.-B. Trebbia, T.~Schumm, A.~Aspect, C.~I. Westbrook, and
  I.~Bouchoule.
\newblock {Observations of Density Fluctuations in an Elongated Bose Gas: Ideal
  Gas and Quasicondensate Regimes}.
\newblock
  \href{http://journals.aps.org/prl/pdf/10.1103/PhysRevLett.96.130403}{\it
  Phys. Rev. Lett. \bf96\rm, 130403 (2006).}

\bibitem{Yuen2015}
B.~Yuen, I.~J.~M. Barr, J.~P. Cotter, E.~Butler, and E.~A. Hinds.
\newblock {Enhanced oscillation lifetime of a Bose-Einstein condensate in the
  3D/1D crossover}.
\newblock
  \href{http://iopscience.iop.org/article/10.1088/1367-2630/17/9/093041/pdf}{\it
  New J. Phys. \bf17\rm, 093041 (2015).}

\bibitem{Rauer2016}
B.~Rauer, P.~Gri{\v s}ins, I.~E. Mazets, T.~Schweigler, W.~Rohringer,
  R.~Geiger, T.~Langen, and J.~Schmiedmayer.
\newblock {Cooling of a one-dimensional Bose gas}.
\newblock
  \href{http://journals.aps.org/prl/pdf/10.1103/PhysRevLett.116.030402}{\it
  Phys. Rev. Lett. \bf116\rm, 030402 (2016).}

\bibitem{Hofferberth2007}
S.~Hofferberth, I.~Lesanovsky, B.~Fischer, T.~Schumm, and J.~Schmiedmayer.
\newblock {Non-equilibrium coherence dynamics in one-dimensional Bose gases}.
\newblock
  \href{http://www.nature.com/nature/journal/v449/n7160/abs/nature06149.html}{\it
  Nature \bf449\rm, 324-327 (2007).}

\bibitem{Gring2012}
M.~Gring, M.~Kuhnert, T.~Langen, T.~Kitagawa, B.~Rauer, M.~Schreitl, I.~Mazets,
  D.~{Adu Smith}, E.~Demler, and J.~Schmiedmayer.
\newblock {Relaxation and Prethermalization in an Isolated Quantum System}.
\newblock \href{http://www.sciencemag.org/content/337/6100/1318.full.pdf}{\it
  Science \bf337\rm, 1318-1322 (2012).}

\bibitem{Langen2015}
T.~Langen, S.~Erne, R.~Geiger, B.~Rauer, T.~Schweigler, M.~Kuhnert,
  W.~Rohringer, I.~E. Mazets, T.~Gasenzer, and J.~Schmiedmayer.
\newblock {Experimental observation of a generalized Gibbs ensemble}.
\newblock \href{https://www.sciencemag.org/content/348/6231/207.full.pdf}{\it
  Science \bf348\rm, 207-211 (2015).}

\bibitem{Riedel2010}
M.~F. Riedel, P.~B\"{o}hi, Y.~Li, T.~W. H\"{a}nsch, A.~Sinatra, and
  P.~Treutlein.
\newblock {Atom-chip-based generation of entanglement for quantum metrology}.
\newblock
  \href{http://www.nature.com/nature/journal/v464/n7292/pdf/nature08988.pdf}{\it
  Nature \bf464\rm, 1170-1173 (2010).}

\bibitem{Haas2014}
F.~Haas, J.~Volz, R.~Gehr, J.~Reichel, and J.~Est{\`e}ve.
\newblock {Entangled States of More Than 40 Atoms in an Optical Fiber Cavity}.
\newblock \href{https://www.sciencemag.org/content/344/6180/180.full.pdf}{\it
  Science \bf344\rm, 180-183 (2014).}

\bibitem{Barontini2015}
G.~Barontini, L.~Hohmann, F.~Haas, J.~Est{\`e}ve, and J.~Reichel.
\newblock {Deterministic generation of multiparticle entanglement by quantum
  Zeno dynamics}.
\newblock \href{http://www.sciencemag.org/content/344/6180/160.full.pdf}{\it
  Science \bf349\rm, 1317-1321 (2015).}

\bibitem{Maussang2010}
K.~Maussang, G.~E. Marti, T.~Schneider, P.~Treutlein, Y.~Li, A.~Sinatra,
  R.~Long, J.~Est{\`e}ve, and J.~Reichel.
\newblock {Enhanced and Reduced Atom Number Fluctuations in a BEC Splitter}.
\newblock
  \href{http://journals.aps.org/prl/pdf/10.1103/PhysRevLett.105.080403}{\it
  Phys. Rev. Lett. \bf105\rm, 080403 (2010).}

\bibitem{Deutsch2010}
C.~Deutsch, F.~Ramirez-Martinez, C.~Lacro{\^u}te, F.~Reinhard, T.~Schneider,
  J.~N. Fuchs, F.~Pi{\'e}chon, F.~Lalo{\"e}, J.~Reichel, and P.~Rosenbusch.
\newblock {Spin Self-Rephasing and Very Long Coherence Times in a Trapped
  Atomic Ensemble}.
\newblock
  \href{http://journals.aps.org/prl/pdf/10.1103/PhysRevLett.105.020401}{\it
  Phys. Rev. Lett. \bf105\rm, 020401 (2010).}

\bibitem{Margalit2015}
Y.~Margalit, Z.~Zhou, S.~Machluf, D.~Rohrlich, Y.~Japha, and R.~Folman.
\newblock {A self-interfering clock as a which-path witness}.
\newblock \href{https://www.sciencemag.org/content/349/6253/1205.full.pdf}{\it
  Science \bf349\rm, 1205-1208 (2015).}

\bibitem{Trotzky2015}
S.~Trotzky, S.~Beattie, C.~Luciuk, S.~Smale, A.~B. Bardon, T.~Enss, E.~Taylor,
  S.~Zhang, and J.~H. Thywissen.
\newblock {Observation of the Leggett-Rice effect in a unitary Fermi gas}.
\newblock
  \href{http://journals.aps.org/prl/pdf/10.1103/PhysRevLett.114.015301}{\it
  Phys. Rev. Lett. \bf114\rm, 015301 (2015).}

\bibitem{vanZoest2010}
T.~{van Zoest}, N.~Gaaloul, Y.~Singh, H.~Ahlers, W.~Herr, S.~T. Seidel,
  W.~Ertmer, E.~Rasel, M.~Eckart, E.~Kajari, S.~Arnold, G.~Nandi, W.~P.
  Schleich, R.~Walser, A.~Vogel, K.~Sengstock, K.~Bongs, W.~Lewoczko-Adamczyk,
  M.~Schiemangk, T.~Schuldt, A.~Peters, T.~K\"onemann, H.~M\"untinga,
  C.~L\"ammerzahl, H.~Dittus, T.~Steinmetz, T.~W. H\"ansch, and J.~Reichel.
\newblock {Bose-Einstein Condensation in Microgravity}.
\newblock \href{http://www.sciencemag.org/content/328/5985/1540.full.pdf/}{\it
  Science \bf328\rm, 1540-1543 (2010).}

\bibitem{Wolf2007}
P.~Wolf, P.~Lemonde, A.~Lambrecht, S.~Bize, A.~Landragin, and A.~Clairon.
\newblock {From optical lattice clocks to the measurement of forces in the
  Casimir regime}.
\newblock \href{http://journals.aps.org/pra/pdf/10.1103/PhysRevA.75.063608}{\it
  Phys. Rev. A \bf75\rm, 063608 (2007).}

\bibitem{Sorrentino2009}
F.~Sorrentino, A.~Alberti, G.~Ferrari, V.~V. Ivanov, N.~Poli, M.~Schioppo, and
  G.~M. Tino.
\newblock {Quantum sensor for atom-surface interactions below $\rm10\,\mum$}.
\newblock \href{http://journals.aps.org/pra/pdf/10.1103/PhysRevA.79.013409}{\it
  Phys. Rev. A \bf79\rm, 013409 (2009).}

\bibitem{Henkel1999}
C.~Henkel, S.~P{\"o}tting, and M.~Wilkens.
\newblock {Loss and heating of particles in small and noisy traps}.
\newblock \href{http://link.springer.com/article/10.1007/s003400050823}{\it
  Appl. Phys. B \bf69\rm, 379-387 (1999).}

\bibitem{Henkel2003}
C.~Henkel, P.~Kr{\"u}ger, R.~Folman, and J.~Schmiedmayer.
\newblock {Fundamental limits for coherent manipulation on atom chips}.
\newblock \href{http://link.springer.com/article/10.1007/s00340-003-1112-z}{\it
  Appl. Phys. B \bf76\rm, 173-182 (2003).}

\bibitem{Wang2004}
D.-W. Wang, M.~D. Lukin, and E.~Demler.
\newblock {Disordered Bose-Einstein Condensates in Quasi-One-Dimensional
  Magnetic Microtraps}.
\newblock
  \href{http://journals.aps.org/prl/pdf/10.1103/PhysRevLett.92.076802}{\it
  Phys. Rev. Lett. \bf92\rm, 076802 (2004).}

\bibitem{Japha2008}
Y.~Japha, O.~Entin-Wohlman, T.~David, R.~Salem, S.~Aigner, J.~Schmiedmayer, and
  R.~Folman.
\newblock {Model for organized current patterns in disordered conductors}.
\newblock \href{http://journals.aps.org/prb/pdf/10.1103/PhysRevB.77.201407}{\it
  Phys. Rev. B \bf77\rm, 201407(R) (2008).}

\bibitem{Dellabetta2012}
B.~Dellabetta, T.~L. Hughes, M.~J. Gilbert, and B.~L. Lev.
\newblock {Imaging topologically protected transport with quantum degenerate
  gases}.
\newblock \href{http://journals.aps.org/prb/pdf/10.1103/PhysRevB.85.205442}{\it
  Phys. Rev. B \bf85\rm, 205442 (2012).}

\bibitem{Haakh2014}
H.~R. Haakh, C.~Henkel, S.~Spagnolo, L.~Rizzuto, and R.~Passante.
\newblock {Dynamical Casimir-Polder interaction between an atom and surface
  plasmons}.
\newblock \href{http://journals.aps.org/pra/pdf/10.1103/PhysRevA.89.022509}{\it
  Phys. Rev. A \bf89\rm, 022509 (2014).}

\bibitem{Burkov2007}
A.~A. Burkov, M.~D. Lukin, and E.~Demler.
\newblock {Decoherence Dynamics in Low-Dimensional Cold Atom Interferometers}.
\newblock
  \href{http://journals.aps.org/prl/pdf/10.1103/PhysRevLett.98.200404}{\it
  Phys. Rev. Lett. \bf98\rm, 200404 (2007).}

\bibitem{LlorenteGarcia2013}
I.~{Llorente Garc{\'i}a}, B.~Darqui{\'e}, C.~D.~J. Sinclair, E.~A. Curtis,
  M.~Tachikawa, J.~J. Hudson, and E.~A. Hinds.
\newblock {Shaking-induced dynamics of cold atoms in magnetic traps}.
\newblock \href{http://journals.aps.org/pra/pdf/10.1103/PhysRevA.88.043406}{\it
  Phys. Rev. A \bf88\rm, 043406 (2013).}

\bibitem{Japha2016x}
Y.~Japha, S.~Zhou, M.~Keil, R.~Folman, C.~Henkel, and A.~Vardi.
\newblock {Suppression and enhancement of decoherence in an atomic Josephson
  junction}.
\newblock \href{http://arxiv.org/pdf/1511.00173v1.pdf}{\tt arXiv:1511.00173v1
  \rm(2015).} ({\it New J. Phys.,} \rm in press (2016).).

\bibitem{Charron2006}
E.~Charron, M.~A. Cirone, A.~Negretti, J.~Schmiedmayer, and T.~Calarco.
\newblock {Theoretical analysis of a realistic atom-chip quantum gate}.
\newblock \href{http://journals.aps.org/pra/pdf/10.1103/PhysRevA.74.012308}{\it
  Phys. Rev. A \bf74\rm, 012308 (2006).}

\bibitem{Treutlein2006}
P.~Treutlein, T.~W. H{\"a}nsch, J.~Reichel, A.~Negretti, M.~A. Cirone, and
  T.~Calarco.
\newblock {Microwave potentials and optimal control for robust quantum gates on
  an atom chip}.
\newblock \href{http://journals.aps.org/pra/pdf/10.1103/PhysRevA.74.022312}{\it
  Phys. Rev. A \bf74\rm, 022312 (2006).}

\bibitem{Muller2011}
M.~M. M{\"u}ller, H.~R. Haakh, T.~Calarco, C.~P. Koch, and C.~Henkel.
\newblock {Prospects for fast Rydberg gates on an atom chip, in: \it Special
  Issue on Neutral Particles, \rm R. Folman, ed.}
\newblock
  \href{http://link.springer.com/article/10.1007%2Fs11128-011-0296-0}{\it
  Quantum Inf. Process. \bf10\rm, 771-792 (2011).}

\bibitem{Petrosyan2009}
D.~Petrosyan, G.~Bensky, G.~Kurizki, I.~Mazets, J.~Majer, and J.~Schmiedmayer.
\newblock {Reversible state transfer between superconducting qubits and atomic
  ensembles}.
\newblock \href{http://journals.aps.org/pra/pdf/10.1103/PhysRevA.79.040304}{\it
  Phys. Rev. A \bf79\rm, 040304(R) (2009).}

\bibitem{Scheel2007}
S.~Scheel, R.~Fermani, and E.~A. Hinds.
\newblock {Feasibility of studying vortex noise in two-dimensional
  superconductors with cold atoms}.
\newblock \href{http://journals.aps.org/pra/pdf/10.1103/PhysRevA.75.064901}{\it
  Phys. Rev. A \bf75\rm, 064901 (2007).}

\bibitem{Hohenester2007}
U.~Hohenester, A.~Eiguren, S.~Scheel, and E.~A. Hinds.
\newblock {Spin-flip lifetimes in superconducting atom chips:
  Bardeen-Cooper-Schrieffer versus Eliashberg theory}.
\newblock \href{http://journals.aps.org/pra/pdf/10.1103/PhysRevA.76.033618}{\it
  Phys. Rev. A \bf76\rm, 033618 (2007).}

\bibitem{Sokolovsky2014a}
V.~Sokolovsky, L.~Prigozhin, and J.~W. Barrett.
\newblock {3D modeling of magnetic atom traps on type-II superconductor chips}.
\newblock
  \href{http://iopscience.iop.org/article/10.1088/0953-2048/27/12/124004/pdf}{\it
  Supercond. Sci. Tech. \bf27\rm, 124004 (2014).}

\bibitem{Sokolovsky2014b}
V.~Sokolovsky, D.~Rohrlich, and B.~Horovitz.
\newblock {Trapping neutral atoms in the field of a vortex pinned by a
  superconducting nano-disc}.
\newblock \href{http://journals.aps.org/pra/pdf/10.1103/PhysRevA.89.053422}{\it
  Phys. Rev. A \bf89\rm, 053422 (2014).}

\bibitem{Dikovsky2005}
V.~Dikovsky, Y.~Japha, C.~Henkel, and R.~Folman.
\newblock {Reduction of magnetic noise in atom chips by material optimization}.
\newblock
  \href{http://link.springer.com/article/10.1140%2Fepjd%2Fe2005-00203-9#page-1}{\it
  Eur. Phys. J. D \bf35\rm, 87-95 (2005).}

\bibitem{Fermani2007}
R.~Fermani, S.~Scheel, and P.~L. Knight.
\newblock {Trapping cold atoms near carbon nanotubes: Thermal spin flips and
  Casimir-Polder potential}.
\newblock \href{http://journals.aps.org/pra/pdf/10.1103/PhysRevA.75.062905}{\it
  Phys. Rev. A \bf73\rm, 062905 (2007).}

\bibitem{David2008}
T.~David, Y.~Japha, V.~Dikovsky, R.~Salem, C.~Henkel, and R~Folman.
\newblock {Magnetic interactions of cold atoms with anisotropic conductors}.
\newblock
  \href{http://link.springer.com/article/10.1140%2Fepjd%2Fe2008-00119-x#/page-1}{\it
  Eur. Phys. J. D \bf148\rm, 321-332 (2008).}

\bibitem{Petrov2009}
P.~G. Petrov, S.~Machluf, S.~Younis, R.~Macaluso, T.~David, B.~Hadad, Y.~Japha,
  M.~Keil, E.~Joselevich, and R.~Folman.
\newblock {Trapping cold atoms using surface-grown carbon nanotubes}.
\newblock \href{http://journals.aps.org/pra/pdf/10.1103/PhysRevA.79.043403}{\it
  Phys. Rev. A \bf79\rm, 043403 (2009).}

\bibitem{Jha2015}
P.~K. Jha, X.~Ni, C.~Wu, Y.~Wang, and X.~Zhang.
\newblock {Metasurface-Enabled Remote Quantum Interference}.
\newblock
  \href{http://journals.aps.org/prl/pdf/10.1103/PhysRevLett.115.025501}{\it
  Phys. Rev. Lett. \bf115\rm, 025501 (2015).}

\bibitem{Japha2007}
Y.~Japha, O.~Arzouan, Y.~Avishai, and R.~Folman.
\newblock {Using Time-Reversal Symmetry for Sensitive Incoherent Matter-Wave
  Sagnac Interferometry}.
\newblock
  \href{http://journals.aps.org/prl/pdf/10.1103/PhysRevLett.99.060402}{\it
  Phys. Rev. Lett. \bf99\rm, 060402 (2007).}

\bibitem{Guarrera2015}
V.~Guarrera, R.~Szmuk, J.~Reichel, and P.~Rosenbusch.
\newblock {Microwave-dressed state-selective potentials for atom
  interferometry}.
\newblock
  \href{http://iopscience.iop.org/article/10.1088/1367-2630/17/8/083022/pdf}{\it
  New J. Phys. \bf17\rm, 083022 (2015).}

\bibitem{Stevenson2015}
R.~Stevenson, M.~R. Hush, T.~Bishop, I.~Lesanovsky, and T.~Fernholz.
\newblock {Sagnac Interferometry with a Single Atomic Clock}.
\newblock
  \href{http://journals.aps.org/prl/pdf/10.1103/PhysRevLett.115.163001}{\it
  Phys. Rev. Lett. \bf115\rm, 163001 (2015).}

\bibitem{Ammar2015}
M.~Ammar, M.~Dupont-Nivet, L.~Huet, J.-P. Pocholle, P.~Rosenbusch,
  I.~Bouchoule, C.~I. Westbrook, J.~Est{\`e}ve, J.~Reichel, C.~Guerlin, and
  S.~Schwartz.
\newblock {Symmetric microwave potentials for interferometry with thermal atoms
  on a chip}.
\newblock \href{http://journals.aps.org/pra/pdf/10.1103/PhysRevA.91.053623}{\it
  Phys. Rev. A \bf91\rm, 053623 (2015).}

\bibitem{Nest2010}
M.~Nest, Y.~Japha, R.~Folman, and R.~Kosloff.
\newblock {Dynamic Matter-Wave Pulse Shaping}.
\newblock \href{http://journals.aps.org/pra/pdf/10.1103/PhysRevA.81.043632}{\it
  Phys. Rev. A \bf81\rm, 043632 (2010).}

\bibitem{Stickney2007}
J.~A. Stickney, D.~Z. Anderson, and A.~A. Zozulya.
\newblock {Transistorlike behavior of a Bose-Einstein condensate in a
  triple-well potential}.
\newblock \href{http://journals.aps.org/pra/pdf/10.1103/PhysRevA.75.013608}{\it
  Phys. Rev. A \bf75\rm, 013608 (2007).}

\bibitem{Pepino2009}
R.~A. Pepino, J.~Cooper, D.~Z. Anderson, and M.~J. Holland.
\newblock {Atomtronic Circuits of Diodes and Transistors}.
\newblock
  \href{http://journals.aps.org/prl/pdf/10.1103/PhysRevLett.103.140405}{\it
  Phys. Rev. Lett. \bf103\rm, 140405 (2009).}

\bibitem{Chow2015}
W.~W. Chow, C.~J.~E. Straatsma, and D.~Z. Anderson.
\newblock {Numerical model for atomtronic circuit analysis}.
\newblock \href{http://journals.aps.org/pra/pdf/10.1103/PhysRevA.92.013621}{\it
  Phys. Rev. A \bf92\rm, 013621 (2015).}

\bibitem{Das2009}
K.~K. Das and S.~Aubin.
\newblock {Quantum Pumping with Ultracold Atoms on Microchips: Fermions versus
  Bosons}.
\newblock
  \href{http://journals.aps.org/prl/pdf/10.1103/PhysRevLett.103.123007}{\it
  Phys. Rev. Lett. \bf103\rm, 123007 (2009).}

\bibitem{Simpson2014}
D.~P. Simpson, D.~M. Gangardt, I.~V. Lerner, and P.~Kr{\"u}ger.
\newblock {One-Dimensional Transport of Bosons between Weakly Linked
  Reservoirs}.
\newblock
  \href{http://journals.aps.org/prl/pdf/10.1103/PhysRevLett.112.100601}{\it
  Phys. Rev. Lett. \bf112\rm, 100601 (2014).}

\bibitem{Wu2010x}
B.~Wu, J.~F. Hulbert, E.~J. Lunt, K.~Hurd, A.~R. Hawkins, and H.~Schmidt.
\newblock {Slow light on a chip via atomic quantum state control}.
\newblock
  \href{http://www.nature.com/nphoton/journal/v4/n11/pdf/nphoton.2010.211.pdf}{\it
  Nature Photon. \bf4\rm, 776-779 (2010).}

\bibitem{Stern2013}
L.~Stern, B.~Desiatov, I.~Goykhman, and U.~Levy.
\newblock {Nanoscale light–matter interactions in atomic cladding
  waveguides}.
\newblock
  \href{http://www.nature.com/ncomms/journal/v4/n3/pdf/ncomms2554.pdf}{\it
  Nature Commun. \bf4\rm, 1548 (2013).}

\bibitem{Knappe2005}
S.~Knappe, P.~D.~D. Schwindt, V.~Shah, L.~Hollberg, J.~Kitching, L.~Liew, and
  J.~Moreland.
\newblock {A chip-scale atomic clock based on \Rb\ with improved frequency
  stability}.
\newblock
  \href{https://www.osapublishing.org/oe/abstract.cfm?uri=oe-13-4-1249}{\it
  Opt. Express \bf13\rm, 1249-1253 (2005).}

\bibitem{Jimenez-Martinez2014}
R.~Jim{\'e}nez-Mart{\'i}nez, D.~J. Kennedy, M.~Rosenbluh, E.~A. Donley,
  S.~Knappe, S.~J. Seltzer, H.~L. Ring, V.~S. Bajaj, and J.~Kitching.
\newblock {Optical hyperpolarization and NMR detection of \Xe\ on a
  microfluidic chip}.
\newblock
  \href{http://www.nature.com/ncomms/2014/140520/ncomms4908/pdf/ncomms4908.pdf}{\it
  Nature Commun. \bf5\rm, 3908 (2014).}

\bibitem{Zhu2014}
X.~Zhu, Y.~Matsuzaki, R.~Ams{\"u}ss, K.~Kakuyanagi, T.~Shimo-Oka, N.~Mizuochi,
  K.~Nemoto, K.~Semba, W.~J. Munro, and S.~Saito.
\newblock {Observation of dark states in a superconductor diamond quantum
  hybrid system}.
\newblock
  \href{http://www.nature.com/ncomms/2014/140408/ncomms4524/pdf/ncomms4524.pdf}{\it
  Nature Commun. \bf5\rm, 3524 (2014).}

\bibitem{You2014}
J.-B. You, W.~L. Yang, Z.-Y. Xu, A.~H. Chan, and C.~H. Oh.
\newblock {Phase transition of light in circuit-QED lattices coupled to
  nitrogen-vacancy centers in diamond}.
\newblock \href{http://journals.aps.org/prb/pdf/10.1103/PhysRevB.90.195112}{\it
  Phys. Rev. B \bf90\rm, 195112 (2014).}

\bibitem{Blatt2008}
R.~Blatt and D.~Wineland.
\newblock {Entangled states of trapped atomic ions}.
\newblock
  \href{http://www.nature.com/nature/journal/v453/n7198/pdf/nature07125.pdf}{\it
  Nature \bf453\rm, 1008-1015 (2008).}

\bibitem{Amini2011x}
J.~M. Armini, J.~Britton, D.~Leibfried, and D.~J. Wineland.
\newblock {Micro-Fabricated Chip Traps for Ions, Chapter~13 in: \it Atom Chips,
  \rm J. Reichel and V. Vuletic, eds.}
\newblock
  \href{http://onlinelibrary.wiley.com/doi/10.1002/9783527633357.ch13/summary}{(Wiley-VCH:
  Hoboken, NJ, 2011) pp.~395-420.}

\bibitem{Monroe2013}
C.~Monroe and J.~Kim.
\newblock {Scaling the Ion Trap Quantum Processor}.
\newblock \href{http://www.sciencemag.org/content/339/6124/1164.full.pdf}{\it
  Science \bf339\rm, 1164-1169 (2013).}

\bibitem{Cho2015}
D.-I.~D. Cho, S.~Hong, M.~Lee, and T.~Kim.
\newblock {A review of silicon microfabricated ion traps for quantum
  information processing}.
\newblock
  \href{http://www.mnsl-journal.com/content/pdf/s40486-015-0013-3.pdf}{\it
  Micro. Nano. Systems Lett. \bf3\rm, 2 (2015).}

\bibitem{Gallagher1994}
T.~F. Gallagher.
\newblock {\it Rydberg Atoms}.
\newblock \href{http://www.book-info.com/isbn/0-521-02166-9.htm}{(Cambridge
  Univ. Press: Cambridge, UK, 1994).}

\bibitem{Carter2012}
J.~D. Carter, O.~Cherry, and J.~D.~D. Martin.
\newblock {Electric-field sensing near the surface microstructure of an atom
  chip using cold Rydberg atoms}.
\newblock \href{http://journals.aps.org/pra/pdf/10.1103/PhysRevA.86.053401}{\it
  Phys. Rev. A \bf86\rm, 053401 (2012).}

\bibitem{McGuirk2004}
J.~M. McGuirk, D.~M. Harber, J.~M. Obrecht, and E.~A. Cornell.
\newblock {Alkali-metal adsorbate polarization on conducting and insulating
  surfaces probed with Bose-Einstein condensates}.
\newblock \href{http://journals.aps.org/pra/pdf/10.1103/PhysRevA.69.062905}{\it
  Phys. Rev. A \bf69\rm, 062905 (2004).}

\bibitem{Tauschinsky2010}
A.~Tauschinsky, R.~M.~T. Thijssen, S.~Whitlock, H.~B. van Linden van~den
  Heuvell, and R.~J.~C. Spreeuw.
\newblock {Spatially resolved excitation of Rydberg atoms and surface effects
  on an atom chip}.
\newblock \href{http://journals.aps.org/pra/pdf/10.1103/PhysRevA.81.063411}{\it
  Phys. Rev. A \bf81\rm, 063411 (2010).}

\bibitem{Hattermann2012}
H.~Hattermann, M.~Mack, F.~Karlewski, F.~Jessen, D.~Cano, and J.~Fort{\'a}gh.
\newblock {Detrimental adsorbate fields in experiments with cold Rydberg gases
  near surfaces}.
\newblock \href{http://journals.aps.org/pra/pdf/10.1103/PhysRevA.86.022511}{\it
  Phys. Rev. A \bf86\rm, 022511 (2012).}

\bibitem{Naber2016}
J.~Naber, S.~Machluf, L.~Torralbo-Campo, M.~L. Soudijn, N.~J. {van Druten},
  H.~B. {van Linden van den Heuvell}, and R.~J.~C. Spreeuw.
\newblock {Adsorbate dynamics on a silica-coated gold surface measured by
  Rydberg Stark spectroscopy}.
\newblock
  \href{http://iopscience.iop.org/article/10.1088/0953-4075/49/9/094005/pdf}{\it
  J. Phys. B \bf49\rm, 094005 (2016).}

\bibitem{Chan2014}
K.~S. Chan, M.~Siercke, C.~Hufnagel, and R.~Dumke.
\newblock {Adsorbate Electric Fields on a Cryogenic Atom Chip}.
\newblock
  \href{http://journals.aps.org/prl/pdf/10.1103/PhysRevLett.112.026101}{\it
  Phys. Rev. Lett. \bf112\rm, 026101 (2014).}

\bibitem{Lukin2001}
M.~D. Lukin, M.~Fleischhauer, R.~Cote, L.~M. Duan, D.~Jaksch, J.~I. Cirac, and
  P.~Zoller.
\newblock {Dipole Blockade and Quantum Information Processing in Mesoscopic
  Atomic Ensembles}.
\newblock
  \href{http://journals.aps.org/prl/pdf/10.1103/PhysRevLett.87.037901}{\it
  Phys. Rev. Lett. \bf87\rm, 037901 (2001).}

\bibitem{Saffman2010}
M.~Saffman, T.~G. Walker, and K.~M{\o}lmer.
\newblock {Quantum information with Rydberg atoms}.
\newblock \href{http://journals.aps.org/rmp/pdf/10.1103/RevModPhys.82.2313}{\it
  Rev. Mod. Phys. \bf82\rm, 2313-2363 (2010).}

\bibitem{Urban2009}
E.~Urban, T.~A. Johnson, T.~Henage, L.~Isenhower, D.~D. Yavuz, T.~G. Walker,
  and M.~Saffman.
\newblock {Observation of Rydberg blockade between two atoms}.
\newblock
  \href{http://www.nature.com/nphys/journal/v5/n2/pdf/nphys1178.pdf}{\it Nature
  Phys. \bf5\rm, 110-114 (2009).}

\bibitem{Gaetan2009}
A.~Ga{\"e}tan, Y.~Miroshnychenko, T.~Wilk, A.~Chotia, M.~Viteau, D.~Comparat,
  P.~Pillet, A.~Browaeys, and P.~Grangier.
\newblock {Observation of collective excitation of two individual atoms in the
  Rydberg blockade regime}.
\newblock
  \href{http://www.nature.com/nphys/journal/v5/n2/pdf/nphys1183.pdf}{\it Nature
  Phys. \bf5\rm, 115-118 (2009).}

\bibitem{Leung2014}
V.~Y.~F. Leung, D.~R.~M. Pijn, H.~Schlatter, L.~Torralbo-Campo, A.~L.~La Rooij,
  G.~B. Mulder, J.~Naber, M.~L. Soudijn, A.~Tauschinsky, C.~Abarbanel,
  B.~Hadad, E.~Golan, R.~Folman, and R.~J.~C. Spreeuw.
\newblock {Magnetic-film atom chip with 10\,$\mum$ period lattices of
  microtraps for quantum information science with Rydberg atoms}.
\newblock
  \href{http://scitation.aip.org/content/aip/journal/rsi/85/5/10.1063/1.4874005}{\it
  Rev. Sci. Instrum. \bf85\rm, 053102 (2014).}

\bibitem{Hermann-Avigliano2014}
C.~Hermann-Avigliano, R.~C. Teixeira, T.~L. Nguyen, T.~Cantat-Moltrecht,
  G.~Nogues, I.~Dotsenko, S.~Gleyzes, J.~M. Raimond, S.~Haroche, and M.~Brune.
\newblock {Long coherence times for Rydberg qubits on a superconducting atom
  chip}.
\newblock \href{http://journals.aps.org/pra/pdf/10.1103/PhysRevA.90.040502}{\it
  Phys. Rev. A \bf90\rm, 040502(R) (2014).}

\bibitem{Carter2013}
J.~D. Carter and J.~D.~D. Martin.
\newblock {Coherent manipulation of cold Rydberg atoms near the surface of an
  atom chip}.
\newblock \href{http://journals.aps.org/pra/pdf/10.1103/PhysRevA.88.043429}{\it
  Phys. Rev. A \bf88\rm, 043429 (2013).}

\bibitem{Teixeira2015}
R.~C. Teixeira, C.~Hermann-Avigliano, T.~L. Nguyen, T.~Cantat-Moltrecht,
  J. M. Raimond, S.~Haroche, S.~Gleyzes, and M.~Brune.
\newblock {Microwaves Probe Dipole Blockade and van der Waals Forces in a Cold
  Rydberg Gas}.
\newblock
  \href{http://journals.aps.org/prl/pdf/10.1103/PhysRevLett.115.013001}{\it
  Phys. Rev. Lett. \bf115\rm, 013001 (2015).}

\bibitem{DeMarco1999}
B.~DeMarco and D.~S. Jin.
\newblock {Onset of Fermi Degeneracy in a Trapped Atomic Gas}.
\newblock \href{http://www.sciencemag.org/content/285/5434/1703.full.pdf}{\it
  Science \bf287\rm, 1703-1706 (1999).}

\bibitem{Truscott2001}
A.~G. Truscott, K.~E. Strecker, W.~I. McAlexander, G.~B. Partridge, and R.~G.
  Hulet.
\newblock {Observation of Fermi Pressure in a Gas of Trapped Atoms}.
\newblock \href{https://www.sciencemag.org/content/291/5513/2570.full.pdf}{\it
  Science \bf291\rm, 2570-2572 (2001).}

\bibitem{Aubin2006}
S.~Aubin, S.~Myrskog, M.~H.~T. Extavour, L.~J. LeBlanc, D.~McKay, A.~Stummer,
  and J.~H. Thywissen.
\newblock {Rapid sympathetic cooling to Fermi degeneracy on a chip}.
\newblock \href{http://www.nature.com/nphys/journal/v2/n6/pdf/nphys309.pdf}{\it
  Nature Phys. \bf2\rm, 384-387 (2006).}

\bibitem{Extavour2009}
M.~H.~T. Extavour.
\newblock {Fermions and Bosons on an Atom Chip}.
\newblock \href{http://arxiv.org/pdf/cond-mat/0609259v1.pdf}{\it Ph.D. Thesis,
  \rm University of Toronto \rm(2009).}

\bibitem{Extavour2011x}
M.~H.~T. Extavour, L.~J. LeBlanc, J.~McKeever, A.~B. Bardon, S.~Aubin,
  S.~Myrskog, T.~Schumm, and J.~H. Thywissen.
\newblock {Fermions on atom chips, Chapter~12 in: \it Atom Chips, \rm V.
  Vuletic and J. Reichel, eds.}
\newblock
  \href{http://eu.wiley.com/WileyCDA/WileyTitle/productCd-3527407553,subjectCd-PHG1.html}{(Wiley-VCH:
  Hoboken, NJ, 2011) pp.~365-394.}

\bibitem{Extavour2006}
M.~H.~T. Extavour, L.~J. LeBlanc, T.~Schumm, B.~Cieslak, S.~Myrskog,
  A.~Stummer, S.~Aubin, and J.~H. Thywissen.
\newblock {Dual-species quantum degeneracy of \K\ and \Rb\ on an atom chip, in:
  \it AtomicPhysics 20, XX International Conference on Atomic Physics, \rm C.
  Roos, H. H{\"a}ffner, and R. Blatt, eds.}
\newblock
  \href{http://scitation.aip.org/content/aip/proceeding/aipcp/10.1063/1.2400654}{AIP
  Conf. Proc. \bf869\rm, 241-249 (2006).}

\bibitem{Bardon2014a}
A.~B. Bardon.
\newblock {Dynamics of a Unitary Fermi Gas}.
\newblock
  \href{http://ultracold.physics.utoronto.ca/reprints/BardonPhDthesis.pdf}{\it
  Ph.D. Thesis, \rm University of Toronto \rm(2014).}

\bibitem{Bardon2014b}
A.~B. Bardon, S.~Beattie, C.~Luciuk, W.~Cairncross, D.~Fine, N.~S. Cheng,
  G.~J.~A. Edge, E.~Taylor, S.~Zhang, S.~Trotzky, and J.~H. Thywissen.
\newblock {Transverse Demagnetization Dynamics of a Unitary Fermi Gas}.
\newblock \href{https://www.sciencemag.org/content/344/6185/722.full.pdf}{\it
  Science \bf344\rm, 722-724 (2014).}

\bibitem{Luciuk2016x}
C.~Luciuk, S.~Trotzky, S.~Smale, Z.~Yu, S.~Zhang, and J.~H. Thywissen.
\newblock {Evidence for universal relations describing a gas with $p$-wave
  interactions}.
\newblock \href{http://arxiv.org/pdf/1505.08151v3.pdf}{\tt arXiv:1505.08151v3
  \rm(2015).} ({\it Nature Phys.,} \rm in press (2016).).

\bibitem{Shuman2010}
E.~S. Shuman, J.~F. Barry, and D.~DeMille.
\newblock {Laser cooling of a diatomic molecule}.
\newblock
  \href{http://www.nature.com/nature/journal/v467/n7317/pdf/nature09443.pdf}{\it
  Nature \bf467\rm, 820-823 (2010).}

\bibitem{Barry2012}
J.~F. Barry, E.~S. Shuman, E.~B. Norrgard, and D.~DeMille.
\newblock {Laser Radiation Pressure Slowing of a Molecular Beam}.
\newblock
  \href{http://journals.aps.org/prl/pdf/10.1103/PhysRevLett.108.103002}{\it
  Phys. Rev. Lett. \bf108\rm, 103002 (2012).}

\bibitem{Zhelyazkova2014}
V.~Zhelyazkova, A.~Cournol, T.~E. Wall, A.~Matsushima, J.~J. Hudson, E.~A.
  Hinds, M.~R. Tarbutt, and B.~E. Sauer.
\newblock {Laser cooling and slowing of CaF molecules}.
\newblock \href{http://journals.aps.org/pra/pdf/10.1103/PhysRevA.89.053416}{\it
  Phys. Rev. A \bf89\rm, 053416 (2014).}

\bibitem{Bethlem2000}
H.~L. Bethlem, G.~Berden, F.~M.~H. Crompvoets, R.~T. Jongma, A.~J.~A. van Roij,
  and G.~Meijer.
\newblock {Electrostatic trapping of ammonia molecules}.
\newblock
  \href{http://www.nature.com/nature/journal/v406/n6795/full/406491A0.html}{\it
  Nature \bf406\rm, 491-494 (2000).}

\bibitem{Rieger2005}
T.~Rieger, T.~Junglen, S.~A. Rangwala, P.~W.~H. Pinkse, and G.~Rempe.
\newblock {Continuous Loading of an Electrostatic Trap for Polar Molecules}.
\newblock
  \href{http://journals.aps.org/prl/pdf/10.1103/PhysRevLett.95.173002}{\it
  Phys. Rev. Lett. \bf95\rm, 173002 (2005).}

\bibitem{Nikitin2003}
E.~Nikitin, E.~Dashevskaya, J.~Alnis, M.~Auzinsh, E.~R.~I. Abraham, B.~R.
  Furneaux, M.~Keil, C.~McRaven, N.~Shafer-Ray, and R.~Waskowsky.
\newblock {Measurement and prediction of the speed-dependent throughput of a
  magnetic octupole velocity filter including nonadiabatic effects}.
\newblock \href{http://journals.aps.org/pra/pdf/10.1103/PhysRevA.68.023403}{\it
  Phys. Rev. A \bf68\rm, 023402 (2003).}

\bibitem{Zeppenfeld2012}
M.~Zeppenfeld, B.~G.~U. Englert, R.~Gl{\"o}ckner, A.~Prehn, M.~Mielenz,
  C.~Sommer, L.~D. {van Buuren}, M.~Motsch, and G.~Rempe.
\newblock {Sisyphus cooling of electrically trapped polyatomic molecules}.
\newblock
  \href{http://www.nature.com/nature/journal/v491/n7425/pdf/nature11595.pdf}{\it
  Nature \bf491\rm, 570-573 (2012).}

\bibitem{Meek2009b}
S.~A. Meek, H.~Conrad, and G.~Meijer.
\newblock {Trapping molecules on a chip}.
\newblock \href{https://www.sciencemag.org/content/324/5935/1699.full.pdf}{\it
  Science \bf324\rm, 1699-1702 (2009).}

\bibitem{Kishimoto2006}
T.~Kishimoto, H.~Hachisu, J.~Fujiki, K.~Nagato, M.~Yasuda, and H.~Katori.
\newblock {Electrodynamic Trapping of Spinless Neutral Atoms with an Atom
  Chip}.
\newblock
  \href{http://journals.aps.org/prl/pdf/10.1103/PhysRevLett.96.123001}{\it
  Phys. Rev. Lett. \bf96\rm, 123001 (2006).}

\bibitem{Meek2008}
S.~A. Meek, H.~L. Bethlem, H.~Conrad, and G.~Meijer.
\newblock {Trapping molecules on a chip in traveling potential wells}.
\newblock
  \href{http://journals.aps.org/prl/pdf/10.1103/PhysRevLett.100.153003}{\it
  Phys. Rev. Lett. \bf100\rm, 153003 (2008).}

\bibitem{Meek2009a}
S.~A. Meek, H.~Conrad, and G.~Meijer.
\newblock {A Stark decelerator on a chip}.
\newblock
  \href{http://iopscience.iop.org/article/10.1088/1367-2630/11/5/055024/pdf}{\it
  New J. Phys. \bf11\rm, 055024 (2009).}

\bibitem{Santambrogio2015}
G.~Santambrogio.
\newblock {Trapping molecules on chips}.
\newblock
  \href{http://link.springer.com/article/10.1140%2Fepjti%2Fs40485-015-0024-8#/page-1}{\it
  Eur. Phys. J. Tech. Instrum. \bf2\rm, 14 (2015).}

\bibitem{Marx2013}
S.~Marx, D.~{Adu Smith}, M.~J. Abel, T.~Zehentbauer, G.~Meijer, and
  G.~Santambrogio.
\newblock {Imaging cold molecules on a chip}.
\newblock
  \href{http://journals.aps.org/prl/pdf/10.1103/PhysRevLett.111.243007}{\it
  Phys. Rev. Lett. \bf111\rm, 243007 (2013).}

\bibitem{Hogan2012}
S.~D. Hogan, P.~Allmendinger, H.~Sa{\ss}mannshausen, H.~Schmutz, and F.~Merkt.
\newblock {Surface-Electrode Rydberg-Stark Decelerator}.
\newblock
  \href{http://journals.aps.org/prl/pdf/10.1103/PhysRevLett.108.063008}{\it
  Phys. Rev. Lett. \bf108\rm, 063008 (2012).}

\bibitem{Schnell2007}
M.~Schnell, P.~L{\"u}tzow, J.~{van Veldhoven}, H.~L. Bethlem, J.~K{\"u}pper,
  B.~Friedrich, M.~Schleier-Smith, H.~Haak, and G.~Meijer.
\newblock {A Linear AC Trap for Polar Molecules in Their Ground State}.
\newblock \href{http://pubs.acs.org/doi/pdf/10.1021/jp070902n}{\it J. Phys.
  Chem. A \bf111\rm, 7411-7419 (2007).}

\bibitem{Chin2010}
C.~Chin, R.~Grimm, P.~Julienne, and E.~Tiesinga.
\newblock {Feshbach resonances in ultracold gases}.
\newblock \href{http://journals.aps.org/rmp/pdf/10.1103/RevModPhys.82.1225}{\it
  Rev. Mod. Phys. \bf82\rm, 1225-1286 (2010).}

\bibitem{Tscherbul2010}
T.~V. Tscherbul, T.~Calarco, I.~Lesanovsky, R.~V. Krems, A.~Dalgarno, and
  J.~Schmiedmayer.
\newblock {rf-field-induced Feshbach resonances}.
\newblock \href{http://journals.aps.org/pra/pdf/10.1103/PhysRevA.81.050701}{\it
  Phys. Rev. A \bf81\rm, 050701 (2010).}

\bibitem{Carr2009}
L.~D. Carr, D.~DeMille, R.~V. Krems, and J.~Ye.
\newblock {Cold and ultracold molecules: science, technology and applications}.
\newblock \href{https://groups.chem.ubc.ca/krems/publications/njp2009b.pdf}{\it
  New J. Phys. \bf11\rm, 055049 (2009).}

\bibitem{Quemener2012}
G.~Qu{\'e}m{\'e}ner and P.~S. Julienne.
\newblock {Ultracold Molecules under Control!}
\newblock \href{http://pubs.acs.org/doi/pdf/10.1021/cr300092g}{\it Chem. Rev.
  \bf112\rm, 4949-5011 (2012).}

\bibitem{Lemeshko2013}
M.~Lemeshko, R.~V. Krems, J.~M. Doyle, and S.~Kaise.
\newblock {Manipulation of molecules with electromagnetic fields}.
\newblock
  \href{http://www.tandfonline.com/doi/pdf/10.1080/00268976.2013.813595}{\it
  Molec. Phys. \bf111\rm, 1648-1682 (2013).}

\bibitem{Villata2011}
M.~Villata.
\newblock {CPT symmetry and antimatter gravity in general relativity}.
\newblock
  \href{http://iopscience.iop.org/article/10.1209/0295-5075/94/20001/pdf}{\it
  Europhys. Lett. \bf94\rm, 20001 (2011).}

\bibitem{Dufour2015}
G.~Dufour, D.~B. Cassidy, P.~Crivelli, P.~Debu, A.~Lambrecht, V.~V.
  Nesvizhevsky, S.~Reynaud, A.~Yu. Voronin, and T.~E. Wall.
\newblock {Prospects for Studies of the Free Fall and Gravitational Quantum
  States of Antimatter}.
\newblock \href{http://www.hindawi.com/journals/ahep/2015/379642/}{\it Adv.
  High Energy Phys. \bf2015\rm, 379642 (2015).}

\bibitem{Indelicato2014}
P.~Indelicato, G.~Chardin, P.~Grandemange, D.~Lunney, V.~Manea, A.~Badertscher,
  P.~Crivelli, A.~Curioni, A.~Marchionni, B.~Rossi, A.~Rubbia, V.~Nesvizhevsky,
  D.~{Brook-Roberge}, P.~Comini, P.~Debu, P.~Dupr{/'e}, L.~Liszkay,
  B.~Mansouli{\'e}, P.~P{\'e}rez, J.-M. Rey, B.~Reymond, N.~Ruiz, Y.~Sacquin,
  B.~Vallage, F.~Biraben, P.~Clad{\'e}, A.~Douillet, G.~Dufour, S.~Guellati,
  L.~Hilico, A.~Lambrecht, R.~Gu{\'e}rout, J.-P. Karr, F.~Nez, S.~Reynaud,
  C.~I. Szabo, V.-Q. Tran, J.~Trapateau, A.~Mohri, Y.~Yamazaki, M.~Charlton,
  S.~Eriksson, N.~Madsen, D.~P. {van der Werf}, N.~Kuroda, H.~Torii,
  Y.~Nagashima, F.~{Schmidt-Kaler}, J.~Walz, S.~Wolf, P.-A. Hervieux,
  G.~Manfredi, A.~Voronin, P.~Froelich, S.~Wronka, and M.~Staszczak.
\newblock {The Gbar project, or how does antimatter fall?}
\newblock \href{http://link.springer.com/article/10.1007/s10751-014-1019-6}{\it
  Hyperfine Interact. \bf228\rm, 141-150 (2014).}

\bibitem{Kellerbauer2015}
A.~Kellerbauer.
\newblock {Why Antimatter Matters}.
\newblock
  \href{http://journals.cambridge.org/action/displayAbstract?fromPage=online&aid=9534318&fileId=S1062798714000532}{\it
  Eur. Rev. \bf23\rm, 45-56 (2015).}

\bibitem{Dufour2013}
G.~Dufour, A.~G{\'e}rardin, R.~Gu{\'e}rout, A.~Lambrecht, V.~V. Nesvizhevsky,
  S.~Reynaud, and A.~Yu. Voronin.
\newblock {Quantum reflection of antihydrogen from the Casimir potential above
  matter slabs}.
\newblock \href{http://journals.aps.org/pra/pdf/10.1103/PhysRevA.87.012901}{\it
  Phys. Rev. A \bf87\rm, 012901 (2013).}

\bibitem{Amole2014}
C.~Amole, G.~B. Andresen, M.~D. Ashkezari, M.~Baquero-Ruiz, W.~Bertsche, P.~D.
  Bowe, E.~Butler, A.~Capra, P.~T. Carpenter, C.~L. Cesar, S.~Chapman,
  M.~Charlton, A.~Deller, S.~Eriksson, J.~Escallier, J.~Fajans, T.~Friesen,
  M.~C. Fujiwara, D.~R. Gill, A.~Gutierrez, J.~S. Hangst, W.~N. Hardy, R.~S.
  Hayano, M.~E. Hayden, A.~J. Humphries, J.~L. Hurt, R.~Hydomako, C.~A. Isaac,
  M.~J. Jenkins, S.~Jonsell, L.~V. J{\o}rgensen, S.~J. Kerrigan,
  L.~Kurchaninov, N.~Madsen, A.~Marone, J.~T.~K. McKenna, S.~Menary, P.~Nolan,
  K.~Olchanski, A.~Olin, B.~Parker, A.~Povilus, P.~Pusa, F.~Robicheaux,
  E.~Sarid, D.~Seddon, S.~{Seif El Nasr}, D.~M. Silveira, C.~Sod, J.~W. Storey,
  R.~I. Thompson, J.~Thornhill, D.~Wells, D.~P. {van der Werf}, J.~S. Wurtele,
  and Y.~Yamazaki.
\newblock {The ALPHA antihydrogen trapping apparatus}.
\newblock
  \href{http://www.sciencedirect.com/science/article/pii/S0168900213012771}{\it
  Nucl. Instrum. Meth. A \bf735\rm, 319-340 (2014).}

\bibitem{Scampoli2014x}
P.~Scampoli and J.~Storey.
\newblock {The AEgIS experiment at CERN for the measurement of antihydrogen
  gravity acceleration}.
\newblock
  \href{http://www.worldscientific.com/action/doSearch?AllField=Scampoli&publication=mpla}{\it
  Mod. Phys. Lett. A \bf29\rm, 1430017 (2014).}

\bibitem{Itano1995}
W.~M. Itano, J.~C. Bergquist, J.~J. Bollinger, and D.~J. Wineland.
\newblock {Cooling methods in ion traps}.
\newblock
  \href{http://iopscience.iop.org/article/10.1088/0031-8949/1995/T59/013/pdf}{\it
  Phys. Scripta \bf T59\rm, 106-120 (1995).}

\bibitem{Leefer2016x}
N.~Leefer, K.~Krimmel, W.~Bertsche, D.~Budker, J.~Fajans, R.~Folman,
  H.~Haeffner, and F.~Schmidt-Kaler.
\newblock {Investigation of two-frequency Paul traps for antihydrogen
  production}.
\newblock \href{http://arxiv.org/pdf/1603.09444v1.pdf}{\tt arXiv:1603.09444v1
  \rm(2016).}

\bibitem{Eltony2015x}
A.~M. Eltony, D.~Gangloff, M.~Shi, A.~Bylinskii, V.~Vuleti{\'c}, and I.~L.
  Chuang.
\newblock {Technologies for trapped-ion quantum information systems}.
\newblock \href{http://arxiv.org/pdf/1502.05739v2.pdf}{\tt arXiv:1502.05739v2
  \rm(2015).}

\bibitem{Hite2013}
D.~A. Hite, Y.~Colombe, A.~C. Wilson, D.~T.~C. Allcock, D.~Leibfried, D.~J.
  Wineland, and D.~P. Pappas.
\newblock {Surface science for improved ion traps}.
\newblock
  \href{http://www.nist.gov/pml/div688/grp10/upload/113-MRSBulletin-Ion-Traps.pdf}{\it
  Mater. Res. Soc. Bull. \bf38\rm, 826-833 (2013).}

\bibitem{Stick2006}
D.~Stick, W.~K. Hensinger, S.~Olmschenk, M.J. Madsen, K.~Schwab, and C.~Monroe.
\newblock {Ion trap in a semiconductor chip}.
\newblock \href{http://www.nature.com/nphys/journal/v2/n1/pdf/nphys171.pdf}{\it
  Nature Phys. \bf2\rm, 36-39 (2006).}

\bibitem{Seidelin2006}
S.~Seidelin, J.~Chiaverini, R.~Reichle, J.~J. Bollinger, D.~Leibfried,
  J.~Britton, J.~H. Wesenberg, R.~B. Blakestad, R.~J. Epstein, D.~B. Hume,
  W.~M. Itano, J.~D. Jost, C.~Langer, R.~Ozeri, N.~Shiga, and D.~J. Wineland.
\newblock {Microfabricated Surface-Electrode Ion Trap for Scalable Quantum
  Information Processing}.
\newblock
  \href{http://journals.aps.org/prl/pdf/10.1103/PhysRevLett.96.253003}{\it
  Phys. Rev. Lett. \bf96\rm, 253003 (2006).}

\bibitem{Moehring2011}
D.~L. Moehring, C.~Highstrete, D.~Stick, K.~M. Fortier, R.~Haltli, C.~Tigges,
  and M.~G. Blain.
\newblock {Design, fabrication and experimental demonstration of junction
  surface ion traps}.
\newblock
  \href{http://iopscience.iop.org/article/10.1088/1367-2630/13/7/075018/pdf}{\it
  New J. Phys. \bf13\rm, 075018 (2011).}

\bibitem{Daniilidis2011}
N.~Daniilidis, S.~Narayanan, S.~A. M{\"o}ller, R.~Clark, T.~E. Lee, P.~J. Leek,
  A.~Wallraff, St. Schulz, F.~Schmidt-Kaler, and H.~H{\"a}ffner.
\newblock {Fabrication and heating rate study of microscopic surface electrode
  ion traps}.
\newblock
  \href{http://iopscience.iop.org/article/10.1088/1367-2630/13/1/013032/pdf}{\it
  New J. Phys. \bf13\rm, 013032 (2011).}

\bibitem{Deslauriers2006}
L.~Deslauriers, S.~Olmschenk, D.~Stick, W.~K. Hensinger, J.~Sterk, and
  C.~Monroe.
\newblock {Scaling and Suppression of Anomalous Heating in Ion Traps}.
\newblock
  \href{http://journals.aps.org/prl/pdf/10.1103/PhysRevLett.97.103007}{\it
  Phys. Rev. Lett. \bf97\rm, 103007 (2006).}

\bibitem{Labaziewicz2008}
J.~Labaziewicz, Y.~Ge, P.~Antohi, D.~Leibrandt, K.~R. Brown, and I.~L. Chuang.
\newblock {Suppression of Heating Rates in Cryogenic Surface-Electrode Ion
  Traps}.
\newblock
  \href{http://journals.aps.org/prl/pdf/10.1103/PhysRevLett.100.013001}{\it
  Phys. Rev. Lett. \bf100\rm, 013001 (2008).}

\bibitem{Hite2012}
D.~A. Hite, Y.~Colombe, A.~C. Guise, K.~R. Brown, U.~Warring, R.~J\"{o}rdens,
  J.~D. Jost, K.~S. McKay, D.~P. Pappas, D.~Leibfried, and D.~J. Wineland.
\newblock {100-Fold Reduction of Electric-Field Noise in an Ion Trap Cleaned
  with In Situ Argon-Ion-Beam Bombardment}.
\newblock
  \href{http://journals.aps.org/prl/pdf/10.1103/PhysRevLett.109.103001}{\it
  Phys. Rev. Lett. \bf109\rm, 103001 (2012).}

\bibitem{Allcock2011}
D.~T.~C. Allcock, L.~Guidoni, T.~P. Harty, C.~J. Ballance, M.~G. Blain, A.~M.
  Steane, and D.~M. Lucas.
\newblock {Reduction of heating rate in a microfabricated ion trap by
  pulsed-laser cleaning}.
\newblock
  \href{http://iopscience.iop.org/article/10.1088/1367-2630/13/12/123023/pdf}{\it
  New J. Phys. \bf13\rm, 123023 (2011).}

\bibitem{Wang2010a}
S.~X. Wang, Y.~Ge, J.~Labaziewicz, E.~Dauler, K.~Berggren, and I.~L. Chuang.
\newblock {Superconducting microfabricated ion traps}.
\newblock
  \href{http://scitation.aip.org/content/aip/journal/apl/97/24/10.1063/1.3526733}{\it
  Appl. Phys. Lett. \bf97\rm, 244102 (2010).}

\bibitem{Eltony2014}
A.~M. Eltony, H.~G. Park, S.~X. Wang, J.~Kong, and I.~L. Chuang.
\newblock {Motional Heating in a Graphene-Coated Ion Trap}.
\newblock \href{http://pubs.acs.org/doi/pdf/10.1021/nl502468g}{\it Nano Lett.
  \bf14\rm, 5712-5716 (2014).}

\bibitem{Mehta2015x}
K.~K. Mehta, C.~D. Bruzewicz, R.~McConnell, R.~J. Ram, J.~M. Sage, and
  J.~Chiaverini.
\newblock {Integrated optical addressing of an ion qubit}.
\newblock \href{http://arxiv.org/pdf/1510.05618v1.pdf}{\tt arXiv:1510.05618v1
  \rm(2015).}

\bibitem{Narayanan2011}
S.~Narayanan, N.~Daniilidis, S.~A. M\"oller, R.~Clark, F.~Ziesel, K.~Singer,
  F.~Schmidt-Kaler, and H.~H\"affner.
\newblock {Electric field compensation and sensing with a single ion in a
  planar trap}.
\newblock
  \href{http://scitation.aip.org/content/aip/journal/jap/110/11/10.1063/1.3665647}{\it
  J. Appl. Phys. \bf110\rm, 114909 (2011).}

\bibitem{Cetina2013}
M.~Cetina, A.~Bylinskii, L.~Karpa, D.~Gangloff, K.~M. Beck, Y.~Ge, M.~Scholz,
  A.~T. Grier, I.~Chuang, and V.~Vuleti{\'c}.
\newblock {One-dimensional array of ion chains coupled to an optical cavity}.
\newblock
  \href{http://iopscience.iop.org/article/10.1088/1367-2630/15/5/053001/pdf}{\it
  New J. Phys. \bf15\rm, 053001 (2013).}

\bibitem{Karpa2013}
L.~Karpa, A.~Bylinskii, D.~Gangloff, M.~Cetina, and V.~Vuleti{\'c}.
\newblock {Suppression of Ion Transport due to Long-Lived Subwavelength
  Localization by an Optical Lattice}.
\newblock
  \href{http://journals.aps.org/prl/pdf/10.1103/PhysRevLett.111.163002}{\it
  Phys. Rev. Lett. \bf111\rm, 163002 (2013).}

\bibitem{Gangloff2015}
D.~Gangloff, A.~Bylinskii, I.~Counts, W.~Jhe, and V.~Vuleti{\'c}.
\newblock {Velocity tuning of friction with two trapped atoms}.
\newblock
  \href{http://www.nature.com/nphys/journal/v11/n11/pdf/nphys3459.pdf}{\it
  Nature Phys. \bf11\rm, 915-919 (2015).}

\bibitem{VanDyck1976x}
R.~{Van Dyck, Jr.}, P.~Ekstrom, and H.~Dehmelt.
\newblock {Axial, magnetron, cyclotron and spin-cyclotron-beat frequencies
  measured on single electron almost at rest in free space (geonium)}.
\newblock
  \href{http://www.nature.com/nature/journal/v262/n5571/abs/262776a0.html}{\it
  Nature \bf262\rm, 776-777 (1976).}

\bibitem{Wineland1973}
D.~Wineland, P.~Ekstrom, and H.~Dehmelt.
\newblock {Monoelectron Oscillator}.
\newblock
  \href{http://journals.aps.org/prl/pdf/10.1103/PhysRevLett.31.1279}{\it Phys.
  Rev. Lett. \bf31\rm, 1279-1282 (1973).}

\bibitem{Brown1986}
L.~S. Brown and G.~Gabrielse.
\newblock {Geonium theory: Physics of a single electron or ion in a Penning
  trap}.
\newblock \href{http://journals.aps.org/rmp/pdf/10.1103/RevModPhys.58.233}{\it
  Rev. Mod. Phys. \bf58\rm, 233-311 (1986).}

\bibitem{Verdu2011}
J.~Verd{\'u}.
\newblock {Theory of the coplanar-waveguide Penning trap}.
\newblock
  \href{http://iopscience.iop.org/article/10.1088/1367-2630/13/11/113029/pdf}{\it
  New J. Phys. \bf13\rm, 113029 (2011).}

\bibitem{AlRjoub2012}
A.~Al-Rjoub and J.~Verd{\'u}.
\newblock {Electronic detection of a single particle in a coplanar-waveguide
  Penning trap}.
\newblock \href{http://link.springer.com/article/10.1007/s00340-012-5069-7}{\it
  Appl. Phys. B \bf107\rm, 955-964 (2012).}

\bibitem{Hoffrogge2011}
J.~Hoffrogge, R.~Fr\"{o}hlich, M.~A. Kasevich, and P.~Hommelhoff.
\newblock {Microwave Guiding of Electrons on a Chip}.
\newblock
  \href{http://journals.aps.org/prl/pdf/10.1103/PhysRevLett.106.193001}{\it
  Phys. Rev. Lett. \bf106\rm, 193001 (2011).}

\bibitem{Hammer2015}
J.~Hammer, S.~Thomas, P.~Weber, and P.~Hommelhoff.
\newblock {Microwave Chip-Based Beam Splitter for Low-Energy Guided Electrons}.
\newblock
  \href{http://journals.aps.org/prl/pdf/10.1103/PhysRevLett.114.254801}{\it
  Phys. Rev. Lett. \bf114\rm, 254801 (2015).}

\bibitem{CCBY}
{Creative Commons License}.
\newblock
  \href{http://creativecommons.org/licenses/by-nc-sa/3.0/}{CC~BY-NC-SA.}

\bibitem{HanschBio2005}
T.~W. H{\"a}nsch.
\newblock
  \href{http://www.nobelprize.org/nobel_prizes/physics/laureates/2005/hansch-bio.html}{Nobel
  Prize biography, Stockholm, 2005.}

\bibitem{Geim2009}
A.~K. Geim.
\newblock {Graphene: Status and Prospects}.
\newblock \href{http://www.sciencemag.org/content/324/5934/1530.full.pdf}{\it
  Science \bf324\rm, 1530-1534 (2009).}

\bibitem{Wang2010b}
X.~Wang and H.~Dai.
\newblock {Etching and narrowing of graphene from the edges}.
\newblock
  \href{http://www.nature.com/nchem/journal/v2/n8/pdf/nchem.719.pdf}{\it Nature
  Chem. \bf2\rm, 661-665 (2010).}

\bibitem{Subramaniam2013}
C.~Subramaniam, T.~Yamada, K.~Kobashi, A.~Sekiguchi, D.~N. Futaba, M.~Yumura,
  and K.~Hata.
\newblock {One hundred fold increase in current carrying capacity in a carbon
  nanotube-copper composite}.
\newblock
  \href{http://www.nature.com/ncomms/2013/130723/ncomms3202/pdf/ncomms3202.pdf}{\it
  Nature Commun. \bf4\rm, 2202 (2013).}

\bibitem{Mehta2015a}
R.~Mehta, S.~Chugh, and Z.~Chen.
\newblock {Enhanced Electrical and Thermal Conduction in Graphene-Encapsulated
  Copper Nanowires}.
\newblock \href{http://pubs.acs.org/doi/pdf/10.1021/nl504889t}{\it Nano Lett.
  \bf15\rm, 2024-2030 (2015).}

\bibitem{Fruchtman2012}
A.~Fruchtman and B.~Horovitz.
\newblock {Single vortex fluctuations in a superconducting chip as generating
  dephasing and spin flips in cold atom traps}.
\newblock
  \href{http://iopscience.iop.org/article/10.1209/0295-5075/99/53002/pdf}{\it
  Europhys. Lett. \bf99\rm, 53002 (2012).}

\bibitem{Jose2014}
S.~Jose, P.~Surendran, Y.~Wang, I.~Herrera, L.~Krzemien, S.~Whitlock,
  R.~McLean, A.~Sidorov, and P.~Hannaford.
\newblock {Periodic array of Bose-Einstein condensates in a magnetic lattice}.
\newblock \href{http://journals.aps.org/pra/pdf/10.1103/PhysRevA.89.051602}{\it
  Phys. Rev. A \bf89\rm, 051602(R) (2014).}

\bibitem{Sinuco-Leon2011}
G.~Sinuco-Le{\'o}n, B.~Kaczmarek, P.~Kr{\"u}ger, and T.~M. Fromhold.
\newblock {Atom chips with two-dimensional electron gases: Theory of
  near-surface trapping and ultracold-atom microscopy of quantum electronic
  systems}.
\newblock \href{http://journals.aps.org/pra/pdf/10.1103/PhysRevA.83.021401}{\it
  Phys. Rev. A \bf83\rm, 021401(R) (2011).}

\bibitem{Salem2010}
R.~Salem, Y.~Japha, J.~Chab{\'e}, B.~Hadad, M.~Keil, K.~A. Milton, and
  R.~Folman.
\newblock {Nanowire atomchip traps for sub-micron atom-surface distances}.
\newblock
  \href{http://m.iopscience.iop.org/article/10.1088/1367-2630/12/2/023039/pdf}{\it
  New J. Phys. \bf12\rm, 023039 (2010).}

\bibitem{Chuang2011a}
H.-C. Chuang, E.~A. Salim, V.~Vuleti{\'c}, D.~Z. Anderson, and V.~M. Bright.
\newblock {Multi-layer atom chips for atom tunneling experiments near the chip
  surface}.
\newblock
  \href{http://www.sciencedirect.com/science/article/pii/S092442471000004X}{\it
  Sensors Actuat. A-Phys \bf165\rm, 101-106 (2011).}

\bibitem{Schmied2010}
R.~Schmied, D.~Leibfried, R.~J.~C. Spreeuw, and S.~Whitlock.
\newblock {Optimized magnetic lattices for ultracold atomic ensembles}.
\newblock
  \href{http://iopscience.iop.org/article/10.1088/1367-2630/12/10/103029/pdf}{\it
  New J. Phys. \bf12\rm, 103029 (2010).}

\bibitem{Herrera2015}
I.~Herrera, Y.~Wang, P.~Michaux, D.~Nissen, P.~Surendran, S.~Juodkazis,
  S.~Whitlock, R.~J. McLean, A.~Sidorov, M.~Albrecht, and P.~Hannaford.
\newblock {Sub-micron period lattice structures of magnetic microtraps for
  ultracold atoms on an atom chip}.
\newblock
  \href{http://iopscience.iop.org/article/10.1088/0022-3727/48/11/115002/pdf}{\it
  J. Phys. D \bf48\rm, 115002 (2015).}

\bibitem{Kruger2003}
P.~Kr{\"u}ger, X.~Luo, M.~W. Klein, K.~Brugger, A.~Haase, S.~Wildermuth,
  S.~Groth, I.~Bar-Joseph, R.~Folman, and J.~Schmiedmayer.
\newblock {Trapping and manipulating neutral atoms with electrostatic fields}.
\newblock
  \href{http://journals.aps.org/prl/pdf/10.1103/PhysRevLett.91.233201}{\it
  Phys. Rev. Lett. \bf91\rm, 233201 (2003).}

\bibitem{Nirrengarten2006}
T.~Nirrengarten, A.~Qarry, C.~Roux, A.~Emmert, G.~Nogues, M.~Brune, J.-M.
  Raimond, and S.~Haroche.
\newblock {Realization of a Superconducting Atom Chip}.
\newblock
  \href{http://journals.aps.org/prl/pdf/10.1103/PhysRevLett.97.200405}{\it
  Phys. Rev. Lett. \bf97\rm, 200405 (2006).}

\bibitem{Mukai2007}
T.~Mukai, C.~Hufnagel, A.~Kasper, T.~Meno, A.~Tsukada, K.~Semba, and
  F.~Shimizu.
\newblock {Persistent Supercurrent Atom Chip}.
\newblock
  \href{http://journals.aps.org/prl/pdf/10.1103/PhysRevLett.98.260407}{\it
  Phys. Rev. Lett. \bf98\rm, 260407 (2007).}

\bibitem{Roux2008}
C.~Roux, A.~Emmert, A.~Lupascu, T.~Nirrengarten, G.~Nogues, M.~Brune1, J.-M.
  Raimond, and S.~Haroche.
\newblock {Bose-Einstein condensation on a superconducting atom chip}.
\newblock
  \href{http://iopscience.iop.org/article/10.1209/0295-5075/81/56004/pdf}{\it
  al., Eur. Phys. Lett. \bf81\rm, 56004 (2008).}

\bibitem{Kasch2010}
B.~Kasch, H.~Hattermann, D.~Cano, T.~E. Judd, S.~Scheel, C.~Zimmermann,
  R.~Kleiner, D.~Koelle, and J.~Fort{\'a}gh.
\newblock {Cold atoms near superconductors: atomic spin coherence beyond the
  Johnson noise limit}.
\newblock
  \href{http://iopscience.iop.org/article/10.1088/1367-2630/12/6/065024/pdf}{\it
  New J. Phys. \bf12\rm, 065024 (2010).}

\bibitem{Muller2010a}
T.~M{\"u}ller, B.~Zhang, R.~Fermani, K.~S. Chan, M.~J. Lim, and R.~Dumke.
\newblock {Programmable trap geometries with superconducting atom chips}.
\newblock \href{http://journals.aps.org/pra/pdf/10.1103/PhysRevA.81.053624}{\it
  Phys. Rev. A \bf81\rm, 053624 (2010).}

\bibitem{Imai2014}
H.~Imai, K.~Inaba, H.~Tanji-Suzuki, M.~Yamashita, and T.~Mukai.
\newblock {Bose-Einstein condensate on a persistent-supercurrent atom chip}.
\newblock \href{http://link.springer.com/article/10.1007/s00340-014-5768-3}{\it
  Appl. Phys. B \bf116\rm, 821-829 (2014).}

\bibitem{Minniberger2014}
S.~Minniberger, F.~Diorico, S.~Haslinger, C.~Hufnagel, Ch. Novotny, N.~Lippok,
  J.~Majer, Ch. Koller, S.~Schneider, and J.~Schmiedmayer.
\newblock {Magnetic conveyor belt transport of ultracold atoms to a
  superconducting atomchip}.
\newblock \href{http://link.springer.com/article/10.1007/s00340-014-5790-5}{\it
  Appl. Phys. B \bf116\rm, 1017-1021 (2014).}

\bibitem{Scheel2005}
S.~Scheel, P.~K. Rekdal, P.~L. Knight, and E.~A. Hinds.
\newblock {Atomic spin decoherence near conducting and superconducting films}.
\newblock \href{http://journals.aps.org/pra/pdf/10.1103/PhysRevA.72.042901}{\it
  Phys. Rev. A \bf72\rm, 042901 (2005).}

\bibitem{Skagerstam2006}
B.-S.~K. Skagerstam, U.~Hohenester, A.~Eiguren, and P.~K. Rekdal.
\newblock {Spin Decoherence in Superconducting Atom Chips}.
\newblock
  \href{http://journals.aps.org/prl/pdf/10.1103/PhysRevLett.97.070401}{\it
  Phys. Rev. Lett. \bf97\rm, 070401 (2006).}

\bibitem{Dikovsky2009}
V.~Dikovsky, V.~Sokolovsky, B.~Zhang, C.~Henkel, and R.~Folman.
\newblock {Superconducting atom chips: advantages and challenges}.
\newblock
  \href{http://link.springer.com/article/10.1140%2Fepjd%2Fe2008-00261-5#/page-1}{\it
  Eur. Phys. J. D \bf51\rm, 247-259 (2009).}

\bibitem{Nogues2009}
G.~Nogues, C.~Roux, T.~Nirrengarten, A.~Lupa{\c s}cu, A.~Emmert, M.~Brune,
  J.-M. Raimond, S.~Haroche, B.~Pla{\c c}ais, and J.-J. Greffet.
\newblock {Effect of vortices on the spin-flip lifetime of atoms in
  superconducting atom-chips}.
\newblock
  \href{http://iopscience.iop.org/article/10.1209/0295-5075/87/13002/pdf}{\it
  Europhys. Lett. \bf87\rm, 13002 (2009).}

\bibitem{Bernon2013}
S.~Bernon, H.~Hattermann, D.~Bothner, M.~Knufinke, P.~Weiss, F.~Jessen,
  D.~Cano, M.~Kemmler, R.~Kleiner, D.~Koelle, and J.~Fort{\'a}gh.
\newblock {Manipulation and coherence of ultra-cold atoms on a superconducting
  atom chip}.
\newblock
  \href{http://www.nature.com/ncomms/2013/130829/ncomms3380/pdf/ncomms3380.pdf}{\it
  Nature Commun. \bf4\rm, 2380 (2013).}

\bibitem{Naides2013}
M.~A. Naides, R.~W. Turner, R.~A. Lai, J.~M. DiSciacca, and B.~L. Lev.
\newblock {Trapping ultracold gases near cryogenic materials with rapid
  reconfigurability}.
\newblock
  \href{http://scitation.aip.org/content/aip/journal/apl/103/25/10.1063/1.4852017}{\it
  Appl. Phys. Lett. \bf103\rm, 251112 (2013).}

\bibitem{Haslinger2011}
S.~Haslinger, R.~Ams{\"u}ss, C.~Koller, C.~Hufnagel, N.~Lippok, J.~Majer,
  J.~Verdu, S.~Schneider, and J.~Schmiedmayer.
\newblock {Electron beam driven alkali metal atom source for loading a
  magneto-optical trap in a cryogenic environment}.
\newblock
  \href{http://link.springer.com/content/pdf/10.1007%2Fs00340-011-4447-x.pdf}{\it
  Appl. Phys. B \bf102\rm, 819-823 (2011).}

\bibitem{Mishra2014}
K.~Mishra, C.~Murade, B.~Carreel, I.~Roghair, J.~M. Oh, G.~Manukyan, D.~Ende,
  and F.~Mugele.
\newblock {Optofluidic lens with tunable focal length and asphericity}.
\newblock \href{http://www.nature.com/articles/srep06378}{\it Sci. Rep.
  \bf4\rm, 6378 (2014).}

\bibitem{Leonard2014}
J.~L{\'e}onard, M.~Lee, A.~Morales, T.~M. Karg, T.~Esslinger, and T.~Donner.
\newblock {Optical transport and manipulation of an ultracold atomic cloud
  using focus-tunable lenses}.
\newblock
  \href{http://iopscience.iop.org/article/10.1088/1367-2630/16/9/093028/pdf}{\it
  New J. Phys. \bf16\rm, 093028 (2014).}

\bibitem{Siercke2012}
M.~Siercke, K.~S. Chan, B.~Zhang, M.~Beian, M.~J. Lim, and R.~Dumke.
\newblock {Reconfigurable self-sufficient traps for ultracold atoms based on a
  superconducting square}.
\newblock \href{http://journals.aps.org/pra/pdf/10.1103/PhysRevA.85.041403}{\it
  Phys. Rev. A \bf85\rm, 041403(R) (2012).}

\bibitem{Muller2010b}
T.~M{\"u}ller, B.~Zhang, R.~Fermani, K.~S. Chan, Z.~W. Wang, C.~B. Zhang, M.~J.
  Lim, and R.~Dumke.
\newblock {Trapping of ultra-cold atoms with the magnetic field of vortices in
  a thin-film superconducting micro-structure}.
\newblock
  \href{http://iopscience.iop.org/article/10.1088/1367-2630/12/4/043016/pdf}{\it
  New J. Phys. \bf12\rm, 043016 (2010).}

\bibitem{Zhang2012}
B.~Zhang, M.~Siercke, K.~S. Chan, M.~Beian, M.~J. Lim, and R.~Dumke.
\newblock {Magnetic confinement of neutral atoms based on patterned vortex
  distributions in superconducting disks and rings}.
\newblock \href{http://journals.aps.org/pra/pdf/10.1103/PhysRevA.85.013404}{\it
  Phys. Rev. A \bf85\rm, 013404 (2012).}

\bibitem{Sokolovsky2016}
V.~Sokolovsky and L.~Prigozhin.
\newblock {Lattices of ultracold atom traps over arrays of nano- and mesoscopic
  superconducting disks}.
\newblock
  \href{http://iopscience.iop.org/article/10.1088/0022-3727/49/16/165006/pdf}{\it
  J. Phys. D \bf49\rm, 165006 (2016).}

\bibitem{Chuang2014a}
H.-C. Chuang, Y.-S. Lin, Y.-H. Lin, and C.-S. Huang.
\newblock {The fabrication of a double-layer atom chip with through silicon
  vias for an ultra-high-vacuum cell}.
\newblock
  \href{http://iopscience.iop.org/article/10.1088/0960-1317/24/4/045013/pdf}{\it
  J. Micromech. Microeng. \bf24\rm, 045013 (2014).}

\bibitem{Trinker2008}
M.~Trinker, S.~Groth, S.~Haslinger, S.~Manz, T.~Betz, S.~Schneider,
  I.~Bar-Joseph, T.~Schumm, and J.~Schmiedmayer.
\newblock {Multilayer atom chips for versatile atom micromanipulation}.
\newblock
  \href{http://scitation.aip.org/content/aip/journal/apl/92/25/10.1063/1.2945893}{\it
  Appl. Phys. Lett. \bf92\rm, 254102 (2008).}

\bibitem{Chuang2011b}
H.-C. Chuang, C.-W. Wend, and H.-F. Li.
\newblock {Design, microfabrication and characterization of planarized
  multilayer atom chips with enhanced heat dissipation}.
\newblock
  \href{http://iopscience.iop.org/article/10.1088/0960-1317/21/12/125009/pdf}{\it
  J. Micromech. Microeng. \bf21\rm, 125009 (2011).}

\bibitem{Birkl2001}
G.~Birkl, F.~B.~J. Buchkremer, R.~Dumke, and W.~Ertmer.
\newblock {Atom optics with microfabricated optical elements}.
\newblock
  \href{http://www.sciencedirect.com/science/article/pii/S0030401801011075}{\it
  Opt. Commun. \bf191\rm, 67-81 (2001).}

\bibitem{Folman2011x}
R.~Folman, P.~Treutlein, and J.~Schmiedmayer.
\newblock {Atom Chip Fabrication, Chapter~3 in: \it Atom Chips, \rm J. Reichel
  and V. Vuletic, eds.}
\newblock
  \href{http://onlinelibrary.wiley.com/doi/10.1002/9783527633357.ch3/summary}{(Wiley-VCH:
  Hoboken, NJ, 2011) pp.~61-117.}

\bibitem{Salim2013}
E.~A. Salim, S.~C. Caliga, J.~B. Pfeiffer, and D.~Z. Anderson.
\newblock {High resolution imaging and optical control of Bose-Einstein
  condensates in an atom chip magnetic trap}.
\newblock
  \href{http://scitation.aip.org/content/aip/journal/apl/102/8/10.1063/1.4793522}{\it
  Appl. Phys. Lett. \bf102\rm, 084104 (2013).}

\bibitem{Caliga2016}
S.~C. Caliga, C.~J.~E. Straatsma, and D.~Z. Anderson.
\newblock {Transport dynamics of ultracold atoms in a triple-well
  transistor-like potential}.
\newblock
  \href{http://iopscience.iop.org/article/10.1088/1367-2630/18/2/025010/pdf}{\it
  New J. Phys. \bf18\rm, 025010 (2016).}

\bibitem{Huet2012}
L.~Huet, M.~Ammar, E.~Morvan, N.~Sarazin, J.~P. Pocholle, J.~Reichel,
  C.~Guerlin, and S.~Schwartz.
\newblock {Experimental investigation of transparent silicon carbide for atom
  chips}.
\newblock
  \href{http://scitation.aip.org/content/aip/journal/apl/100/12/10.1063/1.3689777}{\it
  Appl. Phys. Lett. \bf100\rm, 121114 (2012).}

\bibitem{Chuang2014b}
H.-C. Chuang, C.-S. Huang, H.-P. Chen, C.-S. Huang, and Y.-H. Lin.
\newblock {The Design, Fabrication and Characterization of a Transparent Atom
  Chip}.
\newblock \href{http://www.mdpi.com/1424-8220/14/6/10292}{\it Sensors \bf14\rm,
  10292-10305 (2014).}

\bibitem{Trupke2006}
M.~Trupke, F.~Ramirez-Martinez, E.~A. Curtis, J.~P. Ashmore, S.~Eriksson, E.~A.
  Hinds, Z.~Moktadir, C.~Gollasch, M.~Kraft, G.~Vijaya Prakash, and J.~J.
  Baumberg.
\newblock {Pyramidal micromirrors for microsystems and atom chips}.
\newblock
  \href{http://scitation.aip.org/content/aip/journal/apl/88/7/10.1063/1.2172412}{\it
  App. Phys. Lett. \bf88\rm, 071116 (2006).}

\bibitem{Nshii2013}
C.~C. Nshii, M.~Vangeleyn, J.~P. Cotter, P.~F. Griffin, E.~A. Hinds, C.~N.
  Ironside, P.~See, A.~G. Sinclair, E.~Riis, and A.~S. Arnold.
\newblock {A surface-patterned chip as a strong source of ultracold atoms for
  quantum technologies}.
\newblock
  \href{http://www.nature.com/nnano/journal/v8/n5/pdf/nnano.2013.47.pdf}{\it
  Nature Nanotechnol. \bf8\rm, 321-324 (2013).}

\bibitem{Lee2013a}
J.~Lee, J.~A. Grover, L.~A. Orozco, and S.~L. Rolston.
\newblock {Sub-Doppler cooling of neutral atoms in a grating magneto-optical
  trap}.
\newblock
  \href{https://www.osapublishing.org/josab/abstract.cfm?uri=josab-30-11-2869}{\it
  J. Opt. Soc. Am. B \bf30\rm, 2869-2874 (2013).}

\bibitem{Kohnen2011}
M.~Kohnen, M.~Succo, P.~G. Petrov, R.~A. Nyman, M.~Trupke, and E.~A. Hinds.
\newblock {An array of integrated atom-photon junctions}.
\newblock
  \href{http://www.nature.com/nphoton/journal/v5/n1/pdf/nphoton.2010.255.pdf}{\it
  Nature Photon. \bf5\rm, 35-38 (2011).}

\bibitem{Purdy2008}
T.~P. Purdy and D.~M. Stamper-Kurn.
\newblock {Integrating cavity quantum electrodynamics and ultracold-atom chips
  with on-chip dielectric mirrors and temperature stabilization}.
\newblock
  \href{http://link.springer.com/article/10.1007%2Fs00340-007-2879-0#/page-1}{\it
  Appl. Phys. B \bf90\rm, 401-405 (2008).}

\bibitem{Goldwin2011}
J.~Goldwin, M.~Trupke, J.~Kenner, A.~Ratnapala, and E.A. Hinds.
\newblock {Fast cavity-enhanced atom detection with low noise and high
  fidelity}.
\newblock
  \href{http://www.nature.com/ncomms/journal/v2/n8/pdf/ncomms1428.pdf}{\it
  Nature Commun. \bf2\rm, 418 (2011).}

\bibitem{Heine2010}
D.~Heine, W.~Rohringer, D.~Fischer, M.~Wilzbach, T.~Raub, S.~Loziczky, X.~Liu,
  S.~Groth, B.~Hessmo, and J.~Schmiedmayer.
\newblock {A single-atom detector integrated on an atom chip: fabrication,
  characterization and application}.
\newblock
  \href{http://iopscience.iop.org/article/10.1088/1367-2630/12/9/095005/pdf}{\it
  New J. Phys. \bf12\rm, 095005 (2010).}

\bibitem{Aoki2006}
T.~Aoki, B.~Dayan, E.~Wilcut, W.~P. Bowen, A.~S. Parkins, T.~J. Kippenberg,
  K.~J. Vahala, and H.~J. Kimble.
\newblock {Observation of strong coupling between one atom and a monolithic
  microresonator}.
\newblock
  \href{http://www.nature.com/nature/journal/v443/n7112/abs/nature05147.html}{\it
  Nature \bf443\rm, 671-674 (2006).}

\bibitem{Barclay2006}
P.~E. Barclay, K.~Srinivasan, O.~Painter, B.~Lev, and H.~Mabuchi.
\newblock {Integration of fiber-coupled high-{\it Q} SiN$_x$ microdisks with
  atom chips}.
\newblock
  \href{http://scitation.aip.org/content/aip/journal/apl/89/13/10.1063/1.2356892}{\it
  Appl. Phys. Lett. \bf89\rm, 131108 (2006).}

\bibitem{Rosenblit2006}
M.~Rosenblit, Y.~Japha, P.~Horak, and R.~Folman.
\newblock {Simultaneous optical trapping and detection of atoms by microdisk
  resonators}.
\newblock \href{http://journals.aps.org/pra/pdf/10.1103/PhysRevA.73.063805}{\it
  Phys. Rev. A \bf73\rm, 063805 (2006).}

\bibitem{Rosenblit2007}
M.~Rosenblit, P.~Horak, E.~Fleminger, Y.~Japha, and R.~Folman.
\newblock {Design of microcavity resonators for single-atom detection}.
\newblock
  \href{http://nanophotonics.spiedigitallibrary.org/article.aspx?articleid=1227422}{\it
  J. Nanophoton. \bf1\rm, 011670 (2007).}

\bibitem{Shomroni2014}
I.~Shomroni, S.~Rosenblum, Y.~Lovsky, O.~Bechler, G.~Guendelman, and B.~Dayan.
\newblock {All-optical routing of single photons by a one-atom switch
  controlled by a single photon}.
\newblock \href{https://www.sciencemag.org/content/345/6199/903.full.pdf}{\it
  Science \bf345\rm, 903-906 (2014).}

\bibitem{Rosenblum2016}
S.~Rosenblum, O.~Bechler, I.~Shomroni, Y.~Lovsky, G.~Guendelman, and B.~Dayan.
\newblock {Extraction of a single photon from an optical pulse}.
\newblock
  \href{http://www.nature.com/nphoton/journal/v10/n1/pdf/nphoton.2015.227.pdf}{\it
  Nature Photon. \bf10\rm, 19-22 (2016).}

\bibitem{Lev2004}
B.~Lev, K.~Srinivasan, P.~Barclay, O.~Painter, and H.~Mabuchi.
\newblock {Feasibility of detecting single atoms using photonic bandgap
  cavities}.
\newblock
  \href{http://iopscience.iop.org/article/10.1088/0957-4484/15/10/010/pdf}{\it
  Nanotechnology \bf15\rm, S556-S561 (2004).}

\bibitem{Thompson2013b}
J.~D. Thompson, T.~G. Tiecke, N.~P. {de Leon}, J.~Feist, A.~V. Akimov,
  M.~Gullans, A.~S. Zibrov, V.~Vuleti{\'c}, and M.~D. Lukin.
\newblock {Coupling a Single Trapped Atom to a Nanoscale Optical Cavity}.
\newblock
  \href{http://science.sciencemag.org/content/sci/340/6137/1202.full.pdf}{\it
  Science \bf340\rm, 1202-1205 (2013).}

\bibitem{Goban2014}
A.~Goban, C.-L. Hung, S.-P. Yu, J.~D. Hood, J.~A. Muniz, J.~H. Lee, M.~J.
  Martin, A.~C. McClung, K.~S. Choi, D.~E. Chang, O.~Painter, and H.~J. Kimble.
\newblock {Atom-light interactions in photonic crystals}.
\newblock
  \href{http://www.nature.com/ncomms/2014/140508/ncomms4808/pdf/ncomms4808.pdf}{\it
  Nature Commun. \bf5\rm, 3808 (2014).}

\bibitem{Douglas2015}
J.~S. Douglas, H.~Habibian, C.-L. Hung, A.~V. Gorshkov, H.~J. Kimble, and D.~E.
  Chang.
\newblock {Quantum many-body models with cold atoms coupled to photonic
  crystals}.
\newblock
  \href{http://www.nature.com/nphoton/journal/v9/n5/pdf/nphoton.2015.57.pdf}{\it
  Nature Photon. \bf9\rm, 326-331 (2015).}

\bibitem{Derntl2014}
C.~Derntl, M.~Schneider, J.~Schalko, A.~Bittner, J.~Schmiedmayer, U.~Schmid,
  and M.~Trupke.
\newblock {Arrays of open, independently tunable microcavities}.
\newblock
  \href{https://www.osapublishing.org/oe/abstract.cfm?uri=oe-22-18-22111}{\it
  Opt. Express \bf22\rm, 22111-22120 (2014).}

\bibitem{Abdelrahman2014}
A.~Abdelrahman, T.~Mukai, H.~H{\"a}ffner, and T.~Byrnes.
\newblock {Coherent all-optical control of ultracold atoms arrays in permanent
  magnetic traps}.
\newblock
  \href{https://www.osapublishing.org/oe/abstract.cfm?uri=oe-22-3-3501}{\it
  Opt. Express \bf22\rm, 3501-3513 (2014).}

\bibitem{Straatsma2015}
C.~J.~E. Straatsma, M.~K. Ivory, J.~Duggan, J.~Ramirez-Serrano, D.~Z. Anderson,
  and E.~A. Salim.
\newblock {On-chip optical lattice for cold atom experiments}.
\newblock
  \href{https://www.osapublishing.org/ol/abstract.cfm?uri=ol-40-14-3368}{\it
  Opt. Lett. \bf40\rm, 3368-3371 (2015).}

\bibitem{ColdQuanta2015}
\href{http://coldquanta.com/optical-lattice-atom-chips/}{\tt{http://coldquanta.com}}.

\bibitem{Gullans2012}
M.~Gullans, T.~G. Tiecke, D.~E. Chang, J.~Feist, J.~D. Thompson, J.~I. Cirac,
  P.~Zoller, and M.~D. Lukin.
\newblock {Nanoplasmonic Lattices for Ultracold Atoms}.
\newblock
  \href{http://journals.aps.org/prl/pdf/10.1103/PhysRevLett.109.235309}{\it
  Phys. Rev. Lett. \bf109\rm, 235309 (2012).}

\bibitem{Romero-Isart2013}
O.~Romero-Isart, C.~Navau, A.~Sanchez, P.~Zoller, and J.~I. Cirac.
\newblock {Superconducting Vortex Lattices for Ultracold Atoms}.
\newblock
  \href{http://journals.aps.org/prl/pdf/10.1103/PhysRevLett.111.145304}{\it
  Phys. Rev. Lett. \bf111\rm, 145304 (2013).}

\bibitem{Geraci2009}
A.~A. Geraci and J.~Kitching.
\newblock {Ultracold mechanical resonators coupled to atoms in an optical
  lattice}.
\newblock \href{http://journals.aps.org/pra/pdf/10.1103/PhysRevA.80.032317}{\it
  Phys. Rev. A \bf80\rm, 032317 (2009).}

\bibitem{Sinuco-Leon2015}
G.~A. Sinuco-Le{\'o}n and B.~M. Garraway.
\newblock {Radio-frequency dressed lattices for ultracold alkali atoms}.
\newblock
  \href{http://iopscience.iop.org/article/10.1088/1367-2630/17/5/053037/pdf}{\it
  New J. Phys. \bf17\rm, 053037 (2015).}

\bibitem{Hunger2010a}
D.~Hunger, S.~Camerer, T.~W. H{\"a}nsch, D.~K{\"o}nig, J.~P. Kotthaus,
  J.~Reichel, and P.~Treutlein.
\newblock {Resonant Coupling of a Bose-Einstein Condensate to a Micromechanical
  Oscillator}.
\newblock
  \href{http://journals.aps.org/prl/pdf/10.1103/PhysRevLett.104.143002}{\it
  Phys. Rev. Lett. \bf104\rm, 143002 (2010).}

\bibitem{Treutlein2014}
P.~Treutlein, C.~Genes, K.~Hammerer, M.~Poggio, and P.~Rabl.
\newblock {Hybrid Mechanical Systems, in: \it Cavity Optomechanics \rm M.
  Aspelmeyer, T. J. Kippenberg, and F. Marquardt, eds.}
\newblock
  \href{http://link.springer.com/chapter/10.1007%2F978-3-642-55312-7_14}{(Springer:
  Berlin, 2014) pp.~327-351.}

\bibitem{Kurizki2015}
G.~Kurizki, P.~Bertet, Y.~Kubo, K.~M{\o}lmerc, D.~Petrosyan, P.~Rabl, and
  J.~Schmiedmayer.
\newblock {Quantum technologies with hybrid systems}.
\newblock \href{http://www.pnas.org/content/112/13/3866.full.pdf}{\it Proc.
  Natl. Acad. Sci. USA \bf112\rm, 3866-3873 (2015).}

\bibitem{Montoya2015}
C.~Montoya, J.~Valencia, A.~A. Geraci, M.~Eardley, J.~Moreland, L.~Hollberg,
  and J.~Kitching.
\newblock {Resonant interaction of trapped cold atoms with a magnetic
  cantilever tip}.
\newblock \href{http://journals.aps.org/pra/pdf/10.1103/PhysRevA.91.063835}{\it
  Phys. Rev. A \bf91\rm, 063835 (2015).}

\bibitem{Kalman2011}
O.~K{\'a}lm{\'a}n, T.~Kiss, J.~Fort{\'a}gh, and P.~Domokos.
\newblock {Quantum Galvanometer by Interfacing a Vibrating Nanowire and Cold
  Atoms}.
\newblock \href{http://pubs.acs.org/doi/ipdf/10.1021/nl203762g}{\it Nano Lett.
  \bf12\rm, 435-439 (2012).}

\bibitem{Stehle2011}
C.~Stehle, H.~Bender, C.~Zimmermann, D.~Kern, M.~Fleischer, and S.~Slama.
\newblock {Plasmonically tailored micropotentials for ultracold atoms}.
\newblock
  \href{http://www.nature.com/nphoton/journal/v5/n8/pdf/nphoton.2011.159.pdf}{\it
  Nature Photon. \bf5\rm, 494-498 (2011).}

\bibitem{Shaffer2011}
J.~P. Shaffer.
\newblock {Atom optics: Marriage of atoms and plasmons}.
\newblock
  \href{http://www.nature.com/nphoton/journal/v5/n8/pdf/nphoton.2011.174.pdf}{\it
  Nature Photon. \bf5\rm, 451-452 (2011).}

\bibitem{Bender2014}
H.~Bender, C.~Stehle, C.~Zimmermann, S.~Slama, J.~Fiedler, S.~Scheel, S.~Y.
  Buhmann, and V.~N. Marachevsky.
\newblock {Probing Atom-Surface Interactions by Diffraction of Bose-Einstein
  Condensates}.
\newblock \href{http://journals.aps.org/prx/pdf/10.1103/PhysRevX.4.011029}{\it
  Phys. Rev. X \bf4\rm, 011029 (2014).}

\bibitem{Weiss2015}
P.~Weiss, M.~Knufinke, S.~Bernon, D.~Bothner, L.~S{\'a}rk{\'a}ny,
  C.~Zimmermann, R.~Kleiner, D.~Koelle, J.~Fort{\'a}gh, and H.~Hattermann.
\newblock {Sensitivity of Ultracold Atoms to Quantized Flux in a
  Superconducting Ring}.
\newblock
  \href{http://journals.aps.org/prl/pdf/10.1103/PhysRevLett.114.113003}{\it
  Phys. Rev. Lett. \bf114\rm, 113003 (2015).}

\bibitem{Verdu2009}
J.~Verd{\'u}, H.~Zoubi, Ch. Koller, J.~Majer, H.~Ritsch, and J.~Schmiedmayer.
\newblock {Strong Magnetic Coupling of an Ultracold Gas to a Superconducting
  Waveguide Cavity}.
\newblock
  \href{http://journals.aps.org/prl/pdf/10.1103/PhysRevLett.103.043603}{\it
  Phys. Rev. Lett. \bf103\rm, 043603 (2009).}

\bibitem{Jessen2014}
F.~Jessen, M.~Knufinke, S.~C. Bell, P.~Vergien, H.~Hattermann, P.~Weiss,
  M.~Rudolph, M.~Reinschmidt, K.~Meyer, T.~Gaber, D.~Cano, A.~G{\"u}nther,
  S.~Bernon, D.~Koelle, R.~Kleiner, and J.~Fort{\'a}gh.
\newblock {Trapping of ultracold atoms in a $\rm^3He$/$\rm^4He$ dilution
  refrigerator}.
\newblock
  \href{http://link.springer.com/article/10.1007%2Fs00340-013-5750-5#/page-1}{\it
  Appl. Phys. B \bf116\rm, 665-671 (2014).}

\bibitem{Lee2002}
H.~Lee, P.~Kok, and J.~P. Dowling.
\newblock {A quantum Rosetta stone for interferometry}.
\newblock
  \href{http://www.tandfonline.com/doi/pdf/10.1080/0950034021000011536}{\it J.
  Mod. Optic \bf49\rm, 2325-2338 (2002).}

\bibitem{Reiserer2013}
A.~Reiserer, C.~N{\"o}lleke, S.~Ritter, and G.~Rempe.
\newblock {Ground-State Cooling of a Single Atom at the Center of an Optical
  Cavity}.
\newblock
  \href{http://journals.aps.org/prl/pdf/10.1103/PhysRevLett.110.223003}{\it
  Phys. Rev. Lett. \bf110\rm, 223003 (2013).}

\bibitem{Scheel2006}
S.~Scheel, R.~Fermani, P.~K. Rekdal, P.~L. Knight, and E.~A. Hinds.
\newblock {Atomic spin relaxation and spatial decoherence near metallic and
  superconducting surfaces}.
\newblock
  \href{http://iopscience.iop.org/article/10.1088/1742-6596/36/1/028/pdf}{\it
  J. Phys.: Conf. Ser. \bf36\rm, 188-193 (2006).}

\bibitem{Leanhardt2002}
A.~E. Leanhardt, A.~P. Chikkatur, D.~Kielpinski, Y.~Shin, T.~L. Gustavson,
  W.~Ketterle, and D.~E. Pritchard.
\newblock {Propagation of Bose-Einstein Condensates in a Magnetic Waveguide}.
\newblock
  \href{http://journals.aps.org/prl/pdf/10.1103/PhysRevLett.89.040401}{\it
  Phys. Rev. Lett. \bf89\rm, 040401 (2002).}

\bibitem{Kraft2002}
S.~Kraft, A.~G{\"u}nther, H.~Ott, D.~Wharam, C.~Zimmermann, and J.~Fort{\'a}gh.
\newblock {Anomalous longitudinal magnetic field near the surface of copper
  conductors}.
\newblock
  \href{http://iopscience.iop.org/article/10.1088/0953-4075/35/21/102/pdf}{\it
  J. Phys. B \bf35\rm, L469-L474 (2002).}

\bibitem{Esteve2004}
J.~Est{\`e}ve, C.~Aussibal, T.~Schumm, C.~Figl, D.~Mailly, I.~Bouchoule, C.~I.
  Westbrook, and A.~Aspect.
\newblock {Role of wire imperfections in micromagnetic traps for atoms}.
\newblock \href{http://journals.aps.org/pra/pdf/10.1103/PhysRevA.70.043629}{\it
  Phys. Rev. A \bf70\rm, 043629 (2004).}

\bibitem{Schumm2005b}
T.~Schumm, J.~Est{\`e}ve, C.~Figl, J.-B. Trebbia, C.~Aussibal, H.~Nguyen,
  D.~Mailly, I.~Bouchoule, C.~I. Westbrook, and A.~Aspect.
\newblock {Atom chips in the real world: the effects of wire corrugation}.
\newblock
  \href{http://epjd.epj.org/articles/epjd/abs/2005/02/d04179/d04179.html?mb=0}{\it
  Eur. Phys. J. D \bf32\rm, 171-180 (2005).}

\bibitem{Kruger2007}
P.~Kr{\"u}ger, L.~M. Andersson, S.~Wildermuth, S.~Hofferberth, E.~Haller,
  S.~Aigner, S.~Groth, I.~Bar-Joseph, and J.~Schmiedmayer.
\newblock {Potential roughness near lithographically fabricated atom chips}.
\newblock \href{http://journals.aps.org/pra/pdf/10.1103/PhysRevA.76.063621}{\it
  Phys. Rev. A \bf76\rm, 063621 (2007).}

\bibitem{Trebbia2007}
J.-B. Trebbia, C.~L. {Garrido Alzar}, R.~Cornelussen, C.~I. Westbrook, and
  I.~Bouchoule.
\newblock {Roughness Suppression via Rapid Current Modulation on an Atom Chip}.
\newblock
  \href{http://journals.aps.org/prl/pdf/10.1103/PhysRevLett.98.263201}{\it
  Phys. Rev. Lett. \bf98\rm, 263201 (2007).}

\bibitem{Zhou2015x}
S.~Zhou, D.~Groswasser, M.~Keil, Y.~Japha, and R.~Folman.
\newblock {Robust quantum spatial coherence near a classical environment}.
\newblock \href{http://arxiv.org/abs/1505.02654v2.pdf}{\tt arXiv:1505.02654v2
  \rm(2015).}

\bibitem{Jiang1993}
S.~Jiang, P.~Hallemeier, C.~Surya, and J.~M. Phillips.
\newblock {Low-frequency excess noise in YBCO thin films near the transition
  temperature}.
\newblock
  \href{http://scitation.aip.org/content/aip/proceeding/aipcp/10.1063/1.44584}{AIP
  Conf. Proc. \bf285\rm, 119-122 (1993).}

\bibitem{Khrebtov1993}
I.~A. Khrebtov, V.~N. Leonov, A.~D. Tkachenko, A.~V. Bobyl, V.~Yu. Davydov, and
  V.~I. Kozub.
\newblock {Comparative low-frequency noise studies of YBaCuO films}.
\newblock
  \href{http://scitation.aip.org/content/aip/proceeding/aipcp/10.1063/1.44585}{AIP
  Conf. Proc. \bf285\rm, 123-126 (1993).}

\bibitem{Treutlein2004}
P.~Treutlein, P.~Hommelhoff, T.~Steinmetz, T.~W. H{\"a}nsch, and J.~Reichel.
\newblock {Coherence in Microchip Traps}.
\newblock
  \href{http://journals.aps.org/prl/pdf/10.1103/PhysRevLett.92.203005}{\it
  Phys. Rev. Lett. \bf92\rm, 203005 (2004).}

\bibitem{Wang2005}
Y.-J. Wang, D.~Z. Anderson, V.~M. Bright, E.~A. Cornell, Q.~Diot, T.~Kishimoto,
  M.~Prentiss, R.~A. Saravanan, S.~R. Segal, and S.~Wu.
\newblock {Atom Michelson interferometer on a chip using a Bose-Einstein
  condensate}.
\newblock
  \href{http://journals.aps.org/prl/pdf/10.1103/PhysRevLett.94.090405}{\it
  Phys. Rev. Lett. \bf94\rm, 090405 (2005).}

\bibitem{Schumm2005a}
T.~Schumm, S.~Hofferberth, L.~M. Andersson, S.~Wildermuth, S.~Groth,
  I.~Bar-Joseph, J.~Schmiedmayer, and P.~Kr{\"u}ger.
\newblock {Matter-wave interferometry in a double well on an atom chip}.
\newblock \href{http://www.nature.com/nphys/journal/v1/n1/pdf/nphys125.pdf}{\it
  Nature Phys. \bf1\rm, 57-62 (2005).}

\bibitem{Jo2007}
G.-B. Jo, Y.~Shin, S.~Will, T.~A. Pasquini, M.~Saba, W.~Ketterle, D.~E.
  Pritchard, M.~Vengalattore, and M.~Prentiss.
\newblock {Long phase coherence time and number squeezing of two Bose-Einstein
  condensates on an atom chip}.
\newblock
  \href{http://journals.aps.org/prl/pdf/10.1103/PhysRevLett.98.030407}{\it
  Phys. Rev. Lett. \bf98\rm, 030407 (2007).}

\bibitem{Baumgartner2010}
F.~Baumg\"{a}rtner, R.~J. Sewell, S.~Eriksson, I.~Llorente-Garcia, J.~Dingjan,
  J.~P. Cotter, and E.~A. Hinds.
\newblock {Measuring Energy Differences by BEC Interferometry on a Chip}.
\newblock
  \href{http://journals.aps.org/prl/pdf/10.1103/PhysRevLett.105.243003}{\it
  Phys. Rev. Lett. \bf105\rm, 243003 (2010).}

\bibitem{Gunther2005}
A.~G{\"u}nther, S.~Kraft, M.~Kemmler, D.~Koelle, R.~Kleiner, C.~Zimmermann, and
  J.~Fort{\'a}gh.
\newblock {Diffraction of a Bose-Einstein Condensate from a Magnetic Lattice on
  a Microchip}.
\newblock
  \href{http://journals.aps.org/prl/pdf/10.1103/PhysRevLett.95.170405}{\it
  Phys. Rev. Lett. \bf95\rm, 170405 (2005).}

\bibitem{Sukenik1993}
C.~I. Sukenik, M.~G. Boshier, D.~Cho, V.~Sandoghdar, and E.~A. Hinds.
\newblock {Measurement of the Casimir-Polder Force}.
\newblock \href{http://journals.aps.org/prl/pdf/10.1103/PhysRevLett.70.560}{\it
  Phys. Rev. Lett. \bf70\rm, 560-563 (1993).}

\bibitem{Perreault2005}
J.~D. Perreault and A.~D. Cronin.
\newblock {Observation of Atom Wave Phase Shifts Induced by Van Der Waals
  Atom-Surface Interactions}.
\newblock
  \href{http://journals.aps.org/prl/pdf/10.1103/PhysRevLett.95.133201}{\it
  Phys. Rev. Lett. \bf95\rm, 133201 (2005).}

\bibitem{Lepoutre2009}
S.~Lepoutre, H.~Jelassi, V.~P.~A. Lonij, G.~Tr{\'e}nec, M.~B{\"u}chner, A.~D.
  Cronin, and J.~Vigu{\'e}.
\newblock {Dispersive atom interferometry phase shifts due to atom-surface
  interactions}.
\newblock
  \href{http://iopscience.iop.org/article/10.1209/0295-5075/88/20002/pdf}{\it
  Europhys. Lett. \bf88\rm, 200002 (2009).}

\bibitem{Landragin1996}
A.~Landragin, J.-Y. Courtois, G.~Labeyrie, N.~Vansteenkiste, C.~I. Westbrook,
  and A.~Aspect.
\newblock {Measurement of the van der Waals Force in an Atomic Mirror}.
\newblock
  \href{http://journals.aps.org/prl/pdf/10.1103/PhysRevLett.77.1464}{\it Phys.
  Rev. Lett. \bf77\rm, 1464-1467 (1996).}

\bibitem{Bender2010}
H.~Bender, Ph.~W. Courteille, C.~Marzok, C.~Zimmermann, and S.~Slama.
\newblock {Direct Measurement of Intermediate-Range Casimir-Polder Potentials}.
\newblock
  \href{http://journals.aps.org/prl/pdf/10.1103/PhysRevLett.104.083201}{\it
  Phys. Rev. Lett. \bf104\rm, 083201 (2010).}

\bibitem{Stehle2014}
C.~Stehle, C.~Zimmermann, and S.~Slama.
\newblock {Cooperative coupling of ultracold atoms and surface plasmons}.
\newblock
  \href{http://www.nature.com/nphys/journal/v10/n12/pdf/nphys3129.pdf}{\it
  Nature Phys. \bf10\rm, 937-942 (2014).}

\bibitem{Intravaia2011}
F.~Intravaia, C.~Henkel, and M.~Antezza.
\newblock {Fluctuation-Induced Forces Between Atoms and Surfaces: The
  Casimir-Polder Interaction, in: \it Casimir Physics, \rm D. Dalvit, P.
  Milonni, D. Roberts, and F. da~Rosa, eds.}
\newblock
  \href{http://link.springer.com/chapter/10.1007%2F978-3-642-20288-9_11}{(Springer:
  Berlin, 2011) pp.~345-391.}

\bibitem{Antezza2004}
M.~Antezza, L.~P. Pitaevskii, and S.~Stringari.
\newblock {Effect of the Casimir-Polder force on the collective oscillations of
  a trapped Bose-Einstein condensate}.
\newblock \href{http://journals.aps.org/pra/pdf/10.1103/PhysRevA.70.053619}{\it
  Phys. Rev. A \bf70\rm, 053619 (2004).}

\bibitem{Antezza2005}
M.~Antezza, L.~P. Pitaevskii, and S.~Stringari.
\newblock {New Asymptotic Behavior of the Surface-Atom Force out of Thermal
  Equilibrium}.
\newblock
  \href{http://journals.aps.org/prl/pdf/10.1103/PhysRevLett.95.113202}{\it
  Phys. Rev. Lett. \bf95\rm, 113202 (2005).}

\bibitem{Laliotis2014}
A.~Laliotis, T.~{Passerat de Silans}, I.~Maurin, M.~Ducloy, and D.~Bloch.
\newblock {Casimir-Polder interactions in the presence of thermally excited
  surface modes}.
\newblock
  \href{http://www.nature.com/ncomms/2014/140709/ncomms5364/pdf/ncomms5364.pdf}{\it
  Nature Commun. \bf5\rm, 4364 (2014).}

\bibitem{Schneeweiss2012}
P.~Schneeweiss, M.~Gierling, G.~Visanescu, D.~P. Kern, T.~E. Judd,
  A.~G{\"u}nther, and J.~Fort{\'a}gh.
\newblock {Dispersion forces between ultracold atoms and a carbon nanotube}.
\newblock
  \href{http://www.nature.com/nnano/journal/v7/n8/pdf/nnano.2012.93.pdf}{\it
  Nature Nanotech. \bf7\rm, 515-519 (2012).}

\bibitem{Gierling2011}
M.~Gierling, P.~Schneeweiss, G.~Visanescu, P.~Federsel, M.~H{\"a}ffner, D.~P.
  Kern, T.~E. Judd, A.~G{\"u}nther, and J.~Fort{\'a}gh.
\newblock {Cold-atom scanning probe microscopy}.
\newblock
  \href{http://www.nature.com/nnano/journal/v6/n7/pdf/nnano.2011.80.pdf}{\it
  Nature Nanotech. \bf6\rm, 446-451 (2011).}

\bibitem{Weiss2013}
C.~T. Wei{\ss}, P.~V. Mironova, J.~Fort{\'a}gh, W.~P. Schleich, and R.~Walser.
\newblock {Immersing carbon nanotubes in cold atomic gases}.
\newblock \href{http://journals.aps.org/pra/pdf/10.1103/PhysRevA.88.043623}{\it
  Phys. Rev. A \bf88\rm, 043623 (2013).}

\bibitem{Halif2011}
M.~N.~A. Halif, R.~Messina, and T.~M. Fromhold.
\newblock {Calculation of the Casimir-Polder interaction between Bose-Einstein
  condensates and microengineered surfaces: a pairwise-summation approach}.
\newblock
  \href{http://iopscience.iop.org/article/10.1088/1742-6596/286/1/012045/pdf}{\it
  J. Phys.: Conf. Ser. \bf286\rm, 012045 (2011).}

\bibitem{Rodriguez2011}
A.~W. Rodriguez, F.~Capasso, and S.~G. Johnson.
\newblock {The Casimir effect in microstructured geometries}.
\newblock
  \href{http://www.nature.com/nphoton/journal/v5/n4/pdf/nphoton.2011.39.pdf}{\it
  Nature Photon. \bf5\rm, 211-221 (2011).}

\bibitem{Chan2008}
H.~B. Chan, Y.~Bao, J.~Zou, R.~A. Cirelli, F.~Klemens, W.~M. Mansfield, and
  C.~S. Pai.
\newblock {Measurement of the Casimir Force between a Gold Sphere and a Silicon
  Surface with Nanoscale Trench Arrays}.
\newblock
  \href{http://journals.aps.org/prl/pdf/10.1103/PhysRevLett.101.030401}{\it
  Phys. Rev. Lett. \bf101\rm, 030401 (2008).}

\bibitem{Henkel2005}
C.~Henkel and K.~Joulain.
\newblock {Casimir force between designed materials: What is possible and what
  not}.
\newblock
  \href{http://iopscience.iop.org/article/10.1209/epl/i2005-10344-3/pdf}{\it
  Europhys. Lett. \bf72\rm, 929-935 (2005).}

\bibitem{Leonhardt2007}
U.~Leonhardt and T.~G. Philbin.
\newblock {Quantum levitation by left-handed metamaterials}.
\newblock
  \href{http://iopscience.iop.org/article/10.1088/1367-2630/9/8/254/pdf}{\it
  New J. Phys. \bf9\rm, 254 (2007).}

\bibitem{Yannopapas2009}
V.~Yannopapas and N.~V. Vitanov.
\newblock {First-Principles Study of Casimir Repulsion in Metamaterials}.
\newblock
  \href{http://journals.aps.org/prl/pdf/10.1103/PhysRevLett.103.120401}{\it
  Phys. Rev. Lett. \bf103\rm, 120401 (2009).}

\bibitem{Zhao2009}
R.~Zhao, J.~Zhou, Th. Koschny, E.~N. Economou, and C.~M. Soukoulis.
\newblock {Repulsive Casimir Force in Chiral Metamaterials}.
\newblock
  \href{http://journals.aps.org/prl/pdf/10.1103/PhysRevLett.103.103602}{\it
  Phys. Rev. Lett. \bf103\rm, 103602 (2009).}

\bibitem{Munday2009}
J.~N. Munday, F.~Capasso, and V.~A. Parsegian.
\newblock {Measured long-range repulsive Casimir-Lifshitz forces}.
\newblock
  \href{http://www.nature.com/nature/journal/v457/n7226/pdf/nature07610.pdf}{\it
  Nature \bf457\rm, 170-173 (2009).}

\bibitem{Scheel2011x}
S.~Scheel and E.~A. Hinds.
\newblock {Atoms at Micrometer Distances from a Macroscopic Body, Chapter~4 in:
  \it Atom Chips, \rm J. Reichel and V. Vuletic, eds.}
\newblock
  \href{http://onlinelibrary.wiley.com/doi/10.1002/9783527633357.ch4/summary}{(Wiley-VCH:
  Hoboken, NJ, 2011) pp.~121-146.}

\bibitem{Eberlein2011}
C.~Eberlein and R.~Zietal.
\newblock {Casimir-Polder interaction between a polarizable particle and a
  plate with a hole}.
\newblock \href{http://journals.aps.org/pra/pdf/10.1103/PhysRevA.83.052514}{\it
  Phys. Rev. A \bf83\rm, 052514 (2011).}

\bibitem{Milton2012}
K.~A. Milton, P.~Parashar, N.~Pourtolami, and I.~Brevik.
\newblock {Casimir-Polder repulsion: Polarizable atoms, cylinders, spheres, and
  ellipsoids}.
\newblock \href{http://journals.aps.org/prd/pdf/10.1103/PhysRevD.85.025008}{\it
  Phys. Rev. D \bf85\rm, 025008 (2012).}

\bibitem{Friedrich2002}
H.~Friedrich, G.~Jacoby, and C.~G. Meister.
\newblock {Quantum reflection by Casimir–van der Waals potential tails}.
\newblock \href{http://journals.aps.org/pra/pdf/10.1103/PhysRevA.65.032902}{\it
  Phys. Rev. A \bf65\rm, 032902 (2002).}

\bibitem{Berkhout1989}
J.~J. Berkhout, O.~J. Luiten, I.~D. Setija, T.~W. Hijmans, T.~Mizusaki, and
  J.~T.~M. Walraven.
\newblock {Quantum Reflection: Focusing of Hydrogen Atoms with a Concave
  Mirror}.
\newblock
  \href{http://journals.aps.org/prl/pdf/10.1103/PhysRevLett.63.1689}{\it Phys.
  Rev. Lett. \bf63\rm, 1689-1692 (1989).}

\bibitem{Shimizu2001}
F.~Shimizu.
\newblock {Specular Reflection of Very Slow Metastable Neon Atoms from a Solid
  Surface}.
\newblock \href{http://journals.aps.org/prl/pdf/10.1103/PhysRevLett.86.987}{\it
  Phys. Rev. Lett. \bf86\rm, 987-990 (2001).}

\bibitem{Pasquini2004}
T.~A. Pasquini, Y.~Shin, C.~Sanner, M.~Saba, A.~Schirotzek, D.~E. Pritchard,
  and W.~Ketterle.
\newblock {Quantum reflection from a solid surface at normal incidence}.
\newblock
  \href{http://journals.aps.org/prl/pdf/10.1103/PhysRevLett.93.223201}{\it
  Phys. Rev. Lett. \bf93\rm, 223201 (2004).}

\bibitem{Shimizu2002}
F.~Shimizu and J.~Fujita.
\newblock {Giant Quantum Reflection of Neon Atoms from a Ridged Silicon
  Surface}.
\newblock \href{http://journals.jps.jp/doi/pdf/10.1143/JPSJ.71.5}{\it J. Phys.
  Soc. Jpn. \bf71\rm, 5-8 (2002).}

\bibitem{Zhao2011b}
B.~S. Zhao, G.~Meijer, and W.~Sch{\"o}llkopf.
\newblock {Quantum Reflection of He$_2$ Several Nanometers Above a Grating
  Surface}.
\newblock \href{http://www.sciencemag.org/content/331/6019/892.full.pdf}{\it
  Science \bf331\rm, 892-894 (2011).}

\bibitem{Stickler2015}
B.~A. Stickler, U.~Even, and K.~Hornberger.
\newblock {Quantum reflection and interference of matter waves from
  periodically doped surfaces}.
\newblock \href{http://journals.aps.org/pra/pdf/10.1103/PhysRevA.91.013614}{\it
  Phys. Rev. A \bf91\rm, 013614 (2015).}

\bibitem{Wildermuth2005}
S.~Wildermuth, S.~Hofferberth, I.~Lesanovsky, E.~Haller, L.~M. Andersson,
  S.~Groth, I.~Bar-Joseph, P.~Kr{\"u}ger, and J.~Schmiedmayer.
\newblock {Bose-Einstein condensates: Microscopic magnetic-field imaging}.
\newblock
  \href{http://www.nature.com/nature/journal/v435/n7041/full/435440a.html}{\it
  Nature \bf435\rm, 440-441 (2005).}

\bibitem{Strekalov2011}
D.~V. Strekalov, N.~Yu, and K.~Mansour.
\newblock {Sub-Shot Noise Power Source for Microelectronics}.
\newblock
  \href{http://ntrs.nasa.gov/archive/nasa/casi.ntrs.nasa.gov/20120006513.pdf}{\it
  NASA Tech Briefs \rm(Pasadena, CA, 2011).}

\bibitem{Folman2013}
R.~Folman.
\newblock {On the possibility of a relativistic correction to the E and B
  fields around a current-carrying wire}.
\newblock
  \href{http://iopscience.iop.org/article/10.1088/1742-6596/437/1/012013/pdf}{\it
  J. Phys.: Conf. Ser. \bf437\rm, 012013 (2013).}

\bibitem{Dickerson2013}
S.~M. Dickerson, J.~M. Hogan, A.~Sugarbaker, D.~M.~S. Johnson, and M.~A.
  Kasevich.
\newblock {Multiaxis Inertial Sensing with Long-Time Point Source Atom
  Interferometry}.
\newblock
  \href{http://journals.aps.org/prl/pdf/10.1103/PhysRevLett.111.083001}{\it
  Phys. Rev. Lett. \bf111\rm, 083001 (2013).}

\bibitem{Rosi2014}
G.~Rosi, F.~Sorrentino, L.~Cacciapuoti, M.~Prevedelli, and G.~M. Tino.
\newblock {Precision measurement of the Newtonian gravitational constant using
  cold atoms}.
\newblock
  \href{http://www.nature.com/nature/journal/v510/n7506/pdf/nature13433.pdf}{\it
  Nature \bf510\rm, 518-521 (2014).}

\bibitem{Dimopoulos2003}
S.~Dimopoulos and A.~A. Geraci.
\newblock {Probing submicron forces by interferometry of Bose-Einstein
  condensed atoms}.
\newblock \href{http://journals.aps.org/prd/pdf/10.1103/PhysRevD.68.124021}{\it
  Phys. Rev. D \bf68\rm, 124021 (2003).}

\bibitem{PereiraDosSantos2009}
F.~{Pereira dos Santos}, P.~Wolf, A.~Landragin, M.-C. Angonin, P.~Lemonde,
  S.~Bize, A.~Clairon, A.~Lambrecht, B.~Lamine, and S.~Reynaud.
\newblock {Measurement of Short Range Forces using Cold Atoms, in: \it
  Frequency Standards and Metrology, \rm L. Maleki, ed.}
\newblock
  \href{http://www.worldscientific.com/doi/pdf/10.1142/9789812838223_0004}{(World
  Scientific: Singapore, 2009) pp.~44-52.}

\bibitem{Chiaverini2003}
J.~Chiaverini, S.~J. Smullin, A.~A. Geraci, D.~M. Weld, and A.~Kapitulnik.
\newblock {New Experimental Constraints on Non-Newtonian Forces below
  100\,$\mum$}.
\newblock
  \href{http://journals.aps.org/prl/pdf/10.1103/PhysRevLett.90.151101}{\it
  Phys. Rev. Lett. \bf90\rm, 151101 (2003).}

\bibitem{Ribeiro2013}
S.~Ribeiro and S.~Scheel.
\newblock {Shielding vacuum fluctuations with graphene}.
\newblock \href{http://journals.aps.org/pra/pdf/10.1103/PhysRevA.88.042519}{\it
  Phys. Rev. A \bf88\rm, 042519 (2013).}

\bibitem{Behunin2014}
R.~O. Behunin, D.~A.~R. Dalvit, R.~S. Decca, and C.~C. Speake.
\newblock {Limits on the accuracy of force sensing at short separations due to
  patch potentials}.
\newblock \href{http://journals.aps.org/prd/pdf/10.1103/PhysRevD.89.051301}{\it
  Phys. Rev. D \bf89\rm, 051301(R) (2014).}

\bibitem{Gallego2009}
D.~Gallego, S.~Hofferberth, T.~Schumm, P.~Kr{\"u}ger, and J.~Schmiedmayer.
\newblock {Optical lattice on an atom chip}.
\newblock \href{https://www.osapublishing.org/ol/abstract.cfm?id=188534}{\it
  Opt. Lett. \bf34\rm, 3463-3465 (2009).}

\bibitem{Geraci2008}
A.~A. Geraci, S.~J. Smullin, D.~M. Weld, J.~Chiaverini, and A.~Kapitulnik.
\newblock {Improved constraints on non-Newtonian forces at 10 microns}.
\newblock \href{http://journals.aps.org/prd/pdf/10.1103/PhysRevD.78.022002}{\it
  Phys. Rev. D \bf78\rm, 022002 (2008).}

\bibitem{Geraci2015}
A.~Geraci and H.~Goldman.
\newblock {Sensing short range forces with a nanosphere matter-wave
  interferometer}.
\newblock \href{http://journals.aps.org/prd/pdf/10.1103/PhysRevD.92.062002}{\it
  Phys. Rev. D \bf92\rm, 062002 (2015).}

\bibitem{Tino2014}
G.~M. Tino.
\newblock {Testing gravity with atom interferometry, in: \it Atom
  Interferometry}.
\newblock \href{http://ebooks.iospress.nl/volume/atom-interferometry}{\it Proc.
  Int. Sch. Phys. \bf188\rm, 457-491 (2014)}{\ G. M. Tino and M. A. Kasevich,
  eds.}

\bibitem{Poli2011}
N.~Poli, F.-Y. Wang, M.~G. Tarallo, A.~Alberti, M.~Prevedelli, and G.~M. Tino.
\newblock {Precision Measurement of Gravity with Cold Atoms in an Optical
  Lattice and Comparison with a Classical Gravimeter}.
\newblock
  \href{http://journals.aps.org/prl/pdf/10.1103/PhysRevLett.106.038501}{\it
  Phys. Rev. Lett. \bf106\rm, 038501 (2011).}

\bibitem{Beaufils2011}
Q.~Beaufils, G.~Tackmann, X.~Wang, B.~Pelle, S.~Pelisson, P.~Wolf, and
  F.~{Pereira dos Santos}.
\newblock {Laser Controlled Tunneling in a Vertical Optical Lattice}.
\newblock
  \href{http://journals.aps.org/prl/pdf/10.1103/PhysRevLett.106.213002}{\it
  Phys. Rev. Lett. \bf106\rm, 213002 (2011).}

\bibitem{Zhou2013}
M.-K. Zhou, B.~Pelle, A.~Hilico, and F.~{Pereira dos Santos}.
\newblock {Atomic multiwave interferometer in an optical lattice}.
\newblock \href{http://journals.aps.org/pra/pdf/10.1103/PhysRevA.88.013604}{\it
  Phys. Rev. A \bf88\rm, 013604 (2013).}

\bibitem{Burrage2015}
C.~Burrage, E.~J. Copeland, and E.~A. Hinds.
\newblock {Probing dark energy with atom interferometry}.
\newblock
  \href{http://iopscience.iop.org/article/10.1088/1475-7516/2015/03/042/pdf}{\it
  J. Cosmol. Astropart. Phys. \bf2015\rm, 042 (2015).}

\bibitem{Machluf2013a}
S.~Machluf, Y.~Japha, and R.~Folman.
\newblock {Coherent Stern-Gerlach momentum splitting on an atom chip}.
\newblock
  \href{http://www.nature.com/ncomms/2013/130909/ncomms3424/pdf/ncomms3424.pdf}{\it
  Nature Comm. \bf4\rm, 2424 (2013).}

\bibitem{Cronin2009}
A.~D. Cronin, J.~Schmiedmayer, and D.~E. Pritchard.
\newblock {Optics and interferometry with atoms and molecules}.
\newblock \href{http://journals.aps.org/rmp/pdf/10.1103/RevModPhys.81.1051}{\it
  Rev. Mod. Phys. \bf81\rm, 1051-1129 (2009).}

\bibitem{Schaff2014}
J.-F. Schaff, T.~Langen, and J.~Schmiedmayer.
\newblock {Interferometry with atoms, in: \it Atom Interferometry}.
\newblock \href{http://ebooks.iospress.nl/volume/atom-interferometry}{\it Proc.
  Int. Sch. Phys. \bf188\rm, 1-88 (2014)}{\ G. M. Tino and M. A. Kasevich,
  eds.}

\bibitem{Muntinga2013}
H.~M{\"u}ntinga, H.~Ahlers, M.~Krutzik, A.~Wenzlawski, S.~Arnold, D.~Becker,
  K.~Bongs, H.~Dittus, H.~Duncker, N.~Gaaloul, C.~Gherasim, E.~Giese,
  C.~Grzeschik, T.~W. H{\"a}nsch, O.~Hellmig, W.~Herr, S.~Herrmann, E.~Kajari,
  S.~Kleinert, C.~L{\"a}mmerzahl, W.~{Lewoczko-Adamczyk}, J.~Malcolm, N.~Meyer,
  R.~Nolte, A.~Peters, M.~Popp, J.~Reichel, A.~Roura, J.~Rudolph,
  M.~Schiemangk, M.~Schneider, S.~T. Seidel, K.~Sengstock, V.~Tamma,
  T.~Valenzuela, A.~Vogel, R.~Walser, T.~Wendrich, P.~Windpassinger, W.~Zeller,
  T.~{van Zoest}, W.~Ertmer, W.~P. Schleich, and E.~M. Rasel.
\newblock {Interferometry with Bose-Einstein Condensates in Microgravity}.
\newblock
  \href{http://journals.aps.org/prl/pdf/10.1103/PhysRevLett.110.093602}{\it
  Phys. Rev. Lett. \bf110\rm, 093602 (2013).}

\bibitem{Schuldt2015}
T.~Schuldt, C.~Schubert, M.~Krutzik, L.~{Gesa Bote}, N.~Gaaloul, J.~Hartwig,
  H.~Ahlers, W.~Herr, K.~{Posso-Trujillo}, J.~Rudolph, S.~Seidel, T.~Wendrich,
  W.~Ertmer, S.~Herrmann, A.~{Kubelka-Lange}, A.~Milke, B.~Rievers, E.~Rocco,
  A.~Hinton, K.~Bongs, M.~Oswald, M.~Franz, M.~Hauth, A.~Peters, Ahmad Bawamia,
  Andreas Wicht, B.~Battelier, A.~Bertoldi, P.~Bouyer, A.~Landragin,
  D.~Massonnet, T.~L{\'e}v{\`e}que, A.~Wenzlawski, O.~Hellmig,
  P.~Windpassinger, K.~Sengstock, W.~{von Klitzing}, C.~Chaloner, D.~Summers,
  P.~Ireland, I.~Mateos, C.~F. Sopuerta, F.~Sorrentino, G.~M. Tino,
  M.~Williams, C.~Trenkel, D.~Gerardi, M.~Chwalla, J.~Burkhardt, U.~Johann,
  A.~Heske, E.~Wille, M.~Gehler, L.~Cacciapuoti, N.~G{\"u}rlebeck,
  C.~Braxmaier, and E.~Rasel.
\newblock {Design of a dual species atom interferometer for space}.
\newblock
  \href{http://link.springer.com/article/10.1007%2Fs10686-014-9433-y#/page-1}{\it
  Exp. Astron. \bf39\rm, 167-206 (2015).}

\bibitem{Shin2004}
Y.~Shin, M.~Saba, T.~A. Pasquini, W.~Ketterle, D.~E. Pritchard, and A.~E.
  Leanhardt.
\newblock {Atom Interferometry with Bose-Einstein Condensates in a Double-Well
  Potential}.
\newblock
  \href{http://journals.aps.org/prl/pdf/10.1103/PhysRevLett.92.050405}{\it
  Phys. Rev. Lett. \bf92\rm, 050405 (2004).}

\bibitem{Steffens2015}
A.~Steffens, M.~Friesdorf, T.~Langen, B.~Rauer, T.~Schweigler, R.~H{\"u}bener,
  J.~Schmiedmayer, C.~A. Riofr{\'i}o, and J.~Eisert.
\newblock {Towards experimental quantum-field tomography with ultracold atoms}.
\newblock
  \href{http://www.nature.com/ncomms/2015/150703/ncomms8663/pdf/ncomms8663.pdf}{\it
  Nature Commun. \bf6\rm, 7663 (2015).}

\bibitem{Berrada2013}
T.~Berrada, S.~{van Frank}, R.~B{\"u}cker, T.~Schumm, J.-F. Schaff, and
  J.~Schmiedmayer.
\newblock {Integrated Mach-Zehnder interferometer for Bose-Einstein
  condensates}.
\newblock
  \href{http://www.nature.com/ncomms/2013/130627/ncomms3077/pdf/ncomms3077.pdf}{\it
  Nature Commun. \bf4\rm, 2077 (2013).}

\bibitem{LeBlanc2011}
L.~J. LeBlanc, A.~B. Bardon, J.~McKeever, M.~H.~T. Extavour, D.~Jervis, J.~H.
  Thywissen, F.~Piazza, and A.~Smerzi.
\newblock {Dynamics of a Tunable Superfluid Junction}.
\newblock
  \href{http://journals.aps.org/prl/pdf/10.1103/PhysRevLett.106.025302}{\it
  Phys. Rev. Lett. \bf106\rm, 025302 (2011).}

\bibitem{Scott2009}
R.~G. Scott, D.~A.~W. Hutchinson, T.~E. Judd, and T.~M. Fromhold.
\newblock {Quantifying finite-temperature effects in atom-chip interferometry
  of Bose-Einstein condensates}.
\newblock \href{http://journals.aps.org/pra/pdf/10.1103/PhysRevA.79.063624}{\it
  Phys. Rev. A \bf79\rm, 063624 (2009).}

\bibitem{Judd2008}
T.~E. Judd, R.G. Scott, and T.~M. Fromhold.
\newblock {Atom-chip diffraction of Bose-Einstein condensates: The role of
  interatomic interactions}.
\newblock \href{http://journals.aps.org/pra/pdf/10.1103/PhysRevA.78.053623}{\it
  Phys. Rev. A \bf78\rm, 053623 (2008).}

\bibitem{Grond2011}
J.~Grond, U.~Hohenester, J.~Schmiedmayer, and A.~Smerzi.
\newblock {Mach-Zehnder interferometry with interacting trapped Bose-Einstein
  condensates}.
\newblock \href{http://journals.aps.org/pra/pdf/10.1103/PhysRevA.84.023619}{\it
  Phys. Rev. A \bf84\rm, 023619 (2011).}

\bibitem{Shin2005}
Y.~Shin, C.~Sanner, G.-B. Jo, T.~A. Pasquini, M.~Saba, W.~Ketterle, D.~E.
  Pritchard, M.~Vengalattore, and M.~Prentiss.
\newblock {Interference of Bose-Einstein condensates split with an atom chip}.
\newblock \href{http://journals.aps.org/pra/pdf/10.1103/PhysRevA.72.021604}{\it
  Phys. Rev. A \bf72\rm, 021604(R) (2005).}

\bibitem{Machluf2013b}
S.~Machluf.
\newblock {Coherent Splitting of Matter-Waves on an Atom Chip Using a
  State-Dependent Magnetic Potential}.
\newblock
  \href{http://www.bgu.ac.il/atomchip/Theses/Shimon_Machluf_PhD_2013.pdf}{\it
  Ph.D. Thesis, \rm Ben-Gurion University \rm(2013).}

\bibitem{Bohi2009}
P.~B\"{o}hi, M.~F. Riedel, J.~Hoffrogge, J.~Reichel, T.~W. H\"{a}nsch, and
  P.~Treutlein.
\newblock {Coherent manipulation of Bose-Einstein condensates with
  state-dependent microwave potentials on an atom chip}.
\newblock
  \href{http://www.nature.com/nphys/journal/v5/n8/pdf/nphys1329.pdf}{\it Nature
  Phys. \bf5\rm, 592-597 (2009).}

\bibitem{Fernholz2007}
T.~Fernholz, R.~Gerritsma, P.~Kr{\"u}ger, and R.~J.~C. Spreeuw.
\newblock {Dynamically controlled toroidal and ring-shaped magnetic traps}.
\newblock \href{http://journals.aps.org/pra/pdf/10.1103/PhysRevA.75.063406}{\it
  Phys. Rev. A \bf75\rm, 063406 (2007).}

\bibitem{Gerlach1922a}
W.~Gerlach and O.~Stern.
\newblock {Der experimentelle Nachweis der Richtungsquantelung im Magnetfeld}.
\newblock
  \href{http://link.springer.com/article/10.1007%2FBF01326983#page-1}{\it Z.
  Phys. \bf9\rm, 349-352 (1922).}

\bibitem{Gerlach1922b}
W.~Gerlach and O.~Stern.
\newblock {Das magnetische Moment des Silberatoms}.
\newblock
  \href{http://link.springer.com/article/10.1007%2FBF01326984#page-1}{\it Z.
  Phys. \bf9\rm, 353-355 (1922).}

\bibitem{Scully1989}
M.~O. Scully, B.-G. Englert, and J.~Schwinger.
\newblock {Spin coherence and Humpty-Dumpty. III. The effects of observation}.
\newblock \href{http://journals.aps.org/pra/pdf/10.1103/PhysRevA.40.1775}{\it
  Phys. Rev. A \bf40\rm, 1775-1784 (1989).}

\bibitem{Wu2004}
S.~Wu, W.~Rooijakkers, P.~Striehl, and M.~Prentiss.
\newblock {Bidirectional propagation of cold atoms in a ``stadium''-shaped
  magnetic guide}.
\newblock \href{http://journals.aps.org/pra/pdf/10.1103/PhysRevA.70.013409}{\it
  Phys. Rev. A \bf70\rm, 013409 (2004).}

\bibitem{Baker2009}
P.~M. Baker, J.~A. Stickney, M.~B. Squires, J.~A. Scoville, E.~J. Carlson,
  W.~R. Buchwald, and S.~M. Miller.
\newblock {Adjustable microchip ring trap for cold atoms and molecules}.
\newblock \href{http://journals.aps.org/pra/pdf/10.1103/PhysRevA.80.063615}{\it
  Phys. Rev. A \bf80\rm, 063615 (2009).}

\bibitem{Muller2008}
T.~M{\"u}ller, X.~Wu, A.~Mohan, A.~Eyvazov, Y.~Wu, and R.~Dumke.
\newblock {Towards a guided atom interferometer based on a superconducting atom
  chip}.
\newblock
  \href{http://iopscience.iop.org/article/10.1088/1367-2630/10/7/073006/pdf}{\it
  New J. Phys. \bf10\rm, 073006 (2008).}

\bibitem{Jiang2015}
X.-J. Jiang, X.-L. Li, X.-P. Xu, H.-C. Zhang, and Y.-Z. Wang.
\newblock {Archimedean-Spiral-Based Microchip Ring Waveguide for Cold Atoms}.
\newblock
  \href{http://iopscience.iop.org/article/10.1088/0256-307X/32/2/020301/pdf}{\it
  Chinese Phys. Lett. \bf32\rm, 020301 (2015).}

\bibitem{Morizot2006}
O.~Morizot, Y.~Colombe, V.~Lorent, H.~Perrin, and B.~M. Garraway.
\newblock {Ring trap for ultracold atoms}.
\newblock \href{http://journals.aps.org/pra/pdf/10.1103/PhysRevA.74.023617}{\it
  Phys. Rev. A \bf74\rm, 023617 (2006).}

\bibitem{Lesanovsky2006}
I.~Lesanovsky, T.~Schumm, S.~Hofferberth, L.~M. Andersson, P.~Kr{\"u}ger, and
  J.~Schmiedmayer.
\newblock {Adiabatic radio-frequency potentials for the coherent manipulation
  of matter waves}.
\newblock \href{http://journals.aps.org/pra/pdf/10.1103/PhysRevA.73.033619}{\it
  Phys. Rev. A \bf73\rm, 033619 (2006).}

\bibitem{Lesanovsky2007}
I.~Lesanovsky and W.~{von Klitzing}.
\newblock {Time-Averaged Adiabatic Potentials: Versatile Matter-Wave Guides and
  Atom Traps}.
\newblock
  \href{http://journals.aps.org/prl/pdf/10.1103/PhysRevLett.99.083001}{\it
  Phys. Rev. Lett. \bf99\rm, 083001 (2007).}

\bibitem{Vangeleyn2014}
M.~Vangeleyn, B.~M. Garraway, H.~Perrin, and A.~S. Arnold.
\newblock {Inductive dressed ring traps for ultracold atoms}.
\newblock
  \href{http://iopscience.iop.org/article/10.1088/0953-4075/47/7/071001/pdf}{\it
  J. Phys. B \bf47\rm, 071001 (2014).}

\bibitem{Sinuco-Leon2014}
G.~A. Sinuco-Le{\'o}n, K.~A. Burrows, A.~S. Arnold, and B.~M. Garraway.
\newblock {Inductively guided circuits for ultracold dressed atoms}.
\newblock
  \href{http://www.nature.com/ncomms/2014/141028/ncomms6289/pdf/ncomms6289.pdf}{\it
  Nature Commun. \bf5\rm, 5289 (2014).}

\bibitem{West2012b}
A.~D. West, C.~G. Wade, K.~J. Weatherill, and I.~G. Hughes.
\newblock {Piezoelectrically actuated time-averaged atomic microtraps}.
\newblock
  \href{http://scitation.aip.org/content/aip/journal/apl/101/2/10.1063/1.4736580}{\it
  Appl. Phys. Lett. \bf101\rm, 023115 (2012).}

\bibitem{Sherlock2011}
B.~E. Sherlock, M.~Gildemeister, E.~Owen, E.~Nugent, and C.~J. Foot.
\newblock {Time-averaged adiabatic ring potential for ultracold atoms}.
\newblock \href{http://journals.aps.org/pra/pdf/10.1103/PhysRevA.83.043408}{\it
  Phys. Rev. A \bf83\rm, 043408 (2011).}

\bibitem{Pritchard2012}
J.~D. Pritchard, A.~N. Dinkelaker, A.~S. Arnold, P.~F. Griffin, and E.~Riis.
\newblock {Demonstration of an inductively coupled ring trap for cold atoms}.
\newblock
  \href{http://iopscience.iop.org/article/10.1088/1367-2630/14/10/103047/pdf}{\it
  New J. Phys. \bf14\rm, 103047 (2012).}

\bibitem{Lovecchio2016}
C.~Lovecchio, F.~Sch{\"a}fer, S.~Cherukattil, M.~{{Al\`i} Khan}, I.~Herrera,
  F.~S. Cataliotti, T.~Calarco, S.~Montangero, and F.~Caruso.
\newblock {Optimal preparation of quantum states on an atom chip device}.
\newblock \href{http://journals.aps.org/pra/pdf/10.1103/PhysRevA.93.010304}{\it
  Phys. Rev. A \bf93\rm, 010304(R) (2016).}

\bibitem{vanFrank2014}
S.~{van Frank}, A.~Negretti, T.~Berrada, R.~B{\"u}cker, S.~Montangero, J.-F.
  Schaff, T.~Schumm, T.~Calarco, and J.~Schmiedmayer.
\newblock {Interferometry with non-classical motional states of a Bose-Einstein
  condensate}.
\newblock
  \href{http://www.nature.com/ncomms/2014/140530/ncomms5009/pdf/ncomms5009.pdf}{\it
  Nature Commun. \bf5\rm, 4009 (2014).}

\bibitem{CC}
{Creative Commons License}.
\newblock \href{http://creativecommons.org/licenses/by/3.0/}{CC~BY.}

\bibitem{Egorov2011}
M.~Egorov, R.~P. Anderson, V.~Ivannikov, B.~Opanchuk, P.~Drummond, B.~V. Hall,
  and A.~I. Sidorov.
\newblock {Long-lived periodic revivals of coherence in an interacting
  Bose-Einstein condensate}.
\newblock \href{http://journals.aps.org/pra/pdf/10.1103/PhysRevA.84.021605}{\it
  Phys. Rev. A \bf84\rm, 021605 (2011).}

\bibitem{Petrovic2013}
J.~Petrovic, I.~Herrera, P.~Lombardi, F.~Sch\"{a}fer, and F.~S. Cataliotti.
\newblock {A multi-state interferometer on an atom chip}.
\newblock \href
  {http://iopscience.iop.org/article/10.1088/1367-2630/15/4/043002/pdf}{\it New
  J. Phys. \bf15\rm, 043002 (2013).}

\bibitem{Bucker2011}
R.~B{\"u}cker, J.~Grond, S.~Manz, T.~Berrada, T.~Betz, C.~Koller,
  U.~Hohenester, T.~Schumm, A.~Perrin, and J.~Schmiedmayer.
\newblock {Twin-atom beams}.
\newblock
  \href{http://www.nature.com/nphys/journal/v7/n8/pdf/nphys1992.pdf}{\it Nature
  Phys. \bf7\rm, 608-611 (2011).}

\bibitem{Martinez-Garaot2013}
S.~Mart{\'i}nez-Garaot, E.~Torrontegui, X.~Chen, M.~Modugno,
  D.~Gu{\'e}ry-Odelin, S.-Y. Tseng, and J.~G. Muga.
\newblock {Vibrational Mode Multiplexing of Ultracold Atoms}.
\newblock
  \href{http://journals.aps.org/prl/pdf/10.1103/PhysRevLett.111.213001}{\it
  Phys. Rev. Lett. \bf111\rm, 213001 (2013).}

\bibitem{Charron2002}
E.~Charron, E.~Tiesinga, F.~Mies, and C.~Williams.
\newblock {Optimizing a Phase Gate Using Quantum Interference}.
\newblock
  \href{http://journals.aps.org/prl/pdf/10.1103/PhysRevLett.88.077901}{\it
  Phys. Rev. Lett. \bf82\rm, 077901 (2002).}

\bibitem{Marti2015}
G.~E. Marti, R.~Olf, and D.~M. Stamper-Kurn.
\newblock {Collective excitation interferometry with a toroidal Bose-Einstein
  condensate}.
\newblock \href{http://journals.aps.org/pra/pdf/10.1103/PhysRevA.91.013602}{\it
  Phys. Rev. A \bf91\rm, 013602 (2015).}

\bibitem{Ryu2013}
C.~Ryu, P.~W. Blackburn, A.~A. Blinova, and M.~G. Boshier.
\newblock {Experimental Realization of Josephson Junctions for an Atom SQUID}.
\newblock
  \href{http://journals.aps.org/prl/pdf/10.1103/PhysRevLett.111.205301}{\it
  Phys. Rev. Lett. \bf111\rm, 205301 (2013).}

\bibitem{Rudolph2015}
J.~Rudolph, W.~Herr, C.~Grzeschik, T.~Sternke, A.~Grote, M.~Popp, D.~Becker,
  H.~M{\"u}ntinga, H.~Ahlers, A.~Peters, C.~L\"{a}mmerzahl, K.~Sengstock,
  N.~Gaaloul, W.~Ertmer, and E.~M. Rasel.
\newblock {A high-flux BEC source for mobile atom interferometers}.
\newblock
  \href{http://iopscience.iop.org/article/10.1088/1367-2630/17/6/065001/pdf}{\it
  New J. Phys. \bf17\rm, 065001 (2015).}

\bibitem{Eckel2014}
S.~Eckel, F.~Jendrzejewski, A.~Kumar, C.~J. Lobb, and G.~K. Campbell.
\newblock {Interferometric Measurement of the Current-Phase Relationship of a
  Superfluid Weak Link}.
\newblock \href{http://journals.aps.org/prx/pdf/10.1103/PhysRevX.4.031052}{\it
  Phys. Rev. X \bf4\rm, 031052 (2014).}

\bibitem{NJP_SpecialIssue}
{Focus on Atomtronics-Enabled Quantum Technologies}.
\newblock
  \href{http://iopscience.iop.org/1367-2630/focus/Focus%20on%20Atomtronics-enabled%20Quantum%20Technologies}{Special
  Issue, \it New J. Phys.}

\bibitem{OnACIS2013}
\href{http://www.agence-nationale-recherche.fr/?Project=ANR-13-ASTR-0031}{\tt{http://www.agence-nationale-recherche.fr/en/}}.

\bibitem{deAngelis2011}
M.~{de Angelis}, M.~C. Angonin, Q.~Beaufils, Ch. Becker, A.~Bertoldi, K.~Bongs,
  T.~Bourdel, P.~Bouyer, V.~Boyer, S.~D{\"o}rscher, H.~Duncker, W.~Ertmer,
  T.~Fernholz, T.~M. Fromhold, W.~Herr, P.~Kr{\"u}ger, Ch. K{\"u}rbis, C.~J.
  Mellor, F.~{Pereira Dos Santos}, A.~Peters, N.~Poli, M.~Popp, M.~Prevedelli,
  E.~M. Rasel, J.~Rudolph, F.~Schreck, K.~Sengstock, F.~Sorrentino,
  S.~Stellmer, G.~M. Tino, T.~Valenzuela, T.~J. Wendrich, A.~Wicht,
  P.~Windpassinger, and P.~Wolf.
\newblock {iSense: A Portable Ultracold-Atom-Based Gravimeter}.
\newblock
  \href{http://www.sciencedirect.com/science/article/pii/S1877050911006272}{\it
  Procedia Computer Science \bf7\rm, 334-336 (2011).}

\bibitem{Malcolm2016}
J.~I. Malcolm.
\newblock {Construction of a portable platform for cold atom interferometry}.
\newblock \href{http://etheses.bham.ac.uk/6472/}{\it Ph.D. Thesis, \rm
  University of Birmingham \rm(2016).}

\bibitem{GarridoAlzar2012}
C.~L. {Garrido Alzar}, W.~Yan, and A.~Landragin.
\newblock {Towards High Sensitivity Rotation Sensing Using an Atom Chip, in:
  \it Research in Optical Sciences,}.
\newblock
  \href{https://www.osapublishing.org/abstract.cfm?URI=HILAS-2012-JT2A.10}{\it
  OSA Technical Digest, \rm paper JT2A.10 (2012).}

\bibitem{MatterWave}
\url{http://matterwave.eu/}.

\bibitem{Barrett2014}
B.~Barrett, R.~Geiger, I.~Dutta, M.~Meunier, B.~Canuel, A.~Gauguet, P.~Bouyer,
  and A.~Landragin.
\newblock {The Sagnac effect: 20 years of development in matter-wave
  interferometry}.
\newblock
  \href{http://www.sciencedirect.com/science/article/pii/S1631070514001467}{\it
  Comptes Rendus Phys. \bf15\rm, 875-883 (2014).}

\bibitem{Ramanathan2011}
A.~Ramanathan, K.~C. Wright, S.~R. Muniz, M.~Zelan, W.~T. {Hill,\,III}, C.~J.
  Lobb, K.~Helmerson, W.~D. Phillips, and G.~K. Campbell.
\newblock {Superflow in a Toroidal Bose-Einstein Condensate: An Atom Circuit
  with a Tunable Weak Link}.
\newblock
  \href{http://journals.aps.org/prl/pdf/10.1103/PhysRevLett.106.130401}{\it
  Phys. Rev. Lett. \bf106\rm, 130401 (2011).}

\bibitem{Turpin2015}
A.~Turpin, J.~Polo, Yu.~V. Loiko, J.~K{\"u}ber, F.~Schmaltz, T.~K. Kalkandjiev,
  V.~Ahufinger, G.~Birkl, and J.~Mompart.
\newblock {Blue-detuned optical ring trap for Bose-Einstein condensates based
  on conical refraction}.
\newblock
  \href{https://www.osapublishing.org/oe/abstract.cfm?uri=oe-23-2-1638}{\it
  Opt. Express \bf23\rm, 1638-1650 (2015).}

\bibitem{Zhou2014}
S.~Zhou, J.~Chab{\'e}, R.~Salem, T.~David, D.~Groswasser, M.~Keil, Y.~Japha,
  and R.~Folman.
\newblock {Phase space tomography of cold-atoms dynamics in a weakly corrugated
  potential}.
\newblock \href{http://journals.aps.org/pra/pdf/10.1103/PhysRevA.90.033620}{\it
  Phys. Rev. A \bf90\rm, 033620 (2014).}

\bibitem{Bohi2010}
P.~B{\"o}hi, M.~F. Riedel, T.~W. H{\"a}nsch, and P.~Treutlein.
\newblock {Imaging of microwave fields using ultracold atoms}.
\newblock
  \href{http://scitation.aip.org/content/aip/journal/apl/97/5/10.1063/1.3470591}{\it
  Appl. Phys. Lett. \bf97\rm, 0511010 (2010).}

\bibitem{Ockeloen2013}
C.~F. Ockeloen, R.~Schmied, M.~F. Riedel, and P.~Treutlein.
\newblock {Quantum Metrology with a Scanning Probe Atom Interferometer}.
\newblock
  \href{http://journals.aps.org/prl/pdf/10.1103/PhysRevLett.111.143001}{\it
  Phys. Rev. Lett. \bf111\rm, 143001 (2013).}

\bibitem{Esnault2011}
F.~X. Esnault, N.~Rossettoa, D.~Holleville, J.~Delporteb, and N.~Dimarcqa.
\newblock {HORACE: A compact cold atom clock for Galileo}.
\newblock
  \href{http://www.sciencedirect.com/science/article/pii/S0273117710008124}{\it
  Adv. Space Res. \bf47\rm, 854-858 (2011).}

\bibitem{Vuletic2011}
V.~Vuleti{\'c}, I.~D. Leroux, and M.~H. Schleier-Smith.
\newblock {Microchip-Based Trapped-Atom Clocks, Chapter~8 in: \it Atom Chips,
  \rm J. Reichel and V. Vuletic, eds.}
\newblock
  \href{http://onlinelibrary.wiley.com/doi/10.1002/9783527633357.ch8/summary}{(Wiley-VCH:
  Hoboken, NJ, 2011) pp.~265-282.}

\bibitem{Szmuk2015}
R.~Szmuk, V.~Dugrain, W.~Maineult, J.~Reichel, and P.~Rosenbusch.
\newblock {Stability of a trapped-atom clock on a chip}.
\newblock \href{http://journals.aps.org/pra/pdf/10.1103/PhysRevA.92.012106}{\it
  Phys. Rev. A \bf92\rm, 012106 (2015).}

\bibitem{Nicholson2015}
T.~L. Nicholson, S.~L. Campbell, R.~B. Hutson, G.~E. Marti, B.~J. Bloom, R.~L.
  McNally, W.~Zhang, M.~D. Barrett, M.~S. Safronova, G.~F. Strouse, W.~L. Tew,
  and J.~Ye.
\newblock {Systematic evaluation of an atomic clock at $2\times10^{-18}$ total
  uncertainty}.
\newblock
  \href{http://www.nature.com/ncomms/2015/150421/ncomms7896/pdf/ncomms7896.pdf}{\it
  Nature Commun. \bf6\rm, 6896 (2015).}

\bibitem{Yb_ion_clock2014}
\href{http://www.agence-nationale-recherche.fr/en/anr-funded-project/?tx_lwmsuivibilan_pi2%5BCODE%5D=ANR-14-CE26-0031}
  {\tt{http://www.agence-nationale-recherche.fr/en/}}.

\bibitem{Monroe2008}
C.~M. Monroe and D.~J. Wineland.
\newblock {Quantum Computing with Ions}.
\newblock
  \href{http://www.scientificamerican.com/article/quantum-computing-with-ions/}{\it
  Sci. Am. \bf299\rm, 64-71 (August, 2008).}

\bibitem{Harty2014}
T.~P. Harty, D.~T.~C. Allcock, C.~J. Ballance, L.~Guidoni, H.~A. Janacek, N.~M.
  Linke, D.~N. Stacey, and D.~M. Lucas.
\newblock {High-Fidelity Preparation, Gates, Memory, and Readout of a
  Trapped-Ion Quantum Bit}.
\newblock
  \href{https://physics.aps.org/featured-article-pdf/10.1103/PhysRevLett.113.220501}{\it
  Phys. Rev. Lett. \bf113\rm, 220501 (2014).}

\bibitem{BellPrize2015}
R.~Blatt.
\newblock \href{cqiqc.physics.utoronto.ca/bell_prize/blatt.html}{Bell Prize
  citation, Toronto, 2015.}

\bibitem{Wilson2014}
A.~C. Wilson, Y.~Colombe, K.~R. Brown, E.~Knill, D.~Leibfried, and D.~J.
  Wineland.
\newblock {Tunable spin-spin interactions and entanglement of ions in separate
  potential wells}.
\newblock
  \href{http://www.nature.com/nature/journal/v512/n7512/pdf/nature13565.pdf}{\it
  Nature \bf512\rm, 57-60 (2014).}

\bibitem{Guise2015}
N.~D. Guise, S.~D. Fallek, K.~E. Stevens, K.~R. Brown, C.~Volin, A.~W. Harter,
  J.~M. Amini, R.~E. Higashi, S.~T. Lu, H.~M. Chanhvongsak, T.~A. Nguyen, M.~S.
  Marcus, T.~R. Ohnstein, and D.~W. Youngner.
\newblock {Ball-grid array architecture for microfabricated ion traps}.
\newblock
  \href{http://scitation.aip.org/content/aip/journal/jap/117/17/10.1063/1.4917385}{\it
  J. Appl. Phys. \bf117\rm, 174901 (2015).}

\bibitem{Mielenz2015x}
M.~Mielenz, H.~Kalis, M.~Wittemer, F.~Hakelberg, R.~Schmied, M.~Blain,
  P.~Maunz, D.~Leibfried, U.~Warring, and T.~Schaetz.
\newblock {Freely configurable quantum simulator based on a two-dimensional
  array of individually trapped ions}.
\newblock \href{http://arxiv.org/pdf/1512.03559v2.pdf}{\tt arXiv:1512.03559v2
  \rm(2015).}

\bibitem{Feynman1982}
R.~P. Feynman.
\newblock {Simulating physics with computers}.
\newblock \href{http://link.springer.com/article/10.1007/BF02650179}{\it Int.
  J. Theor. Phys. \bf21\rm, 467-488 (1982).}

\bibitem{Blatt2012}
R.~Blatt and C.~F. Roos.
\newblock {Quantum simulations with trapped ions}.
\newblock
  \href{http://www.nature.com/nphys/journal/v8/n4/pdf/nphys2252.pdf}{\it Nature
  Phys. \bf8\rm, 277-284 (2012).}

\bibitem{Bloch2012}
I.~Bloch, J.~Dalibard, and S.~Nascimb{\`e}ne.
\newblock {Quantum simulations with ultracold quantum gases}.
\newblock
  \href{http://www.nature.com/nphys/journal/v8/n4/pdf/nphys2259.pdf}{\it Nature
  Phys. \bf8\rm, 267-276 (2012).}

\bibitem{Bakr2009}
W.~S. Bakr, J.~I. Gillen, A.~Peng, S.~F{\"o}lling, and M.~Greiner.
\newblock {A quantum gas microscope for detecting single atoms in a
  Hubbard-regime optical lattice}.
\newblock
  \href{http://www.nature.com/nature/journal/v462/n7269/pdf/nature08482.pdf}{\it
  Nature \bf462\rm, 74-77 (2009).}

\bibitem{Gillen2009}
J.~I. Gillen, W.~S. Bakr, A.~Peng, P.~Unterwaditzer, S.~F{\"o}lling, and
  M.~Greiner.
\newblock {Two-dimensional quantum gas in a hybrid surface trap}.
\newblock \href{http://journals.aps.org/pra/pdf/10.1103/PhysRevA.80.021602}{\it
  Phys. Rev. A \bf80\rm, 021602(R) (2009).}

\bibitem{Simon2011}
J.~Simon, W.~S. Bakr, R.~Ma, M.~E. Tai, P.~M. Preiss, and M.~Greiner.
\newblock {Quantum simulation of antiferromagnetic spin chains in an optical
  lattice}.
\newblock
  \href{http://www.nature.com/nature/journal/v472/n7343/pdf/nature09994.pdf}{\it
  Nature \bf472\rm, 307-312 (2011).}

\bibitem{Haller2015}
E.~Haller, J.~Hudson, A.~Kelly, D.~A. Cotta, B.~Peaudecerf, G.~D. Bruce, and
  S.~Kuhr.
\newblock {Single-atom imaging of fermions in a quantum-gas microscope}.
\newblock
  \href{http://www.nature.com/nphys/journal/v11/n9/pdf/nphys3403.pdf}{\it
  Nature Phys. \bf11\rm, 738-742 (2015).}

\bibitem{Farkas2013}
D.~M. Farkas, K.~M. Hudek, S.~Wu, and D.~Z. Anderson.
\newblock {Efficient direct evaporative cooling in an atom-chip magnetic trap}.
\newblock \href{http://journals.aps.org/pra/pdf/10.1103/PhysRevA.87.053417}{\it
  Phys. Rev. A \bf87\rm, 053417 (2013).}

\bibitem{Dupont-Nivet2015}
M.~Dupont-Nivet, M.~Casiulis, T.~Laudat, C.~I. Westbrook, and S.~Schwartz.
\newblock {Microwave-stimulated Raman adiabatic passage in a Bose-Einstein
  condensate on an atom chip}.
\newblock \href{http://journals.aps.org/pra/pdf/10.1103/PhysRevA.91.053420}{\it
  Phys. Rev. A \bf91\rm, 053420 (2015).}

\bibitem{Morgan2013}
T.~Morgan, L.~J. O'Riordan, N.~Crowley, B.~O'Sullivan, and Th. Busch.
\newblock {Coherent transport by adiabatic passage on atom chips}.
\newblock \href{http://journals.aps.org/pra/pdf/10.1103/PhysRevA.88.053618}{\it
  Phys. Rev. A \bf88\rm, 053618 (2013).}

\bibitem{Giovannetti2004}
V.~Giovannetti, S.~Lloyd, and L.~Maccone.
\newblock {Quantum-Enhanced Measurements: Beating the Standard Quantum Limit}.
\newblock \href{http://www.sciencemag.org/content/306/5700/1330.full.pdf}{\it
  Science \bf306\rm, 1330-1336 (2004).}

\bibitem{Gross2010}
C.~Gross, T.~Zibold, E.~Nicklas, J.~Est{\`e}ve, and M.~K. Oberthaler.
\newblock {Nonlinear atom interferometer surpasses classical precision limit}.
\newblock
  \href{http://www.nature.com/nature/journal/v464/n7292/pdf/nature08919.pdf}{\it
  Nature \bf464\rm, 1165-1169 (2010).}

\bibitem{Volz2011}
J.~Volz, R.~Gehr, G.~Dubois, J.~Est{\`e}ve, and J.~Reichel.
\newblock {Measurement of the internal state of a single atom without energy
  exchange}.
\newblock
  \href{http://www.nature.com/nature/journal/v475/n7355/pdf/nature10225.pdf}{\it
  Nature \bf475\rm, 210-213 (2011).}

\bibitem{Briegel1998}
H.-J. Briegel, W.~D{\"u}r, J.~I. Cirac, and P.~Zoller.
\newblock {Quantum Repeaters: The Role of Imperfect Local Operations in Quantum
  Communication}.
\newblock
  \href{http://journals.aps.org/prl/pdf/10.1103/PhysRevLett.81.5932}{\it Phys.
  Rev. Lett. \bf81\rm, 5932-5935 (1998).}

\bibitem{Li2015}
T.~Li and F.-G. Deng.
\newblock {Heralded high-efficiency quantum repeater with atomic ensembles
  assisted by faithful single-photon transmission}.
\newblock \href{http://www.nature.com/articles/srep15610}{\it Sci. Rep.\bf5\rm,
  15610 (2015).}

\bibitem{Kimble2008}
H.~J. Kimble.
\newblock {The quantum internet}.
\newblock
  \href{http://www.nature.com/nature/journal/v453/n7198/pdf/nature07127.pdf}{\it
  Nature \bf453\rm, 1023-1030 (2008).}

\bibitem{Northup2014}
T.~E. Northup and R.~Blatt.
\newblock {Quantum information transfer using photons}.
\newblock
  \href{http://www.nature.com/nphoton/journal/v8/n5/pdf/nphoton.2014.53.pdf}{\it
  Nature Photon. \bf8\rm, 356-363 (2014).}

\end{thebibliography}

\addtocontents{toc}{\protect\vspace{0pt}}

\end{document}